\documentclass[aps,onecolumn,10pt]{revtex4}
\usepackage{amsmath}
\usepackage{amssymb}
\usepackage{hyperref}
\usepackage{graphicx}
\usepackage{float}
\usepackage{slashed}
\numberwithin{equation}{section}

\hypersetup{colorlinks=true}
\begin{document}
\allowdisplaybreaks
\setcounter{equation}{0}
SLAC-PUB-17390
 
\title{Comparing light-front  quantization with instant-time quantization}

\author{Philip D. Mannheim${}^1$, Peter Lowdon${}^2$ and  Stanley J. Brodsky${}^3$}
\affiliation{${}^1$ Department of Physics, University of Connecticut, Storrs, CT 06269, USA \\
${}^2$CPHT, CNRS, Ecole Polytechnique, Institut Polytechnique de Paris, Route de Saclay, 91128 Palaiseau, France\\
${}^3$ SLAC National Accelerator Laboratory, Stanford University, Menlo Park, CA 94025, USA\\
philip.mannheim@uconn.edu,~peter.lowdon@polytechnique.edu,~sjbth@slac.stanford.edu}

\date{June 25 2020}

\begin{abstract}

In this paper we compare light-front quantization and instant-time quantization both at the level of operators and at the level of their Feynman diagram matrix elements. At the level of operators light-front quantization and instant-time quantization lead to equal light-front time commutation (or anticommutation) relations that appear to be quite different from equal instant-time  commutation (or anticommutation) relations.  Despite this we show that at unequal times instant-time and light-front commutation (or anticommutation) relations actually can be transformed into each other, with it only being the restriction to equal times that makes the commutation (or anticommutation) relations appear to be so different. While our results are valid for both bosons and fermions, for fermions there are subtleties associated with tip of the light cone contributions that need to be taken care of. At the level of Feynman diagrams we show for non-vacuum Feynman diagrams that the pole terms in four-dimensional light-front Feynman diagrams reproduce the widely used three-dimensional light-front on-shell Hamiltonian Fock space formulation in which the light-front energy and light-front momentum are on shell. Moreover, we show that the contributions of pole terms in  non-vacuum instant-time and non-vacuum light-front Feynman diagrams are equal. However, because of circle at infinity contributions we show that this equivalence of pole terms fails for four-dimensional light-front vacuum tadpole diagrams. Then, and precisely because of these circle at infinity contributions, we show that light-front vacuum tadpole diagrams are not only nonzero, they quite remarkably are actually equal to the pure pole term instant-time vacuum tadpole diagrams. Light-front vacuum diagrams are not correctly describable by the on-shell Hamiltonian formalism, and thus not by the closely related infinite momentum frame prescription either. Thus for the light-front vacuum sector we must use the off-shell Feynman formalism as it contains information that is not accessible in the on-shell Hamiltonian Fock space approach. We show that light-front quantization is intrinsically nonlocal, and that for fermions this nonlocality is present in Ward identities. One can project fermion spinors into so-called good and bad components, and both of these components contribute in Ward identities. Central to our analysis is that the  transformation from instant-time coordinates and fields to light-front coordinates and fields is a unitary, spacetime-dependent translation. Consequently, not only are instant-time quantization and light-front quantization equivalent, because of general coordinate invariance  they are unitarily equivalent. 
\end{abstract}

\maketitle

\tableofcontents

\section{Introduction}
\label{S1}

Since the original work of Dirac \cite{Dirac:1949cp} there has been continuing interest in light-front (also known as  ``light-cone" or ``front-form") quantization of quantum field theories.  Comprehensive reviews can be found in \cite{Brodsky:1997de,Leutwyler:1977vy,Bakker:2013cea,Burkardt:1995ct}. The light-front approach is  based on three-dimensional Hamiltonian field theory quantized at fixed light-front time $ x^+= x^0+ x^3$.  The rules for calculations for Light-Front Hamiltonian QCD for both perturbative and nonperturbative applications are summarized in \cite{Lepage:1980fj}.  As is the case with the standard four-dimensional covariant Feynman Lagrangian theory,  the light-front formalism is Poincar\'e invariant and causal.  Observables in hadron physics such as form factors, structure functions, and distribution amplitudes are based on the nonperturbative light-front hadronic wave functions, the eigenfunctions of the QCD Light-Front Hamiltonian~\cite{Lepage:1979zb,Brodsky:2003pw}.  In the case of scattering amplitudes, the covariant Feynman diagram approach and the Light-Front Hamiltonian approach give identical results.  One can also replicate the calculation rules for light-front $x^+$-ordered perturbation theory using standard time-ordered perturbation theory based on quantization at fixed time $x^0$ (also known as  ``instant-time" or ``instant-form" quantization)  by choosing a Lorentz frame where the observer moves at infinite momentum \cite{Weinberg:1966jm,Brodsky:1973kb,Chang1969,Yan1973}. 

Two of the key conceptual questions that are raised by the light-front approach are whether or not light-front quantization differs from instant-time quantization, and whether or not the full content of the light-front approach is given by the on-shell Light-Front Hamiltonian formulation alone. In this paper we shall address both of these questions. We shall show that the answer to the first question is that despite their seeming differences the two quantization procedures are in fact equivalent and describe the same canonical quantization procedure; while the answer to the second question is that in the non-vacuum sector the on-shell approach suffices, but in the vacuum sector quantum field theory contains information that the on-shell approach cannot access.

In our discussion of these conceptual aspects of the light-front quantization program we provide both new results and new derivations of results that have been reported before. We identify some key differences between instant-time quantization (canonical commutators or anticommutators of quantum fields evaluated at a common $x^0$) and light-front quantization (common $x^+=x^0+x^3$). Despite these differences, we in general show that if any two directions of quantization can be related by a general coordinate transformation (the spacetime-dependent translation $x^0\rightarrow x^0+x^3$ being a such a general coordinate transformation, and incidentally not being a Lorentz boost), then the general coordinate invariance of unequal-time commutators or anticommutators of q-number fields entails that from instant-time commutators one can construct light-front commutators and vice-versa. Consequently, light-front quantization is a consequence of instant-time quantization and does not need to be independently postulated. By the same token, the general coordinate invariance of both c-number Feynman diagrams and c-number path integrals entails the equality to all orders of matrix elements of products of fields evaluated with either quantization procedure. For bosonic fields this equality holds despite the fact that the instant-time and light-front  canonical commutators have very different light-cone singularity structures. For fermions this equality holds despite the fact that time-ordered products of fields and thus the differential equations obeyed by them in the  instant-time and light-front formulations have very different light-cone singularity structures. Central to our analysis of Feynman diagrams will be a treatment of Green's functions in coordinate space, where these differences cause the $x^+\rightarrow 0$ limit in the light-front case to be singular in a way that the $x^0\rightarrow 0$ limit in the instant-time case is not. These singularities are of relevance only in vacuum tadpole graphs such as the one shown in Fig. \ref{undressedtadpole} below, a graph that we construct as the $x^+\rightarrow 0$ limit of an unequal light-front time-ordered propagator. In instant-time quantization such vacuum tadpole graphs only receive pole contributions in Feynman contours in the complex $p_0$ plane ($p_0$ being the conjugate of $x^0$). However, in the light-front quantization case these vacuum tadpole graphs also receive contributions from circle at infinity terms in Feynman contours in the complex $p_+$ plane ($p_+$ being the conjugate of $x^+$). Remarkably, rather than causing the values of instant-time and light-front vacuum tadpole graphs to differ, as we had noted in \cite{Mannheim2019a},  these circle at infinity terms actually cause them to be the same.

Our intent in this paper is to develop light-front quantization via a first principles approach at the operator level so that we can then compare and contrast the application of this approach to instant-time quantization and light-front  quantization. We shall thus follow a very general quantization procedure. While we shall work in flat spacetime we have found it convenient to formulate everything using a general coordinate invariance approach. Thus starting from an arbitrary general coordinate invariant action $I=\int d^4x(-g)^{1/2}L$ that is a function of some generic field $\chi(x)$, one constructs an automatically conserved energy momentum tensor by varying the action with respect to a background metric as per $T_{\mu\nu}=2(-g)^{-1/2}\delta I/\delta g^{\mu\nu}$, $\delta (-g)^{1/2}=-\tfrac{1}{2}(-g)^{1/2}g_{\mu\nu}\delta g^{\mu\nu}$. One then constructs flat spacetime momentum generators $P_{\mu}=\int d^3x T^{\sigma}_{\phantom{0}\mu}$, where $\sigma$  could be $x^0$ or $x^+$. If one wishes to quantize in some specific coordinate direction $\sigma$,  then for any action that depends on $\partial_{\sigma}\chi(x)$  one constructs a canonical conjugate $\pi(x)=(-g)^{-1/2}\delta I/\delta \partial_{\sigma}{\chi}$ to the field $\chi(x)$. One then fixes the normalization of the $[\chi,\pi]$ commutator at equal $\sigma$ from the requirement that the $P_{\mu}$ generate translations according to $[P_{\mu},\chi(x)]=-i\partial_{\mu}\chi(x)$.

While these remarks are straightforward they actually require some clarification. Specifically, as we show below in (\ref{2.1}), the light-front action $I_S$ for a scalar field $\phi$ contains terms of the form $\int d^4x (-g)^{1/2}[\partial_+\phi\partial_-\phi+\partial_-\phi\partial_+\phi]$ (here $\partial_-=\partial/\partial x^{-}$ where $x^-=x^0-x^3$). If the action is composed of quantum fields that do not necessarily commute with one another, in constructing the canonical conjugate $\Pi(x)=(-g)^{-1/2}\delta I_S/\delta \partial_{\sigma}{\phi}$ we would either have to do the functional variation from the left or from the right. But if we do it from one side alone, say from the left, we obtain $\partial_-\phi$ from the $\partial_+\phi\partial_-\phi$ term, but we would have to commute through the $\partial_-\phi$ term in order to vary the  $\partial_-\phi\partial_+\phi$ term, and thus pick up a $[\partial_-\phi,\partial_+\phi]$ commutator term. Since this commutator need not be a c-number we would be unable to actually determine the canonical conjugate at all. Thus we must initially take the action to be classical, construct the canonical conjugate first and only after doing so then quantize. However, while this would take care of the canonical conjugate issue, it does not address ordering issues in the q-number energy-momentum tensor needed for $[P_{\mu},\phi(x)]=-i\partial_{\mu}\phi(x)$, and we will have to deal with these issues below. While the construction of the canonical conjugate is an issue for light-front quantization, it is not an issue in instant-time quantization since the analogous term is $\int d^4x (-g)^{1/2}[\partial_0\phi\partial_0\phi-\partial_3\phi\partial_3\phi]$, and $\partial_0\phi$ commutes with itself.

For the instant-time quantization of a free scalar field $\phi(x)$ with action 
\begin{eqnarray}
I_S=\int d^4x(-g)^{1/2}\left[\tfrac{1}{2}\partial_{\mu}\phi\partial^{\mu}\phi-\tfrac{1}{2}m^2\phi^2\right],
\label{1.1}
\end{eqnarray}
the canonical conjugate $\Pi(x)$, the conserved energy-momentum tensor, and the equation of motion are given by 
\begin{eqnarray}
\Pi=\partial_0\phi,\quad T_{\mu\nu}=\partial_{\mu}\phi\partial_{\nu}\phi-g_{\mu\nu}\left[\tfrac{1}{2}\partial_{\alpha}\phi\partial^{\alpha}\phi-\tfrac{1}{2}m^2\phi^2\right], \quad [\partial_{\mu}\partial^{\mu}+m^2]\phi=0.
\label{1.2}
\end{eqnarray}
From $[P_{\mu},\phi]=-i\partial_{\mu}\phi$ the normalization of the equal $x^0$ commutator $[\phi,\Pi]$ is given by
\begin{eqnarray}
[\phi(x^0,x^1,x^2,x^3),\partial_0\phi(x^0,y^2,y^2,y^3]=i\delta(x^1-y^1) \delta(x^2-y^2)\delta(x^3-y^3),
\label{1.3}
\end{eqnarray}
to fix the quantization of the theory.

To solve the equations of motion that would then ensue when one adds interactions to $I_S$ one defines a Green's function $D(x^{\mu},{\rm instant})$ that obeys a differential equation of the form
\begin{eqnarray}
\left[\left(\frac{\partial}{\partial x^0}\right)^2-\left(\frac{\partial}{\partial x^1}\right)^2-\left(\frac{\partial}{\partial x^2}\right)^2-\left(\frac{\partial}{\partial x^3}\right)^2+m^2\right]D(x^{\mu},{\rm instant})=-\delta^4(x).
\label{1.4}
\end{eqnarray}
With Feynman boundary conditions this Green's function can be written in the instant-time case as 
\begin{eqnarray}
D(x^{\mu},{\rm instant})&=&\frac{1}{(2\pi)^4}\int dp_0dp_1dp_2dp_3 \frac{e^{-i(p_0x^0+p_1x^1+p_2x^2+p_3x^3)}}{(p_0)^2-(p_1)^2-(p_2)^2-(p_3)^2-m^2+i\epsilon}.
\label{1.5}
\end{eqnarray}
Now there is an alternate procedure for finding solutions to (\ref{1.4}). One introduces the  
time-ordered matrix element $-i\langle \Omega|T^{(0)}[\phi(x)\phi(0)|\Omega\rangle=-i\langle \Omega|\theta(x^0)\phi(x)\phi(0)|\Omega\rangle-i\langle \Omega|\theta(-x^0)\phi(0)\phi(x)]|\Omega\rangle$ ($T^{(0)}$ denotes time ordering with respect to $x^0$), and from the equation of motion and canonical commutator one establishes that it obeys 
\begin{eqnarray}
\left[\left(\frac{\partial}{\partial x^0}\right)^2-\left(\frac{\partial}{\partial x^1}\right)^2-\left(\frac{\partial}{\partial x^2}\right)^2-\left(\frac{\partial}{\partial x^3}\right)^2+m^2\right][-i\langle \Omega|T^{(0)}[\phi(x)\phi(0)]|\Omega\rangle]=-\delta^4(x),
\label{1.6}
\end{eqnarray}
i.e., that it  obeys the Green's function equation (\ref{1.4}). We can thus identify $D(x^{\mu},{\rm instant})=-i\langle \Omega|T^{(0)}[\phi(x)\phi(0)]|\Omega\rangle$.

While the Feynman representation of $D(x^{\mu},{\rm instant})$ in (\ref{1.5}) is off the mass shell since $(p_0)^2$ is not constrained to be equal to $(p_1)^2+(p_2)^2+(p_3)^2+m^2$, which it would be in $\phi=\exp(ip_{\mu}x^{\mu})$ solutions to (\ref{1.2}), the pole term contribution to (\ref{1.5}) is on shell. In (\ref{1.5}) we can close the $p_0$ contour below the real $p_0$ axis in the complex $p_0$ plane when $x^0$ is positive and can close above the real $p_0$ axis when $x^0$ is negative, and in neither case do we get any circle at infinity contribution because the $e^{ip_0x^0}$ terms provides suppression. (While the $(p_0)^2$ term in the denominator in (\ref{1.5}) also gives suppression on the circle at infinity, it would not be sufficient if we were to close below the $p_0$ axis when $x^0<0$.) Performing the contour integration then yields
\begin{eqnarray}
&&D(x^{\mu},{\rm instant},{\rm pole})
\nonumber\\
&&=-\frac{i\theta(x^0)}{(2\pi)^3}\int dp_1dp_2dp_3\frac{e^{-i(E_px^0+p_1x^1+p_2x^2+p_3x^3)}}{2E_p}
-\frac{i\theta(-x^0)}{(2\pi)^3}\int dp_1dp_2dp_3\frac{e^{i(E_px^0+p_1x^1+p_2x^2+p_3x^3)}}{2E_p},
\label{1.7}
\end{eqnarray}
where $E_p=+[(p_1)^2+(p_2)^2+(p_3)^2+m^2]^{1/2}$, and where we have transformed $(p_1,p_2,p_3)$ to $(-p_1,-p_2,-p_3)$ in the second integral. 

The structure given in (\ref{1.7}) can also be obtained by making an on-shell Fock expansion of $\phi$ of the form
\begin{eqnarray}
\phi(x^0,\vec{x})=\int_{-\infty}^{\infty}dp_1\int_{-\infty}^{\infty}dp_2\int_{-\infty}^{\infty}dp_3 \frac{1}{(2\pi)^{3/2}(2E_p)^{1/2}}[a(\vec{p})e^{-iE_p x^0+i\vec{p}\cdot\vec{x}}+a^{\dagger}(\vec{p})e^{+iE_p x^0-i\vec{p}\cdot\vec{x}}],
\label{1.8}
\end{eqnarray}
where the normalization of the creation and annihilation operator algebra, viz.  
\begin{eqnarray}
[a(\vec{p}),a^{\dagger}(\vec{p}^{~\prime})]=\delta^3(\vec{p}-\vec{p}^{~\prime}),
\label{1.9}
\end{eqnarray}
is fixed from the normalization of the canonical commutator given in (\ref{1.3}). Then 
on interpreting the vacuum $|\Omega \rangle$ that appears in (\ref{1.6}) as the one that the $a(\vec{p})$ annihilate according to $a(\vec{p})|\Omega \rangle=0$, on inserting (\ref{1.8}) into  $-i\langle \Omega|T^{(0)}[\phi(x)\phi(0)]|\Omega\rangle$ we find that (\ref{1.7}) immediately follows.

In light-front quantization every one of these steps gets modified, and yet nonetheless the form that we eventually obtain for the corresponding scalar field analog of (\ref{1.6}) is the same as in the instant-time quantization case. That this is possible even in principle is because unlike the q-number fields themselves and their q-number algebras given in (\ref{1.3}) and (\ref{1.9}), quantities such as the instant-time $-i\langle \Omega|T^{(0)}[\phi(x)\phi(0)]|\Omega\rangle$  or the light-front $-i\langle \Omega|T^{(+)}[\phi(x)\phi(0)]|\Omega\rangle$ ($T^{(+)}$ denotes time-ordering with respect to $x^+$) are c-numbers, and c-numbers can transform into each other under a general coordinate transformation,  even though q-numbers need not. To determine how these various relations are modified in the light-front case we shall follow the sequence described above and begin by first identifying canonical conjugates and constructing the momentum generators so as to determine the canonical commutators.

As we show below, for scalar fields we find that the canonical commutator as constructed by this procedure takes the form
\begin{align}
&[\phi(x^+,x^1,x^2,x^-),2\partial_-\phi(x^+,y^1,y^2,y^-)]=i\delta(x^1-y^1)\delta(x^2-y^2)\delta(x^--y^-),
\label{1.10}
\end{align}
which can be integrated to
\begin{align}
&[\phi(x^+,x^1,x^2,x^-),\phi(x^+,y^1,y^2,y^-)]=-\frac{i}{4}\epsilon(x^--y^-)\delta(x^1-y^1)\delta(x^2-y^2),
\nonumber\\
&[\phi(x^+,x^1,x^2,x^-),\phi(x^+,0,0,0)]=-\frac{i}{4}\epsilon(x^-)\delta(x^1)\delta(x^2),
\label{1.11}
\end{align}
where $\epsilon(x)=\theta(x)-\theta(-x)$. In contrast, in instant-time quantization 
\begin{eqnarray}
[\phi(x^0,x^1,x^2,x^3),\phi(x^0,0,0,0)]=0.
\label{1.12a}
\end{eqnarray}
As we see, the instant-time and light-front $[\phi,\phi]$ commutators of q-number $\phi$ fields are radically different. However, despite this quite marked difference at the operator level  (the $[\phi,\phi]$ commutators cannot be transformed into each other since one is zero and the other is not), as we show below, no such difference is  encountered in matrix elements. Moreover, for the commutator itself we can even say something without needing to take matrix elements at all. Specifically, in the instant-time quantization of a free scalar field theory not only is the  equal-time commutator of two q-number scalar fields actually a c-number, so is the unequal-time instant-time ($IT$) commutator $i\Delta(IT;x^0,x^1,x^2,x^3)=[\phi(x^0,\vec{x}),\phi(0)]$, with, as described in detail below,  it behaving on the light cone as
\begin{eqnarray}
i\Delta(IT;x^0,x^1,x^2,x^3)=-\frac{i}{2\pi}\epsilon(x^0)\delta[(x^0)^2-(x^1)^2-(x^2)^2-(x^3)^2].
\label{1.13a}
\end{eqnarray}  
Changing to light-front variables (i.e., just transforming the coordinates, but not changing the direction of quantization) we obtain
\begin{eqnarray}
&&i\Delta(IT;x^+,x^1,x^2,x^-)=[\phi(x^+,x^1,x^2,x^-),\phi(0,0,0,0)]
\nonumber\\
&&=-\frac{i}{2\pi}\epsilon[\tfrac{1}{2}(x^++x^-)]\delta[x^+x^--(x^1)^2-(x^2)^2].
\label{1.14a}
\end{eqnarray}  
By being an unequal-time commutator, $i\Delta(IT;x^0,x^1,x^2,x^3)$ is defined for all $x^{\mu}$. Thus we can evaluate it at $x^+=0$, and obtain 
\begin{align}
&i\Delta(IT;0,x^1,x^2,x^-)=[\phi(0,x^1,x^2,x^-),\phi(0,0,0,0)]=
-\frac{i}{2\pi}\epsilon(\tfrac{1}{2}x^-)\delta((x^1)^2+(x^2)^2)=-\frac{i}{4}\epsilon(x^-)\delta(x^1)\delta(x^2).
\label{1.15a}
\end{align}  
We recognize the right-hand side of (\ref{1.15a}) as the right-hand side of (\ref{1.11}), and thus we can identify 
\begin{eqnarray}
i\Delta(IT;0,x^1,x^2,x^-)=i\Delta(LF;0,x^1,x^2,x^-),
\label{1.16a}
\end{eqnarray}
where $LF$ denotes light front and where $i\Delta(LF;x^+,x^1,x^2,x^-)=[\phi(x^+,x^1,x^2,x^-),\phi(0,0,0,0)]$. Thus, as noted in  \cite{Harindranath1996,Mannheim2019b,Mannheim2020a}, we establish the equivalence of the instant-time and light-front  commutation relations in this particular case. We thus see that despite the fact that the equal $x^0$ instant-time $[\phi,\phi]$ commutators and the equal $x^+$ light-front $[\phi,\phi]$ commutators appear to be radically different, it is only the restriction to equal $x^0$ or equal $x^+$ that makes them appear to be so. In this paper we shall extend this equivalence to fermions and gauge bosons and to time-ordered combinations of field operators.

Beyond commutators, in light-front quantization the Fock space expansion is given by
\begin{align}
\phi(x^+,x^1,x^2,x^-)&=\frac{2}{(2\pi)^{3/2}}\int_{-\infty}^{\infty}dp_1\int_{-\infty}^{\infty}dp_2\int_0^{\infty} \frac{dp_-}{(4p_-)^{1/2}}
\Big{[}e^{-i(F_p^2x^+/4p_-+p_-x^-+p_1x^1+p_2x^2)}a(\vec{p})
\nonumber\\
&+e^{i(F_p^2x^+/4p_-
+p_-x^-+p_1x^1+p_2x^2)}a^{\dagger}(\vec{p})\Big{]},
\label{1.17a}
\end{align}
with $F_p^2=(p_1)^2+(p_2)^2+m^2$, and with $[a(\vec{p}),a^{\dagger}(\vec{p}^{~\prime})]=(1/2)\delta(p_--p_-^{\prime})\delta(p_1-p_1^{\prime})\delta(p_2-p_2^{\prime})$ following from (\ref{1.10}). While (\ref{1.17a}) looks similar to the instant-time (\ref{1.8}), there is one key difference. For instant-time quantization the limit $x^0\rightarrow 0$ can be taken without difficulty, but for the light-front case the limit $x^+\rightarrow 0$ is singular since $x^+$ appears in the combination $x^+/p_-$ and $p_-=0$ is included in the integration range. Thus one gets different answers depending on which of $x^+$ and $p_-$ one sets to zero first. That there is such an ambiguity is because on-shell light-front Fock states obey $4p_+p_--(p_1)^2-(p_2)^2-m^2=0$. Consequently $p_-=0$ corresponds to $p_+=\infty$, and with the conjugate of $p_+$ being $x^+$, $p_-=0$ corresponds to $x^+=0$. 

With the light-front form of the instant-time (\ref{1.5}) being of the form 
\begin{eqnarray}
D(x^{\mu},{\rm front})&=&\frac{2}{(2\pi)^4}\int dp_+dp_1dp_2dp_- \frac{e^{-i(p_+x^++p_1x^1+p_2x^2+p_-x^-)}}{4p_+p_--(p_1)^2-(p_2)^2-m^2+i\epsilon},
\label{1.18a}
\end{eqnarray}
we see that poles in light-front Feynman diagrams are located at $p_+=[(p_1)^2+(p_2)^2+m^2]/4p_-$, and just like the Fock space expansion, thus become undefined at $p_-=0$. As stressed in \cite{Chang1969,Yan1973,
McCartor1988,Heinzl2003,Ji1996,Tomaras2001,Ilderton2014,Strivastava2002,Collins2018},  handling so-called zero-mode $p_-=0$ singularities is one of the main challenges for light-front studies. The novelty of our current study  is in  handling the $p_-=0$ region in a way that does not lead to singularities in the $p_-\rightarrow 0$ limit. Specifically, we use an appropriate cut-off procedure or we rewrite Feynman diagram denominators using the exponential regulator technique, viz. we set $\int_0^{\infty}d\alpha\exp[i\alpha(A+i\epsilon)]=-1/iA$, with the $i\epsilon$ term suppressing the $\alpha =\infty$ contribution when $A$ is real.  The exponential regulator technique is particularly well-suited to handling the $p_-=0$ region, since with $p_-$ now being in an exponent rather than in a denominator the $p_-\rightarrow 0$ limit is no longer singular. On thus being able to control the $p_-\rightarrow 0$ limit,  we find that it is  the contribution of the $p_-=0$ region that  enables light-front vacuum Feynman graphs to be both non-vanishing and equal to their instant-time counterparts. Moreover, since $p_-=0$ entails that the energy $p_+$ is given by $p_+=\infty$, in coordinate space this corresponds to the light-front time being given by $x^+=0$. In consequence, the limit $x^+\rightarrow 0$ is singular, just as we had noted above. Singularities at $p_-=0$ and at $x^+=0$ are thus correlated, and in this paper we study their interplay. Our ability to handle $p_-=0$ singularities is because we have more control in a Feynman diagram than in an on-shell approach since a Feynman diagram involves an off-shell contour. Difficulties in treating the $p_-=0$ region come about because one goes to the pole straight away and reduces the four-dimensional Feynman integral to a three-dimensional one. By starting off the mass shell we can delay  (or, as we shall see below, even avoid) going to the complex $p_+$ plane poles until after we have controlled the $p_-=0$ region. Difficulties associated with the $p_-=0$ region are really difficulties associated with the on-shell three-dimensional formalism, and can be handled without difficulty in an off-shell four-dimensional formalism.

To provide further insight into the nature of the $x^+\rightarrow 0$ limit, we note that before we set $x^+=0$ the light-front matrix element $-i\langle \Omega|T^{(+)}[\phi(x)\phi(0)]|\Omega\rangle$ does coincide with the solution to the light-front Green's function equation 
\begin{eqnarray}
\left[4\left(\frac{\partial}{\partial x^+}\right)\left(\frac{\partial}{\partial x^-}\right)-\left(\frac{\partial}{\partial x^1}\right)^2-\left(\frac{\partial}{\partial x^2}\right)^2+m^2\right]D(x^{\mu},{\rm front})=-2\delta(x^+)\delta(x^1)\delta(x^2)\delta(x^-),
\label{1.19a}
\end{eqnarray}
and thus we can set
\begin{eqnarray}
i\langle \Omega|T^{(+)}[\phi(x)\phi(0)]|\Omega\rangle=D(x^{\mu},{\rm front})=\frac{2}{(2\pi)^4}\int dp_+dp_1dp_2dp_- \frac{e^{-i(p_+x^++p_1x^1+p_2x^2+p_-x^-)}}{4p_+p_--(p_1)^2-(p_2)^2-m^2+i\epsilon}.
\label{1.20a}
\end{eqnarray}
As with the instant-time $e^{-ip_0x^0}$ case, the $e^{-ip_+x^+}$ term provides suppression of the circle at infinity, and as long as $x^+$ and $x^0$ are non-zero, for both $D(x^{\mu},{\rm instant})$ and $D(x^{\mu},{\rm front})$ there are only pole contributions. In both of these cases the on-shell pole terms correspond to the on-shell Fock space expansions. Since the light-front Fock expansion corresponds to the Light-Front Hamiltonian approach, whenever $x^+$ is non-zero we can thus justify the use of the Light-Front Hamiltonian approach that is widely used in light-front studies (an approach that is not just on shell but also only involves scattering processes where $x^+$ is restricted to $x^+>0$).

However, things are different when we set $x^+$ to zero, since given the time-ordered $-i\langle \Omega|T^{(+)}[\phi(x)\phi(0)]|\Omega\rangle$ we can set $x^{\mu}=0$ and construct the vacuum tadpole graph as its $x^{\mu}\rightarrow 0$ limit, to thus include the $x^+\rightarrow 0$ limit. If we take the $x^{\mu}\rightarrow 0$ limit of
\begin{eqnarray}
D(x^{\mu})=-i\langle \Omega|[\theta(\tau)\phi(x)\phi(0)+\theta(-\tau)\phi(0)\phi(x)]|\Omega\rangle
\label{1.21a}
\end{eqnarray}
where $\tau$ denotes $x^0$ or $x^+$,  from (\ref{1.5}) and (\ref{1.20a}) we obtain 
\begin{eqnarray}
D(x^{\mu}=0,{\rm instant})&=&\frac{1}{(2\pi)^4}\int dp_0dp_1dp_2dp_3 \frac{1}{(p_0)^2-(p_1)^2-(p_2)^2-(p_3)^2-m^2+i\epsilon},
\nonumber\\
D(x^{\mu}=0,{\rm front})&=&\frac{2}{(2\pi)^4}\int dp_+dp_1dp_2dp_- \frac{1}{4p_+p_--(p_1)^2-(p_2)^2-m^2+i\epsilon},
\label{1.22a}
\end{eqnarray}
in the respective limits. In constructing the vacuum graphs as the limit in which we set $x^{\mu}=0$ we see that both time orderings contribute since 
\begin{eqnarray}
-i\langle \Omega|[\theta(\tau)\phi(x)\phi(0)+\theta(-\tau)\phi(0)\phi(x)]|\Omega\rangle \rightarrow
-i\langle \Omega|[\theta(0^+)\phi(0)\phi(0)+\theta(0^-)\phi(0)\phi(0)]|\Omega\rangle
=-i\langle \Omega|\phi(0)\phi(0)|\Omega\rangle,
\label{1.23a}
\end{eqnarray}
where, as we discuss in (\ref{14.4}) below, we set $\theta(0)=1/2$.
In this regard light-front vacuum graphs depart from the standard Light-Front Hamiltonian approach, since that approach is designed for scattering processes that propagate forward in time alone, to thus be restricted to $x^+>0$. However in vacuum graphs the limits of both the $x^+>0$ and $x^+<0$ components contribute. Vacuum graphs thus contain information that is not accessible in the Light-Front Hamiltonian approach.  Now as it stands $D(x^{\mu}=0,{\rm instant})$ and $D(x^{\mu}=0,{\rm front})$ as given in (\ref{1.22a}) must be equal to each other since on any given Feynman contour each one is just a momentum transform of the other, with any given complex plane $p_0$ contour transforming into an associated complex plane $p_+$ contour. However, there is still a central difference between the two cases. With there being two powers of $(p_0)^2$ in the denominator of $D(x^{\mu}=0,{\rm instant})$, in closing the $p_0$ contour the circle at infinity in the complex $p_0$ plane is suppressed and only pole terms contribute. However, there is only one power of $p_+$ in the denominator of $D(x^{\mu}=0,{\rm front})$ and thus the circle at infinity in the complex $p_+$ plane is not suppressed. Thus while one can equate $D(x^{\mu}=0,{\rm instant})$ and $D(x^{\mu}=0,{\rm front})$ at the off-shell four-dimensional level one cannot equate their on-shell pole contributions, with instant-time pole terms corresponding not to light-front  pole terms but to light-front pole plus light-front circle contributions combined. That the pole terms could not coincide is because of the singularity associated with the $x^+/p_-$ term in the light-front Fock space expansion given in (\ref{1.17a}). With this singularity being entangled with the circle at infinity in the complex $p_+$ plane (as noted above, $p_-=0$ corresponds to $p_+=\infty$ in the Fock space expansion, and $p_+=\infty$ corresponds to $x^+=0$, just as needed for the tadpole graph), one of the key objectives of this paper will be in taking care of this singularity so as to show explicitly that the $p_+$ circle at infinity term does ensure that $D(x^{\mu}=0,{\rm instant})$ and $D(x^{\mu}=0,{\rm front})$ are indeed equal.

For fermions the light-front anticommutation and time-ordered product relations involve projected fermion states called good and bad fermions. And while the anticommutation relation of two good fermions or that of a good fermion and bad fermion are both well-behaved (see (\ref{4.13a}) and (\ref{4.23a}) below), the anticommutation relation of two bad fermions (see (\ref{4.22a}) below) is very badly-behaved. Since no projected states appear in the instant-time case since  all instant-time fermion anticommutators are well-behaved, and since projection operators are not invertible, again light-front quantization appears to be quite different from instant-time quantization. Nonetheless, just as in the scalar field case, by studying anticommutators at unequal $x^0$ or unequal $x^+$ we shall show that it is only the restriction to equal $x^0$ or equal $x^+$ that causes the instant-time and light-front canonical  anticommutators to appear to be so different. 

Moreover, while we can make the identification of $i\langle \Omega|T^{(+)}[\phi(x)\phi(0)]|\Omega\rangle$ with the momentum integral given in (\ref{1.20a}) in the scalar field case, this is not the case for fermions. Rather, as we discuss below, for light-front fermions one has 
\begin{align}
-i\langle \Omega|T^{(+)}[\psi_{\beta}(x)\bar{\psi}_{\alpha}(0)]|\Omega\rangle
&=2\int dp_+dp_-dp_1dp_2\Big{[} \frac{e^{-i(p_+x^++p_1x^1+p_2x^2+p_-x^-)}}{\gamma^+p_++\gamma^-p_++\gamma^1p_1+\gamma^2p_2-m+i\epsilon}\Big{]}_{\beta\alpha}
\nonumber\\
&+\tfrac{i}{4}\gamma^+_{\beta\alpha}\delta(x^+)\epsilon(x^-)\delta(x^1)\delta(x^2),
\label{1.24a}
\end{align}
with the light-front time-ordered product not being given solely as a momentum integral, but as a momentum integral together with an additional light-cone singularity term. This additional singular term has no instant-time counterpart. So again instant-time quantization and light-front quantization appear to differ. However, as we will show below, this is not in fact the case, with the extra singular term actually decoupling in matrix elements. Some of our results regarding the scalar field sector have already been reported in a short paper \cite{Mannheim2019a}, and in this paper we generalize those results to all orders, and to both the fermion and gauge boson sectors, again finding that in the vacuum sector there are circle at infinity contributions to light-front vacuum sector diagrams. 

Despite the equivalence that we establish in this paper between instant-time $x^0=y^0$ and light-front $x^+=y^+$ commutators and anticommutators, we should note that because this equivalence is between functions defined at $x^0=y^0$ and $x^+=y^+$, it is actually a local one. However in order to define quantities such as momentum generators one needs some global information as well, since even though $T^{\mu\nu}$ is a local operator its integral over the spatial coordinates is global. Thus even though one can transform both  $T^{\mu\nu}$ and $\partial_{\mu}T^{\mu\nu}=0$ from instant time  to light front without difficulty just by covariance, this is not true for their spatial integrals. With the instant-time Hamiltonian being $H(IT)=\int dx^1dx^2dx^3T^0_{~~0}$,  $H(IT)$ will be independent of $x^0$ if $T^{\mu\nu}$ vanishes at large $x^1$, $x^2$ and $x^3$ ($\partial_0\int dx^1dx^2dx^3T^{0}_{~~0}=-\int dx^1dx^2dx^3\partial_iT^{i}_{~~0}=-\int dS_iT^{i}_{~~0}$).  Similarly, the light-front $H(LF)=(1/2)\int dx^1dx^2dx^-T^+_{~~+}$ will be  independent of $x^+$ if $T^{\mu\nu}$ vanishes at large $x^1$, $x^2$ and $x^-$ ($(1/2)\partial_+\int dx^1dx^2dx^-T^{+}_{~~+}=-(1/2)\int dx^1dx^2dx^-\partial_iT^{i}_{~~+}=-(1/2)\int dS_iT^{i}_{~~+}$). In light--front coordinates vanishing at large $x^3$ is vanishing at large $(x^+-x^-)/2$. Thus $H(IT)$ requires vanishing at large $(x^+-x^-)/2$ while $H(LF)$ requires vanishing at large $x^-$. Thus the two cases require asymptotic convergence in different directions, directions that may not be compatible. In this sense then there is an intrinsic global difference between the instant-time and light-front formulations, one that is not seen in the local commutation and anticommutation relations. I.e. the two formulations are identical locally but not necessarily globally. For the purposes of this paper we shall assume that boundary conditions are such that $H(IT)$ is independent of $x^0$ and $H(LF)$ is independent of $x^+$. Moreover, as we will see in Sec. \ref{S4}, even though the fermionic $H(LF)$ is composed of good fermions alone, the spatial surface integral $-(1/2)\int dx^1dx^2dx^-\partial_iT^{i}_{~~+}$ also involves the bad fermions. Then, because the bad fermions are related to the good fermions by the non-local integral constraint given in (\ref{4.16a}) below, to establish the $x^+$ independence of the fermionic light-front $H(LF)$ requires more convergence than is required to establish the $x^0$ independence of the fermionic instant-time $H(IT)$. No similar concerns arise for bosonic fields.

One of the conceptual questions raised by the light-front formulation is whether forward propagation in $x^+$ is the same as or different from forward propagation in $x^0$. Thus in radioactive decay for instance one has to ask whether the Rutherford instant-time $N(x^0)=e^{-Ax^0}$  formula might differ from its light-front $N(x^+)=e^{-Ax^+}$ analog, and if they do differ from each other which one should one use. Even though these two formulae look to be different,  we now show that they are in fact the same. To compare them we introduce the proper time $\tau$ and rewrite the Rutherford formula as $N(\tau)=e^{-A\tau}$. Now in instant-time coordinates we have $\tau^2=(x^0)^2-(x^3)^2-(x^1)^2-(x^2)^2$. Thus for timelike $\tau^2>0$ and all $x^{\mu}$ positive for simplicity we can make boosts in the $x^1$ and $x^2$ directions $x^i\rightarrow (x^i-vx^0)/(1-v^2)^{1/2}$, and with $x^0>x^i$ (since $\tau^2>0$) we can bring $x^1$ and $x^2$ to zero. Also we can set $x^3$ to zero by an $x^3$ boost, and then $\tau^2=(x^0)^2$ and $N(\tau)=e^{-Ax^0}$. To make contact with the light-front case we can make a substitution $x^{\pm}=x^0\pm x^3$, $x^0=(x^++x^-)/2$, $x^3=(x^+-x^-)/2$ to light-front coordinates, and note that this is just a substitution not a transformation of the form $x^0\rightarrow x^0+x^3$. In light-front coordinates we have $\tau^2=x^+x^--(x^1)^2-(x^2)^2$.  We now set $x^1$ and $x^2$ to zero as before. This now leaves $\tau^2=x^+x^-$. We now make the previous $x^3$ boost with  $x^3\rightarrow (x^3-vx^0)/(1-v^2)^{1/2}$ to bring $x^3$ to zero. With $x^3=0$ we have $x^+=x^-$ and thus $\tau^2=(x^+)^2$ and $N(\tau)=e^{-Ax^+}$. Consequently, in the Lorentz frame in which $x^1=0$, $x^2=0$, $x^3=0$, the Lorentz invariant $N(\tau)$ takes two forms, $N(\tau)=e^{-Ax^0}$ and $N(\tau)=e^{-Ax^+}$, and they are equivalent. Thus in this frame both instant-time observers and light-front observers see the same decay rate. For timelike intervals forward in $x^+$ is the same as forward in $x^0$, with causality in $x^+$ thus being the same as causality in $x^0$. For completeness we note that for events with $\tau^2<0$ we can bring $x^0$ to zero via $x^0\rightarrow (x^0-vx^3)/(1-v^2)^{1/2}$ since $x^3>x^0$ for spacelike intervals. Then with  $x^+=-x^-=x^3$ we have $\tau^2=-(x^3)^2-(x^1)^2-(x^2)^2=-(x^-)^2-(x^1)^2-(x^2)^2$, to again establish equivalence.

Underlying our entire analysis and central to it is that the $x^0\rightarrow x^0+x^3=x^+$, $x^3\rightarrow x^0-x^3=x^-$ transformation from instant-time coordinates and fields to light-front coordinates and fields is a spacetime-dependent translation. Now in quantum theory translations are unitary transformations on quantum operators.  Consequently, not only are instant-time quantization and light-front quantization equivalent, because of general coordinate invariance they are unitarily equivalent.  In Secs. \ref{S7} and \ref{S8} we explore this unitary equivalence in detail.

In Secs. \ref{S2} to \ref{S8} of our paper we compare equal instant-time and equal light-front time quantization procedures, studying the nature of canonical conjugates and canonical commutation relations in scalar field, fermion field, Abelian and non-Abelian gauge field, and graviton field theories. While our analysis will enable us to obtain equal light-front time commutation relations that had previously been reported in the literature, for fermions and gauge fields it will also enable us to obtain some that had not. These additional relations involve light-front time derivatives of the fields and cannot be obtained by canonical quantization since these light-front time derivatives are not canonical conjugate variables. In addition, in Sec. \ref{S8}  we follow \cite{Mannheim2019b,Mannheim2020a} and show first how to construct equal light-front time commutators and anticommutators starting from unequal-time instant-time commutators and anticommutators, and then show how to construct equal-time instant-time commutators and anticommutators starting from unequal-time light-front commutators and anticommutators. While this same analysis applies to gauge bosons, in Secs. \ref{S5} and \ref{S9} we present a gauge fixing procedure in which none of zero-mode problems that occur in axial gauge quantization are encountered. In Secs. \ref{S9} to \ref{S17} we compare instant-time and light-front Green's functions for both non-vacuum and vacuum diagrams and show their equivalence, while also showing that the equivalence to both  the on-shell Hamiltonian Fock space prescription and the closely related infinite momentum frame prescription breaks down in the light-front vacuum sector.  In Sec. \ref{S18} we present our conclusions. In an Appendix we present our general notation, analyze good and bad fermion bilinears, and discuss their implications for Ward identities involving such bilinears when symmetries are spontaneously broken.  In particular we find that while one needs the anticommutator of a good fermion with a bad fermion in order to establish the light-front axial-vector Ward identity, one does not need the troublesome, badly-behaved anticommutator between two bad fermions.

\section{Conjugates and Nonlocal Constraints}
\label{S2}

We introduce contravariant light-front variables $x^+=x^0+x^3$, $x^-=x^0- x^3$, and quantize according to equal $x^+$ (light-front formulation) rather than equal $x^0$ (instant-time formulation). (We specify our notation and the structure of coordinate derivatives in the Appendix.) In order to determine the form of canonical commutation relations we first need to identify appropriate light-front canonical conjugates, and as we shall see, for scalars, vectors, or fermion fields there are some substantive subtleties.

In field theory the conjugate of a field is the functional derivative of the action with respect to the relevant time derivative of the field, and for such derivatives we only need to refer to the kinetic energy sector of a field theory if, as is standard, the rest of the Lagrangian density is taken to be free of derivatives. For a Hermitian scalar field the relevant kinetic energy contribution to the action is of the form
\begin{eqnarray}
I_S&=&\int d^4x(-g)^{1/2}\tfrac{1}{2}\partial_{\mu}\phi\partial^{\mu}\phi=\tfrac{1}{2}\int dx^+dx^-dx^1dx^2
\tfrac{1}{2}[\partial_{+}\phi\partial^{+}\phi+\partial_{-}\phi\partial^{-}\phi+\partial_{1}\phi\partial^{1}\phi+
\partial_{2}\phi\partial^{2}\phi]
\nonumber\\
&=&\int d^4x(-g)^{1/2}\tfrac{1}{2}\partial_{\mu}\phi\partial^{\mu}\phi=\tfrac{1}{2}\int dx^+dx^-dx^1dx^2
\tfrac{1}{2}[2\partial_{+}\phi\partial_{-}\phi+2\partial_{-}\phi\partial_{+}\phi-\partial_{1}\phi\partial_{1}\phi-
\partial_{2}\phi\partial_{2}\phi],
\label{2.1}
\end{eqnarray}
where $\partial_+=\partial\phi/\partial x^+$, $\partial_-=\partial\phi/\partial x^-$, $\partial^+\phi=2\partial_-\phi=2\partial\phi/\partial x^-$, $\partial^-\phi=2\partial_+\phi=2\partial\phi/\partial x^+$, and where $g$ is the determinant of the light-front metric introduced in the Appendix. The second form for $I_S$ given in (\ref{2.1}) shows that $I_S$ is Hermitian. The light-front conjugate is thus $(-g)^{-1/2}\delta I_S/\delta \partial_+\phi=\partial^+\phi=2\partial\phi/\partial x^-$. It is thus not a derivative of the field with respect to the light-front time (i.e., not $\partial \phi/\partial x^+$), and thus behaves quite differently than the conjugate in the instant-time form, viz. $\partial \phi/\partial x^0$. We will explore some implications of this distinction below, as it will lead to nonlocal equal $x^+$ light-front time commutators that are nonvanishing for all $x^-$.

For gauge fields with Maxwell Lagrangian density $L=-(1/4)F_{\mu\nu}F^{\mu\nu}$, we have $F_{+-}=\partial_+A_--\partial_-A_+=(1/2)(\partial_+A^+-\partial_-A^-)$. Since there is no light-front time derivative of $A^-$, the field $A_+=g_{+-}A^-=A^-/2$  has no canonical conjugate, though the field $A^+=g^{+-}A_-=2A_-$ does. In the free Maxwell equation $\partial_{\nu}F^{\nu\mu}=0$ with $\mu=+$ we have $\partial_{\nu}F^{\nu+}=0$, i.e., $\partial_-F^{-+}+\partial_1F^{1+}+\partial_2F^{2+}=0$, i.e.,
\begin{eqnarray}
\partial_-(\partial^-A^+ -\partial^+A^-)+\partial_1(\partial^1A^+-\partial^+A^1)+\partial_2(\partial^2A^+-\partial^+A^2)=0. 
\label{2.2}
\end{eqnarray}
Thus in the $A^+=0$ gauge we have $\partial_-\partial^+A^-+\partial_1\partial^+A^1+\partial_2\partial^+A^2=0$, i.e., 
\begin{eqnarray}
&&2\partial_-(\partial_-A^-+\partial_1A^1+\partial_2A^2)=0,\quad A^-=-\frac{1}{(\partial_-)^2}\left(\partial_-(\partial_1A^1+\partial_2A^2)\right),
\nonumber\\
&&A^-=-\int dy^-D_2(x^--y^-)\partial_-(\partial_1A^1+\partial_2A^2),
\label{2.3}
\end{eqnarray}
where $D_2=(\partial_-)^{-2}$ is an inverse propagator. We recognize (\ref{2.3}) as a constraint condition that involves no time derivative, and not only that, with $D_2=(\partial_-)^{-2}$ being an inverse propagator, the constraint condition is even nonlocal. Consequently, in the $A^+=0$ gauge (a gauge that is commonly used in light-front studies of gauge theories) the field $A^-$ that has no conjugate is not an independent dynamical field but a nonlocally constrained one, an issue we shall explore in some detail below.

Comparison with the Coulomb gauge in QED is instructive. In the instant-time case we have $F_{03}=\partial_0A_3-\partial_3A_0$. Thus now it is $A_0$ (and thus $A^0$) that has no conjugate. The zeroth component of the equations of motion is of the form $\partial_{\nu}F^{\nu 0}=0$, i.e., $\partial_i(\partial^iA^0-\partial^0A^i)=0$. Thus in the Coulomb gauge where $\partial_iA^i=0$ we have $\partial_i\partial^iA^0=0$, a constraint condition that also involves no time derivatives. Consequently, the field $A^0$ that has no conjugate is not an independent dynamical field. The light-front $A^+=2A_-=0$ gauge is thus the analog of the instant-time QED Coulomb gauge, and just like Coulomb gauge QED, light-front quantization for gauge fields also has a nonlocal structure. As well as study the $A^+=0$ gauge, we shall also study gauge theories where the gauge freedom is characterized by  gauge fixing terms. In such a case we do not have the nonlocal constraints that occur in the $A^+=0$ gauge.

For the Dirac Lagrangian density $i\bar{\psi}\gamma^{\mu}\partial_{\mu}\psi$, we find that in the instant-time case the conjugate of $\psi$ is $i\bar{\psi}\gamma^0=i\psi^{\dagger}$. In the light-front formulation it is $i\bar{\psi}\gamma^{+}=i\psi^{\dagger}\gamma^0\gamma^{+}$, where $\gamma^+=\gamma^0+\gamma^3$. As we discuss in detail below, the operators $\Lambda^{+}=(1/2)\gamma^0\gamma^{+}$ and $\Lambda^{-}=(1/2)\gamma^0\gamma^{-}$ ($\gamma^-=\gamma^0-\gamma^3$) are projection operators on to what are known as the ``good" and ``bad" fermions $\psi_{(+)}=\Lambda^+\psi$ and $\psi_{(-)}=\Lambda^-\psi$. We thus recognize the conjugate $i\bar{\psi}\gamma^{+}$ of $\psi$ as the good fermion only. With $\Lambda^++\Lambda^-=1$ we can write the light-front time derivative component $i\bar{\psi}\gamma^+\partial_+\psi$ of the Lagrangian density entirely in terms of the good fermion as  $2i\bar{\psi}_{(+)}\partial_+\psi_{(+)}$. The bad fermion thus has no canonical conjugate and is thus not a dynamical field, with the canonical anticommutator being between the good fermions alone. We thus see that for both fermions and gauge fields not all of the degrees of freedom are dynamical, with, as we discuss in more detail below, the nondynamical ones obeying nonlocal constraint conditions.

\section{Determining the Quantization Conditions for Scalar Fields}
\label{S3}

To determine quantization conditions we require that the  momentum operators generically effect $[P_{\mu},\phi]=-i\partial_{\mu}\phi$ for any quantum field, with the momentum generators themselves being constructed from the relevant field energy-momentum tensors. While our procedure will enable us to obtain and provide insight into equal light-front time commutation relations that had previously been reported in the literature, for fermions and gauge fields it will also enable us to obtain some that had not. These additional relations involve $\partial_+$ derivatives of the fields and cannot be obtained by canonical quantization since these light-front time derivatives are not canonical conjugates. 

For free scalar fields first, we can for the moment take them to be massless since that does not affect the kinetic energy operator of relevance for constructing canonical conjugates, with the equation of motion and energy-momentum tensor constructed from a scalar field action that has been covariantized then being given by 
\begin{eqnarray}
g^{\alpha\beta}\partial_{\alpha}\partial_{\beta}\phi=0, \quad T_{\mu\nu}=\partial_{\mu}\phi\partial_{\nu}\phi-g_{\mu\nu}\tfrac{1}{2}\partial^{\alpha}\phi\partial_{\alpha}\phi.
\label{3.1}
\end{eqnarray}
From (\ref{3.1}) it follows that $\partial_{\nu}T^{\mu\nu}=0$, and thus that $\partial_{+}T^{+}_{~~+}=-\partial_{-}T^{-}_{~~+}-\partial_{1}T^{1}_{~~+}-\partial_{2}T^{2}_{~~+}$. Since the $P_{\mu}$ momentum generators have to transform as a covariant four vector, we construct them from $T^{+}_{\phantom{+}\mu}$ as 
\begin{eqnarray}
P_{\mu}=\tfrac{1}{2}\int dx^-dx^1dx^2 T^{+}_{\phantom{+}\mu}({\rm front}),
\label{3.2}
\end{eqnarray}
and they will obey $\partial_{+}P_{\mu}=0$ if the fields in $T_{\mu\nu}$ vanish sufficiently rapidly at spatial infinity, something we assume to be the case. (The factor of one half can be obtained by transforming $dx^3$ to $dx^-$ with $dx^0$ held fixed, or from the Jacobian as described in the Appendix.) We thus construct the momentum generators from
\begin{eqnarray}
&&T^{+}_{\phantom{+}+}({\rm front})=\tfrac{1}{2}[\partial^{+}\phi\partial_{+}\phi -\partial^-\phi\partial_-\phi-\partial^1\phi\partial_1\phi-\partial^2\phi\partial_2\phi],
\nonumber\\
&&T^{+}_{\phantom{+}-}({\rm front})=\partial^{+}\phi\partial_{-}\phi,\quad
T^{+}_{\phantom{+}1}({\rm front})=\partial^{+}\phi\partial_{1}\phi,\quad
T^{+}_{\phantom{+}2}({\rm front})=\partial^{+}\phi\partial_{2}\phi.
\label{3.3}
\end{eqnarray}
However, since $\partial^+\phi=g^{+\mu}\partial_{\mu}\phi=2\partial_-\phi$, $\partial_+\phi=g_{+\mu}\partial^{\mu}\phi=(1/2)\partial^-\phi$, we find that $\partial^{+}\phi\partial_{+}\phi -\partial^-\phi\partial_-\phi=0$. Thus we can replace (\ref{3.3}) by 
\begin{eqnarray}
&&T^{+}_{\phantom{+}+}({\rm front})=\tfrac{1}{2}[\partial_1\phi\partial_1\phi+\partial_2\phi\partial_2\phi],
\nonumber\\
&&T^{+}_{\phantom{+}-}({\rm front})=2\partial_-\phi\partial_-\phi,\quad
T^{+}_{\phantom{+}1}({\rm front})=2\partial_-\phi\partial_{1}\phi,\quad
T^{+}_{\phantom{+}2}({\rm front})=2\partial_-\phi\partial_{2}\phi.
\label{3.4}
\end{eqnarray}
As q-numbers both $T^{+}_{\phantom{+}1}({\rm front})$ and $T^{+}_{\phantom{+}2}({\rm front})$ have an ordering problem, since if $\partial_-\phi$ and $\partial_i\phi$ do not commute $(i=1,2)$, which turns out to actually be the case, neither  $T^{+}_{\phantom{+}1}({\rm front})$ nor $T^{+}_{\phantom{+}2}({\rm front})$ would be Hermitian. However we are actually interested in the integrated $P_{\mu}$, and we can integrate by parts in  $\int dy^-dy^1dy^2\partial_-\phi(x^+,y^1,y^2,y^-)\partial_{i}\phi(x^+,y^1,y^2,y^-)$ and show that it is equal to $\int dy^-dy^1dy^2\partial_i\phi(x^+,y^1,y^2,y^-)\partial_-\phi(x^+,y^1,y^2,y^-)$. Consequently there is no ordering issue for the $P_{\mu}$ and they are Hermitian.

If we were to proceed canonically, since the conjugate of $\phi$ is $2\partial_-\phi$, we would set
\begin{eqnarray}
&&[\phi(x^+,x^1,x^2,x^-),\partial^+\phi(x^+,y^1,y^2,y^-)]
\nonumber\\
&&=[\phi(x^+,x^1,x^2,x^-),2\partial_-\phi(x^+,y^1,y^2,y^-)]
=i\lambda\delta(x^1-y^1)\delta(x^2-y^2)\delta(x^--y^-),
\label{3.5}
\end{eqnarray}
where $\lambda$ is a coefficient that is to be determined from the normalization of the $[P_{\mu},\phi]=-i\partial_{\mu}\phi$ condition. (In (\ref{3.5}) the $\partial^+=2\partial/\partial {y^-}$ derivative acts on the ${y^-}$ coordinate.) However, (\ref{3.5}) is not in the form of a standard commutator since $\partial^+\phi$ is equal to $2\partial_-\phi$ and is thus not a light-front time derivative of $\phi$. Since $x^-$ and $y^-$ are not equal in (\ref{3.5}) we  can therefore integrate (\ref{3.5}) with respect to $\partial_-$ to obtain  (see e.g. \cite{Suzuki1976,Heinzl2001})
\begin{eqnarray}
[\phi(x^+,x^1,x^2,x^-),\phi(x^+,y^1,y^2,y^-)]=-\frac{i\lambda}{4}\epsilon(x^--y^-)\delta(x^1-y^1)\delta(x^2-y^2),
\label{3.6}
\end{eqnarray}
where $\epsilon(x^--y^-)=\theta(x^--y^-)-\theta(-x^-+y^-)$ is nonvanishing for any $x^--y^-$ no matter how large, in consequence of which the  $[\phi(x^+,x^1,x^2,x^-),\phi(x^+,y^1,y^2,y^-)]$ commutator is nonlocal.  Once the commutator of two $\phi$ fields is nonzero,  equal light-front time commutation relations  are then quite different from equal instant-time commutation relations  where $[\phi(x^0,x^1,x^2,x^3),\phi(x^0,y^1,y^2,y^3)]$ is zero. Nonetheless, as we show in Sec. \ref{S8}, this difference is only an apparent one. In evaluating the $[P_{\mu},\phi]$ commutators  in the following not just (\ref{3.5}) but also (\ref{3.6}) will be needed.

Since the $P_{\mu}$ generators are taken to be light-front time independent, we shall evaluate them at the same $x^+$ as that of the $\phi$ that appears in $[P_{\mu},\phi]$. Thus evaluating for $P_+$ at $x^+=y^+$, assuming enough asymptotic convergence so that we can  integrate by parts, and using the scalar field equation of motion yields
\begin{eqnarray}
[P_+,\phi(x)]&=&\frac{1}{4}\int dy^-dy^1dy^2[\partial_1\phi(y)\partial_1\phi(y)+\partial_2\phi(y)\partial_2\phi(y),\phi(x)]
\nonumber\\
&=&-\frac{1}{4}\int dy^-dy^1dy^2[\phi(y)(\partial_1)^2\phi(y)+\phi(y)(\partial_2)^2\phi(y),\phi(x)]
\nonumber\\
&=&-\frac{i\lambda}{8}\int dy^-dy^1dy^2\epsilon(x^--y^-)\phi(y)\left((\partial_1)^2+(\partial_2)^2\right)\delta(x^1-y^1)\delta(x^2-y^2)
\nonumber\\
&=&-\frac{i\lambda}{8}\int dy^-dy^1dy^2\epsilon(x^--y^-)\left((\partial_1)^2\phi(y)+(\partial_2)^2\phi(y)\right)\delta(x^1-y^1)\delta(x^2-y^2)
\nonumber\\
&=&-\frac{i\lambda}{2}\int dy^-dy^1dy^2\epsilon(x^--y^-)\partial_+\partial_-\phi(y)\delta(x^1-y^1)\delta(x^2-y^2)
\nonumber\\
&=&-i\lambda\int dy^-dy^1dy^2\delta(x^--y^-)\partial_+\phi(y)\delta(x^1-y^1)\delta(x^2-y^2)=-i\lambda \partial_+\phi(x).
\label{3.7}
\end{eqnarray}
With $[P_+,\phi]=-i\partial_+\phi$ following when $\lambda=1$, we thus confirm the generic form for the canonical commutator that is given in (\ref{3.5}).  

Evaluating for $P_-$ and $P_i$ $(i=1,2)$ we obtain
\begin{eqnarray}
[P_-,\phi(x)]&=&\int dy^-dy^1dy^2[\partial_-\phi(y)\partial_-\phi(y),\phi(x)]
\nonumber\\
&=&-i\lambda\int dy^-dy^1dy^2\partial_-\phi(y)\delta(x^--y^-)\delta(x^1-y^1)\delta(x^2-y^2)=-i\lambda \partial_-\phi(x),
\label{3.8}
\end{eqnarray}
\begin{eqnarray}
[P_i,\phi(x)]&=&\int dy^-dy^1dy^2\left[\partial_-\phi(y)\partial_i\phi(y)\phi(x) -\phi(x)\partial_-\phi(y)\partial_i\phi(y)\right]
\nonumber\\
&=&\int dy^-dy^1dy^2\Big{[}\partial_-\phi(y)\left[\partial_i\phi(y),\phi(x)\right]-\left[\phi(x),\partial_-\phi(y)\right]\partial_i\phi(y)\Big{]}
\nonumber\\
&=&\int dy^-dy^1dy^2\Big{[}\partial_i\phi(y)\left[\partial_-\phi(y),\phi(x)\right]-\left[\phi(x),\partial_-\phi(y)\right]\partial_i\phi(y)\Big{]}=-i\lambda \partial_i\phi(x).
\label{3.9}
\end{eqnarray}
As we thus see, with $\lambda=+1$ we implement all four $[P_{\mu},\phi]=-i\partial_{\mu}\phi$,  just as required. Finally, with $\lambda=1$  (\ref{3.6}) and (\ref{3.5}) become the scalar field theory equal light-front time commutation relations that date back to  \cite{Neville1971}, viz. 
\begin{align}
&[\phi(x^+,x^1,x^2,x^-),\phi(x^+,y^1,y^2,y^-)]=-\frac{i}{4}\epsilon(x^--y^-)\delta(x^1-y^1)\delta(x^2-y^2),
\nonumber\\
&[\phi(x^+,x^1,x^2,x^-),2\partial_-\phi(x^+,y^1,y^2,y^-)]=i\delta(x^1-y^1)\delta(x^2-y^2)\delta(x^--y^-).
\label{3.10}
\end{align}

The reason why the $[\phi(x),\phi(y)]$ commutator given in (\ref{3.10}) is able to be nonzero is because even though equal light-front time commutators with $x^+-y^+=0$ only take support at the tip of the light cone, that does not restrict $x^--y^-$. Thus if $(x^+-y^+)(x^--y^-)-(x^1-y^1)^2-(x^2-y^2)^2=0$, then if $x^+-y^+$, $x^1-y^1$ and $x^2-y^2$ are all zero, we can be on the light cone with $x^--y^-$ not constrained at all. Now microcausality requires that the two point light-front commutator $\langle \Omega |[\phi(x),\phi(y)]|\Omega\rangle$  vanish outside the light cone. Thus in general at $x^+-y^+=0$ we can write

\begin{eqnarray}
\langle \Omega |[\phi(x),\phi(y)]|\Omega\rangle=\delta(x^+-y^+)[A\epsilon(x^--y^-)+B\delta(x^--y^-)]\delta(x^1-y^1)\delta(x^2-y^2),
\label{3.11}
\end{eqnarray}
with arbitrary coefficients $A$ and $B$ and still not take support outside the light cone. Thus despite the fact that the 
$A\epsilon(x^--y^-)$ term is nonlocal, there is no violation of microcausality. 

The structure given in (\ref{3.11}) has an analog in momentum space. Solutions to the massless wave equation $g^{\mu\nu}\partial_{\mu}\partial_{\nu}\phi=4\partial_+\partial_-\phi-(\partial_1)^2\phi-(\partial_2)^2\phi=0$ can be written as plane waves of the form $\exp(ik\cdot x)=\exp(ik_+x^++ik_-x^-+ik_1x^1+ik_2x^2)$, where 
\begin{eqnarray}
4k_+k_--(k_1)^2-(k_2)^2=0.
\label{3.12}
\end{eqnarray}
We can rewrite this as 
\begin{eqnarray}
4k_+[C\delta(k_+)+k_-]-(k_1)^2-(k_2)^2=0, \quad 4k_-[D\delta(k_-)+k_+]-(k_1)^2-(k_2)^2=0,
\label{3.13}
\end{eqnarray}
with the $C$ or $D$ terms then appearing in the propagator. Given the $D$ term, we see that we will need to treat the $k_-=0$ limit with care in the analysis we provide below.

Light-front causality even has an implication for the nonrelativistic Schr\"odinger equation. If written in instant-time coordinates, viz. $i\partial_t\psi=H\psi$, one then requires that wave functions vanish at asymptotic spatial coordinates $x^1$, $x^2$, $x^3$. However for a local but not necessarily stationary solution at $t=0$ these asymptotic spatial points are outside the forward light cone at $t=0$ since for timelike intervals $(x^0)^2-(x^1)^2-(x^2)^2-(x^3)^2$ is greater than zero. There is of course no violation of causality since one can derive the Schr\"odinger equation as the low energy limit of a relativistically causal Bethe-Salpeter equation. It is just that causality is not manifest. However,  in light-front coordinates timelike intervals obey $x^+x^--(x^1)^2-(x^2)^2>0$, and one can be in the forward light cone of $x^+$ with small $x^+$ and large enough $x^-$, with causality now being manifest.

\section{Good and Bad Fermions}
\label{S4}

For fermions light-front quantization is somewhat different than for scalar field light-front quantization. The free Dirac action is of the form
\begin{align}
I_D&=\tfrac{1}{2}\int d^4x\bar{\psi}(i\gamma^{\mu}\partial_{\mu}-m)\psi +{\rm H.~c.}
\nonumber\\
&=\tfrac{1}{4}\int dx^+dx^-dx^1dx^2\psi^{\dagger}[\gamma^0(i\gamma^+\partial_++i\gamma^-\partial_-+i\gamma^1\partial_1+i\gamma^2\partial_2-m)]\psi+{\rm H.~c.},
\label{4.1}
\end{align}
where $\gamma^{\pm}=\gamma^0\pm\gamma^3$ and ${\rm H.~c.}$ denotes Hermitian conjugate. (On general grounds \cite{Mannheim2018} one should add on the $CPT$ conjugate rather than the Hermitian conjugate, but they coincide here.) With this action the canonical conjugate of $\psi$ is $i\psi^{\dagger}\gamma^0\gamma^+$.  In the construction of the light-front fermion sector we find a rather sharp distinction with the instant-time fermion sector. Specifically, unlike $\gamma^0$ and $\gamma^3$, which obey $(\gamma^0)^2=1$, $(\gamma^3)^2=-1$, the quantities $\gamma^+$ and $\gamma^-$ obey $(\gamma^+)^2=0$, $(\gamma^-)^2=0$, to thus both be non-invertible divisors of zero. In terms of these $\gamma^+$ and $\gamma^-$ we introduce projection operators
\begin{eqnarray}
\Lambda^{+}=\tfrac{1}{2}\gamma^0\gamma^+=\tfrac{1}{2}[1+\gamma^0\gamma^3],\quad \Lambda^{-}=\tfrac{1}{2}\gamma^0\gamma^-=\tfrac{1}{2}[1-\gamma^0\gamma^3],
\label{4.2}
\end{eqnarray}
which obey  
\begin{eqnarray}
\Lambda^{+}+\Lambda^{-}=I,\quad(\Lambda^{+})^2=\Lambda^{+}=[\Lambda^+]^{\dagger},\quad (\Lambda^{-})^2=\Lambda^{-}=[\Lambda^{-}]^{\dagger},\quad \Lambda^{+}\Lambda^{-}=0.
\label{4.3}
\end{eqnarray}
We identify so-called good and bad fermions $\psi_{(+)}=\Lambda^+\psi$, $\psi_{(-)}=\Lambda^-\psi$ (we clarify this designation below), and thus identify the conjugate of $\psi$ as the good fermion $2i\psi_{(+)}^{\dagger}$, where $\psi^{\dagger}_{(+)}$ denotes $[\psi^{\dagger}]_{(+)}=\psi^{\dagger}\Lambda^+$, which is equal to $[\Lambda^+\psi]^{\dagger}=[\psi_{(+)}]^{\dagger}$ since $\Lambda^+$ is Hermitian. These good and bad fermion fields have no instant-time analog, and in fact could not have  since the $\Lambda^+$ and $\Lambda^-$ projection operators are not invertible.  

Since the $2i\psi_{(+)}^{\dagger}$ conjugate is a good fermion, in the canonical anticommutator of $\psi$ with its conjugate only the good component of $\psi$ will contribute since $\Lambda^+\Lambda^-=0$, and guided by the structure of the scalar field $[\phi(x^+,x^1,x^2,x^-),2\partial_-\phi(x^+,y^1,y^2,y^-)]$ commutator given in (\ref{3.10}) we thus anticipate that the canonical equal light-front time fermion field anticommutator will  be of the form
\begin{eqnarray}
&&\Big{\{}\psi_{(+)}^{\dagger}(x^+,x^1,x^2,x^-),\psi_{(+)}(x^+,y^1,y^2,y^-)\Big{\}}=\kappa\Lambda^+\delta(x^--y^-)\delta(x^1-y^1)\delta(x^2-y^2),
\label{4.4}
\end{eqnarray}
where $\kappa$ is a constant. Through use of the relation $[P_{\mu},\psi_{(+)}]=-i\partial_{\mu}\psi_{(+)}$ we will confirm (\ref{4.4}) while determining the value of the constant $\kappa$. On the right-hand side of (\ref{4.4}) we have introduced a factor $\Lambda^+$, since with the left-hand side of (\ref{4.4}) being proportional to  $(\Lambda^+)^2=\Lambda^+$, the right-hand side must be proportional to $\Lambda^+$ too.

Given the Dirac action $I_D$,  the  fermionic energy-momentum tensor as defined via variation with respect to the metric of an $I_D$ that has been covariantized is given by the automatically symmetric (see e.g. \cite{Mannheim2006})
\begin{eqnarray}
T_{\mu\nu}=\tfrac{i}{4}\bar{\psi}\gamma_{\mu}\partial_{\nu}\psi-\tfrac{i}{4}\partial_{\nu}\bar{\psi}\gamma_{\mu}\psi
+\tfrac{i}{4}\bar{\psi}\gamma_{\nu}\partial_{\mu}\psi-\tfrac{i}{4}\partial_{\mu}\bar{\psi}\gamma_{\nu}\psi.
\label{4.5}
\end{eqnarray}
With the fermion field obeying $(i\gamma^{\mu}\partial_{\mu}-m)\psi=0$, the energy momentum tensor obeys $\partial_{\mu}T^{\mu\nu}=0$, $\partial_{\nu}T^{\mu\nu}=0$, though we note that it is only because of the fact that $T_{\mu\nu}$  is symmetric that both of these conservation conditions hold. Rather than use (\ref{4.5}) as is, (\ref{4.5}) can be simplified by noting that the quantity
\begin{eqnarray}
B^{\alpha\mu\nu}=\bar{\psi}[\gamma^{\alpha}(\gamma^{\mu}\gamma^{\nu}-\gamma^{\nu}\gamma^{\mu})
+(\gamma^{\mu}\gamma^{\nu}-\gamma^{\nu}\gamma^{\mu})\gamma^{\alpha}]\psi
\label{4.6a}
\end{eqnarray}
is antisymmetric on all pairs of its three indices, and for fermions that obey the Dirac equation its derivative is given by
\begin{eqnarray}
\partial_{\alpha}B^{\alpha\mu\nu}=-4\bar{\psi}\gamma^{\mu}\partial^{\nu}\psi+4\partial^{\nu}\bar{\psi}\gamma^{\mu}\psi
+4\bar{\psi}\gamma^{\nu}\partial^{\mu}\psi-4\partial^{\mu}\bar{\psi}\gamma^{\nu}\psi.
\label{4.7a}
\end{eqnarray}
Because of the antisymmetry of $B^{\alpha\mu\nu}$ in its $\alpha,\mu$ indices a second derivative of (\ref{4.7a}) gives $\partial_{\mu}\partial_{\alpha}B^{\alpha\mu\nu}=0$. And because of this very same asymmetry it  follows that when $\mu=+$  the integral $\int dy^-dy^1dy^2\partial_{\alpha}B^{\alpha +\nu}=\int dy^-dy^1dy^2\partial_{i}B^{i +\nu}$ is an asymptotic surface term. Consequently, if the fermion fields are such that this asymptotic surface term vanishes (this being a constraint on the behavior of both good and bad fermions as they both contribute to $B^{i+\nu}$), then for evaluating the spatial integrals of $T^{+}_{\phantom{+}\mu}$ we can replace (\ref{4.5}) by
\begin{eqnarray}
T_{\mu\nu}=\tfrac{i}{4}\bar{\psi}\gamma_{\mu}\partial_{\nu}\psi-\tfrac{i}{4}\partial_{\nu}\bar{\psi}\gamma_{\mu}\psi
+\tfrac{i}{4}\bar{\psi}\gamma_{\nu}\partial_{\mu}\psi-\tfrac{i}{4}\partial_{\mu}\bar{\psi}\gamma_{\nu}\psi-\tfrac{i}{16}\partial_{\alpha}B^{\alpha\mu\nu}=\tfrac{i}{2}\bar{\psi}\gamma^{\mu}\partial^{\nu}\psi-\tfrac{i}{2}\partial^{\nu}\bar{\psi}\gamma^{\mu}\psi.
\label{4.8a}
\end{eqnarray}
We recognize the latter form for $T^{\mu\nu}$ as being that of the canonical energy-momentum tensor, and even though it is not symmetric it gives the same $P_{\mu}$ generators as the symmetric $T_{\mu\nu}$ of (\ref{4.5}), and thus we shall use (\ref{4.8a}) in the following. (While we see that we can proceed from the metric-based (\ref{4.5}) to the canonical (\ref{4.8a}), we note in passing that there is an in principle difference, namely while we need to use the Dirac equation in order to establish the canonical (\ref{4.8a}), we do not need the fermions to go on shell in order to establish (\ref{4.5}) as the variation of (\ref{4.1}) with respect to the metric is made without regard to any variation with respect the fermions.)

To determine the appropriate light-front momentum generators we replace the instant-time $T^{(0)}_{\phantom{0}\mu}$ by $(\partial x^+/\partial x^0)T^{(0)}_{\phantom{0}\mu}=T^{(+)}_{\phantom{0}\mu}$. On recalling that $\Lambda^+=\tfrac{1}{2}\gamma^0\gamma^+$ is a projector, we have
\begin{eqnarray}
&&T^{+}_{\phantom{+}\mu}({\rm front})=i\psi_{(+)}^{\dagger}\partial_{\mu}\psi-i\partial_{\mu}\psi_{(+)}^{\dagger}\psi
=i\psi_{(+)}^{\dagger}\partial_{\mu}\psi_{(+)}-i\partial_{\mu}\psi_{(+)}^{\dagger}\psi_{(+)},
\nonumber\\
&&P_{\mu}=\tfrac{1}{2}\int dy^-dy^1dy^2\left[i\psi_{(+)}^{\dagger}(y)\partial_{\mu}\psi_{(+)}(y)-i\partial_{\mu}\psi_{(+)}^{\dagger}(y)\psi_{(+)}(y)\right],
\label{4.9a}
\end{eqnarray}
and thus see that the fermionic momentum generators are composed of good fermions alone. For $\mu=-,1,2$ we can integrate the second term in (\ref{4.9a}) by parts, with the second term then being equal to the first term. We cannot do this for $\mu=+$. However, for the free Dirac equation the charge $\frac{1}{2}\int dy^-dy^1dy^2\bar{\psi}\gamma^+\psi=\int dy^-dy^1dy^2\psi_{(+)}^{\dagger}\psi_{(+)}$ is conserved. Thus we can set $\int dy^-dy^1dy^2[\partial_+\psi_{(+)}^{\dagger}\psi_{(+)}+\psi_{(+)}^{\dagger}\partial_+\psi_{(+)}]=0$. For all four components of $P_{\mu}$ then we have 
\begin{eqnarray}
P_{\mu}=\int dy^-dy^1dy^2i\psi_{(+)}^{\dagger}(y)\partial_{\mu}\psi_{(+)}(y).
\label{4.10a}
\end{eqnarray}

Given (\ref{4.4}) we obtain
\begin{align}
[P_{\mu},\psi_{(+)}(x)]&=i\int dy^-dy^1dy^2\left(\psi_{(+)}^{\dagger}(y)\partial_{\mu}\psi_{(+)}(y)\psi_{(+)}(x)
-\psi_{(+)}(x)\psi_{(+)}^{\dagger}(y)\partial_{\mu}\psi_{(+)}(y)\right)
\nonumber\\
&=i\int dy^-dy^1dy^2\Big{(}-\psi_{(+)}^{\dagger}(y)\psi_{(+)}(x)\partial_{\mu}\psi_{(+)}(y)
-\psi_{(+)}(x)\psi_{(+)}^{\dagger}(y)\partial_{\mu}\psi_{(+)}(y)
\nonumber\\
&+\psi_{(+)}^{\dagger}(y)\psi_{(+)}(x)\partial_{\mu}\psi_{(+)}(y)
+\psi_{(+)}^{\dagger}(y)\partial_{\mu}\psi_{(+)}(y)\psi_{(+)}(x)\Big{)}
\nonumber\\
&=-i\kappa \Lambda^+\partial_{\mu}\psi_{(+)}(x)
+i\int dy^-dy^1dy^2\psi_{(+)}^{\dagger}(y)\Big{\{}\partial_{\mu}\psi_{(+)}(y),\psi_{(+)}(x)\Big{\}}
\nonumber\\
&=-i\kappa\partial_{\mu}\psi_{(+)}(x)
+i\int dy^-dy^1dy^2\psi_{(+)}^{\dagger}(y)\Big{\{}\partial_{\mu}\psi_{(+)}(y),\psi_{(+)}(x)\Big{\}}.
\label{4.11a}
\end{align}
For  the $\mu=-,1,2$ components of $[P_{\mu},\psi_{(+)}(x)]$ we can integrate by parts and obtain 
\begin{align}
[P_{\mu},\psi_{(+)}(x)]\Big{|}_{\mu =-,1,2}&=-i\kappa\partial_{\mu}\psi_{(+)}(x)\Big{|}_{\mu =-,1,2}
-i\int dy^-dy^1dy^2\partial_{\mu}\psi_{(+)}^{\dagger}(y)\Big{\{}\psi_{(+)}(y),\psi_{(+)}(x)\Big{\}}\Big{|}_{\mu =-,1,2}.
\label{4.12a}
\end{align}
To then eliminate the integral in (\ref{4.12a})  we need to set $\Big{\{}\psi_{(+)}(x),\psi_{(+)}(y)\Big{\}}=0$ at $x^+=y^+$. (We could not instead take $\Big{\{}\psi_{(+)}(x),\psi_{(+)}(y)\Big{\}}$ to be equal to $\delta(x^--y^-)\delta(x^1-y^1)\delta(x^2-y^2)$ or equal to the $\epsilon(x^--y^-)\delta(x^1-y^1)\delta(x^2-y^2)$ form found in the scalar field case). For the $\mu=+$ term we cannot integrate by parts in (\ref{4.11a}) and must thus set $\Big{\{}\psi_{(+)}(x),\partial_+\psi_{(+)}(y)\Big{\}}=0$, a relation that had not previously been anticipated in the literature since $\partial_+\psi_{(+)}(y)$ is not a light-front canonical conjugate. Then, with $\kappa=1$ we recover $[P_{\mu},\psi_{(+)}(x)]=-i \partial_{\mu}\psi_{(+)}(x)$ for all $\mu$.

Finally then, the light-front fermion sector anticommutation relations take the form 
\begin{align}
&\Big{\{}\psi_{(+)}^{\dagger}(x^+,x^1,x^2,x^-),\psi_{(+)}(x^+,y^1,y^2,y^-)\Big{\}}
=\Lambda^+\delta(x^--y^-)\delta(x^1-y^1)\delta(x^2-y^2),
\nonumber\\
&\Big{\{}\psi_{(+)}(x^+,x^1,x^2,x^-),\psi_{(+)}(x^+,y^1,y^2,y^-)\Big{\}}=0,
\nonumber\\
&\Big{\{}\psi^{\dagger}_{(+)}(x^+,x^1,x^2,x^-),\psi^{\dagger}_{(+)}(x^+,y^1,y^2,y^-)\Big{\}}=0,
\nonumber\\
&\Big{\{}\psi_{(+)}(x^+,x^1,x^2,x^-),\partial_+\psi_{(+)}(x^+,y^1,y^2,y^-)\Big{\}}=0,
\label{4.13a}
\end{align}
with the first of these relations dating back to \cite{Chang1973}.
Since the last of the  anticommutation relations in (\ref{4.13a}) involves a $\partial_+$ derivative it could not be obtained by applying the $\partial_+$  derivative to any of the other anticommutator relations in (\ref{4.13a}) since all the fields in those anticommutator relations are at the same $x^+$. This last anticommutator relation is thus independent of the others and has independent validity. (Taking the $\partial_{+}$ derivative of $\big{\{}\psi_{(+)}(x^+,x^1,x^2,x^-),\psi_{(+)}(x^+,y^1,y^2,y^-)\big{\}}=0$ allows one only to conclude that $\big{\{}\psi_{(+)}(x^+,x^1,x^2,x^-),\partial_+\psi_{(+)}(x^+,y^1,y^2,y^-)\big{\}}$ is an odd function of $x^--y^-$, of $x^1-y^1$, and of $x^2-y^2$.) We provide a complementary derivation of this last anticommutation relation in Sec. \ref{S8}.

As constructed, these anticommutation relations are initially somewhat puzzling since in instant-time quantization the anticommutator $\Big{\{}\psi^{\dagger},\psi\Big{\}}=\delta^3$ involves all four of the degrees of freedom of the Dirac spinor. And if we insert $\Lambda^++\Lambda^-=1$ into $\Big{\{}\psi^{\dagger},\psi\Big{\}}=\delta^3$, rather than obtain an instant-time analog of (\ref{4.13a})  we would instead obtain $\Big{\{}\psi_{(+)}^{\dagger},\psi_{(+)}\Big{\}}+\Big{\{}\psi_{(-)}^{\dagger},\psi_{(-)}\Big{\}}=\delta^3$, and thus on multiplying this relation by $\Lambda^+$ and on multiplying it by $\Lambda^-$ would obtain the $\Big{\{}\psi_{(+)}^{\dagger},\psi_{(+)}\Big{\}}=\Lambda^+\delta^3$, $\Big{\{}\psi_{(-)}^{\dagger},\psi_{(-)}\Big{\}}=\Lambda^-\delta^3$ relations, but with all fields evaluated at equal $x^0$. Thus we might (erroneously) anticipate that if we do obtain the $\Big{\{}\psi_{(+)}^{\dagger},\psi_{(+)}\Big{\}}=\Lambda^+\delta^3$  relation as evaluated at equal $x^+$, we should equally obtain  the $\Big{\{}\psi_{(-)}^{\dagger},\psi_{(-)}\Big{\}}=\Lambda^-\delta^3$ relation as evaluated at equal $x^+$ as well. We have described this anticipation as erroneous, since as we show in Sec. \ref{S7} one cannot in fact  transform any equal $x^0$ commutator or anticommutator into any equal $x^+$ one in the first place, so that the  equal $x^+$ $\Big{\{}\psi_{(+)}^{\dagger},\psi_{(+)}\Big{\}}=\Lambda^+\delta^3$ relation itself does not in fact follow from the equal $x^0$  $\Big{\{}\psi^{\dagger},\psi\Big{\}}=\delta^3$ relation to begin with. (We elaborate further on this point in Sec. \ref{S8} below, where we show that the equal $x^+$  $\Big{\{}\psi_{(+)}^{\dagger},\psi_{(+)}\Big{\}}=\Lambda^+\delta^3$ relation actually follows not from the equal $x^0$ instant-time $\Big{\{}\psi^{\dagger},\psi\Big{\}}$ anticommutation relation at all, but from its unequal-time generalization.) 

Nonetheless, even though we do not need to anticipate the existence of an equal $x^+$ $\Big{\{}\psi_{(-)}^{\dagger},\psi_{(-)}\Big{\}}=\Lambda^-\delta^3$ relation for the bad fermion, we do need to see what does happen in the bad fermion sector. To this end we manipulate the Dirac equation $[i(\gamma^+\partial_++\gamma^-\partial_-+\gamma^1\partial_1+\gamma^2\partial_2)-m]\psi=0$. We first multiply on the left  by $\gamma^0$ to obtain
\begin{eqnarray}
2i\partial_+\psi_{(+)}+2i\partial_-\psi_{(-)}+i\gamma^0(\gamma^1\partial_1+\gamma^2\partial_2)\psi-m\gamma^0\psi=0.
\label{4.14a}
\end{eqnarray}
Next we multiply (\ref{4.14a}) by $\Lambda^-$ and also multiply it by $\Lambda^+$ to obtain
\begin{eqnarray}
2i\partial_-\psi_{(-)}&=&[-i\gamma^0(\gamma^1\partial_1+\gamma^2\partial_2)+m\gamma^0]\psi_{(+)},
\nonumber\\
2i\partial_+\psi_{(+)}&=&[-i\gamma^0(\gamma^1\partial_1+\gamma^2\partial_2)+m\gamma^0]\psi_{(-)}.
\label{4.15a}
\end{eqnarray}
Since the $\partial_-\psi_{(-)}$ equation contains no time derivatives, $\psi_{(-)}$ is thus a constrained variable, consistent with it having no conjugate.  Since it is a constrained variable it does not appear in any fundamental anticommutation relation, though as we show in (\ref{4.22a}) and (\ref{4.23a}) below, we can use (\ref{4.13a}) and (\ref{4.15a}) to construct  equal $x^+$  $\Big{\{}\psi^{\dagger}_{(-)},\psi_{(-)}\Big{\}}$, $\Big{\{}\psi^{\dagger}_{(-)},\psi_{(+)}\Big{\}}=\Big{\{}\psi^{\dagger}_{(+)},\psi_{(-)}\Big{\}}^{\dagger}$ anticommutators. Through the use of the inverse propagator $D_1=(\partial_-)^{-1}=[\theta(x^-)-\theta(-x^-)]/2=\epsilon(x^-)/2$ we can rewrite the $\partial_-\psi_{(-)}$ equation in (\ref{4.15a}) as  
\begin{align}
\psi_{(-)}(x^+,x^1,x^2,x^-)&=\frac{1}{4i}\int du^-\epsilon(x^--u^-)[-i\gamma^0(\gamma^1\partial_1+\gamma^2\partial_2)+m\gamma^0]\psi_{(+)}(x^+,x^1,x^2,u^-),
\nonumber\\
[\psi_{(-)}]^{\dagger}&=\frac{i}{4}\int du^-\epsilon(x^--u^-)[i\partial_1[\psi_{(+)}]^{\dagger}\gamma^0\gamma^1+
i\partial_2[\psi_{(+)}]^{\dagger}\gamma^0\gamma^2+m[\psi_{(+)}]^{\dagger}\gamma^0],
\label{4.16a}
\end{align}
and recognize $\psi_{(-)}$ as obeying a nonlocal condition. It is because $\psi_{(-)}$ obeys such a nonlocal constraint (one that is interaction dependent when interactions are involved) that it is known as a bad fermion.

Given (\ref{4.16a}) we can determine the $[P_{\mu},\psi_{(-)}(x)]$ commutator. However, it is more convenient to determine the $[P_{\mu},\partial_-\psi_{(-)}(x)]$ commutator. Now even though $P_{\mu}$ is built out of good fermions alone that does not mean that $[P_{\mu},\partial_-\psi_{(-)}(x)]=0$. Rather, on noting that according to (\ref{4.15a})  the relation between $\partial_-\psi_{(-)}(x)$ and $\psi_{(+)}(x)$ only involves kinematic factors, these factors do not affect the commutators of $\partial_-\psi_{(-)}(x)$ and $\psi_{(+)}(x)$ with $P_{\mu}$. And thus from $[P_{\mu},\psi_{(+)}(x)]=-i\partial_{\mu}\psi_{(+)}(x)$ we obtain 
\begin{eqnarray}
[P_{\mu},\partial_-\psi_{(-)}(x)]=-i\partial_{\mu}\partial_-\psi_{(-)}(x).
\label{4.17a}
\end{eqnarray}
Then from (\ref{4.17a}) we can determine $[P_{\mu},\psi_{(-)}(x)]$ via an integration if needed.
Thus while  $\psi_{(-)}$ acquires a time (or space) dependence, it only acquires it indirectly through the time dependence of $\psi_{(+)}$ in the  $\partial_-\psi_{(-)}$  constraint equation given in (\ref{4.15a}). (The quantity that would have been able to give $\psi_{(-)}$ a direct time dependence would have been the bad fermion is $\tfrac{1}{2}\int dy^-dy^1dy^2T^{-}_{\phantom{+}\mu}({\rm front})=i\int dy^-dy^1dy^2\psi_{(-)}^{\dagger}\partial_{\mu}\psi_{(-)}$, but it is not related to a momentum generator.)

With $\psi_{(-)}$ not appearing in $P_{\mu}=\int dy^-dy^1dy^2T^{+}_{\phantom{+}\mu}({\rm front})/2$ it would appear that $\psi_{(-)}$ only plays a minor role with the dynamics being encoded in $\psi_{(+)}$ alone. However, the bad fermion does appear in various key places. In order to show that the $P_{\mu}$ momentum generators actually are independent of $x^+$ one requires the vanishing of the quantity $\partial_+\int dy^-dy^1dy^2T^{+}_{\phantom{+}\mu}({\rm front})/2=-\int dy^-dy^1dy^2[\partial_-T^{-}_{\phantom{+}\mu}({\rm front})+\partial_1T^{1}_{\phantom{+}\mu}({\rm front})+\partial_2T^{2}_{\phantom{+}\mu}({\rm front})]/2={\rm asymptotic~surface~term}$. With the bad fermions appearing in $\partial_-T^{-}_{\phantom{+}\mu}({\rm front})+\partial_1T^{1}_{\phantom{+}\mu}({\rm front})+\partial_2T^{2}_{\phantom{+}\mu}({\rm front})$, they thus influence the behavior of the asymptotic surface term. The bad fermions also appear in the fermion mass term $\bar{\psi}\psi=[\psi_{(+)}]^{\dagger}\gamma^0\psi_{(-)}+[\psi_{(-)}]^{\dagger}\gamma^0\psi_{(+)}$. And though this mass term does not appear in the fermionic $T_{\mu\nu}=\bar{\psi}\gamma_{\mu}\partial_{\nu}\psi/2-\partial_{\nu}\bar{\psi}\gamma_{\mu}\psi/2$, in Sec. \ref{S9} we show that the bad fermion is needed for both light-front fermionic path integrals and light-front fermionic time-ordered products. And in the Appendix we show that $\psi_{(-)}$ does play a quite central role in the structure of both fermion vector and fermion axial vector currents and in their Ward identities. 
Finally, combining the two equations in (\ref{4.15a}) gives
\begin{eqnarray}
2i\partial_+\psi_{(+)}&=&[-i\gamma^0(\gamma^1\partial_1+\gamma^2\partial_2)+m\gamma^0]\psi_{(-)}
=\frac{1}{2i\partial_-}(-(\partial_1)^2-(\partial_2)^2+m^2)\psi_{(+)},
\nonumber\\
2i\partial_-\psi_{(-)}&=&[-i\gamma^0(\gamma^1\partial_1+\gamma^2\partial_2)+m\gamma^0]\psi_{(+)}
=\frac{1}{2i\partial_-}(-(\partial_1)^2-(\partial_2)^2+m^2)\psi_{(-)},
\label{4.18a}
\end{eqnarray}
to thus recover the Klein-Gordon equation for each of $\psi_{(+)}$ and $\psi_{(-)}$ 
\begin{eqnarray}
&&[4\partial_+\partial_--(\partial_1)^2-(\partial_2)^2+m^2]\psi_{(+)}=0,
\nonumber\\
&&[4\partial_+\partial_--(\partial_1)^2-(\partial_2)^2+m^2]\psi_{(-)}=0,
\label{4.19a}
\end{eqnarray}
equations that involve no mixing between them.

Given (\ref{4.15a}) and (\ref{4.13a}) we can now determine an equal light-front time anticommutator for the bad fermions. Noting that
\begin{align}
&\left[-i\gamma^0\gamma^1\frac{\partial}{\partial x^1}-i\gamma^0\gamma^2\frac{\partial}{\partial x^2}+\gamma^0m\right]_{\mu\alpha}\Lambda^+_{\alpha\beta}\left[i\gamma^0\gamma^1\frac{\partial}{\partial y^1}+i\gamma^0\gamma^2\frac{\partial}{\partial x^2}+\gamma^0m\right]_{\beta\nu}
\nonumber\\
&=\Lambda^-_{\mu\nu}\left[\frac{\partial}{\partial x^1}\frac{\partial}{\partial y^1}+\frac{\partial}{\partial x^2}\frac{\partial}{\partial y^2}+m^2\right],
\label{4.20a}
\end{align}
we obtain
\begin{align}
&\Big{\{}\frac{\partial}{\partial x^-}\psi_{\mu}^{(-)}(x^+,x^1,x^2,x^-),\frac{\partial}{\partial y^-}[\psi_{(-)}^{\dagger}]_{\nu}(x^+,y^1,y^2,y^-)\Big{\}}
\nonumber\\
&=\frac{1}{4}\Lambda^-_{\mu\nu}\left[\frac{\partial}{\partial x^1}\frac{\partial}{\partial y^1}+\frac{\partial}{\partial x^2}\frac{\partial}{\partial y^2}+m^2\right]\delta(x^--y^-)\delta(x^1-y^1)\delta(x^2-y^2)
\nonumber\\
&=\frac{1}{4}\Lambda^-_{\mu\nu}\left[-\frac{\partial}{\partial x^1}\frac{\partial}{\partial x^1}-\frac{\partial}{\partial x^2}\frac{\partial}{\partial x^2}+m^2\right]\delta(x^--y^-)\delta(x^1-y^1)\delta(x^2-y^2).
\label{4.21a}
\end{align}
Consequently we obtain
\begin{align}
&\Big{\{}\psi_{\mu}^{(-)}(x^+,x^1,x^2,x^-),[\psi_{(-)}^{\dagger}]_{\nu}(x^+,y^1,y^2,y^-)\Big{\}}
\nonumber\\
&=\frac{1}{16}\Lambda^-_{\mu\nu}\left[-\frac{\partial}{\partial x^1}\frac{\partial}{\partial x^1}-\frac{\partial}{\partial x^2}\frac{\partial}{\partial x^2}+m^2\right]
\int du^-\epsilon(x^--u^-)\int dv^-\epsilon(y^--v^-)\delta(u^--v^-)\delta(x^1-y^1)\delta(x^2-y^2)
\nonumber\\
&=\frac{1}{16}\Lambda^-_{\mu\nu}\left[-\frac{\partial}{\partial x^1}\frac{\partial}{\partial x^1}-\frac{\partial}{\partial x^2}\frac{\partial}{\partial x^2}+m^2\right]
\int du^-\epsilon(x^--u^-)\epsilon(y^--u^-)\delta(x^1-y^1)\delta(x^2-y^2).
\label{4.22a}
\end{align}

For completeness we note that we can also derive the $\Big{\{}\psi_{\mu}^{(+)}(x^+,x^1,x^2,x^-),[\psi_{(-)}^{\dagger}]_{\nu}(x^+,y^1,y^2,y^-)\Big{\}}$ and its Hermitian conjugate by this same procedure, and obtain
\begin{align}
&\Big{\{}[\psi_{(+)}]_{\nu}(x),[\psi^{\dagger}_{(-)}]_{\sigma}(y)\Big{\}}
\nonumber\\
&
=\Big{\{}[\psi_{(+)}]_{\nu}(x^+,x^1,x^2,x^-),\tfrac{i}{4}\int du^-\epsilon(y^--u^-)[i\gamma^0(\gamma^1\partial_1^y+\gamma^2\partial_2^y)+m\gamma^0]_{\tau\sigma}[\psi^{\dagger}_{(+)}]_{\tau}(x^+,y^1,y^2,u^-)\Big{\}}
\nonumber\\
&=\tfrac{i}{4}\int du^-\epsilon(y^--u^-)\Lambda^+_{\nu\tau}[i\gamma^0(\gamma^1\partial_1^y+\gamma^2\partial_2^y)+m\gamma^0]_{\tau\sigma}
\delta(x^--u^-)\delta(x^1-y^1)\delta(x^2-y^2)\
\nonumber\\
&=\tfrac{i}{8}\epsilon(y^--x^-)[i(\gamma^-\gamma^1\partial_1^y+\gamma^-\gamma^2\partial_2^y)+m\gamma^-]_{\nu\sigma}\delta(x^1-y^1)\delta(x^2-y^2)
\nonumber\\
&=\tfrac{i}{8}\epsilon(x^--y^-)[i(\gamma^-\gamma^1\partial_1^x+\gamma^-\gamma^2\partial_2^x)-m\gamma^-]_{\nu\sigma}\delta(x^1-y^1)\delta(x^2-y^2),
\label{4.23a}
\end{align}
where $\partial^y_1$ denotes $\partial/\partial y^1$, etc. 
As we see, the equal $x^+$ bad fermion sector $\Big{\{}\psi_{\mu}^{(-)}(x^+,x^1,x^2,x^-),[\psi_{(-)}^{\dagger}]_{\nu}(x^+,y^1,y^2,y^-)\Big{\}}$ anticommutator is non-vanishing, with its nonlocal nature being apparent. In Sec. \ref{S8} we will show that we can obtain (\ref{4.21a}),  (\ref{4.22a}) and (\ref{4.23a}) as well as the relations  in (\ref{4.13a}) starting from the unequal $x^0$ instant-time $\Big{\{}\psi^{\dagger},\psi\Big{\}}$ anticommutation relation.

\section{Gauge Fields}
\label{S5}
\subsection{Quantization of Abelian Gauge Fields in the $A^+=0$ Gauge}
\label{S5a}

For gauge fields we write the Maxwell Lagrangian density $L=-\tfrac{1}{4}F_{\mu\nu}F^{\mu\nu}$ in light-front form, to obtain
\begin{eqnarray}
F_{\mu\nu}F^{\mu\nu}&=&-8(\partial_+A_--\partial_-A_+)^2+2(\partial_1A_2-\partial_2A_1)^2
\nonumber\\
&-&4(\partial_+A_1-\partial_1A_+)(\partial_-A_1-\partial_1A_-)-4(\partial_-A_1-\partial_1A_-)(\partial_+A_1-\partial_1A_+)
\nonumber\\
&-&4(\partial_+A_2-\partial_2A_+)(\partial_-A_2-\partial_2A_-)-4(\partial_-A_2-\partial_2A_-)(\partial_+A_2-\partial_2A_+),
\label{5.1}
\end{eqnarray}
with the Maxwell equations $\partial_{\nu}F^{\mu\nu}=0$ taking the form
\begin{align}
&[4\partial_+\partial_--(\partial_1)^2-(\partial_2)^2]A_+-\partial_+[2\partial_-A_++2\partial_+A_--\partial_1A_1-\partial_2A_2]=0,
\nonumber\\
&[4\partial_+\partial_--(\partial_1)^2-(\partial_2)^2]A_--\partial_-[2\partial_-A_++2\partial_+A_--\partial_1A_1-\partial_2A_2]=0,
\nonumber\\
&[4\partial_+\partial_--(\partial_1)^2-(\partial_2)^2]A_1-\partial_1[2\partial_-A_++2\partial_+A_--\partial_1A_1-\partial_2A_2]=0,
\nonumber\\
&[4\partial_+\partial_--(\partial_1)^2-(\partial_2)^2]A_2-\partial_2[2\partial_-A_++2\partial_+A_--\partial_1A_1-\partial_2A_2]=0.
\label{5.2}
\end{align}

With there being no light-front time derivative of $A_+$ in (\ref{5.1}), it does not have a canonical conjugate.  For the Maxwell Lagrangian density the respective conjugates of $A_-$, $A_1$ and $A_2$ are 
\begin{align}
&\Pi^-=\frac{\partial L}{\partial(\partial_+A_-)}=4(\partial_+A_--\partial_-A_+)=(\partial^-A^+-\partial^+A^-), 
\nonumber\\
&\Pi^1=\frac{\partial L}{\partial(\partial_+A_1)}=2(\partial_-A_1-\partial_1A_-)=(-\partial^+A^1+\partial^1A^+),
\nonumber\\
&\Pi^2=\frac{\partial L}{\partial(\partial_+A_2)}=2(\partial_-A_2-\partial_2A_-)=(-\partial^+A^2+\partial^2A^+). 
\label{5.3}
\end{align}
In the commonly used $A^+=2A_-=0$ gauge these conjugates respectively reduce to  
\begin{eqnarray}
\Pi^-=-4\partial_-A_+=-\partial^+A^-, \quad \Pi^1=2\partial_-A_1=-\partial^+A^1, \quad \Pi^2=2\partial_-A_2=-\partial^+A^2, 
\label{5.4}
\end{eqnarray}
while the Maxwell equations reduce to
\begin{align}
[4\partial_+\partial_--(\partial_1)^2-(\partial_2)^2]A_+-\partial_+[2\partial_-A_+-\partial_1A_1-\partial_2A_2]=0,&
\nonumber\\
-\partial_-[2\partial_-A_+-\partial_1A_1-\partial_2A_2]=0,&
\nonumber\\
[4\partial_+\partial_--(\partial_1)^2-(\partial_2)^2]A_1-\partial_1[2\partial_-A_+-\partial_1A_1-\partial_2A_2]=0,&
\nonumber\\
[4\partial_+\partial_--(\partial_1)^2-(\partial_2)^2]A_2-\partial_2[2\partial_-A_+-\partial_1A_1-\partial_2A_2]=0.&
\label{5.5}
\end{align}
As we see, not only does $A_+$ not have a conjugate, it obeys the constraint given as the second relation in (\ref{5.5}), a constraint that does not involve the light-front time derivative $\partial_+$ at all. Thus only $A_1$ and $A_2$ are dynamical fields.

Given the canonical conjugates $\Pi^1=2\partial_-A_1$ and $\Pi^2=2\partial_-A_2$ we can construct canonical commutators, and  as with the scalar field, the commutators are at equal $x^+$ and not at equal $x^-$. Thus in analog to (\ref{3.5}) and (\ref{3.6}),  we take the gauge field canonical commutators in the $A_1,A_2$ sector to be of the form
\begin{align}
&[A_1,\Pi^1]=[A_1(x^+,x^1,x^2,x^-),2\partial_-A_1(x^+,y^1,y^2,y^-)]=i\sigma\delta(x^--y^-)\delta(x^1-y^1)\delta(x^2-y^2),
\nonumber\\
&[A_2,\Pi^2]=[A_2(x^+,x^1,x^2,x^-),2\partial_-A_2(x^+,y^1,y^2,y^-)]=i\sigma\delta(x^--y^-)\delta(x^1-y^1)\delta(x^2-y^2),
\nonumber\\
&[A_1(x^+,x^1,x^2,x^-),2A_1(x^+,y^1,y^2,y^-)]=-\tfrac{i}{2}\sigma\epsilon(x^--y^-)\delta(x^1-y^1)\delta(x^2-y^2),
\nonumber\\
&[A_2(x^+,x^1,x^2,x^-),2A_2(x^+,y^1,y^2,y^-)]=-\tfrac{i}{2}\sigma\epsilon(x^--y^-)\delta(x^1-y^1)\delta(x^2-y^2),
\label{5.6}
\end{align}
where $\sigma$ is a numerical constant. We take all $[A_1,A_2]$ type commutators to be zero.

To fix the momentum commutation relations we recall that by variation of the covariantized Maxwell action with respect to the metric the gauge field energy-momentum tensor is given by
\begin{eqnarray}
T_{\mu\nu}=-F_{\mu\alpha}F_{\nu}^{\phantom{\nu}\alpha}+\tfrac{1}{4}g_{\mu\nu}F_{\alpha\beta}F^{\alpha\beta}.
\label{5.7}
\end{eqnarray}
In the light-front $P_1$ sector we can set 
\begin{align}
P_1&=\tfrac{1}{2}\int dy^-dy^1dy^2T^{(+)}_{\phantom{+}1}=-\tfrac{1}{2}\int dy^-dy^1dy^2F^{+}_{\phantom{+\alpha}\alpha}F_{1}^{\phantom{1}\alpha}
=-\tfrac{1}{2}\int dy^-dy^1dy^2(4F_{-+}F_{1-}-2F_{-2}F_{12})
\nonumber\\
&=\int dy^-dy^1dy^2[2\partial_-A_+\partial_-A_1+\partial_-A_2(\partial_1A_2-\partial_2A_1)].
\label{5.8}
\end{align}
Then on integrating by parts, using the second relation in (\ref{5.5}) and  on integrating by parts again, we can rewrite $P_1$ as 
\begin{align}
P_1=\int dy^-dy^1dy^2[\partial_-A_1\partial_1A_1+\partial_-A_2\partial_1A_2].
\label{5.9}
\end{align}
Equally we have
\begin{align}
P_2&=\int dy^-dy^1dy^2[\partial_-A_1\partial_2A_1+\partial_-A_2\partial_2A_2].
\label{5.10}
\end{align}
We can thus realize 
\begin{eqnarray}
[P_1,A_1(x)]&=&\int dy^-dy^1dy^2\Big{(}\partial_-A_1(y)[\partial_1A_1(y),A_1(x)]-[A_1(x),\partial_-A_1(y)]\partial_1A_1(y)\Big{)}
\nonumber\\
&=&-\int dy^-dy^1dy^2\Big{(}\partial_1A_1(y)[A_1(x),\partial_-A_1(y)]+[A_1(x),\partial_-A_1(y)]\partial_1A_1(y)\Big{)}=-i\sigma \partial_1A_1(x).
\label{5.11}
\end{eqnarray}
Thus we obtain 
\begin{eqnarray}
[P_1,A_1]=-i\partial_1A_1,\quad [P_1,A_2]=-i\partial_1A_2,\quad [P_2,A_1]=-i\partial_2A_1, \quad [P_2,A_2]=-i\partial_2A_2,
\label{5.12}
\end{eqnarray}
if we set $\sigma=1$ in (\ref{5.6}).

For the commutation relations involving  $P_-$, in the $A^+=0$ gauge we directly obtain 
\begin{eqnarray}
T^{+}_{\phantom{+}-}&=&2\partial_-A_1\partial_-A_1+2\partial_-A_2\partial_-A_2.
\label{5.13}
\end{eqnarray}
Then with $P_-=\tfrac{1}{2}\int dy^-dy^1dy^2T^{+}_{\phantom{+}-}$ and  $\sigma=1$, we obtain
\begin{eqnarray}
[P_-,A_1]=-i\partial_-A_1, \quad [P_-,A_2]=-i\partial_-A_2,
\label{5.14}
\end{eqnarray}
just as required.

For the $P_+$ sector we try to simplify $T^{+}_{\phantom{+}+}$ by using  the equations of motion given in (\ref{5.5}) and integrating by parts to eliminate the dependence on $A_+$, so as to obtain
\begin{align}
\int d^3yF_{+-}^2&=\int d^3y (\partial_-A_+)^2=-\int d^3yA_+(\partial_-^2A_+)=-\tfrac{1}{2}\int d^3yA_+(\partial_-\partial_1A_1+\partial_-\partial_2A_2)
\nonumber\\
&=-\tfrac{1}{2}\int d^3y(\partial_-\partial_1A_+A_1+\partial_-\partial_2A_+A_2)
\nonumber\\
&=-\tfrac{1}{4}\int d^3y\Big{(}\left(4\partial_+\partial_-A_1-(\partial_2)^2A_1+\partial_1\partial_2A_2\right)A_1
+\left(4\partial_+\partial_-A_2-(\partial_1)^2A_2+\partial_2\partial_1A_1\right)A_2\Big{)}
\nonumber\\
&=\int d^3y\Big{(}\partial_+A_1\partial_-A_1+\partial_+A_2\partial_-A_2-\tfrac{1}{4}F_{12}^2\Big{)},
\nonumber\\
\tfrac{1}{2}\int d^3y[F_{-1},F_{+1}]&=\tfrac{1}{2}\int d^3y[\partial_-A_1,\partial_+A_1-\partial_1A_+]
=\tfrac{1}{2}\int d^3y\Big{(}[\partial_-A_1,\partial_+A_1]+[A_1,\partial_-\partial_1A_+]\Big{)}
\nonumber\\
&=\tfrac{1}{2}\int d^3y\Big{(}[\partial_-A_1,\partial_+A_1]+\tfrac{1}{2}[A_1,4\partial_+\partial_-A_1-(\partial_2)^2A_1+\partial_1\partial_2A_2]\Big{)}
\nonumber\\
&=\tfrac{1}{2}\int d^3y\Big{(}-\partial_-A_1\partial_+A_1+\partial_+A_1\partial_-A_1
-\tfrac{1}{2}[\partial_2A_1,\partial_1A_2]\Big{)},
\nonumber\\
\tfrac{1}{2}\int d^3y[F_{-2},F_{+2}]&=\tfrac{1}{2}\int d^3y\Big{(}-\partial_-A_2\partial_+A_2+\partial_+A_2\partial_-A_2
-\tfrac{1}{2}[\partial_1A_2,\partial_2A_1]\Big{)},
\nonumber\\
P_+&=\tfrac{1}{2}\int dy^-dy^1dy^2T^{+}_{\phantom{+}+}=\tfrac{1}{2}\int dy^-dy^1dy^2\Big{(}-4F_{-+}F_{+-}+2F_{-1}F_{+1}+2F_{-2}F_{+2}+\tfrac{1}{4}F_{\mu\nu}F^{\mu\nu}\Big{)}
\nonumber\\
&=\tfrac{1}{2}\int dy^-dy^1dy^2\Big{(}2F_{+-}^2+F_{-1}F_{+1}+F_{-2}F_{+2}-F_{+1}F_{-1}-F_{+2}F_{-2}+\tfrac{1}{2}F_{12}^2\Big{)},
\nonumber\\
&=\int dy^-dy^1dy^2\left(\tfrac{3}{2}\partial_+A_1\partial_-A_1-\tfrac{1}{2}\partial_-A_1\partial_+A_1+\tfrac{3}{2}\partial_+A_2\partial_-A_2-\tfrac{1}{2}\partial_-A_2\partial_+A_2\right).
\label{5.15}
\end{align}
As we see, $P_+$ as given in (\ref{5.15}) nicely contains $\partial_+A_1$ and   $\partial_+A_2$ just as we might want in order  to establish that $[P_+,A_1]=-i\partial_+A_1$ and $[P_+,A_2]=-i\partial_+A_2$. Unfortunately, we do not have any a priori expectation for the form of equal light-front time commutators that involve the light-front time derivatives $\partial_+A_1$ and $\partial_+A_2$ since they are not conjugate variables. Nonetheless, the theory is covariant and  thus $[P_+,A_1]=-i\partial_+A_1$ and $[P_+,A_2]=-i\partial_+A_2$ must  hold. 

To proceed we recall how we handled the $[P_+,\psi]$ commutator in the fermion case, namely we applied $\partial_+$ to the light-front time independent charge $\int dy^-dy^1dy^2\bar{\psi}\gamma^+\psi$.
Thus using integration by parts we evaluate
\begin{eqnarray}
&&\int d^3y\Big{(}\partial_-A_1\partial_+A_1-\partial_+A_1\partial_-A_1\Big{)}
=\partial_+\int d^3y\Big{(}\partial_-A_1A_1-A_1\partial_-A_1\Big{)}
-\int d^3y\Big{(}\partial_+\partial_-A_1A_1-A_1\partial_+\partial_-A_1\Big{)}
\nonumber\\
&&=-\tfrac{i}{2}\partial_+\int d^3y\delta(y^-=0)\delta(y^1=0)\delta(y^2=0)
+\int d^3y\Big{(}\partial_+A_1\partial_-A_1-\partial_-A_1\partial_+A_1\Big{)}
\nonumber\\
&&=\int d^3y\Big{(}\partial_+A_1\partial_-A_1-\partial_-A_1\partial_+A_1\Big{)},
\label{5.16}
\end{eqnarray}
and likewise for $A_2$, to thus obtain 
\begin{eqnarray}
&\int dy^-dy^1dy^2\Big{(}\partial_+A_1\partial_-A_1-\partial_-A_1\partial_+A_1\Big{)}=0,\quad 
\int dy^-dy^1dy^2\Big{(}\partial_+A_2\partial_-A_2-\partial_-A_2\partial_+A_2\Big{)}=0,
\nonumber\\
&P_+=\int dy^-dy^1dy^2\Big{(}\partial_+A_1\partial_-A_1+\partial_+A_2\partial_-A_2\Big{)}=\int dy^-dy^1dy^2\Big{(}\partial_-A_1\partial_+A_1+\partial_-A_2\partial_+A_2\Big{)}.
\label{5.17}
\end{eqnarray}
We note that (\ref{5.17}) is of interest in its own right since $P_+$ could have been subject to an ordering issue since for q-numbers $\partial_-A_1\partial_+A_1$ is not automatically equal to $\partial_+A_1\partial_-A_1$. However their integrals are equal, and there is thus no ordering issue for $P_+$.

Given (\ref{5.17}) we obtain
\begin{align}
 [P_+,A_1(x)]
 &=\int dy^-dy^1dy^2\Big{(}\partial_-A_1(y)\partial_+A_1(y)A_1(x)-A_1(x)\partial_-A_1(y)\partial_+A_1(y)\Big{)}
\nonumber\\
&=\int dy^-dy^1dy^2\Big{(}\partial_+A_1(y)\partial_-A_1(y)A_1(x)-A_1(x)\partial_-A_1(y)\partial_+A_1(y)\Big{)}\nonumber\\
&=\int dy^-dy^1dy^2\Big{(}\partial_+A_1(y)[\partial_-A_1(y),A_1(x)]+\partial_+A_1(y)A_1(x)\partial_-A_1(y)
\nonumber\\
&-[A_1(x),\partial_-A_1(y)]\partial_+A_1(y)-\partial_-A_1(y)A_1(x)\partial_+A_1(y)\Big{)}
\nonumber\\
&=-i\partial_+A_1(x)+K,
\label{5.18}
\end{align}
where we have used (\ref{5.6}) with $\sigma=1$, and where 
\begin{align}
K=\int dy^-dy^1dy^2\Big{[}\partial_+A_1(y)A_1(x)\partial_-A_1(y)  -  \partial_-A_1(y)A_1(x)\partial_+A_1(y)\Big{]}.
\label{5.19}
\end{align}
The  $[P_+,A_1(x)]=-i\partial_+A_1(x)$ relation required by covariance then follows if $K=0$. 

To determine the condition under which $K$ is zero, on using (\ref{5.16})  we rewrite (\ref{5.19}) as 
\begin{align}
K&=\int dy^-dy^1dy^2\Big{[}\partial_+A_1(y)A_1(x)\partial_-A_1(y) -\partial_-A_1(y) \partial_+A_1(y) A_1(x)
-\partial_-A_1(y)[A_1(x),\partial_+A_1(y)]
\Big{]}
\nonumber\\
&=\int dy^-dy^1dy^2\Big{[}\partial_+A_1(y)A_1(x)\partial_-A_1(y) -\partial_+A_1(y) \partial_-A_1(y) A_1(x)
-\partial_-A_1(y)[A_1(x),\partial_+A_1(y)]\Big{]}
\nonumber\\
&=\int dy^-dy^1dy^2\Big{[}\frac{i}{2}\partial_+A_1(y)\delta(x^--y^-)\delta(x^1-y^1)\delta(x^2-y^2)-\partial_-A_1(y)[A_1(x),\partial_+A_1(y)]\Big{]}.
\label{5.20}
\end{align}
The vanishing of the integrand (or the integral) in the last line in (\ref{5.20}) then enforces $K=0$, with 
the vanishing of the integrand  thus requiring that the light-front time derivatives obey the previously unanticipated
\begin{align}
&\tfrac{i}{2}\delta(x^--y^-)\delta(x^1-y^1)\delta(x^2-y^2)\partial_+A_1(y)=
\partial_-A_1(y)[A_1(x),\partial_+A_1(y)],
\nonumber\\
&\tfrac{i}{2}\delta(x^--y^-)\delta(x^1-y^1)\delta(x^2-y^2)\partial_+A_2(y)=\partial_-A_2(y)[A_2(x),\partial_+A_2(y)]
\label{5.21}
\end{align}
at equal light-front times, with the relations $[P_+,A_1]=-i\partial_+A_1$ and analogously $[P_+,A_2]=-i\partial_+A_2$ then following. 

To complete the analysis, we also need to implement commutation relations that involve $A_+$. However, we note that all of the four $T^{+}_{\phantom{+}+}$, $T^{+}_{\phantom{+}-}$, $T^{+}_{\phantom{+}1}$, and $T^{+}_{\phantom{+}2}$ can be written in a form that does not involve $A_+$ at all. While this confirms that $A_+$ is not an independent degree of freedom, we still need to establish the Poincare commutators  for it. In order to determine the $[P_{\mu},A_+]$ commutators we must proceed as we did above with the bad fermion $\psi_{(-)}$, i.e., we must introduce the inverse operator $D_2=(\partial_-)^{-2}$ and use the nonlocal constraint equation
\begin{eqnarray}
A_+&=&\frac{1}{2(\partial_-)^2}\partial_-\left(\partial_1A_1+\partial_2A_2\right),
\nonumber\\
A_+(x^+,x^1,x^2,x^-)&=&\tfrac{1}{2}\int dy^-D_2(x^--y^-)\partial_-\left(\partial_1A_1(x^+,x^1,x^2,y^-)+\partial_2A_2(x^+,x^1,x^2,y^-)\right).
\label{5.22}
\end{eqnarray}
Then since $[P_{\mu},A_1]=-i\partial_{\mu}A_1$, $[P_{\mu},A_2]=-i\partial_{\mu}A_2$, on integrating by parts for $[P_{-},A_1]$ and $[P_{-},A_2]$  we obtain
\begin{eqnarray}
[P_{\mu},A_+]=-i\partial_{\mu}A_+.
\label{5.23}
\end{eqnarray}
(The structure of (\ref{5.23}) for $A_+$ is analogous to the structure of (\ref{4.17a}) for $\psi_{(-)}$, with both  $A_+$ and $\psi_{(-)}$ obeying constraint equations.) Since we have now successfully obtained all the needed commutators, we thus establish that the field commutators given in (\ref{5.6}) are indeed the correct commutators for a gauge theory. And with $\sigma=1$ they take the form:
\begin{align}
&[A_1(x^+,x^1,x^2,x^-),\partial_-A_1(x^+,y^1,y^2,y^-)]=\tfrac{i}{2}\delta(x^--y^-)\delta(x^1-y^1)\delta(x^2-y^2),
\nonumber\\
&[A_2(x^+,x^1,x^2,x^-),\partial_-A_2(x^+,y^1,y^2,y^-)]=\tfrac{i}{2}\delta(x^--y^-)\delta(x^1-y^1)\delta(x^2-y^2),
\nonumber\\
&[A_1(x^+,x^1,x^2,x^-),A_1(x^+,y^1,y^2,y^-)]=-\tfrac{i}{4}\epsilon(x^--y^-)\delta(x^1-y^1)\delta(x^2-y^2),
\nonumber\\
&[A_2(x^+,x^1,x^2,x^-),A_2(x^+,y^1,y^2,y^-)]=-\tfrac{i}{4}\epsilon(x^--y^-)\delta(x^1-y^1)\delta(x^2-y^2),
\nonumber\\
&\partial_-A_1(x^+,y^1,y^2,y^-)[A_1(x^+,x^1,x^2,x^-),\partial_+A_1(x^+,y^1,y^2,y^-)]
\nonumber\\
&=\tfrac{i}{2}\delta(x^--y^-)\delta(x^1-y^1)\delta(x^2-y^2)\partial_+A_1(x^+,x^1,x^2,x^-),
\nonumber\\
&\partial_-A_2(x^+,y^1,y^2,y^-)[A_2(x^+,x^1,x^2,x^-),\partial_+A_2(x^+,y^1,y^2,y^-)]
\nonumber\\
&=\tfrac{i}{2}\delta(x^--y^-)\delta(x^1-y^1)\delta(x^2-y^2)\partial_+A_2(x^+,x^1,x^2,x^-).
\label{5.24}
\end{align}
In (\ref{5.24}) we have included commutation relations that involve $\partial_+A_1$ and $\partial_+A_2$, relations that had not previously been reported in the literature.

Having obtained the light-front commutation relations that appear in (\ref{5.24}), it is instructive to compare them with the analogous relations that one obtains in instant-time quantization. We discuss the field $A_1$, and analogously to the light-front $A^+=0$ gauge we consider the instant-time $A^0=0$ gauge. The instant-time conjugate to $A_1$ is given by $\Pi^1=\partial L/\partial (\partial_0A_1)=-\partial^0A^1=\partial_0A_1$. In the $A_0=0$ gauge the relevant part of the instant-time $T^{(0)}_{\phantom{0}1}$ for the $[P_1,A_1]$ commutator is given by $-\partial_0 A_2\partial_2A_1-\partial_0 A_3\partial_3A_1$. Thus in $P_1=\int d^3yT^{(0)}_{\phantom{0}1}$ the relevant part is $\int d^3y(\partial_0\partial_2A_2+\partial_0\partial_3 A_3)A_1$. Now the Maxwell equations are of the form $\partial_{\nu}F^{\mu\nu}=\partial_{\nu}(\partial^{\mu}A^{\nu}-\partial^{\nu}A^{\mu})=0$. Thus for $\mu=0$, $A_0=0$  we obtain $\partial_0(\partial_1 A_1+\partial_2A_2+\partial_3 A_3)=0$, with the relevant part of $P_1$ then being $-\int d^3y\partial_0\partial_1A_1A_1=\int d^3y\partial_0A_1\partial_1A_1$. With equal instant-time commutator
\begin{align}
&[A_1(x^0,x^1,x^2,x^3),\partial_0A_1(x^0,y^1,y^2,y^3)]=i\delta(x^1-y^1)\delta(x^2-y^2)\delta(x^3-y^3),
\label{5.25}
\end{align}
we then obtain $[P_1,A_1]=-i\partial_1A_1$. 

Now we can rewrite the first commutator in (\ref{5.24}) as 
\begin{eqnarray}
&[A_1(x^+,x^1,x^2,x^-),\partial^+A_1(x^+,y^1,y^2,y^-)]=i\delta(x^--y^-)\delta(x^1-y^1)\delta(x^2-y^2),
\nonumber\\
&[A_1(x^+,x^1,x^2,x^-),\partial^+A^1(x^+,y^1,y^2,y^-)]=-i\delta(x^--y^-)\delta(x^1-y^1)\delta(x^2-y^2),
\label{5.26}
\end{eqnarray}
and can rewrite (\ref{5.25}) as 
\begin{eqnarray}
&[A_1(x^0,x^1,x^2,x^3),\partial^0A_1(x^0,y^1,y^2,y^3)]=i\delta(x^1-y^1)\delta(x^2-y^2)\delta(x^3-y^3),
\nonumber\\
&[A_1(x^0,x^1,x^2,x^3),\partial^0A^1(x^0,y^1,y^2,y^3)]=-i\delta(x^1-y^1)\delta(x^2-y^2)\delta(x^3-y^3).
\label{5.27}
\end{eqnarray}
As we see, these latter two relations are analogous. And with $g_{11}=-1$ in both the light-front and instant-time cases they generalize to 
\begin{eqnarray}
&&[A_{\mu}(x^+,x^1,x^2,x^-),\partial^+A_{\nu}(x^+,y^1,y^2,y^-)]=-ig_{\mu\nu}\delta(x^--y^-)\delta(x^1-y^1)\delta(x^2-y^2),
\nonumber\\
&&[A_{\mu}(x^0,x^1,x^2,x^3),\partial^0A_{\nu}(x^0,y^1,y^2,y^3)]=-ig_{\mu\nu}\delta(x^1-y^1)\delta(x^2-y^2)\delta(x^3-y^3).
\label{5.28}
\end{eqnarray}
Because of the non-covariance of the $A^0=0$ and $A^+=0$ gauge choices these relations do not hold for all $\mu,\nu$. In the instant-time case they hold for $\mu,\nu=1,2,3$, and for the light-front case they hold for $\mu,\nu=1,2$. As we discuss below, these relations actually can be made to hold for all $\mu,\nu$ if instead of making a gauge choice one adds on a gauge fixing term $-(\partial_{\mu}A^{\mu})^2/2$ to the Maxwell Lagrangian density. 

With (\ref{5.28}) being written in covariant notation we can now raise the $\nu$ index, and recalling that $\Pi^1= -\partial^+A^1$, $\Pi^1=-\partial^0A^1$ in the respective light-front and instant-time cases  we obtain
\begin{eqnarray}
&&[A_{\mu}(x^+,x^1,x^2,x^-),\Pi^{\nu}(x^+,y^1,y^2,y^-)]=i\delta_{\mu}^{\nu}\delta(x^--y^-)\delta(x^1-y^1)\delta(x^2-y^2),
\nonumber\\
&&[A_{\mu}(x^0,x^1,x^2,x^3),\Pi^{\nu}(x^0,y^1,y^2,y^3)]=i\delta_{\mu}^{\nu}\delta(x^1-y^1)\delta(x^2-y^2)\delta(x^3-y^3).
\label{5.29}
\end{eqnarray}

\subsection{Quantization of Abelian Gauge Fields via Gauge Fixing}
\label{S5b}

As well as quantizing in the $A^+=0$ gauge we can instead use gauge fixing by taking the Abelian gauge field action $I_G$ to be of the form 
\begin{eqnarray}
I_G=\int d^4x\left[ -\tfrac{1}{4}F_{\mu\nu}F^{\mu\nu}-\tfrac{1}{2}(\partial_{\mu}A^{\mu})^2\right]=\int d^4x\left[ -\tfrac{1}{2}\partial_{\nu}A_{\mu}\partial^{\nu}A^{\mu}\right],
\label{5.30}
\end{eqnarray}
where $A_{\mu}$ is an Abelian gauge field, where $F_{\mu\nu}=\partial_{\nu}A_{\mu}-\partial_{\mu}A_{\nu}$,  and where we have dropped surface terms in establishing the second form for $I_G$ given in (\ref{5.30}). The presence in the first form for $I_G$ of the $-\chi^2/2$ term where $\chi=\partial_{\mu}A^{\mu}$ causes $I_G$ to be neither gauge invariant  nor equal to the gauge invariant  Maxwell action $I_M=-\tfrac{1}{4}\int d^4xF_{\mu\nu}F^{\mu\nu}$. 

Variation of the $I_G$ action with respect to $A_{\mu}$ yields an equation of motion of the form
\begin{eqnarray}
\partial_{\nu}\partial^{\nu}A_{\mu}=0.
\label{5.31}
\end{eqnarray}
The utility of using (\ref{5.30}) is that the various components of $A_{\mu}$ are decoupled from each other in the equation of motion.  Consequently, we can treat each component of $A_{\mu}$ as an independent degree of freedom, and apply the scalar field analysis given above to each one of them. In this formulation (\ref{5.31}) entails that $\partial_{\nu}\partial^{\nu}\chi=0$. If one imposes the subsidiary conditions $\chi(x^0=0)=0$, $\partial_0\chi(x^0=0)=0$ at the initial time $x^0=0$, then since $\partial_{\nu}\partial^{\nu}\chi=0$ is a second-order derivative equation it follows that the non-gauge-invariant $\chi$ is zero at all times, and in the solution to the equations of motion gauge invariance is then obtained.

To obtain the energy-momentum tensor we covariantize the action  and construct $T_{\mu\nu}=2(-g)^{-1/2}\delta I_G/\delta g^{\mu\nu}$, viz.
\begin{eqnarray}
T_{\mu\nu}=-\partial_{\mu}A^{\alpha}\partial_{\nu}A_{\alpha}+\tfrac{1}{2}g_{\mu\nu}\partial_{\beta}A^{\alpha}\partial^{\beta}A_{\alpha},
\label{5.32}
\end{eqnarray}
and using (\ref{5.31}) can readily check that it obeys $\partial_{\nu}T^{\mu\nu}=0$.
Given (\ref{5.32}) we evaluate 
\begin{eqnarray}
T^{+}_{\phantom{+}1}&=&-2g^{\alpha\sigma}\partial_{-}A_{\sigma}\partial_1A_{\alpha},\quad T^{+}_{\phantom{+}2}=-2g^{\alpha\sigma}\partial_{-}A_{\sigma}\partial_2A_{\alpha},\quad T^{+}_{\phantom{+}-}=-2g^{\alpha\sigma}\partial_{-}A_{\sigma}\partial_-A_{\alpha},
\nonumber\\
 T^{+}_{\phantom{+}+}&=&-2g^{\alpha\sigma}\partial_{-}A_{\sigma}\partial_+A_{\alpha}
+\tfrac{1}{2}\partial_{\beta}A^{\alpha}\partial^{\beta}A_{\alpha}.
\label{5.33}
\end{eqnarray}
With the equation of motion given in (\ref{5.31}) and $T^{\mu\nu}$ as given in (\ref{5.32}) both being diagonal in the $A_{\mu}$ indices, the discussion follows the scalar field case for each component of $A_{\mu}$. Thus with $P_{\mu}=\tfrac{1}{2}\int dy^-dy^1dy^2 T^{+}_{\phantom{+}\mu}$, we enforce $[P_{\mu},A_{\nu}]=-i\partial_{\mu}A_{\nu}$ if we set
\begin{align}
&[A_{\nu}(x^+,x^1,x^2,x^-),\partial_-A_{\mu}(x^+,y^1,y^2,y^-)]=-\frac{i}{2}g_{\mu\nu}\delta(x^1-y^1)\delta(x^2-y^2)\delta(x^--y^-),
\nonumber\\
&[A_{\nu}(x^+,x^1,x^2,x^-),\partial^+A_{\mu}(x^+,y^1,y^2,y^-)]=-ig_{\mu\nu}\delta(x^1-y^1)\delta(x^2-y^2)\delta(x^--y^-),
\nonumber\\
&[A_{\nu}(x^+,x^1,x^2,x^-),\Pi^{\mu}(x^+,y^1,y^2,y^-)]=i\delta^{\mu}_{\nu}\delta(x^1-y^1)\delta(x^2-y^2)\delta(x^--y^-),
\nonumber\\
&[A_{\nu}(x^+,x^1,x^2,x^-), A_{\mu}(x^+,y^1,y^2,y^-)]=\frac{i}{4}g_{\mu\nu}\epsilon(x^--y^-)\delta(x^1-y^1)\delta(x^2-y^2),
\label{5.34}
\end{align}
where $\Pi^{\mu}=-\partial^+A^{\mu}$, to thus define the light-front canonical commutators in  the Abelian case. As constructed, the second relation in (\ref{5.34}) is the light-front analog of the instant-time commutator $[A_{\nu}(x^+,x^1,x^2,x^-),\partial^0A_{\mu}(x^+,y^1,y^2,y^-)]=-ig_{\mu\nu}\delta(x^1-y^1)\delta(x^2-y^2)\delta(x^3-y^3)$ associated with the instant-time quantization of (\ref{5.30}) (see e.g. \cite{Itzykson1980}).

\subsection{Quantization of non-Abelian Gauge Fields via Gauge Fixing}
\label{S5c}

In the non-Abelian Yang-Mills case one has a non-Abelian group with structure coefficients $f_{abc}$. One defines a tensor $G^a_{\mu\nu}=\partial_{\nu}A^a_{\mu}-\partial_{\mu}A^a_{\nu}+gf^{abc}A^b_{\nu}A^c_{\mu}$ where $g$ is the coupling constant. In analog to (\ref{5.30}) one defines an action (see e.g. \cite{Donoghue1992}) 
\begin{eqnarray}
I_{YM}=\int d^4x\left[ -\tfrac{1}{4}G^a_{\mu\nu}G_a^{\mu\nu}-\tfrac{1}{2}\partial_{\mu}A_a^{\mu}\partial_{\nu}A_a^{\nu}
+\partial_{\mu}\bar{c}_a\partial^{\mu}c_a+gf^{abe}A^{\mu}_a\partial_{\mu}\bar{c}_bc_e\right],
\label{5.35}
\end{eqnarray}
where  the $c_a$ and $\bar{c}_a$ are two independent Faddeev-Popov ghost fields that one has to introduce in the non-Abelian case, viz. spin zero Grassmann fields that are to be quantized with  anticommutation relations. To ensure that the action is Hermitian $c_a$ is taken to be Hermitian and $\bar{c}_a$ is taken to be anti-Hermitian. Since the $g$-dependent terms in $I_{YM}$ involve products of either three or four fields they can be treated as part of the interaction. The relevant part of $I_{YM}$ for quantization, viz. the free part,  is, following an integration by parts,  thus of the form 
\begin{eqnarray}
I_{YM}=\int d^4x\left[ -\tfrac{1}{2}\partial_{\nu}A^a_{\mu}\partial^{\nu}A_a^{\mu}+\partial_{\mu}\bar{c}_a\partial^{\mu}c_a\right],
\label{5.36}
\end{eqnarray}
and leads to  equations of motion of the form
\begin{eqnarray}
\partial_{\nu}\partial^{\nu}A_{\mu}^a=0,\quad \partial_{\mu}\partial^{\mu}c_a=0,\quad \partial_{\mu}\partial^{\mu}\bar{c}_a=0.
\label{5.37}
\end{eqnarray}
With both (\ref{5.36}) and (\ref{5.37}) being diagonal in both spacetime and group indices, the discussion thus parallels the Abelian and scalar field cases, with $A_{\mu}^a$ acting the same way as the Abelian $A_{\mu}$, and $c_a$ and $\bar{c}_a$ acting the same way as $\phi$ (so that the canonical conjugate of $c_a$ is $\partial^+\bar{c}_a=2\partial_-\bar{c}_a$, and the canonical conjugate of $\bar{c}_a$ is $2\partial_-c_a$). Thus we obtain

\begin{align}
&[A^a_{\nu}(x^+,x^1,x^2,x^-),\partial_-A^b_{\mu}(x^+,y^1,y^2,y^-)]=-\frac{i}{2}g_{\mu\nu}\delta^{ab}\delta(x^1-y^1)\delta(x^2-y^2)\delta(x^--y^-),
\nonumber\\
&[A^a_{\nu}(x^+,x^1,x^2,x^-), A^b_{\mu}(x^+,y^1,y^2,y^-)]=\frac{i}{4}g_{\mu\nu}\delta^{ab}\epsilon(x^--y^-)\delta(x^1-y^1)\delta(x^2-y^2),
\label{5.38}
\end{align}
\begin{align}
&\Big{\{}c_a(x^+,x^1,x^2,x^-), 2\partial_-\bar{c}_b(x^+,y^1,y^2,y^-)\Big{\}}=i\delta_{ab}\delta(x^1-y^1)\delta(x^2-y^2)\delta(x^--y^-),
\nonumber\\
&\Big{\{}\bar{c}_a(x^+,x^1,x^2,x^-), 2\partial_-c_b(x^+,y^1,y^2,y^-)\Big{\}}=i\delta_{ab}\delta(x^1-y^1)\delta(x^2-y^2)\delta(x^--y^-),
\label{5.39}
\end{align}
to thus define the light-front canonical commutators in  the non-Abelian case.

\section{Gravitational Field}
\label{S6}

For gravity we can establish a  similar pattern. For fluctuations around a flat $\eta_{\mu\nu}$ background with fluctuation metric $\eta_{\mu\nu}+h_{\mu\nu}$, the source-free Einstein equations take the form \cite{Weinberg1972} $\Box h_{\mu\nu}=0$ in the harmonic gauge where $\partial_{\mu}h^{\mu}_{\phantom{\mu}\nu}-\partial_{\nu}[\eta_{\alpha\beta}h^{\alpha\beta}]/2=0$. Similarly, in the fourth-order derivative conformal gravity theory  the fluctuations around flat take the form \cite{Mannheim2006} $\Box^2K_{\mu\nu}=0$ in the transverse gauge $\partial_{\mu}K^{\mu\nu}=0$ where $K_{\mu\nu}=h_{\mu\nu}-\eta_{\mu\nu}\eta_{\alpha\beta}h^{\alpha\beta}/4$ is traceless. Since in both the cases the fluctuation equation is diagonal in the $(\mu,\nu)$ indices, we can treat each component independently. For each component then light-front quantization follows the scalar field  $\Box \phi=0$ case with its characteristic $\epsilon(x^--y^-)$ structure. This is in analog to the use of gauge fixing in the Abelian and non-Abelian gauge theory cases since for both of those cases the wave equations are also diagonal in the indices.

\section{General Assessment of the Commutator and Anticommutator Structure}
\label{S7}

\subsection{General Comments}
\label{S7a}

In trying to assess the commutator and anticommutator structure that we have found, we note that in theories where the canonical conjugate of a field behaves like $\partial_-=\partial/\partial x^-$ acting on the field rather than $\partial_+=\partial/\partial x^+$, the equal light-front time commutation relations can be integrated with respect to $x^-$, to thus lead to a dependence on $\epsilon(x^--y^-)$, a behavior found in both (\ref{3.10}) for scalars and in (\ref{5.24}), (\ref{5.34}) and (\ref{5.38}) for vectors. This occurs because the Lagrangian is quadratic in derivatives. However, this does not occur for half-integral spinors since for them the Lagrangian is only first order in derivatives. The need to involve $\epsilon(x^--y^-)$ will thus be met in any integer spin theory, including electromagnetism, non-Abelian Yang-Mills theories, and gravity. Finally, we note that while the half-integer spin theory only needs standard delta function commutators as exhibited in (\ref{4.13a}), we see for spinors and for vectors in the $A^+=0$ gauge that in light-front quantization there will be constraints and only some of the components of  the fields will be dynamical. Moreover, we recall that in order to show that the momenta $P_{\mu}$ are independent of the light-front time (viz. $\partial_+P_{\mu}=0$), we need to show that the spatial surface terms generated in the $\int dx^-dx^1dx^2$ integrations on both sides of
\begin{eqnarray}
\tfrac{1}{2}\partial_+\int dx^-dx^1dx^2T^{+}_{\phantom{+}\mu}=-\tfrac{1}{2}\int dx^-dx^1dx^2\left(\partial_-T^{-}_{\phantom{-}\mu}+\partial_1T^{1}_{\phantom{1}\mu}
+\partial_2T^{2}_{\phantom{2}\mu}\right)
\label{7.1}
\end{eqnarray}
are convergent at infinity. However, since in the gauge field case $A_+$ appears on the right-hand side of (\ref{7.1}), and since in the $A^+=0$ gauge $A_+$  obeys the nonlocal integral constraint given in (\ref{5.22}) (whose use we took advantage of in integrating by parts in order to eliminate it on the left-hand side of (\ref{7.1})), the light-front gauge fields need to be more convergent at infinity than the instant-time case. (An analogous discussion of this issue from a slightly different perspective may be found in \cite{Schlieder1972}.) A similar situation arises for fermions, with the bad $\psi_{(-)}$ fermion contributing to the surface terms associated with the fermionic version of (\ref{7.1}).

\subsection{Transforming from Instant-time Coordinates to Light-front Coordinates}
\label{S7b}

It is of interest to ask what happens to instant-time quantities when we transform to light-front coordinates. To transform the coordinates it is immediately suggested to consider a Lorentz boost.  
Under a boost with velocity $u$ in the 3-direction one has
\begin{eqnarray}
x^{0\prime}=\frac{x^0+ux^3}{(1-u^2)^{1/2}},\quad x^{3\prime}=\frac{x^3+ux^0}{(1-u^2)^{1/2}},
\label{7.2}
\end{eqnarray}
and thus 
\begin{eqnarray}
x^{+\prime}=x^+\frac{(1+u)^{1/2}}{(1-u)^{1/2}},\quad x^{-\prime}=x^-\frac{(1-u)^{1/2}}{(1+u)^{1/2}},\quad x^{+\prime}x^{-\prime}=x^+x^-.
\label{7.3}
\end{eqnarray}
However, (\ref{7.3}) shows that the light-front variables $x^+$ and $x^-$ are invariant under the boost, and thus the instant-time $x^0$ and $x^3$ cannot be boosted into them. Moreover, even if we approach $u=1$ (the infinite momentum frame)  in (\ref{7.2}) we still do not achieve the desired transformation. Specifically, on setting $u=1-\epsilon^2/2$ with $\epsilon$ small we obtain 
\begin{eqnarray}
x^{0\prime}=\frac{x^+}{\epsilon},\quad x^{3\prime}=\frac{x^+}{\epsilon},\quad x^{+\prime}=\frac{2x^+}{\epsilon},\quad x^{-\prime}=\frac{x^-\epsilon}{2},
\label{7.4}
\end{eqnarray}
and thus while $x^0$ does now transform into $x^+$, so does $x^3$, with it not transforming into $x^-$.

In order to find a transform that does convert instant-time coordinates into light-front coordinates, we note that $x^0\rightarrow x^0+x^3$ and $x^3\rightarrow x^3-x^0$ are actually spacetime-dependent translations, not Lorentz boosts. In classical mechanics one makes these transformations on the coordinates themselves. In quantum mechanics  the translation generators $P_{\mu}$ act on the quantum fields and not on c-numbers and implement $[P_{\mu},\phi]=-i\partial_{\mu}\phi$. Since the momentum generators mutually commute ($[P_{\mu},P_{\nu}]=0$), in the quantum case one can introduce a unitary operator 
\begin{eqnarray}
U(P_0,P_3)=\exp(ix^3P_0)\exp(ix^0P_3)
\label{7.5}
\end{eqnarray}
that effects
\begin{eqnarray}
U\phi(x^0,x^1,x^2,-x^3)U^{-1}= \phi(x^0+x^3,x^1,x^2,x^0-x^3).
\label{7.6}
\end{eqnarray}
To determine how $U(P_0,P_3)$ acts on derivatives of $\phi$ such as $\partial_0\phi$ we write $U\partial_0\phi U^{-1}$ as the limit $\epsilon \rightarrow 0$ of $U[\phi(x^0+\epsilon)-\phi(x^0)]U^{-1}/\epsilon=[\phi(x^0+x^3+\epsilon)-\phi(x^0+x^3)]/\epsilon=\partial_+\phi(x^+)$, and thus obtain
\begin{eqnarray}
U\partial_0\phi(x^0,x^1,x^2,-x^3)U^{-1}= \partial_+\phi(x^0+x^3,x^1,x^2,x^0-x^3).
\label{7.7}
\end{eqnarray}

Given (\ref{7.6}) and (\ref{7.7}) we apply $U(P_0,P_3)$ to the instant-time equal-time commutator, and with the quantum-mechanical $U(P_0,P_3)$ not affecting c-numbers such as delta functions of the coordinates, we obtain
\begin{eqnarray}
&&U[\phi(x^0,x^1,x^2,-x^3),\partial_0\phi(x^0,y^1,y^2,-y^3)]U^{-1}=Ui\delta(x^3-y^3)\delta(x^1-y^1)\delta(x^2-y^2)U^{-1}
\nonumber\\
&&=[\phi(x^0+x^3,x^1,x^2,x^0-x^3),\partial_+(x^0+x^3,y^1,y^2,x^0-y^3)]=i\delta(x^3-y^3)\delta(x^1-y^1)\delta(x^2-y^2).
\label{7.8}
\end{eqnarray}
However, since $y^0=x^0$ at equal instant times,  we can set $x^3-y^3=x^3-x^0+y^0-y^3=y^--x^-$ (this being just a change of classical variables and not a quantum transformation since $U(P_0,P_3)$ does not affect c-numbers). 
Thus we obtain
\begin{eqnarray}
&&[\phi(x^+,x^1,x^2,x^-),\partial_+\phi(x^+,y^1,y^2,y^-)]
=\delta(x^--y^-)\delta(x^1-y^1)\delta(x^2-y^2).
\label{7.9}
\end{eqnarray}
As such, (\ref{7.9}) represents the equal instant-time commutator as transformed to light-front coordinates. However, it is not the equal light-front time commutator that we need for light-front quantization, viz. the one given in (\ref{3.5}) since that one involves $\partial_-\phi$ and not $\partial_+\phi$. Thus (\ref{7.9}) is not a light-front commutator, and instant-time commutators therefore do not transform into light-front commutators (which of course could not be the case since instant-time $[\phi,\phi]$ type commutators are zero while the light-front ones are not). While in general commutation relations transform into some other commutation relations under a unitary transformation, it is the canonical conjugates that do not transform into each other, with equal instant-time quantization relations and equal light-front time quantization relations not being unitarily equivalent. 

Nonetheless, despite this we will show in Sec. \ref{S8} that  instant-time and light-front commutators at unequal times do transform into each other. Specifically, the instant-time matrix element $-i\langle \Omega_I|[\phi(x^0,x^1,x^2,-x^3),\phi(0)]|\Omega_I\rangle$ in the instant-time vacuum $|\Omega_I\rangle$ will transform into 
$-i\langle \Omega_I|U^{\dagger}U[\phi(x^0,x^1,x^2,-x^3),\phi(0)]U^{\dagger}U|\Omega_I\rangle$, i.e. into
$-i\langle \Omega_I|U^{\dagger}[\phi(x^+,x^1,x^2,x^-),\phi(0)]U|\Omega_I\rangle$. Then with the light-front vacuum $|\Omega_F\rangle$ being given by $|\Omega_F\rangle=U|\Omega_I\rangle$ (see (\ref{8.94}) below) we see that the instant time $-i\langle \Omega_I|[\phi(x^0,x^1,x^2,-x^3),\phi(0)]|\Omega_I\rangle$ transforms into the light-front  $-i\langle \Omega_F|[\phi(x^+,x^1,x^2,x^-),\phi(0)]|\Omega_F\rangle$, so that
\begin{eqnarray}
-i\langle \Omega_I|[\phi(x^0,x^1,x^2,-x^3),\phi(0)]|\Omega_I\rangle=-i\langle \Omega_F|[\phi(x^+,x^1,x^2,x^-),\phi(0)]|\Omega_F\rangle,
\label{7.10}
\end{eqnarray}
a relation that we show below to hold to all orders in perturbation theory.

While we cannot transform equal instant-time commutation (and likewise anticommutation) relations into equal light-front time commutation relations, neither at the q-number operator level or  for c-number vacuum matrix elements, nonetheless in Sec. \ref{S8} we show that for the free theory, the starting point for perturbation theory, unequal q-number instant-time commutators of operators (i.e., before we take vacuum matrix elements) actually can be transformed into unequal q-number light-front time commutation relations. Moreover, as we show starting in Sec. \ref{S9}, the two quantization procedures even lead to the same values for Green's functions. In consequence,  the two canonical quantization procedures are equivalent, and one is thus free to start with either.

\section{Transforming from Unequal-time Instant-time Commutators to Equal Light-front Time Commutators And Vice-versa}
\label{S8}
\subsection{Scalar Field Case -- Instant Time to Light Front}
\label{S8a}
In instant-time quantization for a free scalar field one starts with an action
\begin{eqnarray}
I_S&=&\int dx^0dx^1dx^2dx^3\left[\tfrac{1}{2}\partial_{\mu}\phi\partial^{\mu}\phi-\tfrac{1}{2}m^2\phi^2\right]
\nonumber\\
&=&\int dx^0dx^1dx^2dx^3\left[\tfrac{1}{2}\left((\partial_0\phi)^2-(\partial_1\phi)^2-(\partial_2\phi)^2-(\partial_3\phi)^2\right)
-\tfrac{1}{2}m^2\phi^2\right],
\label{8.1}
\end{eqnarray}
and one  identifies a canonical conjugate $\delta I_S/\delta \partial_0\phi=\partial^0\phi=\partial_0\phi$ (one can of course add on interaction terms to $I_S$, but as long as they contain no derivatives they do not affect the identification of the canonical conjugate), and then quantizes the theory according to the equal-time canonical commutation relation
\begin{eqnarray}
&[\phi(x^0,x^1,x^2,x^3), \partial_0\phi(x^0,y^1,y^2,y^3)]=i\delta(x^1-y^1)\delta(x^2-y^2)\delta(x^3-y^3),
\nonumber\\
&[\phi(x^0,x^1,x^2,x^3), \phi(x^0,y^1,y^2,y^3)]=0.
\label{8.2}
\end{eqnarray}
In instant-time quantization one can use the equal-time commutation relation given in (\ref{8.2}) and the wave equation $(\partial_{\mu}\partial^{\mu}+m^2)\phi=0$ associated with $I_S$ to make an on-shell Fock space expansion of $\phi(x)$ of the form
\begin{eqnarray}
\phi(x^0,\vec{x})=\int \frac{d^3p}{(2\pi)^{3/2}(2p)^{1/2}}[a(\vec{p})e^{-iE_p x^0+i\vec{p}\cdot\vec{x}}+a^{\dagger}(\vec{p})e^{+iE_p x^0-i\vec{p}\cdot\vec{x}}],
\label{8.3}
\end{eqnarray}
where $E_p=(\vec{p}^2+m^2)^{1/2}$, and where the normalization of the creation and annihilation operator algebra, viz.  
\begin{eqnarray}
[a(\vec{p}),a^{\dagger}(\vec{q})]=\delta^3(\vec{p}-\vec{q}),
\label{8.4}
\end{eqnarray}
is fixed from the normalization of the canonical commutator given in (\ref{8.2}). Given (\ref{8.3}) and (\ref{8.4}) one can evaluate the unequal-time instant-time ($IT$) commutator between two free scalar fields, to obtain
\begin{eqnarray}
i\Delta(IT;x-y)&=&[\phi(x^0,x^1,x^2,x^3), \phi(y^0,y^1,y^2,y^3)]
\nonumber\\
&=&
\int \frac{d^3pd^3q}{(2\pi)^3(2E_p)^{1/2}(2E_q)^{1/2}}\Big{(}[a(\vec{p}),a^{\dagger}(\vec{q})]e^{-ip\cdot x+iq\cdot y}
+[a^{\dagger}(\vec{p}),a(\vec{q})]e^{ip\cdot x-iq\cdot y}\Big{)}
\nonumber\\
&=&\int_{-\infty}^{\infty}dp_1\int_{-\infty}^{\infty}dp_2\int_{-\infty}^{\infty}dp_3\frac{1}{(2\pi)^3 2E_p}\left(e^{-iE_p(x^0-y^0)+i\vec{p}\cdot(\vec{x}-\vec{y})}-e^{iE_p(x^0-y^0)-i\vec{p}\cdot(\vec{x}-\vec{y})}\right).
\label{8.5}
\end{eqnarray}
We note that this unequal-time commutator is a c-number, and not a q-number. We can also check the validity of (\ref{8.5}) by noting that for $y^{\mu}\neq x^{\mu}$ $i\Delta(IT;x-y)$ obeys the second-order derivative equation $[\partial_{\mu}\partial^{\mu}+m^2]i\Delta(IT;x-y)=0$  since $\phi(x)$ obeys $[\partial_{\mu}\partial^{\mu}+m^2]\phi(x)=0$. At $x^0=y^0$ $i\Delta(IT;x-y)$ vanishes just as required by $[\phi(x^0,x^1,x^2,x^3), \phi(x^0,y^1,y^2,y^3)]=0$, while its $\partial/\partial y_0$ derivative obeys $[\phi(x^0,x^1,x^2,x^3), \partial_0\phi(x^0,y^1,y^2,y^3)]=i\delta(x^1-y^1)\delta(x^2-y^2)\delta(x^3-y^3)$ at $x^0=y^0$. Since the solution to any equation that is a second-order derivative in $x^0$ is uniquely specified once one has fixed the function and its first derivative at the initial time, the expression for $i\Delta(IT;x-y)$ as given in (\ref{8.5}) holds at all times, just as it should.    

Despite its somewhat benign appearance (\ref{8.5}) is actually a highly singular quantity. For $(x-y)^2>0$ (\ref{8.5}) evaluates to (see  \cite{Schwinger1949}, and also \cite{Schweber1964})
\begin{eqnarray}
i\Delta(IT; (x-y)^2>0)=\frac{im}{4\pi}\epsilon(x^0-y^0)\frac{J_1(m[(x-y)^2]^{1/2})}{[(x-y)^2]^{1/2}}.
\label{8.6}
\end{eqnarray}
For $(x-y)^2=0$ (\ref{8.5}) evaluates to
\begin{eqnarray}
i\Delta(IT;(x-y)^2=0)&=&-\frac{i}{2\pi}\epsilon(x^0-y^0)\delta[(x-y)^2].
\label{8.7}
\end{eqnarray}
For $(x-y)^2<0$, $i\Delta(IT;x-y)$ obeys
\begin{eqnarray}
i\Delta(IT;(x-y)^2<0)&=&0,
\label{8.8a}
\end{eqnarray}
to thus vanish for spacelike separated points, just as it should since by microcausality the commutator does not take support outside the light cone. Beyond having $\epsilon(x^0-y^0)$ and $\delta[(x-y)^2]$ singularities, $\Delta(IT;x-y)$ also has a discontinuity at $(x-y)^2=0$, with $\Delta(IT;x-y)$  becoming mass independent at $(x-y)^2=0$ even though the scalar field obeys a massive wave equation. (The mass independence of $\Delta(IT;(x-y)^2=0)$ is just as we would want since the equal instant-time commutator given in (\ref{8.2}) can be obtained from $\Delta(IT;x-y)$, and (\ref{8.2}) is both mass independent  and only takes support on the  light cone).  

In order to see how these various singularities come about we have found it convenient to go off shell and rewrite $\Delta(IT;x-y)$ as a contour integral in a complex $p_0$ plane. We thus set
\begin{align}
i\Delta(IT;x-y)&=-\frac{1}{2\pi i}\frac{1}{8\pi^3}\int_{-\infty }^{\infty}dp_1\int_{-\infty }^{\infty}dp_2\int_{-\infty}^{\infty}dp_3\oint dp_0
\nonumber\\
&\times\bigg{[}
\frac{\theta(x^0-y^0)e^{-ip\cdot (x-y)}-\theta(-x^0+y^0)e^{ip\cdot (x-y)}}{(p_0)^2-(p_3)^2-(p_1)^2-(p_2)^2-m^2+i\epsilon}
+\frac{\theta(x^0-y^0)e^{ip\cdot (x-y)}-\theta(-x^0+y^0)e^{-ip\cdot (x-y)}}{(p_0)^2-(p_3)^2-(p_1)^2-(p_2)^2-m^2-i\epsilon}\bigg{]},
\nonumber\\
&=-\frac{1}{2\pi i}\frac{1}{8\pi^3}\int_{-\infty }^{\infty}dp_1\int_{-\infty }^{\infty}dp_2\int_{-\infty}^{\infty}dp_3\oint dp_0\epsilon(x^0-y^0)
\nonumber\\
&\times\bigg{[}
\frac{e^{-ip\cdot (x-y)}}{(p_0)^2-(p_3)^2-(p_1)^2-(p_2)^2-m^2+i\epsilon}
+\frac{e^{ip\cdot (x-y)}}{(p_0)^2-(p_3)^2-(p_1)^2-(p_2)^2-m^2-i\epsilon}\bigg{]}.
\label{8.9a}
\end{align}
In (\ref{8.9a}) the $+i\epsilon$ term is closed in the lower-half of the complex $p_0$ plane and the $-i\epsilon$ term is closed in the upper-half plane, so that the enclosed poles recover (\ref{8.5}). With the $\theta(x^0-y^0)$ and $\theta(-x^0+y^0)$ factors that have been introduced appearing where they do all circle at infinity contributions are suppressed, with the last line in (\ref{8.9a}) following via a $p_{\mu}\rightarrow -p_{\mu}$ substitution in the $e^{ip\dot(x-y)}$ terms.  However, with there being no circle at infinity suppression if $x^0-y^0=0$, this contour integral representation of $i\Delta(IT;x-y)$ does not hold at $x^0=y^0$, a point we return to below.

On rewriting the denominators in (\ref{8.9a}) in the $\alpha$ regulator form  $\int d\alpha e^{i\alpha(A+i\epsilon)}=-1/i(A+i\epsilon)$ and $\int d\alpha e^{-i\alpha(A-i\epsilon)}=1/i(A-i\epsilon)$ where the $i\epsilon$ factors suppress the $\alpha=\infty$ contributions in both cases, $i\Delta(IT;x-y)$ can now be evaluated as an ordinary integral on the real $p_0$ axis, and with $\int _{-\infty}^{\infty} e^{ia(x-b)^2}dx=(i\pi/a)^{1/2}$ when $b$ is real, $i\Delta(IT;x-y)$ is found to be of the form
\begin{align}
i\Delta(IT;x-y)&=-\frac{1}{2\pi i}\frac{1}{8\pi^3}\int_{-\infty }^{\infty}dp_1\int_{-\infty }^{\infty}dp_2\int_{-\infty}^{\infty}dp_3\int_{-\infty }^{\infty}dp_0\int_0^{\infty}d\alpha\epsilon(x^0-y^0)
\nonumber\\
&\times\bigg{[}
-ie^{-ip\cdot (x-y)}e^{i\alpha[(p_0)^2-(p_3)^2-(p_1)^2-(p_2)^2-m^2+i\epsilon]}
+ie^{ip\cdot (x-y)}e^{-i\alpha[(p_0)^2-(p_3)^2-(p_1)^2-(p_2)^2-m^2-i\epsilon]}\bigg{]}
\nonumber\\
&=-\frac{1}{2\pi i}\frac{1}{8\pi^3}\int_{-\infty }^{\infty}dp_1\int_{-\infty }^{\infty}dp_2\int_{-\infty}^{\infty}dp_3\int_{-\infty }^{\infty}dp_0\int_0^{\infty}d\alpha\epsilon(x^0-y^0)
\nonumber\\
&\times\bigg{[}
-ie^{i\alpha[(p_0-(x^0-y^0)/2\alpha)^2-(p_3-(x^3-y^3)/2\alpha)^2-(p_1-(x^1-y^1)/2\alpha)^2-(p_2-(x^2-y^2)/2\alpha)^2-m^2+i\epsilon]}e^{-i(x-y)^2/4\alpha}
\nonumber\\
&+ie^{-i\alpha[(p_0-(x^0-y^0)/2\alpha)^2-(p_3-(x^3-y^3)/2\alpha)^2-(p_1-(x^1-y^1)/2\alpha)^2-(p_2-(x^2-y^2)/2\alpha)^2-m^2-i\epsilon]}e^{+i(x-y)^2/4\alpha}
\bigg{]}
\nonumber\\
&=-\frac{i}{4\pi^2}\epsilon(x^0-y^0)\int_0^{\infty}\frac{d\alpha}{4\alpha^2}\left[e^{-i(x-y)^2/4\alpha -i\alpha m^2-\alpha\epsilon}
+e^{i(x-y)^2/4\alpha+i\alpha m^2-\alpha \epsilon}\right],
\label{8.10a}
\end{align}
with the required $\epsilon(x^0-y^0)$ singularity nicely emerging. With  
\begin{align}
\int _0^{\infty}\frac{d\alpha}{\alpha^2}e^{i(x-y)^2/4\alpha+i\alpha m^2-\alpha \epsilon}&=-\frac{2m\pi }{[(x-y)^2]^{1/2}} [J_1(m[(x-y)^2]^{1/2})+iY_1(m[(x-y)^2]^{1/2})]
\label{8.11a}
\end{align}
when $(x-y)^2>0$, (\ref{8.6}) follows. 

At $(x-y)^2=0$ we note that the leading divergence  in (\ref{8.10a}) behaves as $\int d\alpha/\alpha^2$ near $\alpha =0$, to thus have a leading term that is both mass independent and singular (just as one would want of a $\delta[(x-y)^2]$ term). To explicitly evaluate (\ref{8.10a}) at $(x-y)^2=0$ we have found it convenient to set $\beta=1/4\alpha$, to thus give 
\begin{align}
i\Delta(IT;x-y)&=-\frac{i}{4\pi^2}\epsilon(x^0-y^0)\int_0^{\infty}d\beta\left[e^{-i\beta (x-y)^2-im^2/4\beta-\beta\epsilon}
+e^{i\beta (x-y)^2+im^2/4\beta-\beta\epsilon}\right].
\label{8.12a}
\end{align}
Since the role of the $i\epsilon$ terms is to indicate how to close the contours in (\ref{8.9a}), we are able to replace $-\alpha\epsilon=-\epsilon/4\beta$ by $-\beta\epsilon$ since $\beta$ is positive everywhere in the integration range. Since at $(x-y)^2=0$ the leading divergence  is independent of $m^2$, we can set $m^2=0$ in the $\beta$ form for the integrals, to then find that at $(x-y)^2=0$ we can integrate the $\beta$ integrals directly, to obtain the singular functions
\begin{align}
i\Delta(IT;(x-y)^2=0)&=-\frac{i}{4\pi^2}\epsilon(x^0-y^0)\left[-\frac{i}{(x-y)^2-i\epsilon}+\frac{i}{(x-y)^2+i\epsilon}\right]
=-\frac{i}{2\pi}\epsilon(x^0-y^0)\delta[(x-y)^2],
\label{8.13a}
\end{align}
with the principal part of $1/[(x-y)^2\pm i\epsilon]=PP[1/(x-y)^2]\mp i\pi \delta[(x-y)^2]$ dropping out. Finally, since $i\Delta(IT;x-y)$ vanishes when we set  $x^0=y^0$ in (\ref{8.9a}), by Lorentz invariance $i\Delta(IT;(x-y)^2<0)$ vanishes for all spacelike separated distances. We thus confirm (\ref{8.6}), (\ref{8.7}) and (\ref{8.8a}). While our off-shell complex  contour integral derivation of (\ref{8.6}), (\ref{8.7}) and (\ref{8.8a}) initially only applies to $x^0\neq y^0$, since $i\Delta(IT;x-y)$ and $\epsilon(x^0-y^0)$ both vanish at $x^0=y^0$, (\ref{8.6}), (\ref{8.7}) and (\ref{8.8a}) actually hold at $x^0=y^0$ as well. While this analysis of the behavior of $i\Delta(IT;x-y)$ at $x^0=y^0$ is straightforward and might even appear to be somewhat pedantic, in the light-front case that we discuss below we will see that such an analysis is necessary.

Since the unequal-time $i\Delta(IT;x-y)$ is defined at all $x^{\mu}$ and $y^{\mu}$, it is equally defined at equal light-front time $x^+=x^0+x^3=y^0+y^3=y^+$. Thus with $x^0=(x^++x^-)/2$, $x^3=(x^+-x^-)/2$, $y^0=(y^++y^-)/2$, $y^3=(y^+-y^-)/2$ we can set 
\begin{align}
(x-y)^2=(x^0-y^0)^2-(x^1-y^1)^2-(x^2-y^2)^2-(x^3-y^3)^2=(x^+-y^+)(x^--y^-)-(x^1-y^1)^2-(x^2-y^2)^2. 
\label{8.14a}
\end{align}
We now note that at $x^+=y^+$ the quantity $(x-y)^2$ would be negative unless $x^1-y^1=0$, $x^2-y^2=0$. However, for $(x-y)^2<0$ the $i\Delta(IT;x-y)$ commutator vanishes. Thus at $x^+=y^+$ the only point of relevance is $(x-y)^2=0$, and at $(x-y)^2=0$ the $i\Delta(IT;(x-y)^2=0)$ commutator is independent of the mass $m$ of the scalar and is given by the mass-independent (\ref{8.7}). When written in light-front coordinates (\ref{8.7}) takes the form
\begin{eqnarray}
i\Delta(IT;(x-y)^2=0)=-\frac{i}{2\pi}\epsilon[\tfrac{1}{2}(x^++x^--y^+-y^-)]\delta[(x^+-y^+)(x^--y^-)-(x^1-y^1)^2-(x^2-y^2)^2].
\label{8.15a}
\end{eqnarray}
Since $\epsilon(x/2)=\epsilon(x)$ for any $x$, at $x^+=y^+$ (\ref{8.15a}) takes the form
\begin{eqnarray}
i\Delta(IT;(x-y)^2=0)\Big{|}_{x^+=y^+}=-\frac{i}{2\pi}\epsilon(x^--y^-)\delta[(x^1-y^1)^2+(x^2-y^2)^2].
\label{8.16a}
\end{eqnarray}
Then since $\delta(a^2+b^2)=\pi \delta(a)\delta(b)/2$ for any $a$ and $b$, we can rewrite (\ref{8.16a}) as 
\begin{eqnarray}
i\Delta(IT;(x-y)^2=0)\Big{|}_{x^+=y^+}&=&[\phi(x^+,x^1,x^2,x^-), \phi(x^+,y^1,y^2,y^-)]
=-\frac{i}{4}\epsilon(x^--y^-)\delta(x^1-y^1)\delta(x^2-y^2).
\label{8.17a}
\end{eqnarray}
We recognize (\ref{8.17a}) as none other than the equal light-front time (\ref{1.11}), with, as noted in \cite{Harindranath1996,Mannheim2019b,Mannheim2020a}, the scalar field equal light-front time commutation relation (\ref{1.11}) thus being derived starting from the unequal instant-time commutation relation (\ref{8.5}). Since the unequal-time instant-time commutation relation (\ref{8.5}) itself follows solely from the equal-time instant-time commutation relation (\ref{8.2}) and the scalar field wave equation, we see that the equal light-front time commutation relation (\ref{1.11}) follows directly from the equal-time instant-time commutation relation (\ref{8.2}) and does not need to be independently postulated.

\subsection{Scalar Field Case -- Light Front to Instant Time}
\label{S8b}

In the light-front case the scalar field canonical commutation relations are of the form
\begin{align}
&[\phi(x^+,x^1,x^2,x^-),2\partial_-\phi(x^+,y^1,y^2,y^-)]=i\delta(x^1-y^1)\delta(x^2-y^2)\delta(x^--y^-),
\nonumber\\
&[\phi(x^+,x^1,x^2,x^-),\phi(x^+,y^1,y^2,y^-)]=-\frac{i}{4}\epsilon(x^--y^-)\delta(x^1-y^1)\delta(x^2-y^2)
\label{8.18a}
\end{align}
given earlier, and the on-shell Fock expansion is of the form
\begin{align}
\phi(x^+,x^1,x^2,x^-)&=\frac{2}{(2\pi)^{3/2}}\int_{-\infty}^{\infty}dp_1\int_{-\infty}^{\infty}dp_2\int_0^{\infty} \frac{dp_-}{(4p_-)^{1/2}}
\Big{[}e^{-i(F_p^2x^+/4p_-+p_-x^-+p_1x^1+p_2x^2)}a(\vec{p})
\nonumber\\
&+e^{i(F_p^2x^+/4p_-
+p_-x^-+p_1x^1+p_2x^2)}a^{\dagger}(\vec{p})\Big{]},
\label{8.19a}
\end{align}
where $F_p^2=(p_1)^2+(p_2)^2+m^2$, where the integration range for $p_-$ is only over $p_-\geq 0$, and where the light-front $[a(\vec{p}),a^{\dagger}(\vec{p}^{~\prime})]$ commutator  is normalized to 
\begin{eqnarray}
[a(\vec{p}),a^{\dagger}(\vec{p}^{~\prime})]=\tfrac{1}{2}\delta(p_--p_-^{\prime})\delta(p_1-p_1^{\prime})\delta(p_2-p_2^{\prime}), 
\label{8.20a}
\end{eqnarray}
as fixed via the normalization of the equal light-front time canonical commutator given in (\ref{8.18a}). Given (\ref{8.19a}) we construct the unequal light-front time ($LF$) commutator $i\Delta(LF;x-y)$, and obtain
\begin{align}
i\Delta(LF;x-y)&=[\phi(x^+,x^1,x^2,x^-),\phi(y^+,y^1,y^2,y^-)]
\nonumber\\
&=\frac{1}{4\pi^3}\int_{-\infty}^{\infty}dp_1\int_{-\infty}^{\infty}dp_2\int_{0}^{\infty}\frac{dp_-}{4p_-}
\big{[}e^{-i[F_p^2(x^+-y^+)/4p_-+p_-(x^--y^-)+p_1(x^1-y^1)+p_2(x^2-y^2)]}
\nonumber\\
&-e^{i[F_p^2(x^+-y^+)/4p_-+p_-(x^--y^-)+p_1(x^1-y^1)+p_2(x^2-y^2)]}\big{]}.
\label{8.21a}
\end{align}
Since (\ref{8.21a}) itself is based on (\ref{8.18a}), both of the relations in (\ref{8.18a})  can be recovered from (\ref{8.21a}). And with  $i\Delta(LF;x-y)$ obeying the first-order derivative in $x^+$ equation $[4\partial_+\partial_--(\partial_1)^2-(\partial_2)^2+m^2]i\Delta(LF;x-y)=0$ at $x^{\mu}\neq y^{\mu}$ and recovering (\ref{8.18a}) at $x^+=y^+$, we see that (\ref{8.21a}) holds at all $x^{\mu}-y^{\mu}$, just as it should.

In order to compare instant-time quantization with light-front quantization it would convenient if we could transform $i\Delta(IT;x-y)$ as given in (\ref{8.5}) to $i\Delta(LF;x-y)$ as given in (\ref{8.21a}) via a transformation of the coordinates and momenta. However, this cannot be done as is since in $i\Delta(IT;x-y)$ the variable $p_3$ ranges between $-\infty$ and $\infty$ while in $i\Delta(LF;x-y)$ the variable $p_-$ only ranges between $0$ and $\infty$. That $p_-$ cannot be negative originates in the fact that in the on-shell light-front Fock expansion given in (\ref{8.19a}) there is a $1/(4p_-)^{1/2}$ term, and it has to be real. A second reason that (\ref{8.21a}) cannot be used as is is because it has a singularity at $p_-=0$, the zero-mode singularity that commonly appears in on-shell light-front studies and challenges them. As noted for instance in \cite{Mannheim2019a,Mannheim2019b} and as will be discussed in detail below when we study some light-front Feynman diagrams, the way to handle on-shell zero-mode singularities such as these and give them a meaning is to go off shell. We shall thus rewrite $i\Delta(LF;x-y)$ as a contour integral in a complex $p_+$ plane, with no zero-mode singularity problem then being found to occur. Going into the complex plane is necessary anyway since the $\epsilon(x^0-y^0)$ and $\delta[(x-y)^2]$ singularities in $i\Delta(IT;x-y)$ must be reflected in and reproduced in $i\Delta(LF;x-y)$, with it precisely being the $(x-y)^2=0$ region where canonical commutators take support. 

To this end we set  
\begin{align}
i\Delta(LF;x-y)&=-\frac{1}{2\pi i}\frac{1}{4\pi^3}\int_{-\infty }^{\infty}dp_1\int_{-\infty }^{\infty}dp_2\int_{0}^{\infty}dp_-\oint dp_+
\nonumber\\
&\times\bigg{[}
\frac{\theta(x^+-y^+)e^{-ip\cdot (x-y)}-\theta(-x^++y^+)e^{ip\cdot (x-y)}}{4p_+p_--(p_1)^2-(p_2)^2-m^2+i\epsilon}
+\frac{\theta(x^+-y^+)e^{ip\cdot (x-y)}-\theta(-x^++y^+)e^{-ip\cdot (x-y)}}{4p_+p_--(p_1)^2-(p_2)^2-m^2-i\epsilon}\bigg{]},
\label{8.22a}
\end{align}
with the $+i\epsilon$ terms being closed in the lower-half complex $p_+$ plane and the $-i\epsilon$ terms being closed in the upper-half plane. As introduced,  the $\theta(x^+-y^+)$ and $\theta(-x^++y^+)$ terms suppress all circle at infinity contributions. In analog to the instant-time analysis this contour integral representation of $i\Delta(LF;x-y)$ does not hold at $x^+=y^+$, a point we return to below.

In order to be able to compare $i\Delta(IT;x-y)$ and $i\Delta(LF;x-y)$ we need to evaluate (\ref{8.22a}) so as to potentially obtain an analog to (\ref{8.10a}). With the only non-zero contributions to the $p_+$ integration being on the real $p_+$ axis we again introduce the $\alpha$ regulators  and obtain
\begin{align}
i\Delta(LF;x-y)&
=-\frac{1}{2\pi i}\frac{1}{4\pi^3}\int_{-\infty }^{\infty}dp_1\int_{-\infty }^{\infty}dp_2\int_{0}^{\infty}dp_-\int_{-\infty }^{\infty}dp_+\int_0^{\infty}d\alpha
\nonumber\\
&\times\bigg{[}
-i[\theta(x^+-y^+)e^{-ip\cdot (x-y)}-\theta(-x^++y^+)e^{ip\cdot (x-y)}]e^{i\alpha[4p_+p_--(p_1)^2-(p_2)^2-m^2+i\epsilon]}
\nonumber\\
&+i[\theta(x^+-y^+)e^{ip\cdot (x-y)}-\theta(-x^++y^+)e^{-ip\cdot (x-y)}]e^{-i\alpha[4p_+p_--(p_1)^2-(p_2)^2-m^2-i\epsilon]}\bigg{]}
\nonumber\\
&
=-\frac{2\pi}{2\pi i}\frac{1}{4\pi^3}\int_{-\infty }^{\infty}dp_1\int_{-\infty }^{\infty}dp_2\int_{0}^{\infty}dp_-\int_0^{\infty}d\alpha
\nonumber\\
&\times\bigg{[}
-i[\theta(x^+-y^+)e^{-ip_-(x^--y^-)-ip_1(x^1-y^1)-ip_2 (x^2-y^2)}\delta(4\alpha p_--x^++y^+)
\nonumber\\
&
-\theta(-x^++y^+)e^{ip_-(x^--y^-)+ip_1 (x^1-y^1)+ip_2 (x^2-y^2)}\delta(4\alpha p_-+x^+-y^+)]e^{i\alpha[-(p_1)^2-(p_2)^2-m^2+i\epsilon]}
\nonumber\\
&+i[\theta(x^+-y^+)e^{ip_-(x^--y^-)+ip_1(x^1-y^1)+ip_2 (x^2-y^2)}\delta(-4\alpha p_-+x^+-y^+)
\nonumber\\
&
-\theta(-x^++y^+)e^{-ip_-(x^--y^-)-ip_1 (x^1-y^1)-ip_2 (x^2-y^2)}\delta(-4\alpha p_--x^++y^+)]e^{-i\alpha[-(p_1)^2-(p_2)^2-m^2-i\epsilon]}
\bigg{]}.
\label{8.23a}
\end{align}
Because the range of $p_-$ is restricted to $(0,\infty)$,  the contributions of the delta functions depend on the sign of $x^+-y^+$. Thus we obtain 
\begin{align}
i\Delta(LF;x-y)&
=-\frac{2\pi}{2\pi i}\frac{1}{4\pi^3}\int_{-\infty }^{\infty}dp_1\int_{-\infty }^{\infty}dp_2\int_0^{\infty}\frac{d\alpha}{4\alpha}
\nonumber\\
&\times\bigg{[}
-i[\theta(x^+-y^+)e^{-i(x^+-y^+)(x^--y^-)/4\alpha-ip_1(x^1-y^1)-ip_2 (x^2-y^2)}
\nonumber\\
&-\theta(-x^++y^+)e^{-i(x^+-y^+)(x^--y^-)/4\alpha+ip_1(x^1-y^1)+ip_2 (x^2-y^2)}]e^{i\alpha[-(p_1)^2-(p_2)^2-m^2+i\epsilon]}
\nonumber\\
&+i[\theta(x^+-y^+)e^{i(x^+-y^+)(x^--y^-)4\alpha+ip_1(x^1-y^1)+ip_2 (x^2-y^2)}
\nonumber\\
&
-\theta(-x^++y^+)e^{i(x^+-y^+)(x^--y^-)4\alpha-ip_1 (x^1-y^1)-ip_2 (x^2-y^2)}]e^{-i\alpha[-(p_1)^2-(p_2)^2-m^2-i\epsilon]}
\bigg{]}.
\label{8.24a}
\end{align}
On substituting $p_1\rightarrow -p_1$,  $p_2\rightarrow -p_2$ in the appropriate places we obtain
\begin{align}
i\Delta(LF;x-y)&
=i\frac{2\pi}{2\pi i}\frac{1}{4\pi^3}\int_{-\infty }^{\infty}dp_1\int_{-\infty }^{\infty}dp_2\int_0^{\infty}\frac{d\alpha}{4\alpha}
\epsilon(x^+-y^+)
\nonumber\\
&\times\bigg{[}e^{-i(x^+-y^+)(x^--y^-)/4\alpha-ip_1(x^1-y^1)-ip_2 (x^2-y^2)}e^{i\alpha[-(p_1)^2-(p_2)^2-m^2+i\epsilon]}
\nonumber\\
&-e^{i(x^+-y^+)(x^--y^-)/4\alpha +ip_1(x^1-y^1)+ip_2 (x^2-y^2)}e^{-i\alpha[-(p_1)^2-(p_2)^2-m^2-i\epsilon]}
\bigg{]}
\nonumber\\
&=i\frac{2\pi}{2\pi i}\frac{1}{4\pi^3}\int_{-\infty }^{\infty}dp_1\int_{-\infty }^{\infty}dp_2\int_0^{\infty}\frac{d\alpha}{4\alpha}
\epsilon(x^+-y^+)
\nonumber\\
&\times\bigg{[}e^{-i(x-y)^2/4\alpha-i\alpha(p_1+(x^1-y^1)/2\alpha)^2-i\alpha(p_2+(x^2-y^2)/2\alpha)^2-i\alpha m^2-\alpha\epsilon}
\nonumber\\
&-e^{i(x-y)^2/4\alpha+i\alpha(p_1+(x^1-y^1)/2\alpha)^2+i\alpha(p_2+(x^2-y^2)/2\alpha)^2+i\alpha m^2-\alpha\epsilon}\bigg{]}.
\label{8.25a}
\end{align}
Then, with a final integration we obtain 
\begin{align}
i\Delta(LF;x-y)&=-\frac{i}{4\pi^2}\epsilon(x^+-y^+)\int_0^{\infty}\frac{d\alpha}{4\alpha^2}\left[e^{-i(x-y)^2/4\alpha -i\alpha m^2-\alpha\epsilon}
+e^{i(x-y)^2/4\alpha+i\alpha m^2-\alpha \epsilon}\right]
\nonumber\\
&=-\frac{i}{4\pi^2}\epsilon(x^+-y^+)\int_0^{\infty}d\beta\left[e^{-i\beta (x-y)^2-im^2/4\beta-\beta\epsilon}
+e^{i\beta (x-y)^2+im^2/4\beta-\beta\epsilon}\right].
\label{8.26a}
\end{align}
This is our main result.

Comparing (\ref{8.26a}) with (\ref{8.10a}) we see that under the transformation  $x^0\rightarrow x^+$, $x^3\rightarrow x^-$, $y^0\rightarrow y^+$, $y^3\rightarrow y^-$, viz. $(x-y)^2=(x^0-y^0)^2-(x^3-y^3)^2-(x^1-y^1)^2-(x^2-y^2)^2\rightarrow (x^+-y^+)(x^--y^-)-(x^1-y^1)^2-(x^2-y^2)^2$ the instant-time $i\Delta(IT;x-y)$ transforms into the light-front $ i\Delta(LF;x-y)$.  We have thus achieved our main objective, showing that $i\Delta(IT;x-y)$ and $i\Delta(LF;x-y)$ are related by a coordinate transformation, being in fact related for arbitrary $(x-y)^2$, i.e., for timelike, lightlike or spacelike $(x-y)^2$. The quantities $i\Delta(IT;x-y)$ and $ i\Delta(LF;x-y)$ are thus equivalent, and we thereby establish that for scalar fields light-front commutation relations and instant-time commutation relations are equivalent, and each  can be considered as being a consequence of the other. In establishing this equivalence we note that even though the instant-time $p_3$ varies between $-\infty$ and $\infty$, the light-front $p_-$ only varies between $0$ and $\infty$. Thus the lack of modes with  negative $p_-$ is not an impediment to establishing the equivalence of  $i\Delta(IT;x-y)$ and $i\Delta(LF;x-y)$.

Given the equivalence of (\ref{8.10a}) and (\ref{8.26a}) we see immediately that $i\Delta(LF;x-y)$ evaluates to
\begin{align}
i\Delta(LF; (x-y)^2>0)&=\frac{im}{4\pi}\epsilon(x^+-y^+)\frac{J_1(m[(x-y)^2]^{1/2})}{[(x-y)^2]^{1/2}}=\frac{im}{4\pi}\epsilon(x^--y^-)\frac{J_1(m[(x-y)^2]^{1/2})}{[(x-y)^2]^{1/2}},
\nonumber\\
i\Delta(LF;(x-y)^2=0)&=-\frac{i}{2\pi}\epsilon(x^+-y^+)\delta[(x-y)^2]=-\frac{i}{2\pi}\epsilon(x^--y^-)\delta[(x-y)^2],
\nonumber\\
i\Delta(LF;(x-y)^2<0)&=0
\label{8.27a}
\end{align}
in the various cases, and like its instant-time counterpart the light-front $i\Delta(LF;x-y)$ has a delta function singularity on the light cone and vanishes identically for spacelike separated points, just as required by microcausality.

For both timelike and lightlike intervals the sign of $x^+-y^+$ is the same as the sign of  $x^--y^-$, and so we can replace the $\epsilon(x^+-y^+)$ factors by $\epsilon(x^--y^-)$ in (\ref{8.27a}). (For spacelike separated points no such replacement is allowed, but since $\Delta(LF;(x-y)^2<0)=0$ none is anyway needed.) We have made this replacement since according to (\ref{8.16a}) $i\Delta(LF;x-y)=[\phi(x^+,x^1,x^2,x^-),\phi(y^+,y^1,y^2,y^-)]$ does not vanish at $x^+=y^+$ while $\epsilon(x^+-y^+)$ does. There is no contradiction here since our contour integral representation of $i\Delta(LF;x-y)$ does not hold at $x^+=y^+$. However, by replacing $\epsilon(x^+-y^+)$ by $\epsilon(x^--y^-)$ we obtain expressions for $i\Delta(LF;x-y)$ which now do hold at $x^+=y^+$. And one can readily check that starting with $i\Delta(LF;(x-y)^2=0)=-\tfrac{i}{2\pi}\epsilon(x^--y^-)\delta[(x-y)^2]$ one can obtain (\ref{8.18a}).

Having derived equal light-front time commutators from unequal-time instant-time commutators in the above, we can equally  derive equal-time instant-time commutators from unequal light-front time commutators. To this end we note that the argument of the delta function term in $i\Delta(LF;(x-y)^2=0)$ is $(x^+-y^+)(x^--y^-)-(x^1-y^1)^2-(x^2-y^2)^2$. Thus on the light cone $x^+-y^+$ and $x^--y^-$ have the same sign. On setting $x^+=x^0+x^3$, $x^-=x^0-x^3$,  $y^+=y^0+y^3$, $y^-=y^0-y^3$, it follows that $2(x^0-y^0)=x^+-y^++x^--y^-$ has the same sign as $x^+-y^+$. Consequently we can rewrite $i\Delta(LF;(x-y)^2=0)$ as
\begin{align}
i\Delta(LF;(x-y)^2=0)&=-\frac{i}{2\pi}\epsilon(x^0-y^0)\delta[(x^0-y^0)^2-(x^3-y^3)^2-(x^1-y^1)^2-(x^2-y^2)^2].
\label{8.28a}
\end{align}
On taking the derivative with respect to $y^0$ and then setting $x^0=y^0$ the instant-time commutation relations given in (\ref{8.2}) then follow.

Moreover, comparing with (\ref{8.7}) we see that under the change of variables $x^+=x^0+x^3$, $x^-=x^0-x^3$,  $y^+=y^0+y^3$, $y^-=y^0-y^3$ we obtain 
\begin{align}
i\Delta(IT;(x-y)^2=0)&=i\Delta(LF;(x-y)^2=0),
\label{8.29a}
\end{align}
with $\Delta(IT;(x-y)^2=0)$ and $i\Delta(LF;(x-y)^2=0)$ thus being equal all over the light cone. In this sense they are completely equivalent with all the seeming differences between the commutation relations given in (\ref{8.2}) and (\ref{8.18a}) only arising because of the restriction to equal times (instant-time or light-front) in the two cases.  Thus while we cannot transform (\ref{8.2}) into (\ref{8.18a}), we can transform $\Delta(IT;(x-y)^2=0)$ into $i\Delta(LF;(x-y)^2=0)$, and that is all we need in order to be able to establish the equivalence of the instant-time and light-front quantization procedures  in the scalar field case.

\subsection{Fermion Field Case -- Instant Time to Light Front}
\label{S8c}

In instant-time quantization with a  free fermionic Dirac action of the form
\begin{eqnarray}
I_D=\int d^4x\bar{\psi}(i\gamma^{\mu}\partial_{\mu}-m)\psi +{\rm H.~c.},
\label{8.30a}
\end{eqnarray}
the canonical conjugate of $\psi$ is $i\psi^{\dagger}$, and the canonical anticommutation relations are of the form
\begin{align}
&\Big{\{}\psi_{\alpha}(x^0,x^1,x^2,x^3),\psi_{\beta}^{\dagger}(x^0,y^1,y^2,y^3)\Big{\}}
=\delta_{\alpha\beta}\delta(x^1-y^1)\delta(x^2-y^2)\delta(x^3-y^3),
\nonumber\\
&\Big{\{}\psi_{\alpha}(x^0,x^1,x^2,x^3),\psi_{\beta}(x^0,y^1,y^2,y^3)\Big{\}}=0.
\label{8.31a}
\end{align}
In solutions to the fermion field Dirac equation $(i\gamma^{\mu}\partial_{\mu}-m)\psi=0$, the instant-time on-shell Fock space expansion of the fermion field is of the form (see e.g. \cite{Bjorken1965})
\begin{eqnarray}
\psi(\vec{x},x^0)=\sum_{s=\pm}\int \frac{d^3p}{(2\pi)^{3/2}}\left(\frac{m}{E_p}\right)^{1/2}[b(\vec{p},s)u(\vec{p},s)e^{-ip\cdot x}+d^{\dagger}(\vec{p})v(\vec{p},s)e^{+ip\cdot x}],
\label{8.32a}
\end{eqnarray}
where $E_p=+[(p_1)^2+(p_2)^2+(p_3)^2]^{1/2}$, where $s$ denotes the spin projection, where the Dirac spinors $u(\vec{p},s)$ and $v(\vec{p},s)$ obey $(\slashed{p}-m)u(\vec{p},s)=0$, $(\slashed{p}+m)v(\vec{p},s)=0$, and where the non-trivial creation and annihilation operator anticommutation relations are of the form 
\begin{eqnarray}
\{b(\vec{p},s),b^{\dagger}(\vec{q},s^{\prime})\}=\delta_{s,s^{\prime}}\delta^3(\vec{p}-\vec{q}),\quad
\{d(\vec{p},s),d^{\dagger}(\vec{q},s^{\prime})\}=\delta_{s,s^{\prime}}\delta^3(\vec{p}-\vec{q}).
\label{8.33a}
\end{eqnarray}
With these relations  the free fermion unequal-time instant-time anticommutator is given by the c-number (see e.g. \cite{Bjorken1965})
\begin{eqnarray}
\Big{\{}\psi_{\alpha}(x^0,x^1,x^2,x^3), \psi_{\beta}^{\dagger}(y^0,y^1,y^2,y^3)\Big{\}}=\left[(i\gamma^{\mu}\frac{\partial}{\partial x^{\mu}}+m)\gamma^0\right]_{\alpha\beta}i\Delta(IT;x-y),
\label{8.34a}
\end{eqnarray}
where $i\Delta(IT;x-y)$ is given in (\ref{8.5}).

In  light-front coordinates the same Dirac action takes the form 
\begin{eqnarray}
I_D=\frac{1}{2}\int dx^+dx^1dx^2dx^-\psi^{\dagger}[i\gamma^0(\gamma^+\partial_++\gamma^-\partial_-+\gamma^1\partial_1+\gamma^2\partial_2)-\gamma^0m]\psi +{\rm H.~c.},
\label{8.35a}
\end{eqnarray}
where $\gamma^{\pm}=\gamma^0\pm \gamma^3$. With this action the light-front  canonical conjugate of $\psi$ is $i\psi^{\dagger}\gamma^0\gamma^+$. In the construction of the light-front fermion sector we find a rather sharp contrast with the instant-time fermion sector. First, unlike $\gamma^0$ and $\gamma^3$, which obey $(\gamma^0)^2=1$, $(\gamma^3)^2=-1$, as had been noted earlier the matrices $\gamma^+$ and $\gamma^-$ obey $(\gamma^+)^2=0$, $(\gamma^-)^2=0$, to thus both be non-invertible divisors of zero. Secondly, the quantities 
\begin{eqnarray}
\Lambda^{+}=\tfrac{1}{2}\gamma^0\gamma^{+}=\tfrac{1}{2}(1+\gamma^0\gamma^3),\quad \Lambda^{-}=\tfrac{1}{2}\gamma^0\gamma^{-}=\tfrac{1}{2}(1-\gamma^0\gamma^3)
\label{8.36a}
\end{eqnarray}
introduced in (\ref{4.2}) obey the relations
\begin{eqnarray}
\Lambda^{+}+\Lambda^{-}=I,\quad(\Lambda^{+})^2=\Lambda^{+}=[\Lambda^+]^{\dagger},\quad (\Lambda^{-})^2=\Lambda^{-}=[\Lambda^{-}]^{\dagger},\quad \Lambda^{+}\Lambda^{-}=0
\label{8.37a}
\end{eqnarray}
given in (\ref{4.3}). We recognize (\ref{8.37a}) as a projector algebra, with $\Lambda^{+}$ and $\Lambda^{-}$ thus being non-invertible projection operators. On identifying  $\psi_{(+)}=\Lambda^+\psi$, $\psi_{(-)}=\Lambda^-\psi$, we identify the conjugate of $\psi$ as $2i\psi_{(+)}^{\dagger}$, where by $\psi^{\dagger}_{(+)}$ we mean $[\psi^{\dagger}]_{(+)}=\psi^{\dagger}\Lambda^+$, which is equal to $[\Lambda^+\psi]^{\dagger}=[\psi_{(+)}]^{\dagger}$ since $\Lambda^+$ is Hermitian. Since the conjugate is built purely out of the good fermion, in the anticommutator of $\psi$ with its conjugate only the good component of $\psi$ will contribute since $\Lambda^+\Lambda^-=0$, with the equal light-front time canonical anticommutator being of the form given in (\ref{4.13a}), viz.
\begin{eqnarray}
&&\Big{\{}[\psi_{(+)}]_{\alpha}(x^+,x^1,x^2,x^-),[\psi_{(+)}^{\dagger}]_{\beta}(x^+,y^1,y^2,y^-)\Big{\}}=\Lambda^+_{\alpha\beta}\delta(x^--y^-)\delta(x^1-y^1)\delta(x^2-y^2).
\label{8.38a}
\end{eqnarray}

In this construction the bad fermion $\psi_{(-)}$ has no canonical conjugate and is thus not a dynamical variable. And through use of the Dirac equation $\psi_{(-)}$ is found to obey the constraint 
\begin{align}
2i\partial_-\psi_{(-)}&=[-i\gamma^0(\gamma^1\partial_1+\gamma^2\partial_2)+m\gamma^0]\psi_{(+)},
\nonumber\\
\psi_{(-)}(x^+,x^1,x^2,x^-)&=\frac{1}{4i}\int du^-\epsilon(x^--u^-)[-i\gamma^0(\gamma^1\partial_1+\gamma^2\partial_2)+m\gamma^0]\psi_{(+)}(x^+,x^1,x^2,u^-)
\label{8.39a}
\end{align}
given in (\ref{4.15a}) and (\ref{4.16a}). From this constraint we find that the bad fermion obeys the anticommutation relations given in (\ref{4.21a}) and (\ref{4.22a}), viz.
\begin{align}
&\Big{\{}\frac{\partial}{\partial x^-}\psi_{\mu}^{(-)}(x^+,x^1,x^2,x^-),\frac{\partial}{\partial y^-}[\psi_{(-)}^{\dagger}]_{\nu}(x^+,y^1,y^2,y^-)\Big{\}}
\nonumber\\
&=\frac{1}{4}\Lambda^-_{\mu\nu}\left[-\frac{\partial}{\partial x^1}\frac{\partial}{\partial x^1}-\frac{\partial}{\partial x^2}\frac{\partial}{\partial x^2}+m^2\right]\delta(x^--y^-)\delta(x^1-y^1)\delta(x^2-y^2),
\nonumber\\
&\Big{\{}\psi_{\mu}^{(-)}(x^+,x^1,x^2,x^-),[\psi_{(-)}^{\dagger}]_{\nu}(x^+,y^1,y^2,y^-)\Big{\}}
\nonumber\\
&=\frac{1}{16}\Lambda^-_{\mu\nu}\left[-\frac{\partial}{\partial x^1}\frac{\partial}{\partial x^1}-\frac{\partial}{\partial x^2}\frac{\partial}{\partial x^2}+m^2\right]
\int du^-\epsilon(x^--u^-)\epsilon(y^--u^-)\delta(x^1-y^1)\delta(x^2-y^2).
\label{8.40a}
\end{align}

In analog to our discussion of the scalar field given above, we now derive the light-front (\ref{8.38a}) and (\ref{8.40a})  starting from the unequal instant-time relation (\ref{8.34a}). To this end we first multiply both sides of (\ref{8.34a}) by $\Lambda^+$ on both the right- and the left-hand sides. Noting that
\begin{align}
&\Lambda^+\gamma^0\Lambda^+=0,\quad \Lambda^+\gamma^1\gamma^0\Lambda^+=0,\quad \Lambda^+\gamma^2\gamma^0\Lambda^+=0,\quad \Lambda^+\gamma^+\gamma^0\Lambda^+=0,\quad 
\Lambda^+\gamma^-\gamma^0\Lambda^+=2\Lambda^+,
\nonumber\\
&\gamma^0\partial_0+\gamma^3\partial_3=\gamma^0(\partial_++\partial_-)+\gamma^3(\partial_+-\partial_-)=\gamma^+\partial_++\gamma^-\partial_-,
\label{8.41a}
\end{align}
from the right-hand side of (\ref{8.34a}) we then obtain 
\begin{eqnarray}
\Lambda^+_{\alpha\gamma}\left[(i\gamma^{\mu}\partial_{\mu}+m)\gamma^0\right]_{\gamma\delta}i\Delta(IT;x-y)\Lambda^+_{\delta\beta}
=2i\Lambda^{+}_{\alpha\beta}\partial_{-}i\Delta(IT;x-y).
\label{8.42a}
\end{eqnarray}
We now substitute $x^0=(x^++x^-)/2$, $x^3=(x^+-x^-)/2$, $y^0=(y^++y^-)/2$, $y^3=(y^+-y^-)/2$, and on the light cone use of (\ref{8.15a}) enables us to rewrite the right-hand side of (\ref{8.42a}) as
\begin{align}
&2i\Lambda^{+}_{\alpha\beta}\frac{\partial}{\partial x^-}i\Delta(IT;(x-y)^2=0)
\nonumber\\
&=2i\Lambda^{+}_{\alpha\beta}\frac{\partial}{\partial x^-}\left[-\frac{i}{2\pi}\epsilon[\tfrac{1}{2}(x^++x^--y^+-y^-)]\delta[(x^+-y^+)(x^--y^-)-(x^1-y^1)^2-(x^2-y^2)^2\right]
\nonumber\\
&=\frac{1}{\pi}\Lambda^{+}_{\alpha\beta}\delta[\tfrac{1}{2}(x^++x^--y^+-y^-)]\delta[(x^+-y^+)(x^--y^-)-(x^1-y^1)^2-(x^2-y^2)^2]
\nonumber\\
&+2i\Lambda^{+}_{\alpha\beta}\left[-\frac{i}{2\pi}\epsilon[\tfrac{1}{2}(x^++x^--y^+-y^-)](x^+-y^+)\delta^{\prime}[(x^+-y^+)(x^--y^-)-(x^1-y^1)^2-(x^2-y^2)^2\right].
\label{8.43a}
\end{align}
At $x^+=y^+$ (\ref{8.43a}) takes the form
\begin{eqnarray}
&&2i\Lambda^{+}_{\alpha\beta}\frac{\partial}{\partial x^-}i\Delta(IT;(x-y)^2=0)\Big{|}_{x^+=y^+}=\Lambda^+_{\alpha\beta}\delta(x^--y^-)\delta(x^1-y^1)\delta(x^2-y^2).
\label{8.44a}
\end{eqnarray}
Equating with the good fermion projection of the left-hand side of (\ref{8.34a}) thus yields
\begin{eqnarray}
&&\Lambda^+_{\alpha\gamma}\Big{\{}\psi_{\gamma}(x^+,x^1,x^2,x^-),\psi_{\delta}(x^+,y^1,y^2,y^-)\Big{\}}\Lambda^+_{\delta\beta}
\nonumber\\
&&=\Big{\{}[\psi_{(+)}(x^+,x^1,x^2,x^-)]_{\alpha},[\psi_{(+)}^{\dagger}]_{\beta}(x^+,y^1,y^2,y^-)\Big{\}}=\Lambda^+_{\alpha\beta}\delta(x^--y^-)\delta(x^1-y^1)\delta(x^2-y^2).
\label{8.45a}
\end{eqnarray}
We recognize (\ref{8.45a}) as the good fermion equal light-front time anticommutation relation given in (\ref{8.38a}). Thus as noted in \cite{Harindranath1996,Mannheim2019b,Mannheim2020a} the good fermion equal light-front time anticommutation relation follows from the equal-time instant-time fermion anticommutation relation.

For the bad fermion anticommutation relation given in (\ref{8.40a}) we note that 
\begin{align}
\Lambda^-\gamma^0\Lambda^-=0,\quad \Lambda^-\gamma^1\gamma^0\Lambda^-=0,\quad \Lambda^-\gamma^2\gamma^0\Lambda^-=0,\quad \Lambda^-\gamma^+\gamma^0\Lambda^-=2\Lambda^-,\quad 
\Lambda^-\gamma^-\gamma^0\Lambda^-=0.
\label{8.46a}
\end{align}
Recalling that $x^0=(x^++x^-)/2$, $x^3=(x^+-x^-)/2$, $y^0=(y^++y^-)/2$, $y^3=(y^+-y^-)/2$, 
from the last line in (\ref{8.5}) we obtain
\begin{align}
\frac{\partial}{\partial x^+}\frac{\partial}{\partial y^-}i\Delta(IT;x-y)
&=\frac{\partial}{\partial x^+}\frac{\partial}{\partial y^-}\Big{[}\int_{-\infty}^{\infty}d^3p\frac{1}{(2\pi)^3 2E_p}[e^{-iE_p(x^0-y^0)}-e^{iE_p(x^0-y^0)}]e^{i\vec{p}\cdot(\vec{x}-\vec{y})}\Big{]}
\nonumber\\
&=\int \frac{d^3p}{(2\pi)^3 2E_p}\Big{(}e^{-iE_p(x^0-y^0)}-e^{iE_p (x^0-y^0)}\Big{)}e^{i\vec{p}\cdot (\vec{x}-\vec{y})}
\frac{i}{2}\Big{[}-E_p+p_3\Big{]}\frac{i}{2}\Big{[}E_p+p_3\Big{]}
\nonumber\\
&=\int \frac{d^3p}{(2\pi)^3 2E_p}\Big{(}e^{-iE_p(x^0-y^0)}-e^{iE_p (x^0-y^0)}\Big{)}e^{i\vec{p}\cdot (\vec{x}-\vec{y})}
\frac{1}{4}\Big{[}E_p^2-(p_3)^2\Big{]}
\nonumber\\
&=\int \frac{d^3p}{(2\pi)^3 2E_p}\Big{(}e^{-iE_p(x^0-y^0)}-e^{iE_p (x^0-y^0)}\Big{)}e^{i\vec{p}\cdot (\vec{x}-\vec{y})}
\frac{1}{4}\Big{[}(p_1)^2+(p_2)^2+m^2\Big{]}
\nonumber\\
&=\frac{1}{4}\Big{[}-\Big{(}\frac{\partial}{\partial x^1}\Big{)}^2-\Big{(}\frac{\partial}{\partial x^2}\Big{)}^2+m^2\Big{]}i\Delta(IT;x-y).
\label{8.47a}
\end{align}
Thus on projecting (\ref{8.34a}) with $\Lambda^-$ we obtain
\begin{align}
&\Big{\{}\frac{\partial}{\partial x^-}\psi_{\alpha}^{(-)}(x^+,x^1,x^2,x^-),\frac{\partial}{\partial y^-}[\psi_{(-)}^{\dagger}]_{\beta}(y^+,y^1,y^2,y^-)\Big{\}}
\nonumber\\
&=2i\Lambda^-_{\alpha\beta}\frac{\partial}{\partial x^-}\frac{\partial}{\partial y^-}\frac{\partial}{\partial x^+}i\Delta(IT;x-y)
\nonumber\\
&=\frac{i}{2}\Lambda^-_{\alpha\beta}\Big{[}-\Big{(}\frac{\partial}{\partial x^1}\Big{)}^2-\Big{(}\frac{\partial}{\partial x^2}\Big{)}^2+m^2\Big{]}\frac{\partial}{\partial x^-}i\Delta(IT;x-y).
\label{8.48a}
\end{align}
Thus at $x^+=y^+$, from (\ref{8.17a}) we obtain 
\begin{align}
&\Big{\{}\frac{\partial}{\partial x^-}\psi_{\alpha}^{(-)}(x^+,x^1,x^2,x^-),\frac{\partial}{\partial y^-}[\psi_{(-)}^{\dagger}]_{\beta}(x^+,y^1,y^2,y^-)\Big{\}}
\nonumber\\
&=\frac{1}{4}\Lambda^-_{\alpha\beta}\left[-\frac{\partial}{\partial x^1}\frac{\partial}{\partial x^1}-\frac{\partial}{\partial x^2}\frac{\partial}{\partial x^2}+m^2\right]\delta(x^--y^-)\delta(x^1-y^1)\delta(x^2-y^2).
\label{8.49a}
\end{align}
We recognize (\ref{8.49a}) as (\ref{8.40a}). With a similar analysis enabling us recover the $\Big{\{}\psi_{(+)},\psi^{\dagger}_{(-)}\Big{\}}$ anticommutator given in (\ref{4.23a}), we see that in both the good and the bad fermion sectors one can derive equal light-front time anticommutators starting from unequal-time instant-time anticommutators. Since these unequal-time instant-time anticommutators themselves follow directly from the Dirac equation and equal instant-time anticommutators, the equal light-front time anticommutators do not need to be independently postulated. In passing we note that the $\Big{\{}\psi_{\alpha}^{(-)}(x^+,x^1,x^2,x^-),[\psi_{(-)}^{\dagger}]_{\beta}(x^+,y^1,y^2,y^-)\Big{\}}$ anticommutator given in (\ref{4.22a}) is highly divergent. However, the $\partial_-\psi_{(-)}$ version given in (\ref{8.49a}) is well-behaved (as it must be because (\ref{4.15a}) constrains the bad fermion derivative $\partial_-\psi_{(-)}$ to behave as the well-behaved good fermion $\psi_{(+)}$). The well-behaved (\ref{8.49a}) suffices for our purposes here.

\subsection{Fermion Field Case -- Light Front to Instant Time}
\label{S8d}

For the light-front case given that only the good fermion is dynamical, initially it is suggested to generalize the equal light-front time good fermion anticommutator given in (\ref{8.38a}) to unequal light-front time. However, starting from a projected light-front relation we could not derive an instant-time relation from it precisely because projectors are not invertible. However, since $\Lambda^++\Lambda^-=I$, it is only together that the good and bad fermion sectors form a complete basis. Thus to derive instant-time anticommutators starting from light-front ones, we must start on the fermion light-front side with something that contains all four of the components of the fermion field, and which in addition is invertible. To this end we thus seek an analog of (\ref{8.34a}). We simply suggest that the light-front analog of (\ref{8.34a}) be given by  
\begin{eqnarray}
\big{\{}\psi_{\alpha}(x^+,x^1,x^2,x^-), \psi_{\beta}^{\dagger}(y^+,y^1,y^2,y^-)\big{\}}=\left[i(\gamma^+\partial_+^x+\gamma^-\partial_-^x+\gamma^1\partial_1^x+\gamma^2\partial_2^x)\gamma^0+m\gamma^0\right]_{\alpha\beta}i\Delta(LF;x-y)
\label{8.50a}
\end{eqnarray}
($\partial^x_+$ denotes $\partial/\partial x^+$ etc.), and then test for whether or not this might in fact be the case.
We note that in (\ref{8.50a}) we have not transformed $\gamma^0$ into $\gamma^+$, since in going from the instant-time $\gamma^0\partial_0+\gamma^3\partial_3$ to the light-front $\gamma^+\partial_++\gamma^-\partial_-$ in (\ref{8.41a})  we only transformed the coordinate derivative operators and not the Dirac gamma matrices. 

To establish the validity of (\ref{8.50a}) we note that since $\psi(x)$ itself obeys the Dirac equation, so does $\big{\{}\psi_{\alpha}(x), \psi_{\beta}^{\dagger}(y)\big{\}}$. However, the Dirac equation is a first-order equation in $\partial/\partial x^+$, so (\ref{8.50a}) will be valid at all $x^+$ if it is valid at one value of $x^+$, which here we take to be $x^+=0$. To check if it is valid at $x^+=0$ we apply $\Lambda^+$ to both sides of (\ref{8.50a}) and also apply $\Lambda^-$ to both sides of (\ref{8.50a}). This yields
\begin{eqnarray}
&&\big{\{}[\psi_{(+)}]_{\alpha}(x^+,x^1,x^2,x^-),[\psi_{(+)}^{\dagger}]_{\beta}(y^+,y^1,y^2,y^-)\big{\}}=2\Lambda^+_{\alpha\beta}i\frac{\partial}{\partial x^-}i\Delta(LF;x-y),
\label{8.51a}
\end{eqnarray}
\begin{eqnarray}
&&\big{\{}[\psi_{(-)}]_{\alpha}(x^+,x^1,x^2,x^-),[\psi_{(-)}^{\dagger}]_{\beta}(y^+,y^1,y^2,y^-)\big{\}}=2\Lambda^-_{\alpha\beta}i\frac{\partial}{\partial x^+}i\Delta(LF;x-y).
\label{8.52a}
\end{eqnarray}
On the light cone $i\Delta(LF;(x-y)^2=0)$ is given by (\ref{8.27a}), and in it we can replace $\epsilon(x^+-y^+)$ by $\epsilon(x^--y^-)$ since the delta function in (\ref{8.27a}) requires that $x^+-y^+$ and $x^--y^-$ have the same sign. Thus from (\ref{8.51a})  we obtain 
\begin{align}
&\big{\{}[\psi_{(+)}]_{\alpha}(x^+,x^1,x^2,x^-),[\psi_{(+)}^{\dagger}]_{\beta}(y^+,y^1,y^2,y^-)\big{\}}|_{(x-y)^2=0}
\nonumber\\
&=2i\Lambda^+_{\alpha\beta}\left(\frac{-i}{2\pi}\right)\bigg{[}2\delta(x^--y^-)
\delta[(x^+-y^+)(x^--y^-)-(x^1-y^1)^2-(x^2-y^2)^2]
\nonumber\\
&+(x^+-y^+)\epsilon(x^--y^-)
\delta^{\prime}[(x^+-y^+)(x^--y^-)-(x^1-y^1)^2-(x^2-y^2)^2
\bigg{]}
\label{8.53a}
\end{align}
in the good fermion sector.
At $x^+=y^+$ we obtain
\begin{eqnarray}
&&\big{\{}[\psi_{(+)}]_{\alpha}(x^+,x^1,x^2,x^-),[\psi_{(+)}^{\dagger}]_{\beta}(x^+,y^1,y^2,y^-)\big{\}}
=\Lambda^+_{\alpha\beta}\delta(x^--y^-)\delta(x^1-y^1)\delta(x^2-y^2).
\label{8.54a}
\end{eqnarray}
We thus obtain the good fermion (\ref{8.38a}).

In order to evaluate the right-hand side of (\ref{8.52a}) in the bad fermion case we have found it convenient to use the contour integral representation of $i\Delta(LF;x-y)$ given in (\ref{8.22a}). On applying $\partial/\partial x^+$ to (\ref{8.22a}) we obtain
\begin{align}
&\frac{\partial}{\partial x^+}i\Delta(LF;x-y)=-\frac{1}{2\pi i}\frac{1}{4\pi^3}\int_{-\infty }^{\infty}dp_1\int_{-\infty }^{\infty}dp_2\int_{0}^{\infty}dp_-\int_{-\infty }^{\infty}dp_+
\nonumber\\
&\times\bigg{[}
\delta(x^+-y^+)\frac{e^{-ip\cdot (x-y)}+e^{ip\cdot (x-y)}}{4p_+p_--(p_1)^2-(p_2)^2-m^2+i\epsilon}
+\delta(x^+-y^+)\frac{e^{ip\cdot (x-y)}+e^{-ip\cdot (x-y)}}{4p_+p_--(p_1)^2-(p_2)^2-m^2-i\epsilon}
\nonumber\\
&+\frac{\theta(x^+-y^+)(-ip_+)e^{-ip\cdot (x-y)}-\theta(-x^++y^+)(ip_+)e^{ip\cdot (x-y)}}{4p_+p_--(p_1)^2-(p_2)^2-m^2+i\epsilon}
\nonumber\\
&+\frac{\theta(x^+-y^+)(ip_+)e^{ip\cdot (x-y)}-\theta(-x^++y^+)(-ip_+)e^{-ip\cdot (x-y)}}{4p_+p_--(p_1)^2-(p_2)^2-m^2-i\epsilon}\bigg{]}.
\label{8.55a}
\end{align}
In each of the terms that contain a delta function the delta functions cause all the $\pm ip_+(x^+-y^+)$ terms in the exponents to vanish identically. However that then causes the residues at the poles in the $+i\epsilon$ and $-i\epsilon$ terms to be equal. Then since the $+i\epsilon$ term is closed below the real $p_+$ axis in a clockwise direction while the $-i\epsilon$ term is closed above the real $p_+$ axis in a counter-clockwise direction the two delta function terms cancel each other identically.

In order to make contact with (\ref{4.21a}), from which (\ref{4.22a}) follows, we apply $\partial/\partial y^-$ to the 
theta-function-dependent terms in (\ref{8.55a}). This yields
\begin{align}
&\frac{\partial}{\partial y^-}\frac{\partial}{\partial x^+}i\Delta(LF;x-y)=-\frac{1}{2\pi i}\frac{1}{4\pi^3}\int_{-\infty }^{\infty}dp_1\int_{-\infty }^{\infty}dp_2\int_{0}^{\infty}dp_-\int_{-\infty }^{\infty}dp_+
\nonumber\\
&\times\bigg{[}
\frac{\theta(x^+-y^+)p_+p_-e^{-ip\cdot (x-y)}-\theta(-x^++y^+)p_+p_-e^{ip\cdot (x-y)}}{4p_+p_--(p_1)^2-(p_2)^2-m^2+i\epsilon}
\nonumber\\
&+\frac{\theta(x^+-y^+)p_+p_-e^{ip\cdot (x-y)}-\theta(-x^++y^+)p_+p_-e^{-ip\cdot (x-y)}}{4p_+p_--(p_1)^2-(p_2)^2-m^2-i\epsilon}\bigg{]}.
\label{8.56a}
\end{align}
Since the only contributions to the contour integrals are poles we can replace the $p_+p_-$ terms in the numerators by $[(p_1)^2+(p_2)^2+m^2]/4$. We can then replace these terms by derivatives with respect to $x^1$ and $x^2$, and obtain
\begin{align}
&\frac{\partial}{\partial y^-}\frac{\partial}{\partial x^+}i\Delta(LF;x-y)=-\frac{1}{2\pi i}\frac{1}{4\pi^3}\int_{-\infty }^{\infty}dp_1\int_{-\infty }^{\infty}dp_2\int_{0}^{\infty}dp_-\int_{-\infty }^{\infty}dp_+\frac{1}{4}\left[-\frac{\partial}{\partial x^1}\frac{\partial}{\partial x^1}-\frac{\partial}{\partial x^2}\frac{\partial}{\partial x^2}+m^2\right]
\nonumber\\
&\times\bigg{[}
\frac{\theta(x^+-y^+)e^{-ip\cdot (x-y)}-\theta(-x^++y^+)e^{ip\cdot (x-y)}}{4p_+p_--(p_1)^2-(p_2)^2-m^2+i\epsilon}
\nonumber\\
&+\frac{\theta(x^+-y^+)e^{ip\cdot (x-y)}-\theta(-x^++y^+)e^{-ip\cdot (x-y)}}{4p_+p_--(p_1)^2-(p_2)^2-m^2-i\epsilon}\bigg{]}.
\label{8.57a}
\end{align}
Comparing with (\ref{8.22a}) we thus obtain 
\begin{align}
&\frac{\partial}{\partial y^-}\frac{\partial}{\partial x^+}i\Delta(LF;x-y)=\frac{1}{4}\left[-\frac{\partial}{\partial x^1}\frac{\partial}{\partial x^1}-\frac{\partial}{\partial x^2}\frac{\partial}{\partial x^2}+m^2\right]i\Delta(LF;x-y).
\label{8.58a}
\end{align}
Finally, on taking a $\partial/\partial x^-$ derivative and applying the $\Lambda^-$ projection operator to (\ref{8.58a}), from (\ref{8.52a}) we obtain 
\begin{align}
&\Big{\{}\frac{\partial}{\partial x^-}\psi_{\alpha}^{(-)}(x^+,x^1,x^2,x^-),\frac{\partial}{\partial y^-}[\psi_{(-)}^{\dagger}]_{\beta}(y^+,y^1,y^2,y^-)\Big{\}}
\nonumber\\
&=2i\Lambda^-_{\alpha\beta}\frac{1}{4}\left[-\frac{\partial}{\partial x^1}\frac{\partial}{\partial x^1}-\frac{\partial}{\partial x^2}\frac{\partial}{\partial x^2}+m^2\right]\frac{\partial}{\partial x^-}i\Delta(LF;x-y).
\label{8.59a}
\end{align}

At $x^+-y^+=0$ the quantity $(x-y)^2$ could only be lightlike or spacelike. But $i\Delta(LF;x-y)$ vanishes for spacelike separations, and thus at $x^+=y^+$ the quantity $(x-y)^2$ must be zero. With $i\Delta(LF;(x-y)^2=0)$ being given in  (\ref{8.27a}), and with (\ref{8.27a}) being rewritable as 
$i\Delta(LF;(x-y)^2=0)=-(i/2\pi)\epsilon(x^--y^-)\delta[(x^+-y^+)(x^--y^-)-(x^1-y^1)^2-(x^2-y^2)^2]$ since the delta function forces $x^+-y^+$ and $x^--y^-$ to have the same sign, on the light cone the equal light-front time bad fermion anticommutator evaluates to
\begin{align}
&\Big{\{}\frac{\partial}{\partial x^-}\psi_{\alpha}^{(-)}(x^+,x^1,x^2,x^-),\frac{\partial}{\partial y^-}[\psi_{(-)}^{\dagger}]_{\beta}(x^+,y^1,y^2,y^-)\Big{\}}
\nonumber\\
&=2i\Lambda^-_{\alpha\beta}\frac{1}{4}\left[-\frac{\partial}{\partial x^1}\frac{\partial}{\partial x^1}-\frac{\partial}{\partial x^2}\frac{\partial}{\partial x^2}+m^2\right]\frac{\partial}{\partial x^-}\left[-\frac{i}{2\pi}\epsilon(x^--y^-)\frac{\pi}{2}\delta(x^1-y^1)\delta(x^2-y^2)\right]
\nonumber\\
&=\frac{1}{4}\Lambda^-_{\alpha\beta}\left[-\frac{\partial}{\partial x^1}\frac{\partial}{\partial x^1}-\frac{\partial}{\partial x^2}\frac{\partial}{\partial x^2}+m^2\right]\delta(x^--y^-)\delta(x^1-y^1)\delta(x^2-y^2).
\label{8.60a}
\end{align}
We recognize (\ref{8.60a}) as (\ref{4.21a}). Finally,  we state without proof that the  $\big{\{}[\psi_{(+)}]_{\alpha}(x^+,x^1,x^2,x^-),[\psi_{(-)}^{\dagger}]_{\beta}(x^+,y^1,y^2,y^-)\big{\}}$ anticommutator given in (\ref{4.23a}) (and thus also its conjugate $\big{\{}[\psi_{(-)}]_{\beta}(x^+,y^1,y^2,y^-),[\psi_{(+)}^{\dagger}]_{\alpha}(x^+,x^1,x^2,x^-)\big{\}}$) can also be derived this way. With the good and bad fermions together being complete since $\Lambda^++\Lambda^-=I$, we thus confirm that the expression for the light-front $\big{\{}\psi_{\alpha}(x^+,x^1,x^2,x^-), \psi_{\beta}^{\dagger}(y^+,y^1,y^2,y^-)\big{\}}$ given in (\ref{8.50a}) is indeed valid. Moreover, not only do we confirm the validity of (\ref{8.50a}), we see how it is possible to generate the derivative terms that appear in (\ref{8.60a}) even though no derivative terms appear in (\ref{8.50a}) itself. Given the now established validity of (\ref{8.50a}) we note in passing that the fourth relation given in (\ref{4.13a}), viz. $\Big{\{}\psi_{(+)}(x^+,x^1,x^2,x^-),\partial_+\psi_{(+)}(x^+,y^1,y^2,y^-)\Big{\}}=0$, the one in (\ref{4.13a}) that is not given by the equal $x^+$ anticommutators, can directly be obtained from (\ref{8.50a}). Finally, comparing (\ref{8.50a}) with the instant-time $\big{\{}\psi_{\alpha}(x^0,x^1,x^2,x^3), \psi_{\beta}^{\dagger}(y^0,y^1,y^2,y^3)\big{\}}$ given in (\ref{8.34a}), we see that the discussion can now completely parallel the scalar field case. We thus establish that even with non-invertible projection operators, just as in the scalar field case, analogously for fermions  light-front anticommutators are equivalent to  instant-time anticommutators.

\subsection{Abelian Gauge Field Case}
\label{S8e}

With gauge fixing the instant-time Abelian gauge field equal-time commutator takes the form (see e.g. \cite{Itzykson1980})
\begin{align}
&[A_{\nu}(x^+,x^1,x^2,x^-),\partial_0A_{\mu}(x^0,y^1,y^2,y^3)]=-ig_{\mu\nu}\delta(x^1-y^1)\delta(x^2-y^2)\delta(x^3-y^3),
\label{8.61a}
\end{align}
while the equation of motion is given by $\partial_{\nu}\partial^{\nu}A_{\mu}=0$.
Consequently, in analog to the scalar field case, one has unequal-time instant-time commutation relations of the form (see e.g. \cite{Itzykson1980})
\begin{align}
[A_{\nu}(x^0,x^1,x^2,x^3),A_{\mu}(y^0,y^1,y^2,y^3)]&=-ig_{\mu\nu}\Delta(IT;x-y)
\nonumber\\
&=\frac{i}{2\pi}g_{\mu\nu}\epsilon(x^0-y^0)\delta[(x^0-y^0)^2-(x^1-y^1)^2-(x^2-y^2)^2-(x^3-y^3)^2],
\label{8.62a}
\end{align}
where $g_{\mu\nu}$ is the instant-time metric and $i\Delta(IT;x-y)$ is the scalar field $i\Delta(IT;x-y)$ as given in the mass-independent (\ref{8.7}).

Since  the instant-time commutation relation given in  (\ref{8.61a}) and the instant-time wave equation are both diagonal in the gauge field indices,  we can proceed just as in the scalar field case, and from (\ref{8.62a}) we obtain equal light-front time commutation relations of the form 
\begin{align}
&[A_{\nu}(x^+,x^1,x^2,x^-),\partial_-A_{\mu}(x^+,y^1,y^2,y^-)]=-\frac{i}{2}g_{\mu\nu}\delta(x^1-y^1)\delta(x^2-y^2)\delta(x^--y^-),
\nonumber\\
&[A_{\nu}(x^+,x^1,x^2,x^-), A_{\mu}(x^+,y^1,y^2,y^-)]=\frac{i}{4}g_{\mu\nu}\epsilon(x^--y^-)\delta(x^1-y^1)\delta(x^2-y^2).
\label{8.63a}
\end{align}
We recognize (\ref{8.63a}) as (\ref{5.34}), with, as noted in \cite{Mannheim2019b,Mannheim2020a},  equal light-front time gauge field commutators thus being derivable starting from unequal-time instant-time gauge field commutators.

\subsection{Non-Abelian Gauge Field Case}
\label{S8f}

In a non-Abelian Yang-Mills case with gauge fixing one simply duplicates the Abelian case for the $A_{\mu}^a$, since as well as being diagonal in the Lorentz indices the equations of motion and equal-time instant-time $A_{\mu}^a$ commutators are diagonal in the group indices. Thus with (\ref{8.62a}) generalizing to 
\begin{align}
&[A^a_{\nu}(x^0,x^1,x^2,x^3),A^b_{\mu}(y^0,y^1,y^2,y^3)]=-ig_{\mu\nu}\delta_{ab}\Delta(IT;x-y),
\label{8.64a}
\end{align}
the equal light-front time commutators given in (\ref{5.38}), viz.
\begin{align}
&[A^a_{\nu}(x^+,x^1,x^2,x^-),\partial_-A^b_{\mu}(x^+,y^1,y^2,y^-)]=-\frac{i}{2}g_{\mu\nu}\delta^{ab}\delta(x^1-y^1)\delta(x^2-y^2)\delta(x^--y^-),
\nonumber\\
&[A^a_{\nu}(x^+,x^1,x^2,x^-), A^b_{\mu}(x^+,y^1,y^2,y^-)]=\frac{i}{4}g_{\mu\nu}\delta^{ab}\epsilon(x^--y^-)\delta(x^1-y^1)\delta(x^2-y^2),
\label{8.65a}
\end{align}
then follow.

For the Faddeev-Popov ghosts the canonical equal-time instant-time anticommutators are of the form
\begin{align}
&\Big{\{}c_a(x^+,x^1,x^2,x^-),\partial_0\bar{c}_b(x^+,y^1,y^2,y^-)\Big{\}}=i\delta_{ab}\delta(x^1-y^1)\delta(x^2-y^2)\delta(x^--y^-),
\nonumber\\
&\Big{\{}\bar{c}_a(x^+,x^1,x^2,x^-),\partial_0c_b(x^+,y^1,y^2,y^-)\Big{\}}=i\delta_{ab}\delta(x^1-y^1)\delta(x^2-y^2)\delta(x^--y^-),
\label{8.66a}
\end{align}
with the unequal-time instant-time anticommutator being of the form
\begin{align}
&\Big{\{}c_a(x^+,x^1,x^2,x^-),\bar{c}_b(y^+,y^1,y^2,y^-)\Big{\}}=i\delta_{ab}\Delta(IT;x-y).
\label{8.67a}
\end{align}
Consequently, the equal light-front time anticommutators are given just as in (\ref{5.39}), viz.
\begin{align}
&\Big{\{}c_a(x^+,x^1,x^2,x^-), 2\partial_-\bar{c}_b(x^+,y^1,y^2,y^-)\Big{\}}=i\delta_{ab}\delta(x^1-y^1)\delta(x^2-y^2)\delta(x^--y^-),
\nonumber\\
&\Big{\{}\bar{c}_a(x^+,x^1,x^2,x^-), 2\partial_-c_b(x^+,y^1,y^2,y^-)\Big{\}}=i\delta_{ab}\delta(x^1-y^1)\delta(x^2-y^2)\delta(x^--y^-).
\label{8.68a}
\end{align}
Thus, as noted in \cite{Mannheim2019b},  in the non-Abelian case equal light-front time commutators and anticommutators can be derived starting from unequal-time instant-time commutators and anticommutators.
Thus in all the scalar, fermion, and gauge field cases we see that canonical light-front commutation or anticommutation relations follow from  instant-time commutation or anticommutation relations and do not need to be independently postulated. 

To conclude this section, we note that while we have only discussed the equivalence of instant-time and light-front  canonical quantization procedures  for free fields, our results immediately apply to interacting theories. Specifically, since perturbative interactions cannot change a Hilbert space, once we show that the free instant-time and free light-front theories are in the same Hilbert space (as their (unequal time) commutators and anticommutators are related purely by kinematic coordinate transformations),  it follows that the interacting theories are in the same Hilbert space too. In fact in general we note that because of the general coordinate invariance of quantum theory, any two directions of quantization that are related by a general coordinate transformation must be equivalent. Since the transformation $x^0\rightarrow x^0+x^3=x^+$ is one such transformation, it follows that light-front quantization is instant-time quantization. We now examine this equivalence in the interacting case in more detail by studying the Lehmann representation. 

\subsection{Extension to Interacting Theories and the Lehmann Representation}
\label{Sg}

Having now established the equivalence of instant-time and light-front commutators in the free field theory case we comment briefly on how these results generalize to the interacting case.  For commutators (and analogously for anticommutators) the generalization is given by the Lehmann representation, an exact all-order relation in quantum field theory. The Lehmann representation  is derived using only some very basic requirements that are thought to occur in any quantum field theory, requirements that do not get modified by the renormalization procedure. These requirements are Poincare invariance, Hermiticity of the momentum generators and thus the completeness of their eigenstates and the reality of their eigenvalues, the  existence of a lowest energy vacuum state, and the uniqueness of that vacuum state. As such, these requirements make no reference to any dynamical equation that might be obeyed by the quantum fields of interest, while also not being restricted to perturbation theory. Moreover, for the purposes of deriving the Lehmann representation there is no need to actually determine any of the eigenvalues of the momentum generators or construct any of the associated eigenstates. All that matters is that they exist, and that must be the case if the momentum generators are Hermitian.

For the instant-time case first, following the discussion of the Lehmann representation given in \cite{Bjorken1965} we introduce the momentum generators $P_{\mu}=\int dx^1dx^2dx^3 T^{0}_{\phantom{0}\mu}$, with a Hermitian scalar field then transforming according to 
\begin{eqnarray}
[P_{\mu},\phi(x)]=-i\partial_{\mu}\phi(x),\quad \phi(x)=e^{iP\cdot x}\phi(0)e^{-iP\cdot x}.
\label{8.69}
\end{eqnarray}
We introduce a complete set of eigenstates $|p^n_{\mu}\rangle$ of the $P_{\mu}$ momentum generators, and other than the vacuum $|\Omega\rangle$ (an eigenstate of $P_{\mu}$ with all $p^n_{\mu}=0$), all the other states lie above the vacuum and have eigenvalues $p^n_{\mu}$ with  positive $p_0^n$ and non-negative $p^n_{\mu}p^{\mu}_n$, with the matrix element of $\phi(x)$ between $\langle \Omega|$ and  $|p_{\mu}^n\rangle$ being of the form
\begin{eqnarray}
&&\langle \Omega|\phi(x)|p^n_{\mu}\rangle=\langle \Omega|\phi(0)|p^n_{\mu}\rangle e^{-ip_{\mu}^n\cdot x},\quad \langle p_{\mu}^n|\phi(x)|\Omega\rangle=\langle p_{\mu}^n|\phi(0)|\Omega\rangle e^{ip_{\mu}^n\cdot x}.
\label{8.70}
\end{eqnarray}
With the states obeying the closure relation 
\begin{eqnarray}
\sum_n|n\rangle\langle n|=I,
\label{8.71}
\end{eqnarray}
we can then write the two-point function as
\begin{eqnarray}
\langle \Omega |\phi(x)\phi(y)|\Omega\rangle=\sum_n|\langle \Omega |\phi(0)|p_{\mu}^n\rangle |^2e^{-ip^n\cdot(x-y)}.
\label{8.72}
\end{eqnarray}
While the closure sum on $n$ given in (\ref{8.71}) must include the vacuum (in order to enforce $\sum_n|n\rangle\langle n|\Omega\rangle=|\Omega\rangle$), the vacuum does not contribute in the sum on $n$ in (\ref{8.72}), since the uniqueness of the vacuum requires that $\langle \Omega |\phi(0)|\Omega\rangle$ be zero as one otherwise would have  spontaneous symmetry breaking and a degenerate vacuum.

We now  introduce the instant time ($IT$) spectral function
\begin{eqnarray}
\rho(q_{\mu},IT)=(2\pi)^3\sum_n\delta^4(p_{\mu}^n-q_{\mu})|\langle \Omega |\phi(0)|p_{\mu}^n\rangle |^2=\rho(q^2,IT)\theta(q_0).
\label{8.73}
\end{eqnarray}
Because of Lorentz invariance the spectral function is only a function of $q^2$, and with all the $p_0^n$ being positive, $\rho(q_{\mu},IT)$ can be written as $\rho(q_{\mu},IT)=\rho(q^2,IT)\theta(q_0)$, with $\rho(q^2,IT)$  vanishing for $q^2<0$ since all the $p^n_{\mu}$ have non-negative $p_{\mu}^np^{\mu}_n$. In terms of $\rho(q_{\mu},IT)$ we obtain the two-point function
Lehmann representation 
\begin{eqnarray}
\langle \Omega |\phi(x)\phi(y)|\Omega\rangle&=&\frac{1}{(2\pi)^3}\int d^4q \rho(q^2,IT)\theta(q_0)e^{-iq\cdot(x-y)}
\nonumber\\
&=&\frac{1}{(2\pi)^3}\int_0^{\infty}d\sigma^2\rho(\sigma^2,IT)\int d^4q \theta(q_0)\delta(q^2-\sigma^2)e^{-iq\cdot(x-y)},
\label{8.74}
\end{eqnarray}
where $\sigma$ is a mass parameter. Thus with $\epsilon(q_0)=\theta(q_0)-\theta(-q_0)$, the vacuum matrix element of the commutator is given by 
\begin{eqnarray}
\langle \Omega |[\phi(x),\phi(y)]|\Omega\rangle=\frac{1}{(2\pi)^3}\int_0^{\infty}d\sigma^2\rho(\sigma^2,IT)\int d^4q \epsilon(q_0)\delta(q^2-\sigma^2)e^{-iq\cdot(x-y)}.
\label{8.75}
\end{eqnarray}

Now in (\ref{8.5}) we expressed the free theory $i\Delta(IT;x-y)$ for a scalar field of mass $m$ as an on shell three-dimensional integral. On introducing a delta function we can rewrite it as a still on-shell four-dimensional integral. And relabeling it as $i\Delta(IT,FREE;x-y,m^2)$ we rewrite (\ref{8.5}) as
\begin{eqnarray}
i\Delta(IT,FREE;x-y,m^2)=\frac{1}{(2\pi)^3}\int d^4q\epsilon(q_0)\delta(q^2-m^2)e^{-iq\cdot(x-y)},
\label{8.76}
\end{eqnarray}
with its evaluation for $(x-y)^2>0$, $(x-y)^2=0$ and $(x-y)^2<0$ being given in (\ref{8.6}), (\ref{8.7}) and (\ref{8.8a}).
Labeling the full commutator term on the left-hand side of (\ref{8.75}) as $i\Delta(IT,FULL;x-y)$, we can thus rewrite (\ref{8.75}) as
\begin{eqnarray}
i\Delta(IT,FULL;x-y)=\int_0^{\infty} d\sigma^2 \rho(\sigma^2,IT)i\Delta(IT,FREE;x-y,\sigma^2),
\label{8.77}
\end{eqnarray}
with the range of $\sigma^2$ being restricted to $(0,\infty)$ since $\rho(\sigma^2,IT)$ vanishes outside the light cone.
With recognize (\ref{8.77}) as  the Lehmann representation for the $x^0\neq y^0$ commutator  \cite{Bjorken1965}.

For the light-front ($LF$) case we again assume Poincare invariance and Hermiticity of the momentum generators. With the requirement that all $p_{\mu}^np_n^{\mu}$ be non-negative, it follows that the instant-time momenta obey $(p^n_0)^2-(p^n_3)^2-(p^n_1)^2-(p^n_2)^2\geq 0$, with light-front momenta thus obeying $4p^n_+p^n_--(p^n_1)^2-(p^n_2)^2\geq 0$. Then with $p_0^n>|p_3^n|$ it follows that $p_+^n=p_0^n+p_3^n$ and $p_-^n=p_0^n-p_3^n$ are both positive. Boundedness from below of the instant-time Hamiltonian thus entails boundedness from below of the light-front Hamiltonian  as well.  

To now establish the Lehmann representation  in the light-front case we  need to write the three-dimensional free light-front theory $x^+\neq y^+$ commutator $ i\Delta(LF,FREE;x-y,m^2)$ given in (\ref{8.21a}) in a four-dimensional form. We anticipate that it will be the analog of (\ref{8.76}) and check to see if this is the case. We thus set
\begin{eqnarray}
&&i\Delta(LF,FREE;x-y,m^2)=\frac{2}{(2\pi)^3}\int_{-\infty}^{\infty} dq_1dq_2dq_+dq_-\epsilon(q_+)\delta(4q_+q_--F_q^2)
\nonumber\\
&&\times e^{-iq_+(x^+-y^+)-iq_-(x^--y^-)-iq_1(x^1-y^1)-iq_2(x^2-y^2)},
\label{8.78}
\end{eqnarray}
where $F_q^2=(q_1)^2+(q_2)^2+m^2$. With the delta function term requiring that $q_+q_-$ be positive, then given the $\epsilon(q_+)$ term  the integration breaks up into two pieces:
\begin{eqnarray}
i\Delta(LF,FREE;x-y,m^2)&=&\frac{2}{(2\pi)^3}\int_{-\infty}^{\infty} dq_1dq_2\int_0^{\infty} dq_+dq_-\delta(4q_+q_--F_q^2)e^{-iq\cdot(x-y)}
\nonumber\\
&-&\frac{2}{(2\pi)^3}\int_{-\infty}^{\infty} dq_1dq_2\int _{-\infty}^0dq_+dq_-\delta(4q_+q_--F_q^2)e^{-iq\cdot(x-y)}.
\label{8.79}
\end{eqnarray}
Setting $q_{\mu}=-q_{\mu}$ in the second integral then yields
\begin{eqnarray}
i\Delta(LF,FREE;x-y,m^2)&=&\frac{2}{(2\pi)^3}\int_{-\infty}^{\infty} dq_1dq_2\int_0^{\infty} dq_+dq_-\delta(4q_+q_--F_q^2)e^{-iq\cdot(x-y)}
\nonumber\\
&-&\frac{2}{(2\pi)^3}\int_{-\infty}^{\infty} dq_1dq_2\int _0^{\infty}dq_+dq_-\delta(4q_+q_--F_q^2)e^{iq\cdot(x-y)}.
\label{8.80}
\end{eqnarray}
Finally, doing the $q_+$ integration yields
\begin{eqnarray}
i\Delta(LF,FREE;x-y,m^2)&=&\frac{2}{(2\pi)^3}\int_{-\infty}^{\infty}dq_1\int_{-\infty}^{\infty}dq_2\int_{0}^{\infty}\frac{dq_-}{4q_-}
\nonumber\\
&&\times
\left[e^{-i(F_q^2x^+/4q_-+q_-x^-+q_1x^1+q_2x^2)}-e^{i(F_q^2x^+/4q_-+q_-x^-+q_1x^1+q_2x^2)}\right].
\label{8.81}
\end{eqnarray}
We recognize (\ref{8.81}) as (\ref{8.21a}), and thus confirm the validity of (\ref{8.78}), with its evaluation for $(x-y)^2>0$ , $(x-y)^2=0$ and $(x-y)^2<0$ being given in (\ref{8.27a}). 

Defining now a light-front spectral function
\begin{eqnarray}
\rho(q_{\mu},LF)=\frac{(2\pi)^3}{2}\sum_n\delta^4(p_{\mu}^n-q_{\mu})|\langle \Omega |\phi(0)|p_{\mu}^n\rangle |^2=\rho(q^2,LF)\theta(q_+),
\label{8.82}
\end{eqnarray}
then with (\ref{8.72}) also holding in light-front coordinates, the $x^+\neq y^+$ light-front coordinate commutator is given by
\begin{eqnarray}
\langle \Omega |[\phi(x),\phi(y)]|\Omega\rangle=\frac{2}{(2\pi)^3}\int_0^{\infty}d\sigma^2\rho(\sigma^2,LF)\int d^4q \epsilon(q_+)\delta(q^2-\sigma^2)e^{-iq\cdot(x-y)}.
\label{8.83}
\end{eqnarray}
Thus in the light-front case the Lehmann representation takes the form 
\begin{eqnarray}
i\Delta(LF,FULL;x-y)=\int_0^{\infty} d\sigma^2 \rho(\sigma^2,LF)i\Delta(LF,FREE;x-y,\sigma^2).
\label{8.84}
\end{eqnarray}

We recognize the light-front (\ref{8.84}) as being completely analogous to the instant-time Lehmann representation given in (\ref{8.77}), and thus anticipate that one can be transformed into the other. We will need to transform both the free $i\Delta(IT,FREE;x-y,\sigma^2)$ and the spectral function. For the free $i\Delta(IT,FREE;x-y,\sigma^2)$ given in (\ref{8.76}) we  set $p_0=p_++p_-$, $p_3=p_+-p_-$. This transforms $(p_0)^2-(p_3)^2-(p_1)^2-(p_2)^2-m^2$ into $4p_+p_--(p_1)^2-(p_2)^2-m^2$. From the delta function constraint it follows that $p_+p_-$ is positive, with $p_+$ and $p_-$ thus having  the same sign. Consequently, with  $\epsilon(p_0)$ transforming into $\epsilon(p_++p_-)$ it follows that $\epsilon(p_++p_-)\delta(p^2-m^2)$ is equal to $\epsilon(p_+)\delta(p^2-m^2)$ alone. Then, after making an analogous transformation on $x^{\mu}-y^{\mu}$ of the form $x^0-y^0=(x^+-y^++x^--y^-)/2$, $x^3-y^3=(x^+-y^+-x^-+y^-)/2$, the equivalence of $i\Delta(IT,FREE;x-y,\sigma^2)$ as given in (\ref{8.76}) and $i\Delta(LF,FREE;x-y,\sigma^2)$ as given in (\ref{8.78}) is established. 

To compare the spectral functions we first write the free theory Fock expansions of (\ref{1.8}) and (\ref{1.17a}) in a four-dimensional form. We introduce $A(IT;\vec{p}=(p_1,p_2,p_3))=(2E_p)^{1/2}a(IT;\vec{p})$ in the instant-time case \cite{Bjorken1965} and $A(LF;\vec{p}=(p_1,p_2,p_-))=(4p_-)^{1/2}a(LF;\vec{p})$ in the light-front case, both as confined to their respective mass shells, $(p_0)^2=E_p^2$, $4p_+p_-=F_p^2$. These creation and annihilation operators obey
\begin{eqnarray}
&&[A(IT;\vec{p}),A^{\dagger}(IT;\vec{p}^{\prime})]=2E_p\delta(p_1-p_1^{\prime})\delta(p_2-p_2^{\prime})\delta(p_3-p_3^{\prime}),
\nonumber\\
&&[A(LF;\vec{p}),A^{\dagger}(LF;\vec{p}^{\prime})]=4p_-\tfrac{1}{2}\delta(p_1-p_1^{\prime})\delta(p_2-p_2^{\prime})\delta(p_--p_-^{\prime}).
\label{8.85}
\end{eqnarray}
The Fock space expansions thus take the form 
\begin{eqnarray}
\phi(IT;x^0,x^1,x^2,x^3)&=&\frac{1}{(2\pi)^{3/2}}\int d^4p \delta((p_0)^2-E_p^2)\theta(p_0)\left[A(IT;\vec{p})e^{-ip\cdot x}
+A^{\dagger}(IT;\vec{p})e^{+ip\cdot x}\right]
\nonumber\\
&=&\frac{1}{(2\pi)^{3/2}}\int d^4p \delta((p_0)^2-E_p^2)\left[\theta(p_0)A(IT;\vec{p})
+\theta(-p_0)A^{\dagger}(IT;-\vec{p})\right]e^{-ip\cdot x}.
\label{8.86}
\end{eqnarray}
\begin{eqnarray}
\phi(LF;x^+,x^1,x^2,x^-)&=&\frac{2}{(2\pi)^{3/2}}\int_{-\infty}^{\infty}dp_1dp_2\int_0^{\infty}dp_+dp_- \delta(4p_+p_--F_p^2)A(LF;\vec{p})e^{-ip\cdot x}
\nonumber\\
&+&\frac{2}{(2\pi)^{3/2}}\int_{-\infty}^{\infty}dp_1dp_2\int_0^{\infty}dp_+dp_- \delta(4p_+p_--F_p^2)A^{\dagger}(LF;\vec{p})e^{ip\cdot x}
\nonumber\\
&=&\frac{2}{(2\pi)^{3/2}}\int_{-\infty}^{\infty}dp_1dp_2\int_0^{\infty}dp_+dp_- \delta(4p_+p_--F_p^2)A(LF;\vec{p})e^{-ip\cdot x}
\nonumber\\
&+&\frac{2}{(2\pi)^{3/2}}\int_{-\infty}^{\infty}dp_1dp_2\int_{-\infty}^0 dp_+dp_- \delta(4p_+p_--F_p^2)A^{\dagger}(LF;-\vec{p})e^{-ip\cdot x}
\nonumber\\
&=&\frac{2}{(2\pi)^{3/2}}\int d^4p \delta(4p_+p_--F_p^2)\left[\theta(p_+)A(LF;\vec{p})
+\theta(-p_+)A^{\dagger}(LF;-\vec{p})\right]e^{-ip\cdot x}.
\label{8.87}
\end{eqnarray}
The utility of these expressions is that they show that $A(IT;\vec{p})$ and $A(LF;\vec{p})$ are Lorentz scalars, as must be the case since the two delta function terms in (\ref{8.85}) are themselves Lorentz scalars  ($\int d^3p/2E_p\times 2E_p\delta^3(\vec{p}-\vec{p}^{\prime})=1$, $2\int dp_1dp_2\int_0^{\infty}dp_-/4p_-\times 4p_-\tfrac{1}{2}\delta(p_1-p_1^{\prime})\delta(p_2-p_2^{\prime})\delta(p_--p_-^{\prime})=1$.) In Sec. \ref{S16} below we will show that $dp_3/2E_p$  can be boosted into $dp_-/2p_-$ by an infinite momentum frame boost.

With the transformation $x^0\rightarrow x^0+x^3$, $x^3\rightarrow x^0-x^3$ being a general coordinate translation on the coordinates,  at the operator level we can transform $\phi(IT;x)$ to $\phi(LF;x)$ via the translation operator introduced in (\ref{7.5}), viz. 
\begin{eqnarray}
U(P_0,P_3)=\exp(ix^3P_0)\exp(ix^0P_3),
\label{8.88}
\end{eqnarray}
where the $P_{\mu}$ are momentum generators that effect $[P_{\mu},\phi(x)]=-i\partial_{\mu}\phi$. In order to apply this transformation to (\ref{8.86}),  we note that because we use $x^3\rightarrow x^0-x^3=x^-$ rather than $x^3\rightarrow x^3-x^0=-x^-$, then rather than reformulate everything in terms of   $x^3\rightarrow x^3-x^0$, we shall restrict to theories in which the action is invariant under $x^3\rightarrow -x^3$ (i.e., actions with only even powers of $\phi$), so that we can set  $\phi(IT;x^0,x^1,x^2,x^3)=\phi(IT;x^0,x^1,x^2,-x^3)$. On now applying (\ref{8.88})  to the left-hand side of (\ref{8.86}) we obtain
\begin{align}
U\phi(IT;x^0,x^1,x^2,x^3)U^{-1}=U\phi(IT;x^0,x^1,x^2,-x^3)U^{-1}=\phi(IT;x^0+x^3,x^1,x^2,x^0-x^3)=\phi(LF;x^+,x^1,x^2,x^-).
\label{8.89}
\end{align}
On substituting $p_0=p_++p_-$, $p_3=p_+-p_-$ into the right-hand side of (\ref{8.86}) we obtain 
\begin{eqnarray}
\phi(LF;x^+,x^1,x^2,x^-)&=&\frac{2}{(2\pi)^{3/2}}\int d^4p \delta(4p_+p_--F_p^2)
\nonumber\\
&&\times\left[\theta(p_++p_-)UA(IT;\vec{p})U^{-1}
+\theta(-p_+-p_-)UA^{\dagger}(IT;-\vec{p})U^{-1}\right]e^{-ip\cdot x},
\label{8.90}
\end{eqnarray}
as now written in light-front coordinates. Then, since $\delta(4p_+p_--F_p^2)\theta(\pm (p_++ p_-))=\delta(4p_+p_--F_p^2)\theta(\pm p_+)$, we obtain
\begin{align}
\phi(LF;x^+,x^1,x^2,x^-)=\frac{2}{(2\pi)^{3/2}}\int d^4p \delta(4p_+p_--F_p^2)\left[\theta(p_+)UA(IT;\vec{p})U^{-1}
+\theta(-p_+)UA^{\dagger}(IT;-\vec{p})U^{-1}\right]e^{-ip\cdot x}.
\label{8.91}
\end{align}
Finally, comparing with (\ref{8.87}) and recalling that $U$ is unitary, we obtain
\begin{eqnarray}
UA(IT; \vec{p})U^{-1}=A(LF;\vec{p}),\quad UA^{\dagger}(IT; -\vec{p})U^{-1}=A^{\dagger}(LF;-\vec{p}),\quad UA^{\dagger}(IT; \vec{p})U^{-1}=A^{\dagger}(LF;\vec{p}).
\label{8.92}
\end{eqnarray}
The instant-time and light-front scalar field and the associated creation and annihilation operators are thus unitarily equivalent. With 
\begin{align}
A(IT; \vec{p})|\Omega(IT)\rangle=0,&\quad A^{\dagger}(IT;\vec{p})|\Omega(IT)\rangle=|p_{\mu}(IT)\rangle,
\nonumber\\
A(LF; \vec{p})|\Omega(LF)\rangle=0,&\quad A^{\dagger}(LF;\vec{p})|\Omega(LF)\rangle=|p_{\mu}(LF)\rangle,
\label{8.93}
\end{align}
we obtain
\begin{eqnarray}
&&U|\Omega(IT)\rangle=|\Omega(LF)\rangle,\quad U|p_{\mu}(IT)\rangle=|p_{\mu}(LF)\rangle,\quad
U\phi(IT;0)U^{-1}=\phi(LF;0).
\label{8.94}
\end{eqnarray}
Given (\ref{8.94}) we thus establish the unitary equivalence of the instant-time and light-front spectral functions given in (\ref{8.73}) and (\ref{8.82}) and thus of the Lehmann representations in the free theory case. Since translations are general coordinate transformations, our result follows from the general coordinate invariance of the scalar field theory. And as is standard in quantum theory, invariance under translations entails invariance under the unitary transformations associated with the translation generators, with instant-time quantization and light--front quantization thus being unitarily equivalent. 

Finally, for the full interacting Lehmann representation, we can transform with the unitary $U(P_0,P_3)$ operator to all orders since the full fields and full momentum operators obey the relation $[P_{\mu},\phi]=-i\partial_{\mu}\phi$ given in (\ref{8.69}) that was initially  used for establishing the Lehmann representation in the first place. The general coordinate invariance of the interacting theory, which we take to be the case,  then establishes the unitary equivalence of the interacting instant-time and light-front theories, with (\ref{8.94}) actually holding to all orders and not just for the free theory. Consequently, we see that 
\begin{align}
&\langle \Omega(IT)|\phi(IT;x^0,x^1,x^2,x^3)|p^n_{\mu}(IT)\rangle=\langle \Omega(IT)|\phi(IT;x^0,x^1,x^2,-x^3)|p^n_{\mu}(IT)\rangle
\nonumber\\
&=\langle \Omega(IT)|U^{\dagger}U\phi(IT;x^0,x^1,x^2,-x^3)U^{\dagger}U|p^n_{\mu}(IT)\rangle=\langle \Omega(LF)|\phi(LF;x^+,x^1,x^2,x^-)|p^n_{\mu}(LF)\rangle.
\label{8.95}
\end{align}
We thus establish the equivalence of the interacting spectral functions, and thus the equivalence of the full Lehmann representations. We thus see that to all orders the vacuum expectation values of the commutators in the instant-time and light-front cases are the same. Thus by working with the commutators at unequal $x^0-y^0$ and unequal $x^+-y^+$ we can establish the equivalence of instant-time and light-front quantization to all orders. This generalizes the free theory result that had been given above, and again shows the centrality of the general unequal time commutators. An analogous Lehmann representation also exists for anticommutators but we do not discuss it here. Discussion of the large $\sigma^2$ behavior of the light-front Lehmann representation spectral function may be found in \cite{Schlieder1972}.

\subsection{Comparing the Instant-time and Light-front Hamiltonians}
\label{S8h}

Unlike commutators, which are local objects that are defined at local coordinates $x^{\mu}$ and $y^{\mu}$, as we had noted in Sec. \ref{S1},  Hamiltonian operators are global objects as they are integrals over all space of energy-momentum tensor operators, with spatially asymptotic boundary conditions being needed in order to show that they are time independent. Comparison of instant-time and light-front Hamiltonians thus requires some global information. To see what is needed we construct the global Hamiltonians from the local energy-momentum tensors. Transformations from instant-time to light-front coordinates of the form  $x^0\rightarrow x^0+x^3$,   $x^3\rightarrow x^0-x^3$ are spacetime-dependent translations, and are thus general coordinate transformations. To construct the energy-momentum tensor one varies a covariantized action with respect to the metric. Since the action is general coordinate invariant, the instant-time and light-front energy-momentum tensors constructed this way are coordinate equivalent, and in this local sense they are completely equivalent. As constructed, both of the instant-time and light-front energy-momentum tensors obey $\partial_{\mu}T^{\mu\nu}=0$. The respective Hamiltonians are  then constructed as 
\begin{eqnarray}
H(IT)=\int dx^1dx^2dx^3T^0_{\phantom{0}0},\quad H(LF)=\tfrac{1}{2}\int dx^1dx^2dx^-T^+_{\phantom{+}+}.
\label{8.96}
\end{eqnarray}
Since $\partial_{\mu}T^{\mu}_{\phantom{\mu}\nu}=0$ the Hamiltonians obey
\begin{eqnarray}
&&\partial_0H(IT)=-\int dx^1dx^2dx^3\left[\partial_1T^{1}_{\phantom{1}0}+\partial_2T^{2}_{\phantom{2}0}+\partial_3T^{3}_{\phantom{3}0}\right],
\nonumber\\
&&\partial_+H(IT)=-\tfrac{1}{2}\int dx^1dx^2dx^-\left[\partial_1T^{1}_{\phantom{1}+}+\partial_2T^{2}_{\phantom{2}+}+\partial_-T^{-}_{\phantom{-}+}\right].
\label{8.97}
\end{eqnarray}
With both of the integrals on the right-hand sides of (\ref{8.97}) being asymptotic surface terms, to show that $H(IT)$ and $H(LF)$ are respectively independent of $x^0$ and $x^+$ requires different global boundary conditions (conditions that we assume to hold), namely asymptotic vanishing in $x^1$, $x^2$ and $x^3$ for $H(IT)$ and asymptotic vanishing in $x^1$, $x^2$ and $x^-$ for $H(LF)$. Now, as such, these  boundary conditions do not transform into each other because $x^1$, $x^2$ and $x^3$ can be reexpressed as $x^1$, $x^2$ and $(x^+-x^-)/2$ and not as $x^1$, $x^2$ and $x^-$. Moreover, the functions $T^{1}_{\phantom{1}0}$, $T^{2}_{\phantom{2}0}$ and  $T^{3}_{\phantom{3}0}$ are different from $T^{1}_{\phantom{1}+}$, $T^{2}_{\phantom{2}+}$ and  $T^{-}_{\phantom{-}+}$. (As noted in Sec. \ref{S4}, for fermions these light-front components even contain the non-local bad fermions.) Thus even though one can transform the local energy-momentum tensors into each other one cannot transform the Hamiltonians into each other. In this respect then $H(IT)$ and $H(LF)$ are intrinsically different. 

As we noted in Sec. \ref{S1}, the canonical equal instant-time and equal light-front time commutators are also intrinsically different. However, as we had shown in Sec. \ref{S3} above, it is this very difference that actually enables both instant-time and light-front momentum generators to obey $[P_{\mu},\phi]=-i\partial_{\mu}\phi$. Thus despite there being both global differences (asymptotic boundary conditions) and local differences (canonical commutators), nonetheless one does find some commonality, namely $H(IT)$ generates translations in $x^0$ and its light-front $H(LF)$ counterpart generates translations in $x^+$, just as they should. Moreover, for the other momentum generators $P_3(IT)$ generates translations in $x^3$ and $P_-(LF)$ generates translations in $x^-$, while both $P_1(IT)$ and $P_1(LF)$ generate translations in $x^1$ and  both $P_2(IT)$ and $P_2(LF)$ generate translations in $x^2$.

To see just how equivalent $H(IT)$ and $H(LF)$ might be, we note that for a free scalar theory with $T_{\mu\nu}=\partial_{\mu}\phi\partial_{\nu}\phi-\tfrac{1}{2}g_{\mu\nu}(\partial^{\alpha}\phi\partial_{\alpha}\phi-m^2\phi^2)$, we can use the Fock expansions given in (\ref{1.8}) and (\ref{1.17a}) to evaluate $H(IT)$ and $H(LF)$, and with the form for $T^{+}_{\phantom{+}+}$ given in Sec. \ref{S3} obtain
\begin{align}
H(IT)&=\int dx^1dx^2dx^3\tfrac{1}{2}\left[(\partial_0\phi)^2+\vec{\nabla}\phi\cdot \vec{\nabla} \phi+m^2\phi^2\right]
=\tfrac{1}{2}\int_{-\infty}^{\infty}dp_1dp_2dp_3\left[a^{\dagger}(\vec{p})a(\vec{p})+a(\vec{p})a^{\dagger}(\vec{p})\right]p_0,
\nonumber\\
H(LF)&=\tfrac{1}{2}\int dx^1dx^2dx^-\tfrac{1}{2}\left[(\partial_1\phi)^2+(\partial_2\phi)^2+m^2\phi^2\right]
=\tfrac{1}{2}\int_{-\infty}^{\infty} dp_1dp_2\int_{0}^{\infty}dp_-\left[a^{\dagger}(\vec{p})a(\vec{p})+a(\vec{p})a^{\dagger}(\vec{p})\right]p_+,
\label{8.98}
\end{align}
with the allowed values of $p_0=E_p$ and $p_+=F_p^2/4p_-$ being positive. In terms of the $A(IT;\vec{p})=(2E_p)^{1/2}a(IT;\vec{p})$ and $A(LF;\vec{p})=(4p_-)^{1/2}a(LF;\vec{p})$ operators introduced above we can write both the Hamiltonian and the other momentum generators in the covariant forms
\begin{align}
&P_{\mu}(IT)=\tfrac{1}{2}\int d^4p\delta((p_0)^2-E_p^2)\theta(p_0)\left[A^{\dagger}(IT;\vec{p})A(IT;\vec{p})+A(IT;\vec{p})A^{\dagger}(IT;\vec{p})\right]p_{\mu},\quad p_{\mu}=(p_0,p_1,p_2,p_3),
\label{8.99}
\end{align}
\begin{align}
&P_{\mu}(LF)=\int d^4p\delta(4p_+p_--F_p^2)\theta(p_+)\left[A^{\dagger}(LF;\vec{p})A(LF;\vec{p})+A(LF;\vec{p})A^{\dagger}(LF;\vec{p})\right]p_{\mu},\quad p_{\mu}=(p_+,p_1,p_2,p_-),
\label{8.100}
\end{align}
with the form for $P_{\mu}(IT)$ being given in \cite{Bjorken1965}.

We now apply $U(P_0,P_3)=\exp(ix^3P_0)\exp(ix^0P_3)$ to both sides of the $P_{\mu}(IT)$ equation given in (\ref{8.99}). Now the four-momentum generators commute with each other according to $[P_{\mu},P_{\nu}]=0$. Thus $UP_{\mu}(IT)U^{-1}=P_{\mu}(IT)$, with the left-hand side being unchanged. On the right-hand side  we apply (\ref{8.92}) and obtain
\begin{eqnarray}
&&P_{\mu}(IT)=\tfrac{1}{2}\int d^4p\delta((p_0)^2-E_p^2)\theta(p_0)\left[A^{\dagger}(LF;\vec{p})A(LF;\vec{p})+A(LF;\vec{p})A^{\dagger}(LF;\vec{p})\right]p_{\mu}.
\label{8.101}
\end{eqnarray}
Changing the variables according to $p_0=p_++p_-$, $p_3=p_+-p_-$ yields
\begin{eqnarray}
&&P_{0}(IT)=\int d^4p\delta(4p_+p_--E_p^2)\theta(p_++p_-)\left[A^{\dagger}(LF;\vec{p})A(LF;\vec{p})+A(LF;\vec{p})A^{\dagger}(LF;\vec{p})\right](p_++p_-),
\nonumber\\
&&P_{3}(IT)=\int d^4p\delta(4p_+p_--E_p^2)\theta(p_++p_-)\left[A^{\dagger}(LF;\vec{p})A(LF;\vec{p})+A(LF;\vec{p})A^{\dagger}(LF;\vec{p})\right](p_+-p_-),
\nonumber\\
&&P_{i}(IT)=\int d^4p\delta(4p_+p_--E_p^2)\theta(p_++p_-)\left[A^{\dagger}(LF;\vec{p})A(LF;\vec{p})+A(LF;\vec{p})A^{\dagger}(LF;\vec{p})\right]p_i,\quad i=1,2.
\label{8.102}
\end{eqnarray}
Because of the delta function term we can replace the $\theta(p_++p_-)$ term by $\theta(p_+)$. We thus obtain
\begin{eqnarray}
&&P_{0}(IT)=\int d^4p\delta(4p_+p_--E_p^2)\theta(p_+)\left[A^{\dagger}(LF;\vec{p})A(LF;\vec{p})+A(LF;\vec{p})A^{\dagger}(LF;\vec{p})\right](p_++p_-),
\nonumber\\
&&P_{3}(IT)=\int d^4p\delta(4p_+p_--E_p^2)\theta(p_+)\left[A^{\dagger}(LF;\vec{p})A(LF;\vec{p})+A(LF;\vec{p})A^{\dagger}(LF;\vec{p})\right](p_+-p_-),
\nonumber\\
&&P_{i}(IT)=\int d^4p\delta(4p_+p_--E_p^2)\theta(p_+)\left[A^{\dagger}(LF;\vec{p})A(LF;\vec{p})+A(LF;\vec{p})A^{\dagger}(LF;\vec{p})\right]p_i,\quad i=1,2.
\label{8.103}
\end{eqnarray}
Finally, comparing with  (\ref{8.100}) we obtain
\begin{eqnarray}
P_0(IT)=P_+(LF)+P_-(LF),\quad P_3(IT)=P_+(LF)-P_-(LF),\quad P_1(IT)=P_1(LF),\quad  P_2(IT)=P_2(LF).
\label{8.104}
\end{eqnarray}
As we see, the relationship between the free theory instant-time momentum operators and the free theory light-front momentum operators tracks the relationship between their eigenvalues. 

Now in constructing the all-order instant-time and light-front Lehmann representations given above the basic input was the existence of a complete set of all-order momentum eigenstates. Since that basis is complete the all-order momentum generators are given by 
\begin{align}
P_{\mu}(IT)=\sum|p^n(IT)\rangle p^n_{\mu}(IT)\langle p^n(IT)|,\quad P_{\mu}(LF)=\sum|p^n(LF)\rangle p^n_{\mu}(LF)\langle p^n(LF).
 \label{8.105}
 \end{align}
With eigenvalues not changing under a unitary transformation, we obtain
\begin{align}
P_0(IT)=UP_0(IT)U^{-1}=U\sum| p^n(IT)\rangle p^n_{0}\langle p^n(IT)|U^{\dagger}=\sum|p^n(LF)\rangle (p^n_++p^n_-)\langle p^n(LF)|=P_+(LF)+P_-(LF).
 \label{8.106}
 \end{align}
The relations given in (\ref{8.104}) thus hold to all orders in interactions. In addition, we note that
from (\ref{8.104}) we obtain the  Lorentz invariant operator identity
\begin{eqnarray}
P^2_0(IT)- P^2_3(IT)-P^2_1(IT)-P^2_2(IT)=4P_+(LF)P_-(LF)-P^2_1(LF)-P^2_2(LF),
\label{8.107}
\end{eqnarray}
a relation that also holds to all orders in interactions. Now as written, each operator in (\ref{8.107}) is infinite-dimensional, with each possessing an infinite number of momentum eigenstates. However, when acting on any particular  set of momentum eigenstates $|p_{\mu}\rangle$ with eigenvalues $p_{\mu}$ that obey $p^2=m^2$ both sides of (\ref{8.107}) have eigenvalue $m^2$. 

Given (\ref{8.104}) and (\ref{8.106}), there initially appears to be a mismatch between the eigenstates of $P_0(IT)$ and $P_+(LF)$. However, for any timelike set of instant-time momentum eigenvalues we can Lorentz boost $p_1$, $p_2$ and $p_3$ to zero, to then leave $p_0=m$. If we impose this same $p_1=0$, $p_2=0$, $p_3=0$ condition on the light-front momentum eigenvalues we would set $p_+=p_-$, $p^2=4p_+^2=m^2$, and thus obtain $p_0=2p_+=m$. When written in terms of contravariant vectors with $p^{\mu}=g^{\mu\nu}p_{\nu}$ this condition takes the form $p^0=p^-$. Thus in the instant-time rest frame the eigenvalues of $P^0(IT)$ and $P^-(LF)$ coincide. In this sense then instant-time and light-front Hamiltonians are equivalent. Having now established the equivalence of commutators and the equivalence of Hamiltonian operators, in Sec. \ref{S9} we will proceed to establish the same equivalence for both free and interacting instant-time and light-front Green's functions.

However, before doing so in detail we note that use of the Lehmann representation actually provides us with a direct  way of establishing that instant-time and light-front Green's functions are indeed equivalent. With the Lehmann representation that we derived above holding for the two-point function and not just for the commutator, the Lehmann representation also holds for time-ordered products, relating the full interacting  $D(x^{\mu}-y^{\mu})=-i\langle \Omega |T[\phi(x)\phi(y)]|\Omega\rangle$ propagator (as time ordered with $x^0$ or $x^+$) to the free propagator according to
\begin{align}
D(IT,FULL;x-y)&=\int_0^{\infty} d\sigma^2 \rho(\sigma^2,IT)D(IT,FREE;x-y,\sigma^2),
\nonumber\\
D(LF,FULL;x-y)&=\int_0^{\infty} d\sigma^2 \rho(\sigma^2,LF)D(LF,FREE;x-y,\sigma^2),
\label{8.108}
\end{align}
with the same instant-time and light-front spectral functions as given in (\ref{8.73}) and (\ref{8.82}), and with
\begin{align}
&D(IT,FREE;x,\sigma^2)=\frac{1}{(2\pi)^4}\int dp_0dp_1dp_2dp_3 \frac{e^{-i(p_0x^0+p_1x^1+p_2x^2+p_3x^3)}}{(p_0)^2-(p_1)^2-(p_2)^2-(p_3)^2-\sigma^2+i\epsilon},
\nonumber\\
&D(LF,FREE;x,\sigma^2)=\frac{2}{(2\pi)^4}\int dp_+dp_1dp_2dp_- \frac{e^{-i(p_+x^++p_1x^1+p_2x^2+p_-x^-)}}{4p_+p_--(p_1)^2-(p_2)^2-\sigma^2+i\epsilon}.
\label{8.109}
\end{align}
With the free propagators transforming into each other under the substitutions $p_0=p_++p_-$, $p_3=p_+-p_-$, $x^0=(x^++x^-)/2$, $x^3=(x^+-x^-)/2$, and with the spectral functions transforming into each other, we are thus able to establish the equivalence of all-order instant-time and light-front time-ordered products, and are even  able to  do so without needing to make any reference to perturbation theory. We shall explore this connection in more detail in Sec. \ref{S9} below, but note now that while there are $p_-=0$ zero-mode singularity concerns in Feynman diagrams  on the light-front side there are no such concerns on the instant-time side to which it is equivalent. Moreover, since we can represent time-ordered products as path integrals on fields in coordinate space (in Sec. \ref{S9} we show this explicitly in the light-front case), in such path integrals no momentum integrations are involved at all and thus there cannot be any zero-mode difficulties. Consequently, all zero-mode concerns in momentum space Feynman diagrams  must be resolvable, though as we shall see below, this will require some care.

\subsection{General Coordinate Invariance and Renormalization}
\label{S8i}

To end this section we note that the feature that underlies our analysis is general coordinate invariance.  Since in general in quantum theory invariance under translations entails  invariance under unitary transformations, we thus establish not only that instant-time quantization and light-front quantization are equivalent procedures, they are unitarily equivalent procedures. The general coordinate invariance and unitary equivalence of the theory ensures that the instant-time and light-front Lehmann representations for commutators transform into each other. Moreover, by the same token  the general coordinate invariance and unitary equivalence of the theory equally ensures that the instant-time and light-front Lehmann representations for time-ordered products transform into each other, to thereby be equal to each other in any order of perturbation. With these Green's function being equal they thus have the same asymptotic behavior, and thus the same divergence structure. Using general coordinate invariant counterterms, the cancellation of infinities is then identical in the instant-time and light-front cases, so that the instant-time and light-front renormalized Green's functions also coincide. This same reasoning ensures that the renormalized instant-time and light-front Lehmann representations coincide as well. Thus not only do we have an instant-time and light-front equivalence in the unrenormalized theory, we have one in the renormalized theory too. And as long as we take renormalization counterterms to equally be general coordinate invariant, which we do, all of our results survive renormalization.

Moreover, this same reasoning ensures that the short-distance behaviors of matrix elements of instant-time and light-front products of fields at the same spacetime points are also the same (we explicitly confirm this in our study of vacuum tadpole graphs that is given below.)  Then with the (\ref{8.94}) unitary equivalence relation  $U|\Omega(IT)\rangle=|\Omega(LF)\rangle$ holding to all orders, normal ordering instant-time products of fields with respect to $|\Omega(IT)\rangle$ is the same as normal ordering light-front products of fields with respect to $|\Omega(LF)\rangle$. In general then, the underlying general coordinate invariance of the theory ensures the equivalence of the instant-time and light-front quantization procedures both in the unrenormalized and renormalized, normal-ordered theories.

\section{Equivalence of Instant-time and Light-front Green's Functions}
\label{S9}

\subsection{Scalar Field Case}
\label{S9a}

By examining the lowest order perturbation graphs in  \cite{Mannheim2019a} we had shown the equivalence of light-front and instant-time Green's functions in the scalar field case in both the non-vacuum and the vacuum sectors. In the present paper we extend the analysis to all orders in scalar field graphs, and also extend the analysis to the fermion and gauge boson cases. In \cite{Mannheim2019a} we had found that it was through circle at infinity contributions in light-front Feynman diagram contours that one could establish the equivalence in the vacuum sector. Since this same effect is present in the fermion and gauge boson cases as well, below  we will review the results of \cite{Mannheim2019a} and  establish the same results in the fermion and gauge boson cases. And in the process of doing so we identify an additional effect that occurs in the fermion case alone, namely a tip of the light cone contribution. However, this effect  does not prevent us from establishing the equivalence of  light-front and instant-time Green's functions in the fermion case.

Since Green's functions are themselves c-numbers (matrix elements of quantum fields) rather than q-numbers, one can investigate the equivalence of instant-time and light-front Green's functions of interacting theories by working with c-numbers alone. There are two formulations of quantum field theory that specifically only involve c-numbers, namely path integrals and Feynman diagrams. Path integrals are integrals over classical paths in coordinate space, while Feynman diagrams are integrals over momenta in momentum space, and the coordinates and momenta are just c-number integration variables. To go from instant-time to light-front path integrals one only has to make a change of variables $x^0=\tfrac{1}{2}(x^++x^-)$, $x^3=\tfrac{1}{2}(x^+-x^-)$, and because of general coordinate invariance the path integral cannot change, as all one has done is make a change of basis. Similarly, for Feynman diagrams one makes the change of variables $p_0=p_++p_-$, $p_3=p_+-p_-$ as given in the Appendix, and, as noted in \cite{Yan1973}, again this is just a change of variable that cannot change the value of any diagram. Since one can discuss both Feynman diagrams and path integrals to all orders, through coordinate invariance proofs of the equivalence of instant-time and light-front  Green's functions can then be established to all orders.

To see how things work in detail it suffices to consider the individual propagators that appear order by order in perturbation theory. Both path integrals and Feynman diagrams calculate propagators, and these are c-number quantities that obey differential equations. If we consider the time-ordered (instant-time or light-front) two-point free massive scalar field Green's function $D(x^{\mu})=-i\langle \Omega|T[\phi(x)\phi(0)]|\Omega\rangle$, we can represent it as the path integral
\begin{eqnarray}
D(x^{\mu})=\int {{\cal D}}[\phi]\phi(x)\phi(0)e^{iS[\phi]},
\label{9.1}
\end{eqnarray}
where $S[\phi]=\int d^4x (-g)^{1/2}\frac{1}{2}[\partial_{\mu}\phi\partial^{\mu}\phi-m^2\phi^2+i\epsilon\phi^2]$, or as the Feynman diagram 
\begin{eqnarray}
D(x^{\mu})=\frac{1}{(2\pi)^4}\int d^4p \frac{e^{-ip\cdot x}}{p^2-m^2+i\epsilon}.
\label{9.2}
\end{eqnarray}
And for either the path integral or the Feynman representation we have
\begin{eqnarray}
\left(\partial_{\alpha}\partial^{\alpha}+m^2\right)D(x^{\mu})
=\frac{1}{(2\pi)^4}\int d^4p e^{-ip_{\alpha}x^{\alpha}}\frac{-p_{\beta}p^{\beta}+m^2}
{p_{\gamma}p^{\gamma}-m^2+i\epsilon}=-\delta^4(x).
\label{9.3}
\end{eqnarray}
And moreover, because of covariance (\ref{9.2}) and (\ref{9.3}) hold in both the instant-time and light-front cases. 

In the instant-time case for fields that obey the free field equation of motion, via use of the instant-time equal-time canonical commutator  for the q-number time-ordered product of the fields (i.e., not its c-number vacuum matrix element)  we obtain
\begin{align}
&\left[(\partial_0)^2-(\partial_3)^2-(\partial_1)^2-(\partial_2)^2+m^2\right]\left[-i\theta(x^0)\phi(x)\phi(0)-i\theta(-x^0)\phi(0)\phi(x)\right] 
\nonumber\\
&=-i\delta(x^0)\left[\partial_0\phi(x)\phi(0)-\phi(0)\partial_0\phi(x)\right]=-\delta(x^0)\delta(x^3)\delta(x^1)\delta(x^2).
\label{9.4}
\end{align}
Similarly, in the light-front case, using (\ref{3.10}) we obtain
\begin{align}
&\left[4\partial_-\partial_+-(\partial_1)^2-(\partial_2)^2+m^2\right]\left[-i\theta(x^+)\phi(x)\phi(0)-i\theta(-x^+)\phi(0)\phi(x)\right] 
\nonumber\\
&=-4i\delta(x^+)\left[\partial_-\phi(x)\phi(0)-\phi(0)\partial_-\phi(x)\right]=-2\delta(x^+)\delta(x^-)\delta(x^1)\delta(x^2),
\label{9.5}
\end{align}
with the factor of $\frac{1}{2}$ in the $\tfrac{1}{2}\int dx^+dx^-dx^1dx^2$ measure being compensated for by a factor of $2$ in the delta function product. We emphasize that even though the free theory time-ordered products are themselves q-numbers, their derivatives with respect to the  equation of motion operators are actually c-numbers, being so even though we have not taken vacuum matrix elements. Also we emphasize that even though the time ordering in (\ref{9.5}) is with respect to $x^+$, we are automatically led to the presence not of $\partial_+\phi(x)$ but of $\partial_-\phi(x)$ instead. 

Now we had noted in (\ref{7.8}) and (\ref{7.9})  that equal instant-time and equal light-front time commutators cannot be transformed into each other by a change of variable and use of the operator $U(P_0,P_3)$ that we introduced in Sec. \ref{S7}, as $\partial_0\phi$ transforms into $\partial_+\phi$  rather than into the needed $\partial_-\phi$. Initially this would equally appear to the case with time-ordered products since even though we can transform the fields via $U(P_0,P_3)$, under $x^0=(x^++x^-)/2$ the  $\theta(x^0)$ factor in (\ref{9.4}) is rewritten as $\theta[(x^++x^-)/2]$ and not as the factor  $\theta(x^+)$ that is needed in (\ref{9.5}). However, since the time-ordered product is a Lorentz scalar it is only a function of $x^2=(x^0)^2-(x^3)^2-(x^1)^2-(x^2)^2=x^+x^--(x^1)^2-(x^2)^2$. Now for timelike or lightlike separated points, but not for spacelike separated points,  $x^+$ and $x^-$ have the same sign. Thus for such points we can set $\theta[(x^++x^-)/2]=\theta(x^+)$. Noting that we can write the time-ordered product $-i\theta(x^+)\phi(x)\phi(0)-i\theta(-x^+)\phi(0)\phi(x)$ as $-(i/2)\Delta_{(1)}(x)+(1/2)\epsilon(x^0)\Delta(x)$ where $\Delta_{(1)}(x)=\phi(x)\phi(0)+\phi(0)\phi(x)$, $i\Delta(x)=\phi(x)\phi(0)-\phi(0)\phi(x)$, recalling that $\Delta(x)$ vanishes for spacelike separations,  and noting that the only quantity that does not vanish for spacelike separations, viz.  $\Delta_{(1)}(x)$, is not multiplied by any $\epsilon(x^0)$ factor, it follows that the free theory instant-time and light-front time-ordered products transform into each other.

To check that this in fact the case,  we need to obtain an explicit  relation between the instant-time and light-front time-ordered products. Thus  with (\ref{9.4}) and (\ref{9.5}) being differential equations, we see that they formally admit of solutions of the form 
\begin{align}
&-i\theta(x^0)\phi(x)\phi(0)-i\theta(-x^0)\phi(0)\phi(x)=\frac{1}{(2\pi)^4}\int dp_0dp_1dp_2dp_3 \frac{e^{-i(p_0x^0+p_1x^1+p_2x^2+p_3x^3)}}{(p_0)^2-(p_1)^2-(p_2)^2-(p_3)^2-m^2+i\epsilon},
\nonumber\\
&-i\theta(x^+)\phi(x)\phi(0)-i\theta(-x^+)\phi(0)\phi(x)=\frac{2}{(2\pi)^4}\int dp_+dp_1dp_2dp_- \frac{e^{-i(p_+x^++p_1x^1+p_2x^2+p_-x^-)}}{4p_+p_--(p_1)^2-(p_2)^2-m^2+i\epsilon}.
\label{9.6}
\end{align}
However, the validity of such solutions would require that  both $-i\theta(x^0)\phi(x)\phi(0)-i\theta(-x^0)\phi(0)\phi(x)$ and $-i\theta(x^+)\phi(x)\phi(0)-i\theta(-x^+)\phi(0)\phi(x)$ be c-numbers, and this cannot be the case since, unlike commutators, these time-ordered products are expressly q-number operators, with the insertion into them of the Fock expansions given in (\ref{1.8}) and (\ref{1.17a}) leading to the presence of $a(\vec{p})a(\vec{p})$ and $a^{\dagger}(\vec{p})a^{\dagger}(\vec{p})$ type terms. To eliminate such terms we take the vacuum matrix elements of (\ref{9.4}) and (\ref{9.5}), and now we can integrate (\ref{9.4}) and (\ref{9.5}) since now all terms are then c-numbers, and thus obtain 
\begin{align}
&-i\langle \Omega|[\theta(x^0)\phi(x)\phi(0)+\theta(-x^0)\phi(0)\phi(x)]|\Omega\rangle=\frac{1}{(2\pi)^4}\int dp_0dp_1dp_2dp_3 \frac{e^{-i(p_0x^0+p_1x^1+p_2x^2+p_3x^3)}}{(p_0)^2-(p_1)^2-(p_2)^2-(p_3)^2-m^2+i\epsilon},
\nonumber\\
&-i\langle \Omega|[\theta(x^+)\phi(x)\phi(0)+\theta(-x^+)\phi(0)\phi(x)]|\Omega\rangle=\frac{2}{(2\pi)^4}\int dp_+dp_1dp_2dp_- \frac{e^{-i(p_+x^++p_1x^1+p_2x^2+p_-x^-)}}{4p_+p_--(p_1)^2-(p_2)^2-m^2+i\epsilon}.
\label{9.7}
\end{align}
Since the two momentum space integrals are just integrals over c-number variables we can apply a coordinate transformation to them with the instant-time momentum integral transforming into the light-front momentum integral under the change of variables $x^0=\tfrac{1}{2}(x^++x^-)$, $x^3=\tfrac{1}{2}(x^+-x^-)$, $p_0=p_++p_-$, $p_3=p_+-p_-$ given in the Appendix. Consequently, we are able to establish the equivalence of the free instant-time  and light-front Green's functions in the scalar field case.  And not only that, we confirm that precisely because the light-front commutation relation is not the transform of the instant-time commutation relation but is instead  given by (\ref{3.10}),  the representation of the propagator as $D(x^{\mu})=-i\langle \Omega|T^{(+)}[\phi(x)\phi(0)]|\Omega\rangle$ is the correct representation of the scalar field Green's function in the light-front case. In other words, and this is the key point, as we change the quantization condition, the commutators readjust so as to leave the Green's functions invariant. Moreover, since for interacting theories we can develop a perturbative Wick expansion via free propagators and vertices, our results immediately generalize to interacting theories as well. And as noted in Sec. \ref{S8} above, the equivalence is even maintained following renormalization. 

Moreover, we can even turn the argument around. Suppose we had not known ahead of time what the form of the light-front canonical commutator might be. Then by requiring that $D(x^{\mu})=-i\langle \Omega|T^{(+)}[\phi(x)\phi(0)]|\Omega\rangle$ obey $\left(\partial_{\alpha}\partial^{\alpha}+m^2\right)D(x^{\mu})=-\delta^4(x)$ we would automatically be led to (\ref{3.10}). Now with (\ref{3.10}) we find $[\phi,\phi]$ type commutators that do not appear in instant-time quantization. However, to show that $D(x^{\mu})=-i\langle \Omega|T^{(+)}[\phi(x)\phi(0)]|\Omega\rangle$ obeys the wave equation we only need the $[\phi,\partial_-\phi]$ commutator given in (\ref{3.10}), with $D(x^{\mu})$ then obeying (\ref{9.7}). Consequently, despite the nonvanishing of $[\phi,\phi]$ type commutators,  the light-front Feynman rules in the interacting case are the same as in the instant-time  case, save only that we evaluate propagators using light-front momenta.

Now in regard to path integrals the reader might be concerned about the validity of and use of the path integral representation of the propagator in the light-front case. Thus in order to establish the equivalence of instant-time and light-front path integrals, we need to establish the actual validity of the path integral representation for  propagators in the light-front case. To this end  it suffices to consider the  five-dimensional formulation of relativistic quantum mechanics introduced by Feynman \cite{Feynman1950} and Nambu \cite{Nambu1950}. (For a review see e.g. \cite{Mannheim1985}.) In nonrelativistic quantum mechanics $\vec{x}$ is an operator and $t$ is a parameter. To implement covariance one can either demote $\vec{x}$ to also be a parameter and thus use quantized fields $\phi(x)$ that depend on all four components of a c-number $x^{\mu}$. Or one can promote $t$ to an operator. However, if one does promote $t$ to an operator one then needs a new quantity to serve as a parameter. So one introduces an $SO(3,1)$ Lorentz scalar $\tau$ and takes four q-number $\hat{x}^{\mu}(\tau)$ to be operators in a five-dimensional space labelled by $x^{\mu}$ and $\tau$. One treats $\tau$ nonrelativistically, and in the free theory case takes the $SO(3,1)$ invariant classical action to be $S=\int d\tau[\dot{\vec{x}}^2-\dot{t}^2]/2=\int d\tau[- \dot{x}_{\mu}(\tau)\dot{x}^{\mu}(\tau)]/2$, i.e., $S=\int d\tau[\bar{p}\cdot \dot{\bar{x}}-p_0\dot{t}-H]=\int d\tau[-p_{\mu}(\tau)\dot{x}^{\mu}(\tau)-H]$, where the dot denotes $d/d\tau$, where $p_{\mu}(\tau)=\dot{x}_{\mu}(\tau)$, and where $H=[\vec{p}^2-(p_0)^2]/2=-p_{\mu}(\tau)p^{\mu}(\tau)/2$. One takes the quantum Hamiltonian operator to be given by the four-dimensional $SO(3,1)$ invariant $\hat{H}=-\hat{p}_{\mu}(\tau)\hat{p}^{\mu}(\tau)/2$, and considers a Schr\"odinger equation of the form $i\partial_{\tau}\psi(\tau)=i\partial_{\tau}\langle \tau |\psi\rangle=\hat{H}\psi(\tau)=\langle \tau |\hat{H}|\psi\rangle$, i.e., of the form $i\partial_{\tau}\langle \tau |=\langle \tau |\hat{H}$, with solution $\langle\tau |=\langle \tau_0|\exp(-i\hat{H}(\tau-\tau_0))$. Subsequently one integrates $\tau$ out and obtains the same four-dimensional Feynman propagator as the one that one obtains with second-quantized fields. The five-dimensional formulation is thus a first-quantized approach that is equivalent to the standard second-quantized field formalism.

To establish the path integral formalism using the five-dimensional formulation of relativistic quantum mechanics we first consider the instant-time case with $SO(3,1)$ coordinates $(x^0, x^1,x^2,x^3)$. We propagate an initial 
wave function $\psi(x^{\mu}_i,\tau_i)=\langle x^{\mu}_i,\tau_i|\psi \rangle$ to a final wave function $\psi(x^{\mu}_f,\tau_f)=\langle x^{\mu}_f,\tau_f|\psi\rangle=i\int d^4x_iG(x^{\mu}_f,\tau_f;x^{\mu}_i,\tau_i)\psi(x^{\mu}_i,\tau_i)$, and can thus set $iG(x^{\mu}_f,\tau_f;x^{\mu}_i,\tau_i)=\langle x^{\mu}_f,\tau_f|x^{\mu}_i,\tau_i\rangle=\langle x^{\mu}_f,\tau_i|e^{-i\hat{H}(\tau_f-\tau_i)}|x^{\mu}_i,\tau_i\rangle$. We now insert a complete set of four-momentum eigenstates at an infinite set of time slices between $\tau_i$ and $\tau_f$, with $\sum |p_{\mu}\rangle \langle p_{\mu}|=I$, where each of the $p_{\mu}$ variables varies between $-\infty$ and $+\infty$. Since for one-dimensional nonrelativistic physics one sets $\langle x|p\rangle=e^{ipx}$, for four-vectors $p_{\mu}$ and $x^{\mu}$ propagating in the five space, on allowing for the Minkowski signature in the $SO(3,1)$ space, one sets $\langle x|p\rangle=e^{-ip_{\mu}x^{\mu}}$. Thus on putting in such intermediate states, on setting $q_{\mu}=p_{\mu}(\epsilon/2)^{1/2}$ and $\delta^{\mu}=x^{\mu}-x_i^{\mu}$, on shifting the $q_{\mu}$ integration, and using $\int _{-\infty}^{\infty} e^{ia(x-b)^2}dx=(i\pi/a)^{1/2}$ when $b$ is real, the first such time slicing at $\tau_i+\epsilon$ gives
\begin{align}
iG(x^{\mu}-x^{\mu}_i,\epsilon)&=\frac{1}{(2\pi)^4}\int dp_0dp_1dp_2dp_3\langle x^{\mu}|\exp(i\hat{p}^2\epsilon/2)|p\rangle\langle p|x^{\mu}_i\rangle
=\frac{1}{(2\pi)^4}\int dp_0dp_1dp_2dp_3\exp[ip^2\epsilon/2-ip_{\mu}\delta^{\mu}]
\nonumber\\
&=\frac{1}{4\pi^4\epsilon^2}\int dq_0dq_1dq_2dq_3 \exp[iq^2-2^{1/2}iq_{\mu}\delta^{\mu}/\epsilon^{1/2}]
=\frac{1}{4\pi^4\epsilon^2}\int dq_0dq_1dq_2dq_3\exp(iq^2)\exp\left(\frac{-i\delta_{\mu}\delta^{\mu}}{2\epsilon}\right)
\nonumber\\
&=-\frac{i}{4\pi^2\epsilon^2}\exp\left(\frac{-i(x-x_i)^2}{2\epsilon}\right)=-\frac{i}{4\pi^2\epsilon^2}\exp(iS_{\rm STAT}),
\label{9.8}
\end{align}
with the phase being the stationary classical action between $x_i,\tau_i$ and $x,\tau_i+\epsilon$ (the Lagrangian $-p_{\mu}(\tau)\dot{x}^{\mu}(\tau)+p_{\mu}(\tau)p^{\mu}(\tau)/2$ being the Legendre transform of the Hamiltonian $H=-p_{\mu}(\tau)p^{\mu}(\tau)/2$).  Iterating to all time slices then gives the instant-time path integral, and up to an overall normalization it takes the form
\begin{eqnarray}
iG(x^f-x^i,\tau)=\exp\left(\frac{-i[(x_f^0-x_i^0)^2-(x_f^1-x_i^1)^2-(x_f^2-x_i^2)^2-(x_f^3-x_i^3)^2}{2\tau}\right).
\label{9.9}
\end{eqnarray}

Finally, to integrate on $\tau$ we need $\tfrac{1}{2}\int _0^{\infty}d\tau iG$ to be convergent at $\tau=\infty$. We thus replace $H$ by $H-i\epsilon=-p_{\mu}p^{\mu}/2-i\epsilon$ so that $-iH\tau$ is replaced by $ip^2\tau/2-\epsilon\tau$. Then, with $\tfrac{1}{2}\int_0^{\infty}d\tau\int_0^{\infty}(d^4p/(2\pi)^4)\exp[ip^2\tau/2-\epsilon\tau-ip\cdot x]=i\int (d^4p/(2\pi)^4)\exp[-ip\cdot x]/(p^2+i\epsilon]$, on integrating on $\tau$ we find that $\tfrac{1}{2}\int _0^{\infty}d\tau iG(x^{\mu}_f-x^{\mu}_i,\tau)$ gives the ($m^2=0$) four-dimensional $iD(x^{\mu}_f-x^{\mu}_i)=\langle \Omega|T^{(+)}[\phi(x^{\mu}_f)\phi(x^{\mu}_i)]|\Omega\rangle$ given in (\ref{9.2}) as written in the instant-time basis, just as it should. 

Now in evaluating four-dimensional Feynman diagrams as contour integrals it is very convenient to introduce an exponential regulator (and in fact we will explicitly do so below in our light-front Feynman diagram analysis), and rewrite the four-dimensional Feynman integral $i\int d^4p\exp[-ip\cdot x]/[p^2+i\epsilon]$ as the five-dimensional $\tfrac{1}{2}\int_0^{\infty}d\alpha\int_0^{\infty}d^4p\exp[ip^2\alpha/2-\epsilon\alpha-ip\cdot x]$ (i.e., as  $\int_0^{\infty}d\alpha\int_0^{\infty}d^4p\exp[ip^2\alpha-\epsilon\alpha-ip\cdot x]$). As we see, the five-dimensional $\tau$ formalism is identical to the $\alpha$ regulator technique, and it is for this reason that the Feynman diagram and path integral formalisms coincide.  

Since we can (and in fact below will) also use the $\alpha$ regulator technique for light-front Feynman diagrams we can thus immediately anticipate that the Feynman diagram and path integral approaches will coincide in the light-front case as well. Thus for $SO(3,1)$ coordinates $(x^+,x^1,x^2,x^-)$ we again time slice in $\tau$, and obtain 
\begin{eqnarray}
iG(x^{\mu}-x^{\mu}_i,\epsilon)&=&\frac{1}{4\pi^4\epsilon^2}\int 2dq_+dq_1dq_2dq_- \exp[i(4q_+q_--(q_1)^2-(q_2)^2)-2^{1/2}i(q_+\delta^++q_-\delta^-+q_1\delta^1+q_2\delta^2)/\epsilon^{1/2}]
\nonumber\\
&&=\frac{1}{\pi^3\epsilon^2}\int dq_-dq_1dq_2 \exp[-i(q_1)^2-i(q_2)^2)-2^{1/2}i(
q_-\delta^-+q_1\delta^1+q_2\delta^2)/\epsilon^{1/2}]\delta(4q_--2^{1/2}\delta^+/\epsilon^{1/2})
\nonumber\\
&&=\frac{1}{4\pi^3\epsilon^2}\int dq_1dq_2 \exp[-i(q_1)^2-i(q_2)^2-i
\delta^+\delta^-/2\epsilon-2^{1/2}i(q_1\delta^1+q_2\delta^2)/\epsilon^{1/2}]
\nonumber\\
&&=\frac{1}{4\pi^3\epsilon^2}\int dq_1dq_2 \exp[-i(q_1)^2-i(q_2)^2]\exp\left(\frac{-i[\delta^+\delta^--(\delta^1)^2-(\delta^2)^2]}{2\epsilon}\right)
\nonumber\\
&&=-\frac{i}{4\pi^2\epsilon^2}\exp\left(\frac{-i(x-x_i)^2}{2\epsilon}\right)
=-\frac{i}{4\pi^2\epsilon^2}\exp(iS_{\rm STAT}).
\label{9.10}
\end{eqnarray}
We thus recover exactly the same form as in the instant-time case. Iterating to all time slices then gives the light-front path integral, and up to an overall normalization it takes the form
\begin{eqnarray}
iG(x^f{\mu}_f-x^{\mu}_i,\tau)=\exp\left(\frac{-i[(x_f^+-x_i^+)(x_f^--x_i^-)-(x_f^1-x_i^1)^2-(x_f^2-x_i^2)^2}{2\tau}\right),
\label{9.11}
\end{eqnarray}
to thus establish its equivalence to the instant-time path integral. Finally, on integrating on $\tau$, $\tfrac{1}{2}\int _0^{\infty}d\tau iG(x^{\mu}_f-x^{\mu}_i,\tau)$ gives the ($m^2=0$) four-dimensional $iD(x^{\mu}_f-x^{\mu}_i)=\langle \Omega|T^{(+)}[\phi(x^{\mu}_f)\phi(x^{\mu}_i)]|\Omega\rangle$ given in (\ref{9.2}) as written in the light-front basis.

As constructed, we note that the presence of the $\delta(4q_--2^{1/2}\delta^+/\epsilon^{1/2})$ term in the light-front path integral entails that $q_-$ and $x^+-x^+_i$ have the same sign. Thus both could be positive or both could be negative, and both possibilities are allowed since the initial $dq_-$ integration was from $-\infty$ to $+\infty$. Since this same relation holds for every time slice, we see that in going from one time slice in $\tau$ to the next the propagation of the light-front $x^+$ could be either forward or backward. This is typical of the five-dimensional formalism. In this formalism one is nonrelativistic in $\tau$ but relativistic in $x^{\mu}$. Thus while one can only go forward in $\tau$, one can forward or backward in $x^+$, just as in the nonrelativistic case where one can only go forward in $t$ but can go forward or backward in the spatial coordinates. Since one can go forward or backward in $x^+$ in the five-dimensional formalism and thus have particles propagate backward in $x^+$ or antiparticles propagate forward in $x^+$, in the five-dimensional formalism one is one-body in $\tau$ but many-body in $x^+$. That the path integral would contain both forward and backward propagation in $x^+$ is to be expected since the path integral corresponds to the Feynman propagator (as it must since it needs the specific Feynman diagram $i\epsilon$ prescription to exist), and the Feynman propagator has both forward and backward time orderings. And below in (\ref{12.4}) and (\ref{13.1}) we explicitly evaluate each of the two time orderings in a specific case. Moreover, if we symbolically write the above light-front path integral as $\int_{-\infty}^{\infty} dq_-F(q_-)\delta(q_--x^+)$ we can rewrite it as $\int_{-\infty}^0dq_-F(q_-)\theta(-x^+)+\int_0^{\infty}F(q_-)\theta(x^+)$, i.e., as $\int_0^{\infty}dq_-[F(-q_-)\theta(-x^+)+F(q_-)\theta(x^+)]$. Thus as is common in light-front studies (cf. (\ref{13.3}) below), we can reformulate the Feynman integral as one that involves $q_-\geq 0$ alone. However, we will still have both time orderings, something that will prove to be central when we evaluate vacuum light-front Feynman diagrams below.

Since our analysis is based on c-numbers our free theory analysis immediately generalizes to interacting theories as one can transform instant-time path integrals and momentum space Feynman diagrams into light-front  path integrals and momentum-space Feynman diagrams just by changes of variables. Instant-time quantization and light-front  quantization thus give the same scalar field Green's functions and scattering amplitudes to all orders in interactions. This had long been known to be the case for specific Feynman diagrams that had been evaluated. Here, and in agreement with \cite{Yan1973},  we derive the equivalence to all orders. Moreover, we also note that an instant-time path integral represents a quantum theory with equal instant-time commutation relations. If we now make a coordinate transformation to some general, not necessarily light-front, transformed time, the path integral will not change but it will now correspond to a quantum theory whose commutation relations are at equal transformed time, even as, and this is the key point, the commutation relations themselves do not transform into each other. For practical purposes one might want to restrict to transformations between bases that do not give the transformed metric (as defined in the Appendix) any coordinate dependence, with the most practical of all being those such as the normalized basis described in the Appendix that leave the determinant of the metric unchanged altogether.

\subsection{Fermion Field Case}
\label{S9b}

In order to show the equivalence of instant-time and light-front Green's function for fermions and gauge fields, we first need to determine whether the light-front propagators are to be written in terms of the full sets of fields (i.e., including the constrained $\psi_{(-)}$ and $A_+$) or only in terms of the dynamical $\psi_{(+)}$ and $A_1$ and $A_2$. As we now show, they are in fact to be written in  terms of the full sets of fields, with the light-front propagators being given directly by $-i\langle \Omega|T^{(+)}[\psi\bar{\psi}]|\Omega\rangle$ and $-i\langle \Omega|T^{(+)}[A^{\mu}A^{\nu}]|\Omega\rangle$ without any specific reference to good or bad fermions or to constrained gauge fields. To see this we would only need to show that these Green's functions as written with full sets of fields respectively obey the light-front Dirac and Maxwell equations with a delta function source. While this is the case for gauge bosons if we use the gauge fixing technique described earlier, it is not in fact the case for fermions \cite{Yan1973} as one obtains the additional terms shown in (\ref{9.16}) and (\ref{9.18}) below, and for fermions the further analysis that we provide below is required. (While one does obtain additional terms for gauge bosons as well if one quantizes in the axial gauge, as we show below, these additional terms are gauge artifacts that do not occur if we use the gauge fixing technique described above.)

To see what is involved for fermions we follow the scalar field discussion and apply the Dirac operator not to the c-number matrix element $-i\langle \Omega|T^{(+)}[\psi\bar{\psi}]|\Omega\rangle$ but to the light-front q-number $-i T^{(+)}[\psi\bar{\psi}]=-i[\theta(x^+-y^+)\psi_{\beta}(x)\bar{\psi}_{\alpha}(y)-\theta(-x^++y^+)\bar{\psi}_{\alpha}(y)\psi_{\beta}(x)]
$ itself, to obtain 
\begin{align}
&[i\gamma^{+}\partial_++i\gamma^{-}\partial_-+i\gamma^{1}\partial_1+i\gamma^{2}\partial_2 -m]_{\lambda\beta}(-i)[\theta(x^+-y^+)\psi_{\beta}(x)\bar{\psi}_{\alpha}(y)-\theta(-x^++y^+)\bar{\psi}_{\alpha}(y)\psi_{\beta}(x)]
\nonumber\\
&=\gamma^+_{\lambda\beta}\delta(x^+-y^+)[\psi_{\beta}(x)\psi^{\dagger}_{\sigma}(y)\gamma^0_{\sigma\alpha}+\psi^{\dagger}_{\sigma}(y)\gamma^0_{\sigma\alpha}\psi_{\beta}(x)]
\nonumber\\
&=2\gamma^0_{\lambda\nu}\Lambda^+_{\nu\beta}\delta(x^+-y^+)[\psi_{\beta}(x)\psi^{\dagger}_{\sigma}(y)\gamma^0_{\sigma\alpha}+\psi^{\dagger}_{\sigma}(y)\gamma^0_{\sigma\alpha}\psi_{\beta}(x)]
\nonumber\\
&=2\gamma^0_{\lambda\nu}\delta(x^+-y^+)[[\psi_{(+)}]_{\nu}(x)\psi^{\dagger}_{\sigma}(y)\gamma^0_{\sigma\alpha}+\psi^{\dagger}_{\sigma}(y)\gamma^0_{\sigma\alpha}[\psi_{(+)}]_{\nu}(x)].
\label{9.12}
\end{align}
Inserting $\Lambda^++\Lambda^-=1$ where $\Lambda^{\pm}=\gamma^0\gamma^{\pm}/2$,  we obtain
\begin{align}
& [i\gamma^+\partial_++i\gamma^-\partial_-+i\gamma^1\partial_1+i\gamma^2\partial_2-m]_{\lambda\beta}(-i)[\theta(x^+-y^+)\psi_{\beta}(x)\bar{\psi}_{\alpha}(y)-\theta(-x^++y^+)\bar{\psi}_{\alpha}(y)\psi_{\beta}(x)]
\nonumber\\
&=2\gamma^0_{\lambda\nu}\delta(x^+-y^+) \Big{\{}[\psi_{(+)}]_{\nu}(x),[\psi^{\dagger}_{(+)}]_{\sigma}(y)\Big{\}}\gamma^0_{\sigma\alpha}
+2\gamma^0_{\lambda\nu}\delta(x^+-y^+) \Big{\{}[\psi_{(+)}]_{\nu}(x),[\psi^{\dagger}_{(-)}]_{\sigma}(y)\Big{\}}\gamma^0_{\sigma\alpha},
\label{9.13}
\end{align}
and note that because of the $\delta(x^+-y^+)$ term the two anticommutator terms that appear in (\ref{9.13}) are to be evaluated at $y^+=x^+$. On using the anticommutation relation given in (\ref{4.13a}) the first anticommutator term in (\ref{9.13}) evaluates to 
\begin{align}
&2\gamma^0_{\lambda\nu}\delta(x^+-y^+)\Big{\{}[\psi_{(+)}]_{\nu}(x),[\psi^{\dagger}_{(+)}]_{\sigma}(y)\Big{\}}\gamma^0_{\sigma\alpha}
=2\delta(x^+-y^+)\delta(x^--y^-)\delta(x^1-y^1)\delta(x^2-y^2)\gamma^0_{\lambda\nu}\Lambda^+_{\nu\sigma}\gamma^0_{\sigma\alpha}
\nonumber\\
&=2\delta(x^+-y^+)\delta(x^--y^-)\delta(x^1-y^1)\delta(x^2-y^2)\Lambda^-_{\lambda\alpha}
\nonumber\\
&=2(\delta_{\lambda\alpha}-[\Lambda^+]_{\lambda\alpha})\delta(x^+-y^+)\delta(x^--y^-)\delta(x^1-y^1)\delta(x^2-y^2).
\label{9.14}
\end{align}
Similarly,  from (\ref{4.23a}) the second anticommutator term evaluates to
\begin{align}
&2\gamma^0_{\lambda\nu}\delta(x^+-y^+)\Big{\{}[\psi_{(+)}]_{\nu}(x),[\psi^{\dagger}_{(-)}]_{\sigma}(y)\Big{\}}\gamma^0_{\sigma\alpha}
\nonumber\\
&=\tfrac{i}{4}\delta(x^+-y^+)\epsilon(x^--y^-)[i(\gamma^1\gamma^+\partial_1^x+\gamma^2\gamma^+\partial_2^x)-m\gamma^+]_{\lambda\alpha}\delta(x^1-y^1)\delta(x^2-y^2),
\label{9.15}
\end{align}
where $\partial^x_1$ denotes $\partial/\partial x^1$, etc. Finally, combining terms and noting that $\Lambda _+=\frac{1}{4}\gamma^-\gamma^+$, we obtain 
\begin{align}
& [i\gamma^+\partial_++i\gamma^-\partial_-+i\gamma^1\partial_1+i\gamma^2\partial_2-m]_{\lambda\beta}(-i)T^{(+)}[\psi_{\beta}(x)\bar{\psi}_{\alpha}(y)]
\nonumber\\
&=2\delta_{\lambda\alpha}\delta(x^+-y^+)\delta(x^--y^-)\delta(x^1-y^1)\delta(x^2-y^2)-\tfrac{1}{2}[\gamma^-\gamma^+]_{\lambda\alpha}\delta(x^+-y^+)\delta(x^--y^-)\delta(x^1-y^1)\delta(x^2-y^2)
\nonumber\\
&+\tfrac{i}{4}\delta(x^+-y^+)\epsilon(x^--y^-)[i(\gamma^1\gamma^+\partial^x_1+\gamma^2\gamma^+\partial^x_2)-m\gamma^+]_{\lambda\alpha}\delta(x^1-y^1)\delta(x^2-y^2).
\label{9.16}
\end{align}
Thus we see that even though $-iT^{(+)}[\psi_{\beta}(x)\bar{\psi}_{\alpha}(y)]$ is a q-number, its Dirac-operator derivative is a c-number. This in direct analog to the $i\Delta(LF;x-y)$ commutator as it is also a c-number. Also we see that in deriving (\ref{9.16}) we had to incorporate both good and bad fermions. Thus for a formulation of light-front time-ordered products of fermions both good and bad fermions are needed, and we cannot be limited to good fermions alone.

As constructed, (\ref{9.16}) has an exact solution, giving for the $(x^+-y^+)$-ordered product itself 
\begin{align}
-iT^{(+)}[\psi_{\beta}(x)\bar{\psi}_{\alpha}(y)]
&=\frac{2}{(2\pi)^4}\int_{-\infty}^{\infty} dp_+dp_-dp_1dp_2\Big{[}\frac{e^{-i[p_+(x^+-y^+)+p_1(x^1-y^1)+p_2(x^2-y^2)+p_-(x^--y^-)]}}{\gamma^+p_++\gamma^-p_-+\gamma^1p_1+\gamma^2p_2-m+i\epsilon}\Big{]}_{\beta\alpha}
\nonumber\\
&+\tfrac{i}{4}\gamma^+_{\beta\alpha}\delta(x^+-y^+)\epsilon(x^--y^-)\delta(x^1-y^1)\delta(x^2-y^2).
\label{9.17}
\end{align}
However, just as in the scalar field case, (\ref{9.17}) could not hold since  $-iT^{(+)}[\psi_{\beta}(x)\bar{\psi}_{\alpha}(y)]$ is a q-number (i.e., if  we expand $\psi(x)$ in terms of creation and annihilation operators $b$ and $d^{\dagger}$, and expand $\psi^{\dagger}$ in terms of $b^{\dagger}$ and $d$, the time-ordered product $-iT^{(+)}[\psi_{\beta}(x)\bar{\psi}_{\alpha}(y)]$ will contain $d^{\dagger}b^{\dagger}$ and $bd$ terms). However, since $-i\langle \Omega|T^{(+)}[\psi_{\beta}(x)\bar{\psi}_{\alpha}(y)]|\Omega\rangle$ is a c-number we can take the vacuum matrix element of (\ref{9.16}) with the $b^{\dagger}d^{\dagger}$ and $bd$ terms then decoupling, and then we do have a pure c-number equation to integrate. This leads to
\begin{align}
-i\langle \Omega|T^{(+)}[\psi_{\beta}(x)\bar{\psi}_{\alpha}(y)]|\Omega\rangle
&=\frac{2}{(2\pi)^4}\int_{-\infty}^{\infty} dp_+dp_1dp_2dp_- \Big{[}\frac{e^{-i[p_+(x^+-y^+)+p_1(x^1-y^1)+p_2(x^2-y^2)+p_-(x^--y^-)]}}{\gamma^+p_++\gamma^-p_-+\gamma^1p_1+\gamma^2p_2-m+i\epsilon}\Big{]}_{\beta\alpha}
\nonumber\\&+\tfrac{i}{4}\gamma^+_{\beta\alpha}\delta(x^+-y^+)\epsilon(x^--y^-)\delta(x^1-y^1)\delta(x^2-y^2)
\nonumber\\
&=S^{LF}_F(x-y)_{\beta\alpha}+\tfrac{i}{4}\gamma^+_{\beta\alpha}\delta(x^+-y^+)\epsilon(x^--y^-)\delta(x^1-y^1)\delta(x^2-y^2),
\label{9.18}
\end{align}
with (\ref{9.18}) serving to define  the light-front Feynman propagator $S^{LF}_F(x-y)$ (as written in light-front coordinates). 

As constructed here, and in contrast to the light-front $i\Delta (LF;x-y)$ commutator given in (\ref{8.21a}), we note that for $S^{LF}_F(x-y)$ the integration range for $p_-$ is not $(0,\infty)$ but $(-\infty,\infty)$, as required since the action of the Dirac operator derivative on $S^{LF}_F(x-y)$ has to generate the four-dimensional delta function exhibited in (\ref{9.16}). 
The light-front $S^{LF}_F(x-y)$ that appears in (\ref{9.18}) can be obtained directly from the instant-time 
\begin{eqnarray}
S^{IT}_F(x-y)=\frac{1}{(2\pi)^4}\int d^4p\frac{e^{-i[p_0(x^0-y^0)+p_1(x^1-y^1)+p_2(x^2-y^2)+p_3(x^3-y^3)]}}{\gamma^0p_0+\gamma^3p_3+\gamma^1p_1+\gamma^2p_2-m+i\epsilon}
\label{9.19a}
\end{eqnarray}
by a general coordinate transformation from instant-time time to light-front coordinates and momenta, a transformation that leaves the c-number Feynman propagator invariant. And with the light-front $-i\langle \Omega|T^{(+)}[\psi_{\beta}(x)\bar{\psi}_{\alpha}(y)]|\Omega\rangle$ only depending on $x^{\mu}-y^{\mu}$, just as required by the translation invariance of the vacuum, we can simplify by setting $y^{\mu}=0$ in (\ref{9.18}) and for light-front time-ordered product obtain
\begin{align}
&-i\langle \Omega|T^{(+)}[\psi_{\beta}(x)\bar{\psi}_{\alpha}(0)]\Omega \rangle
\nonumber\\
&=\frac{2}{(2\pi)^4}\int_{-\infty}^{\infty} dp_+dp_1dp_2dp_-\Big{[}\frac{e^{-i(p_+x^++p_1x^1+p_2x^2+p_-x^-)}}{\gamma^+p_++\gamma^-p_-+\gamma^1p_1+\gamma^2p_2-m+i\epsilon}\Big{]}_{\beta\alpha}
+\tfrac{i}{4}\gamma^+_{\beta\alpha}\delta(x^+)\epsilon(x^-)\delta(x^1)\delta(x^2).
\label{9.20a}
\end{align}

The expression we have derived for the light-front $-i\langle \Omega|T^{(+)}[\psi_{\beta}(x)\bar{\psi}_{\alpha}(0)]|\Omega\rangle$ in (\ref{9.20a}) is a standard light-front literature expression. It  can be derived via the insertion  of the light-front Fock space expansion for $\psi(x)$  into $-i\langle \Omega|T^{(+)}[\psi_{\beta}(x)\bar{\psi}_{\alpha}(0)]|\Omega\rangle$ \cite{Harindranath1996}, with it being obtained in \cite{Yan1973} using functional source techniques. Our approach here is to obtain it by first solving for the Green's function differential equation obeyed by $-iT^{(+)}[\psi_{\beta}(x)\bar{\psi}_{\alpha}(0)]$ using the equations of motion, the constraint equation, and the anticommutation relations, with our derivation serving to complement those of \cite{Yan1973} and \cite{Harindranath1996}. As we see from (\ref{9.20a}), we do not in fact find that the light-front $-i\langle \Omega|T^{(+)}[\psi_{\beta}(x)\bar{\psi}_{\alpha}(0)]|\Omega\rangle$ is equal to the  light-front Feynman propagator $S^{LF}_F(x)_{\beta\alpha}$. Rather, there is an additional term, one that  only takes support on the light-front light cone. Since the vanishing of $x^+$, $x^1$ and $x^2$ entails that $x^+x^--(x^1)^2-(x^2)^2$ is zero for any choice of $x^-$, one can regard $\delta(x^+)\epsilon(x^-)\delta(x^1)\delta(x^2)$ as representing the tip of the light-front light cone, and we shall refer to the singular term in (\ref{9.20a}) as being a tip of the light cone singularity. This singular term has no instant-time quantization counterpart as there $-i\langle \Omega|T^{(0)}[\psi_{\beta}(x)\bar{\psi}_{\alpha}(0)]|\Omega\rangle=-i\langle \Omega|[\theta(x^0)\psi_{\beta}(x)\bar{\psi}_{\alpha}(0)-\theta(-x^0)\bar{\psi}_{\alpha}(0)\psi_{\beta}(x)]\Omega \rangle$ is equal to the standard instant-time $S^{IT}_F(x)_{\beta\alpha}$. The tip of light cone singularity is thus a special feature of light-front quantization. 

In (\ref{9.20a}) we have written the $S^{LF}_F(x)_{\beta\alpha}$ Feynman propagator term in covariant form. Thus under a general coordinate transformation $x^0=\tfrac{1}{2}(x^++x^-)$, $x^3=\tfrac{1}{2}(x^+-x^-)$, $p_0=p_++p_-$, $p_3=p_+-p_-$, the standard instant-time fermion Feynman propagator $S^{IT}_F(x)_{\beta\alpha}$ transforms into the light-front fermion Feynman propagator $S^{LF}_F(x)_{\beta\alpha}$ (just as in the scalar field case). However, the matrix element  $-i\langle \Omega|T^{(+)}[\psi_{\beta}(x)\bar{\psi}_{\alpha}(0)]|\Omega\rangle$ itself does not because of the tip of the light cone singularity. Thus initially it looks as though light-front quantization  for fermions is not equivalent to instant-time quantization. However, the difference  only contributes at the tip of the light-front light cone. Thus for any processes that are not at the tip of the light cone (such as scattering from one spacetime point to another, processes where $x^+\neq 0$) the equivalence of light-front and instant-time fermionic Green's functions is established. 

However, the one place where the tip of the light cone will contribute is in light-front vacuum graphs, and as noted in  \cite{Mannheim2019a} and as discussed in detail below, we can construct vacuum graphs from time-ordered products of fields. We thus use the spacetime coordinate $x^{\mu}$ as a regulator and take the $x^{\mu}\rightarrow 0$ limit of 
\begin{eqnarray}
-i\langle \Omega|T^{(+)}[\psi_{\beta}(x)\bar{\psi}_{\alpha}(0)]|\Omega\rangle&=&-i\langle \Omega |[\theta(x^+)\psi_{\beta}(x^{\mu})\bar{\psi}_{\alpha}(0)-\theta(-x^+)\bar{\psi}_{\alpha}(0)\psi_{\beta}(x^{\mu})]|\Omega\rangle
\nonumber\\
&=&-i\langle \Omega |[-\theta(x^+)\bar{\psi}_{\alpha}(0)\psi_{\beta}(x^{\mu})-\theta(-x^+)\bar{\psi}_{\alpha}(0)\psi_{\beta}(x^{\mu})]|\Omega\rangle
\nonumber\\
&&-i\langle \Omega |[\theta(x^+)\psi_{\beta}(x^{\mu})\bar{\psi}_{\alpha}(0)+\theta(x^+)\bar{\psi}_{\alpha}(0)\psi_{\beta}(x^{\mu})]|\Omega\rangle.
\label{9.21a}
\end{eqnarray}
On inserting $\Lambda^++\Lambda^-=I$ into (\ref{9.21a}) in the limit we obtain 
\begin{align}
&-i\langle \Omega|T^{(+)}[\psi_{\beta}(x)\bar{\psi}_{\alpha}(0)]|\Omega\rangle \rightarrow
i\langle \Omega|\bar{\psi}_{\alpha}(0)\psi_{\beta}(0)]|\Omega\rangle
-i\theta(0^+)\langle \Omega|[\psi_{\beta}(0)\bar{\psi}_{\alpha}(0)+\bar{\psi}_{\alpha}(0)\psi_{\beta}(0)]\Omega\rangle
\nonumber\\
&=i\langle \Omega |\bar{\psi}_{\alpha}(0)\psi_{\beta}(0)|\Omega\rangle
-i\theta(0^+)\gamma^0_{\nu\alpha}\langle \Omega |[\psi_{\beta}(0)(\Lambda^++\Lambda^-)\psi^{\dagger}_{\nu}(0)+\psi^{\dagger}_{\nu}(0)(\Lambda^++\Lambda^-)\psi_{\beta}(0)]|\Omega\rangle
\nonumber\\
&=i\langle \Omega |\bar{\psi}_{\alpha}(0)\psi_{\beta}(0)|\Omega\rangle
-i\theta(0^+)\gamma^0_{\nu\alpha}\langle \Omega |[\Big{\{}\psi^{(+)}_{\beta}(0),[\psi_{(+)}^{\dagger}]_{\nu}(0)\Big{\}}+\Big{\{}\psi^{(-)}_{\beta}(0),[\psi_{(-)}^{\dagger}]_{\nu}(0)\Big{\}}]|\Omega\rangle.
\label{9.22a}
\end{align}
In (\ref{9.22a}) the first anticommutator term  evaluates to 
\begin{align}
-i\theta(0^+)\gamma^0_{\nu\alpha}\langle \Omega |\Big{\{}\psi^{(+)}_{\beta}(0),[\psi_{(+)}^{\dagger}]_{\nu}(0)\Big{\}}|\Omega\rangle=-i\theta(0^+)\gamma^0_{\nu\alpha}\Lambda^+_{\beta\nu}[\delta(0)]^3=-\tfrac{i}{2}(\gamma^0-\gamma^3)_{\beta\alpha}\theta(0^+)[\delta(0)]^3,
\label{9.23a}
\end{align}
while according to (\ref{4.22a}) the second anticommutator term evaluates 
\begin{align}
&-i\theta(0^+)\gamma^0_{\nu\alpha}\langle \Omega |\Big{\{}\psi^{(-)}_{\beta}(0),[\psi_{(-)}^{\dagger}]_{\nu}(0)\Big{\}}|\Omega\rangle
\nonumber\\
&=-\frac{i}{32}(\gamma^0+\gamma^3)_{\beta\alpha}\theta(0^+)
\left[-\frac{\partial}{\partial x^1}\frac{\partial}{\partial x^1}-\frac{\partial}{\partial x^2}\frac{\partial}{\partial x^2}+m^2\right]
\int du^-[\epsilon(-u^-)]^2\delta(x^1=0)\delta(x^2=0).
\label{9.24a}
\end{align}

Now the quantity $\bar{\psi}_{\alpha}(0)\psi_{\beta}(0)$ has 16 components. We can thus develop it in terms of irreducible representations of the Lorentz group as a scalar, a pseudoscalar, a vector, an axial vector, and a rank two antisymmetric tensor (i.e., as  $I$, $\gamma^5$, $\gamma^{\mu}$, $\gamma^{\mu}\gamma^5$ and $\gamma^{\mu}\gamma^{\nu}-\gamma^{\nu}\gamma^{\mu}$). However if Lorentz invariance is not to be broken in the vacuum, we can only allow the scalar and pseudoscalar.  Both involve taking a trace over the spinor indices, and since the discussion is equivalent for both we shall restrict to the scalar. On now taking the Dirac spinor space traces we see that the contributions of both of the anticommutator terms in (\ref{9.22a}) actually vanish since $\gamma^0$ and $\gamma^3$ are both  traceless. Consequently, the tip of the light cone singularity drops out of the trace, leaving us with 
\begin{align}
&-i\eta^{\alpha\beta}\langle \Omega|T^{(+)}[\psi_{\beta}(x)\bar{\psi}_{\alpha}(0)]|\Omega\rangle \rightarrow i\eta^{\alpha\beta}\langle \Omega |\bar{\psi}_{\alpha}(0)\psi_{\beta}(0)|\Omega\rangle
=i\langle \Omega |\bar{\psi}(0)\psi(0)|\Omega\rangle
\nonumber\\
&=\frac{2}{(2\pi)^4}\int dp_+dp_1dp_2dp_-\frac{4m}{4p_+p_--(p_1)^2-(p_2)^2-m^2+i\epsilon}.
\label{9.25a}
\end{align}
a so-called vacuum tadpole graph of the type exhibited in Fig. \ref{undressedtadpole} below.
By coordinate invariance this is exactly the same expression as the one associated with the fermion loop tadpole graph in instant-time quantization, and their equivalence is thus established. However, as noted in \cite{Mannheim2019a} and as is discussed in detail below, the way that the equivalence is actually established is due to a circle at infinity contribution in the complex light-front energy $p_+$ plane, a contribution that has no counterpart in the instant-time case as the circle in the complex instant-time energy $p_0$ plane is suppressed. Since (\ref{9.25a}) only differs from the analogous scalar field case though the factor of $4m$, discussion of the fermion loop tadpole graph is identical to discussion of the scalar field tadpole loop, and we present the discussion of the scalar field tadpole below.

With the tip of the light cone singularity (where $x^+=0$) not contributing to the vacuum graph, and with it not contributing to non-vacuum graphs (where $x^+\neq 0$), we see that despite its presence,  it is not observable. Consequently, just as in the scalar sector, in the fermionic sector light-front quantized diagrams and instant-time quantized diagrams are equal.

Since we need to consider both good and bad fermions in the full $-i\langle \Omega|T^{(+)}[\bar{\psi}(x)\psi(0)]|\Omega\rangle$, we equally would need to consider both good and bad fermions in a path integral representation $\int D[\bar{\psi}]D[\psi]\psi(x)\bar{\psi}(0) \exp(iS[\bar{\psi}, \psi])$ of the same propagator, where the fermion mass term $m\bar{\psi}\psi=m[\psi_{(+)}]^{\dagger}\gamma^0\psi_{(-)}+m[\psi_{(-)}]^{\dagger}\gamma^0\psi_{(+)}$ for instance directly couples the good and bad fermions in the action $S[\bar{\psi}, \psi]$. That we would need to consider both good and bad fermions anyway is because the bad fermion only obeys the constraint equation given in (\ref{4.15a}) in solutions to the Dirac equation of motion, while the path integral is over all paths, stationary and non-stationary combined, with the bad fermion not obeying any constraint in any non-stationary path. With both the good and the bad fermions contributing in the path integral measure, the light-front path integral measure has the same number of degrees of freedom as the instant-time path integral measure (i.e., both 4-component fermions not 2-component ones). And thus by a general coordinate transformation we can show that the two path integrals are  equal, with the light-front and instant-time Green's functions then being the same. 

However, we just established that in fact light-front and instant-time Green's functions are not equal because of the tip of the light cone contribution given in (\ref{9.20a}), and thus we initially have a puzzle. The resolution of the puzzle is that the path integral representation of a quantity such as $-i\langle \Omega|T^{(+)}[\psi(x^+,x^1,x^2,x^-)\bar{\psi}(0)]|\Omega\rangle$ can only be derived if $x^+$ is nonzero since one has to break $-i\langle \Omega|T^{(+)}[\psi(x^+,x^1,x^2,x^-)\bar{\psi}(0)]|\Omega\rangle$ up into infinitesimal time slices and then identify all the paths that evolve from each time slice to the next. Technically, the behavior of $-i\langle \Omega|T^{(+)}[\psi(x^+,x^1,x^2,x^-)\bar{\psi}(0)]|\Omega\rangle$ at $x^+=0$ is outside of the path integral formalism, and whether or not one can construct an expression for $-i\langle \Omega|T^{(+)}[\psi(0,x^1,x^2,x^-)\bar{\psi}(0)]|\Omega\rangle$ as the $x^+\rightarrow 0$ limit of $-i\langle \Omega|T^{(+)}[\psi(x^+,x^1,x^2,x^-)\bar{\psi}(0)]|\Omega\rangle$ depends on each individual case. However, since the tip of the light cone issue does not arise for scattering processes, i.e., processes where the light-front time $x^+$  is nonzero, for scattering the path integral representation is valid, and by general coordinate invariance one can not only show the equivalence of light-front and instant-time scattering processes for the free theory, one can equally show the equivalence to all orders in interactions and needed renormalization counterterms since the all-order path integral and needed counterterms are just as general coordinate invariant as the lowest order contribution. Thus even though there is a tip of the light cone issue for fermions, as we have seen, the equivalence still holds since the tip of the light cone singularity decouples from the fermion loop tadpole graph. Thus for fermions the equivalence of light-front and instant-time Green's functions holds in both the non-vacuum and vacuum sectors.

\subsection{Gauge Field Case}
\label{S9c}

For gauge bosons one can work in the $A^+=0$ gauge or use gauge fixing. If we work with $A^+=0$,  
gauge bosons  obey the wave equation $(\eta_{\mu\nu}\Box-\partial_{\mu}\partial_{\nu})A^{\nu}=0$ given in (\ref{5.5}). 
Consequently in this gauge the light-front propagator obeys
\begin{eqnarray}
(\eta_{\mu\nu}\Box-\partial_{\mu}\partial_{\nu})[-i
\langle \Omega|[\theta(x^+)A^{\nu}(x)A_{\sigma}(0)+\theta(-x^+)A_{\sigma}(0)A^{\nu}(x)]|\Omega\rangle]
=2\eta_{\mu\sigma}\delta(x^+)\delta(x^-)\delta(x^1)\delta(x^2).
\label{9.26a}
\end{eqnarray}
As noted in \cite{Harindranath1996} this equation conveniently  has an exact solution 
\begin{eqnarray}
D^{\mu\nu}(x)=-i\langle \Omega|T^{(+)}[A^{\mu}(x)A^{\nu}(0)]|\Omega\rangle=2\int \frac{dp_+dp_-dp_1dp_2}{(2\pi)^4}\frac{e^{-ip\cdot x}}{p^2+i\epsilon}\left(g^{\mu\nu}-\frac{n^{\mu}p^{\nu}+n^{\nu}p^{\mu}}{n\cdot p}
+\frac{p^2}{(n\cdot p)^2}n^{\mu}n^{\nu}\right),
\label{9.27a}
\end{eqnarray}
where the only non-zero components of the reference vector $n^{\mu}$ are $n_+=1$, $n^-=2$, with the axial gauge condition being given by $n_{\mu}A^{\mu}=A^+=0$. Not only is the axial gauge $D^{\mu\nu}(x)$  not diagonal in the spacetime indices, but as had been noted in  \cite{Yan1973,Leibbrandt1987}, it also contain terms that are singular at $n\cdot p=0$. i.e., at $p_-=0$, the familiar zero-mode problem of light-front quantization.

The way that (\ref{9.27a}) satisfies (\ref{9.26a}) differs for its various components. Thus for $\mu=1$ (and analogously for $\mu=2$) we immediately obtain a propagator equation of the form
\begin{eqnarray}
&&[\eta_{1\nu}(4\partial_{+}\partial_{-}-(\partial_1)^2-(\partial_2)^2)-\partial_{1}\partial_{\nu}][-i
\langle \Omega|[\theta(x^+)A^{\nu}(x)A_{1}(0)+\theta(-x^+)A_{1}(0)A^{\nu}(x)]|\Omega\rangle
\nonumber\\
&&=-i\langle \Omega|4\delta(x^+)[\partial_{-}A_{1}(x)A_{1}(0)-A_{1}(0)\partial_-A_{1}(x)]\Omega\rangle
+i\langle \Omega|\delta(x^+)[\partial_{1}A^{+}(x)A_{1}(0)-A_{1}(0)\partial_{1}A^{+}(x)]|\Omega\rangle.
\label{9.28}
\end{eqnarray}
In the $A^+=0$ gauge the last term vanishes (to thus show how convenient this gauge is), and, just as required,  through use of (\ref{5.24})  not only do we obtain 
\begin{eqnarray}
&&[\eta_{1\nu}\partial_{\alpha}\partial^{\alpha}-\partial_{1}\partial_{\nu}]D^{\nu }_{\phantom{\nu}1}
=[\eta_{2\nu}\partial_{\alpha}\partial^{\alpha}-\partial_{2}\partial_{\nu}]D^{\nu}_{\phantom{\nu} 2}
=2\delta(x^+)\delta(x^-)\delta(x^1)\delta(x^2)
\label{9.29}
\end{eqnarray}
for the propagating $A_1$ and $A_2$ modes, we obtain (\ref{9.29}) regardless of the constraint on $A_+$ given in (\ref{5.22}). As we see  from (\ref{9.29}), there are no zero-mode issues for the propagating $A_1$ and $A_2$ modes,

In regard to the singular terms that appear in (\ref{9.27a}), as noted in \cite{Harindranath1996} these  terms arise because in the $A^+=0$ gauge $A^-=2A_+$ is a constrained field that obeys the nonlocal constraint given in (\ref{5.22}). In particular, we note that the factor $g^{\mu\nu}-(n^{\mu}p^{\nu}+n^{\nu}p^{\mu})/n\cdot p+p^2n^{\mu}n^{\nu}/(n\cdot p)^2$ on the right-hand side of (\ref{9.27a}) evaluates to $-(p_1/p_-)^2-(p_2/p_-)^2$ when $\mu=-$, $\nu=-$, precisely as is needed to verify the validity of (\ref{9.27a}) in the $(-,-)$ sector by inserting the nonlocal (\ref{5.22})  into $D^{--}(x)=-i\langle \Omega|T^{(+)}[A^{-}(x)A^{-}(0)]|\Omega\rangle$. With $A^+$, $g^{++}$ and $n^+$ all being zero, (\ref{9.27a}) holds for all components of  $D^{\mu\nu}(x)$. And with $A^-$ being constrained, only $A^1$ and $A^2$ propagate, and for them we can use $D^{\mu\nu}(p)=g^{\mu\nu}/(p^2+i\epsilon)$. Singularities do appear in $D^{--}$, $D^{-1}$ and $D^{-2}$, and with (\ref{9.27a}) being local in momentum space, $D^{--}$, $D^{-1}$ and $D^{-2}$ are nonlocal in coordinate space, just as $A^-$ is. While this nonlocality is analogous to the bad fermion anticommutator given in (\ref{4.22a}), which is also nonlocal in coordinate space, the two situations are not comparable since the bad fermion obeys a nonlocal constraint due to the intrinsic structure of the light-front Dirac equation ($\gamma^+$ being a divisor of zero). However, $A^+$ only obeys a nonlocal constraint because of the gauge choice, and with the gauge fixing $I_G$ of (\ref{5.30}) leading to $D^{\mu\nu}(p)=g^{\mu\nu}/(p^2+i\epsilon)$ for all $\mu$, $\nu$, the associated singularities in (\ref{9.27a}) would even appear to be avoidable since using $I_G$ apparently leads to no singularities at all. Thus it would appear that the treatments of these $n\cdot p=0$ singularities by Mandelstam \cite{Mandelstam1983} and Leibbrandt \cite{Leibbrandt1984} might not be needed, because with $I_G$ they do not appear. Also we note that in establishing the equivalence of instant-time and light-front vacuum tadpole graphs given in \cite{Mannheim2019a,Mannheim2019b} and below it is necessary to deal with the $p^+=2p_-=0$ region, as even in the scalar field case where there are no gauge issues at all, this zero-mode region puts singularities into Feynman diagrams (the $p_+$ pole term in (\ref{9.25a}) below generates an on-shell  $1/4p_-$ term, just like the one in the on-shell (\ref{8.21a})). These zero-mode  Feynman diagram singularities are distinct from those in (\ref{9.27a}). They are characteristic of light-front studies, and have been treated quite extensively in \cite{Mannheim2019a,Mannheim2019b} and will be discussed in detail below.

As we see, in (\ref{9.29}) there is no tip of the light cone singularity in the $A_1$ and $A_2$ sectors. With (\ref{9.29}) being a c-number relation, we can make a coordinate transformation on it to then recover the instant-time propagator equation in the $A^0=0$ gauge since the gauge condition $A^+=0$ would transform into $A^0=0$. We thus establish that instant-time and light-front gauge field Green's functions are identical. This same lack of any need to consider the constraints on $A_+$ occurs for path integrals as well, as we must do the path integration over all four components of $A_{\mu}$, since constraints on $A_+$ are properties of the solutions to the equations of motion and thus not properties of the unconstrained arbitrary variational path. From this it follows immediately that for gauge bosons the free and all-order  Green's functions calculated via path integration are the same in both instant-time $A^0=0$ quantization and light-front $A^+=0$ quantization since all one has to do is make a coordinate transformation on the coordinates and on the classical fields.

When we use gauge fixing the gauge bosons and Faddeev-Popov ghost fields obey the wave equations given in (\ref{5.31}) and (\ref{5.37}). Since these equations are diagonal in Lorentz and internal symmetry indices, the discussion parallels the scalar field case with there being no tip of the light cone issues, and with the instant-time and light-front gauge boson and ghost propagators transforming into each other just as in the scalar field case. Finally, with both the Feynman diagram and  path integral analyses holding for any theory (there being no tip of the light cone issues for integer spin fields, while the tip of the light cone singularity decouples from the fermion loop tadpole graph), we establish that in all cases the equivalence of light-front and instant-time Green's functions holds.

\subsection{Why There are  Only Tip of the Light Cone Issues for Fermions}
\label{S9d}

As we have seen, tip of the light cone issues occur for fermions but not for bosons. To understand why this is so it is instructive to construct light-front vacuum diagrams (the place where tip of the light cone issues are relevant) by boosting instant-time vacuum diagrams to the infinite momentum frame.  In general, the infinite momentum frame is a very convenient frame to use in quantum field theory for scattering processes (viz. $x^+>0$), since many Feynman diagrams are suppressed if an observer makes a Lorentz boost with a velocity at or close to the velocity of light, and under this boost instant-time diagrams with $x^0>0$ transform into light-front diagrams with $x^+>0$. However, we will show in Sec. \ref{S16}  that use of the infinite momentum frame is actually inadequate for constructing light-front vacuum graphs (viz. $x^+=0$) from instant-time ones ($x^0=0$) because of circle at infinity contributions. However for addressing the vacuum sector tip of the light cone issue the infinite momentum frame still is useful because the tip of the light cone issue appears in the pole term contribution, and as we also show in Sec. \ref{S16}, for the pole contribution the infinite momentum frame is adequate.

Under a Lorentz boost with velocity $u$ in the 3-direction the contravariant and covariant components of a general four-vector $A^{\mu}$ transform as 
\begin{eqnarray}
A^0\rightarrow\frac{A^0+uA^3}{(1-u^2)^{1/2}}, \quad A^3\rightarrow \frac{A^3+uA^0}{(1-u^2)^{1/2}},\quad
A_0\rightarrow\frac{A_0-uA_3}{(1-u^2)^{1/2}}, \quad A_3\rightarrow \frac{A_3-uA_0}{(1-u^2)^{1/2}}.
\label{9.30}
\end{eqnarray}
If we set $(1-u)=\epsilon^2/2$, then with $\epsilon$ small, to leading order we  obtain
\begin{eqnarray}
&&A^0\rightarrow\frac{A^0+A^3}{\epsilon} +O(\epsilon), \quad A^3\rightarrow \frac{A^3+A^0}{\epsilon}+O(\epsilon),\quad
A_0\rightarrow\frac{A_0-A_3}{\epsilon}+O(\epsilon), \quad A_3\rightarrow \frac{A_3-A_0}{\epsilon}+O(\epsilon),
\nonumber\\
&&(A^0)^2-(A^3)^2\rightarrow A^+A^-+O(\epsilon),
\label{9.31}
\end{eqnarray}
where $A^{\pm}=A^0\pm A^3$. This leads to
\begin{eqnarray}
p^3\rightarrow \frac{p^+}{\epsilon}= \frac{2p_-}{\epsilon}, \quad
E_p\rightarrow \frac{2p_-}{\epsilon},\quad 
\frac{dp^3}{E_p}\rightarrow \frac{dp_-}{p_-},
\label{9.32}
\end{eqnarray}
where $E_p=((p_3)^2+(p_1)^2+(p_2)^2+m^2)^{1/2}$.

As well as transform energies and momenta we also have to transform the ranges of integration in Feynman graphs. To this end we recall that under a Lorentz boost the velocity transforms as
\begin{eqnarray}
v \rightarrow \frac{v+u}{1+vu}=\frac{v+1-\epsilon^2/2}{1+v-v\epsilon^2/2}.
\label{9.33}
\end{eqnarray}
Thus with $u=1-\epsilon^2/2$, $v=1-\epsilon^2/2$ transforms into $v^{\prime}=1$, while $v=-1+\epsilon^2/2$ transforms into $v^{\prime}= -1$. With the quantity $p^3+p^0$ being given by $m(v+1)/(1-v^2)^{1/2}=m(1+v)^{1/2}(1-v)^{1/2}$, the range $p^3=-\infty$ to $p^3=+\infty$, viz. $v=-1+\epsilon^2/2$ to $v=1-\epsilon^2/2$, transforms into the $p^+=2p_-$ range $m\epsilon^2/4$ to $\infty$, and thus to the range $0$ to $\infty$ when we set $\epsilon=0$.

As we discuss in detail below, if we now look at an instant-time scalar field vacuum tadpole, we can close the contour below the real $p_0$ axis and only pick up poles, to obtain 
\begin{eqnarray}
D(x^{\mu}=0,{\rm instant},{\rm pole})=\frac{1}{(2\pi)^4}\int d^4p \frac{1}{[(p_0)^2-(p_3)^2-(p_1)^2-(p_2)^2-m^2+i\epsilon]}\Big{|}_{\rm pole}=\frac{-i}{(2\pi)^3}\int_{-\infty}^{\infty} \frac{d^3p}{2E_p}.
\label{9.34}
\end{eqnarray}
We can write this expression in the infinite momentum frame and obtain 
\begin{eqnarray}
D(x^{\mu}=0,{\rm instant},{\rm pole}) \rightarrow
\frac{-i}{(2\pi)^3}\int_{-\infty}^{\infty} dp_1\int_{-\infty}^{\infty}dp_2\int_0^{\infty}\frac{dp_-}{2p_-},
\label{9.35}
\end{eqnarray}
and in Sec. \ref{S15} we will show that the right-hand side of (\ref{9.35})  is the light-front complex $p_+$ plane pole contribution ($D(x^{\mu}=0,{\rm front},{\rm pole})$) to the light-front vacuum tadpole. With this being the case, there are no tip of the light cone issues for scalars. 

For fermions we have a pole term contribution of the form
\begin{eqnarray}
\frac{1}{(2\pi)^4}\int d^4p \frac{1}{[\gamma^{\mu}p_{\mu}-m+i\epsilon]}\Big{|}_{\rm pole}=\frac{-i}{(2\pi)^3}\int_{-\infty}^{\infty} \frac{d^3p}{2E_p}(\gamma^0E_p+\gamma^1p_1+\gamma^2p_2+\gamma^3p_3+m).
\label{9.36}
\end{eqnarray}
To apply the infinite momentum frame we treat the Dirac $\gamma^{\mu}$ matrices as fixed quantities as they act in the spinor space not in spacetime (it is $\bar{\psi}(x)\gamma^{\mu}\psi(x)$ that is a spacetime vector), and obtain
\begin{align}
&\frac{-i}{(2\pi)^3}\int_{-\infty}^{\infty} \frac{d^3p}{2E_p}(
\gamma^0E_p+\gamma^1p_1+\gamma^2p_2+\gamma^3p_3+m)
\nonumber\\
&\rightarrow \frac{-i}{(2\pi)^3}\int_{-\infty}^{\infty} dp_1\int_{-\infty}^{\infty}dp_2\int_0^{\infty}\frac{dp_-}{2p_-}(
\gamma^0E_p+\gamma^1p_1+\gamma^2p_2+\gamma^3p_3+m)
\nonumber\\
&+ \frac{i}{(2\pi)^3}\int_{-\infty}^{\infty} dp_1\int_{-\infty}^{\infty}dp_2\int_0^{\infty}\frac{dp_-}{2p_-}\frac{\gamma^0p_3+\gamma^3E_p}{\epsilon},
\label{9.37}
\end{align}
with both the numerator and denominator having to transform. We recognize (\ref{9.37}) as being in the same form as (\ref{9.20a}), with the $1/\epsilon$ term corresponding to the $\delta(x^+)$ term. Thus in transforming the fermionic instant-time vacuum Feynman diagram to the infinite momentum frame we generate not just the light-front vacuum Feynman propagator but also the tip of the light cone singularity, to thus shed some insight into its origin. 

For Abelian and non-Abelian gauge bosons and for Faddeev-Popov ghosts we can evaluate the instant-time vacuum diagrams using the gauge fixing procedure described in Secs. \ref{S5b} and \ref{S5c}, with the pole terms being of the form 
\begin{eqnarray}
D({\rm Abelian})&=&\frac{1}{(2\pi)^4}\int d^4p \frac{g_{\mu\nu}({\rm instant})}{[(p_0)^2-(p_3)^2-(p_1)^2-(p_2)^2-m^2+i\epsilon]}\Big{|}_{\rm pole}=\frac{-ig_{\mu\nu}({\rm instant)}}{(2\pi)^3}\int_{-\infty}^{\infty} \frac{d^3p}{2E_p},
\nonumber\\
D({\rm non-Abelian})&=&\frac{1}{(2\pi)^4}\int d^4p \frac{g_{\mu\nu}({\rm instant})\delta_{ab}}{[(p_0)^2-(p_3)^2-(p_1)^2-(p_2)^2-m^2+i\epsilon]}\Big{|}_{\rm pole}=\frac{-ig_{\mu\nu}({\rm instant)}\delta_{ab}}{(2\pi)^3}\int_{-\infty}^{\infty} \frac{d^3p}{2E_p},
\nonumber\\
D({\rm ghost})&=&\frac{1}{(2\pi)^4}\int d^4p \frac{\delta_{ab}}{[(p_0)^2-(p_3)^2-(p_1)^2-(p_2)^2-m^2+i\epsilon]}\Big{|}_{\rm pole}=\frac{-i\delta_{ab}}{(2\pi)^3}\int_{-\infty}^{\infty} \frac{d^3p}{2E_p},
\label{9.38}
\end{eqnarray}
where $g_{\mu\nu}({\rm instant})$ is the instant-time metric given in (\ref{A1}).  We can write (\ref{9.38}) in the infinite momentum frame and obtain
\begin{eqnarray}
\frac{-ig_{\mu\nu}({\rm instant})}{(2\pi)^3}\int_{-\infty}^{\infty} \frac{d^3p}{2E_p} &\rightarrow&
\frac{-ig_{\mu\nu}({\rm front})}{(2\pi)^3}\int_{-\infty}^{\infty} dp_1\int_{-\infty}^{\infty}dp_2\int_0^{\infty}\frac{dp_-}{2p_-},
\nonumber\\
\frac{-ig_{\mu\nu}({\rm instant})\delta_{ab}}{(2\pi)^3}\int_{-\infty}^{\infty} \frac{d^3p}{2E_p} &\rightarrow&
\frac{-ig_{\mu\nu}({\rm front})\delta_{ab}}{(2\pi)^3}\int_{-\infty}^{\infty} dp_1\int_{-\infty}^{\infty}dp_2\int_0^{\infty}\frac{dp_-}{2p_-},
\nonumber\\
\frac{-i\delta_{ab}}{(2\pi)^3}\int_{-\infty}^{\infty} \frac{d^3p}{2E_p} &\rightarrow&
\frac{-i\delta_{ab}}{(2\pi)^3}\int_{-\infty}^{\infty} dp_1\int_{-\infty}^{\infty}dp_2\int_0^{\infty}\frac{dp_-}{2p_-},
\label{9.39}
\end{eqnarray}
where $g_{\mu\nu}({\rm front})$ is the light-front metric given in (\ref{A1}). ($g_{\mu\nu}({\rm instant})$ is actually left invariant by the boost but can be written in the light-front coordinate basis as $g_{\mu\nu}({\rm front})$.)  With gauge fixing there is thus no tip of the light cone singularity in the gauge boson case. Now it was noted in \cite{Yan1973} that while one would get zero-mode singularities of the type exhibited in (\ref{9.27a}) in the gauge boson case if one quantized in the axial gauge, they decouple if the gauge boson propagator is coupled to conserved currents. Thus by working with the Secs. \ref{S5b} and \ref{S5c} gauge fixing procedures no tip of the light cone singularities or zero-mode singularities are encountered in the gauge boson case.

The reason why fermions behave differently than bosons is because they link spacetime with the spinor space while bosons are defined in spacetime alone. This is familiar in differential geometry since one has to introduce vierbeins $V^{\mu}_a$ to describe half-integral spin, as one sets $g^{\mu\nu}=\eta^{ab}V^{\mu}_aV^{\nu}_b$  where the vierbeins carry both a spacetime index $\mu$ and a fixed frame spinor index $a$. One replaces $\gamma^{\mu}\partial_{\mu}$ by $\gamma^aV^{\mu}_a\partial_{\mu}$, and under a coordinate transformation the $V^{\mu}_a$ transform but not the fixed frame $\gamma^a$. For integer spin it is not necessary to introduce vierbeins. Consequently, it is only for half-integer spin that one has to deal with tip of the light cone singularities.

\section{Instant-time Feynman Contours for Scalar Fields}
\label{S10}

To study Green's function issues involved in more detail it is instructive to examine some specific Feynman diagrams, and as we noted above, we only need to consider the scalar field case, since with gauge fixing the gauge boson case is analogous, and once we have taken care of tip of the light cone singularities the fermion sector graphs are analogous too. In the instant-time formalism the scalar field $D(x^{\mu})$ Feynman diagram given in (\ref{9.2}) is of the form
\begin{eqnarray}
D(x^{\mu},{\rm instant})=\frac{1}{(2\pi)^4}\int dp_0dp_1dp_2dp_3 \frac{e^{-i(p_0x^0+p_3x^3+p_1x^1+p_2x^2)}}{[(p_0)^2-(p_3)^2-(p_1)^2-(p_2)^2-m^2+i\epsilon]},
\label{10.1}
\end{eqnarray}
and has poles in the lower right-hand and upper left-hand quadrants in the complex $p_0=p_0^R+ip_0^I$ plane. With $\exp(-ip_0x^0)=\exp(-ip_0^Rx^0+p_0^Ix^0)$, for $x^0>0$ the circle at infinity contribution will be suppressed if we close in the lower half complex $p_0$ plane. With the instant-time denominator behaving as $1/(p_0)^2$ at large $p_0$, it would also give suppression of the circle at infinity contribution, but the $\exp(p_0^Ix^0)$ suppression of the lower half circle at infinity contribution in the instant-time case will suffice if $x^0$ is positive definite. We can therefore set $\int_{-\infty}^{\infty}dp_0={\rm pole~contribution}$. There are then two ways to evaluate the Feynman diagram. The first way is to close the contour and evaluate the pole contribution, and the second is to regulate the integral $\int_{-\infty}^{\infty}dp_0$ and evaluate it directly along the real $p_0$ axis. If these two procedures agree, one then confirms that there indeed are no circle at infinity contributions.

For the instant-time pole contribution to forward in time propagation, we note that  if we close the contour below the real $p_0$ axis in a clockwise-directed contour we only pick up the poles in the lower right-hand quadrant in the complex $p_0$ plane, and they all have positive energy (positive energy states propagate forward in time). In instant-time quantization the contribution of the poles is 
\begin{eqnarray}
D(x^0>0,{\rm instant},{\rm pole})=-\frac{i}{(2\pi)^3}\int_{-\infty}^{\infty} \frac{d^3p}{2E_p}\exp(-iE_px^0+i\vec{p}\cdot \vec{x})
=\frac{1}{8\pi}\left(\frac{m^2}{x^2}\right)^{1/2}H^{(2)}_1(m(x^2)^{1/2}),
\label{10.2}
\end{eqnarray}
where $E_p=+(\vec{p}^2+m^2)^{1/2}$, and where the $d^3p$ integral can be done analytically. 

To evaluate the instant-time integral directly along the real $p_0$ axis, we introduce the exponential regulator given by $\int_0^{\infty}d\alpha\exp[i\alpha(A+i\epsilon)]=-1/iA$, with the $i\epsilon$ term suppressing the $\alpha =\infty$ contribution when $A$ is real. With $p^2-m^2$ being real on the real $p_0$ axis we thus obtain
\begin{eqnarray}
D(x^0>0,{\rm instant},{\rm regulator})&=&-\frac{i}{(2\pi)^4}\int_{-\infty}^{\infty} d^4pe^{-ip\cdot x}\int_0^{\infty}
d\alpha e^{i\alpha(p^2-m^2+i\epsilon)}
\nonumber\\
&=&-\frac{i}{(2\pi)^4}\int_{-\infty}^{\infty} d^4p\int_0^{\infty}d\alpha e^{i\alpha(p-x/2\alpha)^2-ix^2/4\alpha-i\alpha m^2-\alpha\epsilon}.
\label{10.3}
\end{eqnarray}
On shifting the  $p_{\mu}$ integration and  setting $q_{\mu}=\alpha^{1/2} p_{\mu}$ we obtain 
\begin{align}
&D(x^0>0,{\rm instant},{\rm regulator})
\nonumber\\
&=-\frac{i}{(2\pi)^4}\int_0^{\infty}\frac{d\alpha}{\alpha^2}e^{-ix^2/4\alpha-i\alpha m^2-\alpha\epsilon}\int_{-\infty}^{\infty} dq_0dq_1dq_2dq_3e^{i[(q_0)^2-(q_1)^2-(q_2)^2-(q_3)^2]}
\nonumber\\
&=-\frac{1}{16\pi^2}\int_0^{\infty}\frac{d\alpha}{\alpha^2}e^{-ix^2/4\alpha-i\alpha m^2-\alpha\epsilon}
=\frac{1}{8\pi}\left(\frac{m^2}{x^2}\right)^{1/2}H^{(2)}_1(m(x^2)^{1/2}).
\label{10.4}
\end{align}
With $D(x^0>0,{\rm instant},{\rm pole})$ and $D(x^0>0,{\rm instant},{\rm regulator})$ being equal, we thus confirm that there is no circle at infinity contribution just as anticipated.

\section{Instant-time Fock Space Formalism in the Scalar Field Non-vacuum Sector}
\label{S11}

In the instant-time case one can take an instant-time forward in time Green's function such as $D(x^0>0,{\rm instant})=-i\langle \Omega_I|\theta(x^0)\phi(x^0,x^1,x^2,x^3)\phi(0)|\Omega_I\rangle$ as evaluated in the instant-time vacuum $|\Omega_I\rangle$, and expand the field in terms of instant-time creation and annihilation operators that create and annihilate particles out of that vacuum as  
\begin{eqnarray}
\phi(x^0,\vec{x})=\int \frac{d^3p}{(2\pi)^{3/2}(2E_p)^{1/2}}[a(\vec{p})\exp(-iE_p t+i\vec{p}\cdot\vec{x})+a^{\dagger}(\vec{p})\exp(+iE_p t-i\vec{p}\cdot\vec{x})],
\label{11.1}
\end{eqnarray}
where $E_p=(\vec{p}^2+m^2)^{1/2}$ and $[a(\vec{p}),a^{\dagger}(\vec{p}^{~\prime})]=\delta^3(\vec{p}-\vec{p}^{~\prime})$. The insertion of $\phi(\vec{x},x^0)$ into $D(x^0>0,{\rm instant})$ immediately leads to the on-shell three-dimensional integral
\begin{eqnarray}
D(x^0>0,{\rm instant},{\rm Fock})=-\frac{i\theta(x^0)}{(2\pi)^3}\int_{-\infty}^{\infty} \frac{d^3p}{2E_p} e^{-iE_p x^0+i\vec{p}\cdot\vec{x}}.
\label{11.2}
\end{eqnarray}
We recognize (\ref{11.2}) as (\ref{10.2}), to thus establish the equivalence of the instant-time off-shell Feynman and on-shell Fock space prescriptions.

As an alternative way of understanding this equivalence, one can look for solutions to $(\partial_{\alpha}\partial^{\alpha}+m^2)D(x^{\mu},{\rm instant})=-\delta^4(x)$, and obtain the off-shell four-dimensional integral
\begin{eqnarray}
D(x^{\mu},{\rm instant})=\frac{1}{(2\pi)^4}\int d^4p \frac{e^{-ip\cdot x}}{p^2-m^2+i\epsilon}=
\frac{1}{(2\pi)^4}\int \frac{d^4p}{2E_p} e^{-ip\cdot x}\left[\frac{1}{p_0-E_p+i\epsilon}
-\frac{1}{p_0+E_p-i\epsilon}\right].
\label{11.3}
\end{eqnarray}
One can then proceed from (\ref{11.3}) to (\ref{11.2}) by closing the Feynman contour below the real $p_0$ axis, to yield a contour integral in which the lower-half $p_0$ plane circle at infinity makes no contribution when the instant-time $x^0$ is positive, while the pole term yields (\ref{11.2}). 

Similarly, one can proceed from (\ref{11.2}) to (\ref{11.3}) by writing the theta function as
\begin{eqnarray}
\theta(x^0)=-\frac{1}{2\pi i}\int_{-\infty}^{\infty} d\omega \frac{e^{-i\omega x^0}}{\omega+i\epsilon},
\label{11.4}
\end{eqnarray}
so that (\ref{11.2}) takes the form 
\begin{eqnarray}
D(x^0>0,{\rm instant})=\frac{1}{(2\pi)^4}\int \frac{d^3p}{2E_p}\int_{-\infty}^{\infty}   d\omega \frac{e^{-i\omega x^0}}{\omega+i\epsilon}  e^{-iE_p x^0+i\vec{p}\cdot\vec{x}}.
\label{11.5}
\end{eqnarray}
On setting $p_0=\omega+E_p$, we can rewrite (\ref{11.5}) as 
\begin{eqnarray}
D(x^0>0,{\rm instant})=\frac{1}{(2\pi)^4}\int \frac{d^4p}{2E_p} \frac{e^{-ip_0 x^0+i\vec{p}\cdot\vec{x}}}{(p_0-E_p+i\epsilon)}.
\label{11.6}
\end{eqnarray}
We recognize (\ref{11.6}) as the forward in time, positive frequency component of (\ref{11.3}), and thus establish the equivalence of the instant-time off-shell four-dimensional Feynman and on-shell three-dimensional Hamiltonian (Fock space) formalisms, and see that the equivalence occurs because the four-dimensional Feynman contour is given by on-shell poles alone. However, as we shall also see, in the light-front vacuum sector that we discuss below this is not in fact the case, with there being a circle at infinity contribution as well.

\section{Light-front Feynman Contours in the Scalar Field Non-vacuum Sector}
\label{S12}

With there being only one power of $p_+$ in the denominator in 
\begin{eqnarray}
D(x^{\mu},{\rm front})=\frac{2}{(2\pi)^4}\int dp_+dp_1dp_2dp_- \frac{e^{-i(p_+x^++p_1x^1+p_2x^2+p_-x^-)}}{4p_+p_--(p_1)^2-(p_2)^2-m^2+i\epsilon},
\label{12.1}
\end{eqnarray}
in the light-front case the circle at infinity contribution cannot be suppressed by the denominator, though with $\exp(-ip_+x^+)$ being of the form $\exp(-iR\cos \theta x^++R\sin\theta x^+)$ when $p_+=Re^{i\theta}$,  in analog to the instant-time case, for $x^+>0$ the circle at infinity contribution will still be suppressed in the lower half complex $p_+$ plane.

We can evaluate $\int_{-\infty}^{\infty}dp_+$ directly along the real axis using an exponential regulator, or via the pole contribution alone when we take advantage of the suppression on the lower half plane circle at infinity and close below the real $p_+$ axis. In regard to the pole terms we note that in (\ref{12.1}) the denominator is given by
\begin{eqnarray}
4p_+p_--(p_1)^2-(p_2)^2-m^2+i\epsilon=4p_-\left[p_+-((p_1)^2+(p_2)^2+m^2)/4p_-+i\epsilon/4p_-\right].
\label{12.2}
\end{eqnarray}
In terms of $F_p=(p_1)^2+(p_2)^2+m^2$, we see that poles in the complex $p_+$ plane occur at $p_+=F_p^2/4p_--i\epsilon/4p_-$. Poles with $p_-\geq 0^+$ thus all lie below the real $p_+$ axis and have positive $F_p^2/4p_-$, while poles with $p_-\leq 0^-$ all lie above the real $p_+$ axis and have negative $F_p^2/4p_-$. For $x^+>0$ closing the $p_+$ contour below the real axis then restricts the poles to $F_p^2/4p_->0$, $p_-\geq 0^+$, just as is considered in the on-shell Light-Front Hamiltonian treatment (see e.g. \cite{Brodsky:1997de}). However, in order to establish that the pole terms are all that one needs, one has to deal with the fact that the pole at $p_-=0^+$ has $F_p^2/4p_-=\infty$ and thus, as noted in \cite{Yan1973}, gets intertwined with the circle at infinity (and in addition one has to show that the $p_-<0^-$ contribution is zero). To see whether this might be a concern, we compare the regulator and pole determinations of the Feynman contour.

To address these issues we need to evaluate $D(x^+>0,{\rm front})$ in a procedure in which there is no ambiguity when we set $p_-=0$. To this end we note that the exponential regulator that we used in (\ref{10.3}) can be used in the light-front case as well as in the instant-time case, with it being well-behaved when $4p_+p_--(p_1)^2-(p_2)^2-m^2$ is real (it again being the $i\epsilon$ term that provides the large $\alpha$ suppression), while also remaining well-behaved if $p_-=0$, just as we want. Through use of the exponential regulator we can determine $D(x^+>0,{\rm front})$ by direct integration along the real $p_+$ axis without regard to any contour contribution, and with $q_1=\alpha^{1/2}p_1$, $q_2=\alpha^{1/2}p_2$ obtain
\begin{eqnarray}
&&D(x^+>0,{\rm front},{\rm regulator})
\nonumber\\
&&=-\frac{2i}{(2\pi)^4}\int_{-\infty}^{\infty} dp_+\int_{-\infty}^{\infty} dp_-\int_{-\infty}^{\infty} dp_1\int_{-\infty}^{\infty} dp_2e^{-i(p_+x^++p_-x^-+p_1x^1+p_2x^2)}
\int_0^{\infty}
d\alpha e^{i\alpha(4p_+p_--(p_1)^2-(p_2)^2-m^2+i\epsilon)}
\nonumber\\
&&=-\frac{2i}{(2\pi)^3}\int_{-\infty}^{\infty} dp_-\int_{-\infty}^{\infty} dp_1\int_{-\infty}^{\infty} dp_2e^{-i(p_-x^-+p_1x^1+p_2x^2)}
\int_0^{\infty}d\alpha e^{i\alpha(-(p_1)^2-(p_2)^2-m^2+i\epsilon)}\delta(4\alpha p_--x^+)
\nonumber\\
&&=-\frac{2i}{4(2\pi)^3}\int_{-\infty}^{\infty} dp_1\int_{-\infty}^{\infty} dp_2e^{-i(p_1x^1+p_2x^2)}
\int_0^{\infty}
\frac{d\alpha}{\alpha} e^{i\alpha(-(p_1)^2-(p_2)^2-m^2+i\epsilon)}e^{-ix^+x^-/4\alpha}
\nonumber\\
&&=-\frac{2i}{4(2\pi)^3}\int_{-\infty}^{\infty} dq_1\int_{-\infty}^{\infty} dq_2e^{-i[(q_1)^2+(q_2)^2]}\int_0^{\infty}\frac{d\alpha}{\alpha^2}e^{-i(x^+x^--(x^1)^2-(x^2)^2)/4\alpha-i\alpha m^2-\alpha\epsilon}
\nonumber\\
&&=-\frac{1}{16\pi^2}\int_0^{\infty}\frac{d\alpha}{\alpha^2}e^{-ix^2/4\alpha-i\alpha m^2-\alpha\epsilon}
=\frac{1}{8\pi}\left(\frac{m^2}{x^2}\right)^{1/2}H^{(2)}_1(m(x^2)^{1/2}).
\label{12.3}
\end{eqnarray}
Comparing with (\ref{10.4}) we thus establish the equivalence of the instant-time and light-front non-vacuum sector Feynman diagrams, just as desired, and just as we had discussed in general in Sec. \ref{S9}. Moreover, we in addition note the appearance of the $\delta(4\alpha p_--x^+)$ term. With $x^+$ being positive for forward in time propagation, we thus establish that for forward in time propagation the only allowed values of $p_-$ are $x^+/4\alpha$. With $\alpha$ varying between $0$ and $\infty$, for forward in time propagation we thus restrict to all $p_-\geq 0^+$, just as required in the on-shell Light-Front Hamiltonian treatment. (For backward in time propagation we would restrict to all $p_-\leq 0^-$.)

To evaluate the complex $p_+$ plane pole contribution when $x^+>0$, as can be seen from (\ref{12.2}), poles below the real axis are restricted to $p_-\geq 0$, and occur at a value of $F_p^2/4p_-$ that is necessarily positive. However, as noted above there  is an ambiguity since the range includes the point at $p_-=0$ where $F_p^2/4p_-=\infty$. We shall thus momentarily exclude the region around $p_-=0$, and thus in (\ref{12.1}) only consider poles below the real $p_+$ axis that have $p_-\geq \delta$ where $\delta$ is a small number. Evaluating the contour integral in the complex $p_+$ plane thus gives
\begin{align}
D(x^+>0,{\rm front},{\rm pole})&=-\frac{2i}{(2\pi)^3}\int_{\delta}^{\infty}\frac{dp_-}{4p_-}\int_{-\infty}^{\infty} dp_1\int_{-\infty}^{\infty}dp_2 
e^{-i(F_p^2x^+/4p_-+p_-x^-+p_1x^1+p_2x^2)-\epsilon x^+/4p_-}
\nonumber\\
&=-\frac{1}{4\pi^2x^+}\int_{\delta}^{\infty} dp_-e^{-ip_-x^-+i[(x^1)^2+(x^2)^2]p_-/x^+-im^2x^+/4p_--\epsilon x^+/4p_-}
\nonumber\\
&=-\frac{1}{4\pi^2x^+}\int_{\delta}^{\infty}dp_-e^{-ip_-x^2/x^+-im^2x^+/4p_--\epsilon x^+/4p_-}.
\label{12.4}
\end{align}
If we now set $\alpha=x^+/4p_-$, we obtain
\begin{align}
D(x^+>0,{\rm front},{\rm pole})=-\frac{1}{16\pi^2}\int_{0}^{x^+/4\delta} \frac{d\alpha}{\alpha^2}e^{-ix^2/4\alpha-i\alpha m^2-\alpha\epsilon}.
\label{12.5}
\end{align}
In (\ref{12.5}) we can now take the limit $\delta \rightarrow 0$, $x^+/4\delta\rightarrow \infty$ without encountering any ambiguity as long as $x^+$ is nonzero, and with $x^+>0$ thus obtain
\begin{align}
D(x^+>0,{\rm front},{\rm pole})=-\frac{1}{16\pi^2}\int_{0}^{\infty} \frac{d\alpha}{\alpha^2}e^{-ix^2/4\alpha-i\alpha m^2-\alpha\epsilon}.
\label{12.6}
\end{align}
With  (\ref{12.6}) being equal to (\ref{12.3}) we thus confirm that $D(x^+>0,{\rm front})$ can be written as 
\begin{eqnarray}
D(x^+>0,{\rm front})=D(x^+>0,{\rm front},{\rm regulator})=D(x^+>0,{\rm front},{\rm pole})=\frac{1}{8\pi}\left(\frac{m^2}{x^2}\right)^{1/2}H^{(2)}_1(m(x^2)^{1/2}),
\label{12.7}
\end{eqnarray}
just as needed. Since we have established that the pole contribution and the exponential regulator calculation are equal, we confirm that there is no circle at infinity contribution.

If we include interactions the propagator acquires a self energy $\Sigma(p)$ and the denominator in the propagator becomes 
\begin{eqnarray}
4p_+p_--(p_1)^2-(p_2)^2-m^2-\Sigma(p)+i\epsilon=4p_-\left(p_+-((p_1)^2+(p_2)^2+m^2+\Sigma(p))/4p_-+i\epsilon/4p_-\right).
\label{12.8}
\end{eqnarray}
Now if the interacting Hamiltonian is Hermitian the poles would still be at real $p_+$, and they would still be below the real $p_+$ axis if $p_-> 0$. However, the poles could move into the negative $p_+$ region. As long as the shift in $p_+$ is finite (after renormalization if necessary), one can always redefine the zero of energy so as to make all poles be at nonnegative $p_+$. Thus even in the interacting case we must still restrict to $p_-> 0$ so as to ensure that all poles stay below the real $p_+$ axis, and even in the interacting case we recover the $p_+>0$, $p_-> 0$ conditions that are used in the on-shell Light-Front Hamiltonian formalism. 

\section{Light-front Fock Space Formalism in the Scalar Field Non-vacuum Sector}
\label{S13}

With the lower half $p_+$ plane pole contribution to the light-front non-vacuum $x^+>0$ forward in time propagator being given in (\ref{12.4}), to obtain the non-vacuum light-front backward propagator we need to close the contour above the real $p_+$ axis since the circle at infinity is then suppressed when $x^+<0$. And in this case the poles that lie above the real $p_+$ axis will have negative $p_-$. Doing the $p_+$ contour integral in (\ref{12.1}) gives 
\begin{eqnarray}
D(x^+<0,{\rm front},{\rm pole})=\frac{2i\theta(-x^+)}{(2\pi)^3}\int_{-\infty}^{\infty}dp_1\int_{-\infty}^{\infty}dp_2\int_{-\infty}^0 \frac{dp_-}{4p_-}e^{-i(F_p^2x^+/4p_-+p_-x^-+p_1x^1+p_2x^2)},
\label{13.1}
\end{eqnarray}
where $F_p^2=(p_1)^2+(p_2)^2+m^2$.
Changing the variables from $p_1$, $p_2$, $p_-$ to $-p_1$, $-p_2$, $-p_-$ gives 
\begin{eqnarray}
D(x^+<0,{\rm front},{\rm pole})&=&-\frac{2i\theta(-x^+)}{(2\pi)^3}\int_{-\infty}^{\infty}dp_1\int_{-\infty}^{\infty}dp_2\int_{0}^{\infty}\frac{dp_-}{4p_-}e^{-i(-F_p^2x^+/4p_--p_-x^--p_1x^1-p_2x^2)}
\nonumber\\
&=& -\frac{2i\theta(-x^+)}{(2\pi)^3}\int_{-\infty}^{\infty}dp_1\int_{-\infty}^{\infty}dp_2\int_{0}^{\infty}\frac{dp_-}{4p_-}e^{i(F_p^2x^+/4p_-+p_-x^-+p_1x^1+p_2x^2)}.
\label{13.2}
\end{eqnarray}
Combining with (\ref{12.4}) the full $D(x^{\mu},{\rm front})$ is thus of the form 
\begin{eqnarray}
D(x^{\mu},{\rm front})&=&-\frac{2i\theta(x^+)}{(2\pi)^3}\int_{-\infty}^{\infty}dp_1\int_{-\infty}^{\infty}dp_2\int_0^{\infty} \frac{dp_-}{4p_-}e^{-i(F_p^2x^+/4p_-+p_-x^-+p_1x^1+p_2x^2)}
\nonumber\\
 &&-\frac{2i\theta(-x^+)}{(2\pi)^3}\int_{-\infty}^{\infty}dp_1\int_{-\infty}^{\infty}dp_2\int_{0}^{\infty}\frac{dp_-}{4p_-}e^{i(F_p^2x^+/4p_-+p_-x^-+p_1x^1+p_2x^2)}.
\label{13.3}
\end{eqnarray}
In (\ref{13.3}) there are no $p_-<0$ contributions. With (\ref{13.3}) $D(x^{\mu},{\rm front})$ is now written in a form that we can compare with the Light-Front Hamiltonian expectation.

To obtain the non-vacuum Light-Front Hamiltonian expectation for $D(x^{\mu},{\rm front})$ we expand the $\phi$ field in terms of on-shell light-front creation and annihilation operators as  
\begin{align}
\phi(x^+,x^-,x^1,x^2)&=\frac{2}{(2\pi)^{3/2}}\int_{-\infty}^{\infty}dp_1\int_{-\infty}^{\infty}dp_2\int_0^{\infty} \frac{dp_-}{(4p_-)^{1/2}}
\Big{[}e^{-i(F_p^2x^+/4p_-+p_-x^-+p_1x^1+p_2x^2)}a(\vec{p})
\nonumber\\
&+e^{i(F_p^2x^+/4p_-
+p_-x^-+p_1x^1+p_2x^2)}a^{\dagger}(\vec{p})\Big{]},
\label{13.4}
\end{align}
and note that the only dependence on $m$ is in the $F_p^2=(p_1)^2+(p_2)^2+m^2$ term in the exponentials, and that the integration range for $p_-$ is only over $p_-\geq 0$. The light-front $[a(\vec{p}),a^{\dagger}(\vec{p}^{~\prime})]$ commutator  is normalized to $[a(\vec{p}),a^{\dagger}(\vec{p}^{~\prime})]=(1/2)\delta(p_--p_-^{\prime})\delta(p_1-p_1^{\prime})\delta(p_2-p_2^{\prime})$ (the factor of $1/2$ being required in light-front coordinates), as fixed via the normalization of the equal light-front time $[\phi(x),\partial_-\phi(y)]$ canonical commutator given in (\ref{3.10}). With this normalization we can then insert this on-shell form for $\phi(x)$ into $D(x^{\mu},{\rm front})=-i\langle \Omega_F |[\theta(x^+)\phi(x^{\mu})\phi(0)+\theta(-x^+)\phi(0)\phi(x^{\mu})]|\Omega_F\rangle$ as evaluated in the light-front vacuum $|\Omega_F\rangle$ that the light-front $a(\vec{p})$ annihilate. With the insertion of this $\phi(x)$ then precisely giving (\ref{13.3}), we establish the equivalence of the Feynman and on-shell Fock space prescriptions in the non-vacuum light-front case. As we see, just as with the instant-time case, since there are only pole terms and no circle at infinity contributions  in the non-vacuum case, we are able to establish the equivalence of the on-shell Light-Front Hamiltonian and Feynman diagram prescriptions in such cases, while thus validating the standard non-vacuum  on-shell Light-Front Hamiltonian prescription that is widely used in light-front studies.

\section{Scalar Field  Vacuum  Sector Feynman Diagrams}
\label{S14}

\subsection{Defining the Vacuum Sector Diagrams}
\label{S14a}

\begin{figure}[H]
\begin{center}
\includegraphics[scale=0.15]{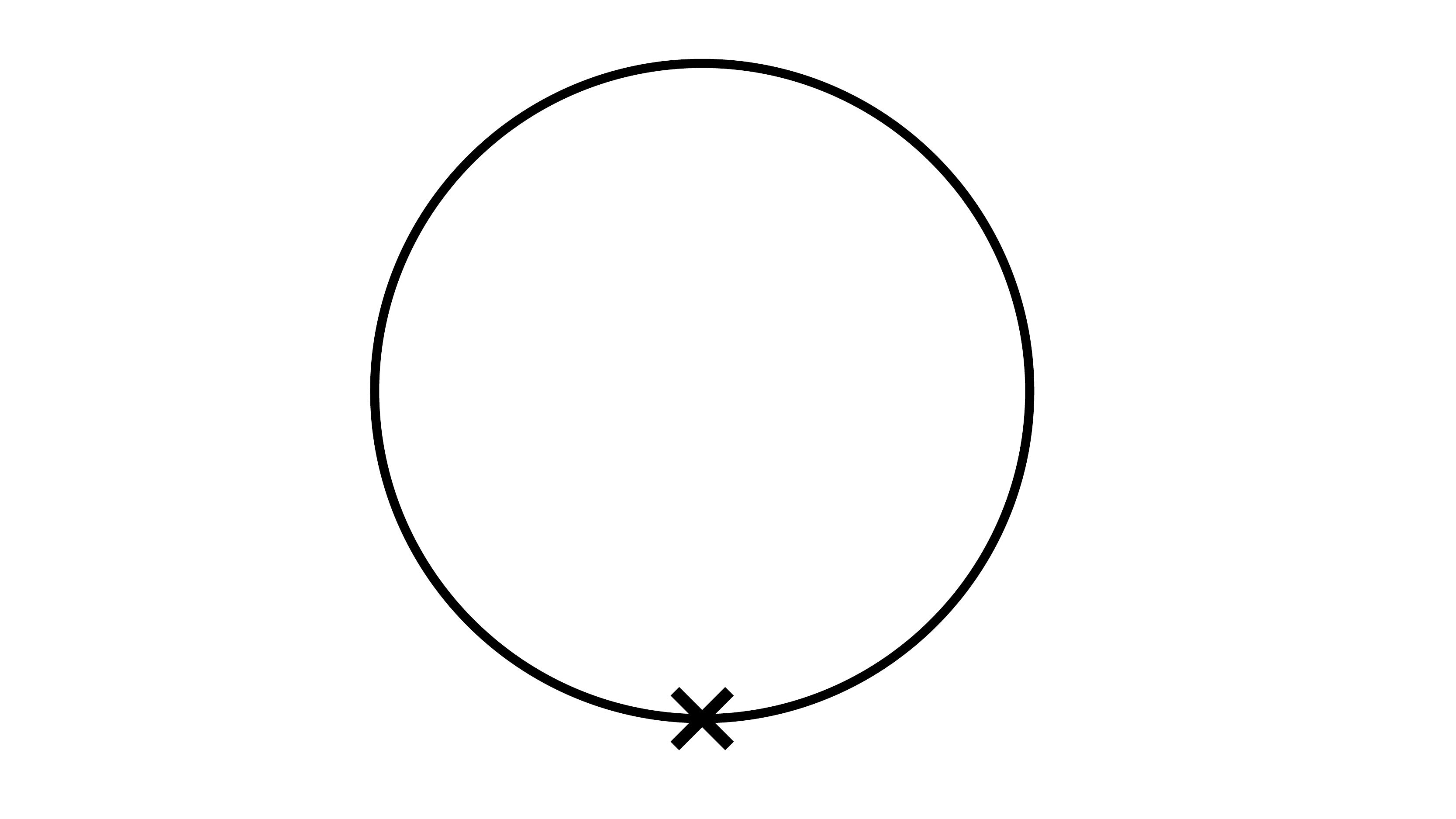}
\caption{Connected $\langle\Omega|\phi(0)\phi(0)|\Omega\rangle$}
\label{undressedtadpole}
\end{center}
\end{figure}

To obtain and give a meaning to a vacuum bubble diagram by point-splitting, we have found it more convenient to set $x^{\mu}=0$ in the time-ordered product of fields $D(x^{\mu})$ rather than in the two-point function $-i\langle \Omega |\phi(x^{\mu})\phi(0)|\Omega\rangle$, and this yields  
\begin{eqnarray}
D(x^{\mu})&=&-i\langle \Omega |[\theta(\tau)\phi(x^{\mu})\phi(0)+\theta(-\tau)\phi(0)\phi(x^{\mu})]|\Omega\rangle
\nonumber\\
&\rightarrow& -i\langle \Omega |[\theta(0^+)\phi(0)\phi(0)+\theta(0^-)\phi(0)\phi(0)]|\Omega\rangle
=-i\langle \Omega |\phi(0)\phi(0)|\Omega\rangle,
\label{14.1}
\end{eqnarray}
where $\tau=x^0$ in the instant-time case and $\tau=x^+$ in the light-front case, and where, following (\ref{14.4}) below, we have set $\theta(0^+)=\theta(0^-)=1/2$. When $x^{\mu}$  is nonzero, $D(x^{\mu})$ can be represented as a single line in coordinate space propagating from the origin of coordinates to $x^{\mu}$. In momentum space $D(x^{\mu})$ contains all momenta. The vacuum bubble is then obtained in coordinate space by linking $x^{\mu}$ back to the origin of coordinates, to give a (zero radius) circle in coordinate space. At $x^{\mu}=0$ the time ordering in $D(x^{\mu})$ drops out, with the respective $D(x^{\mu}=0)$ being given as the Feynman diagrams
\begin{eqnarray}
D(x^{\mu}=0,{\rm instant})&=&\frac{1}{(2\pi)^4}\int dp_0dp_1dp_2dp_3\frac{1}{(p_0)^2-(p_1)^2-(p_2)^2-(p_3)^2-m^2+i\epsilon},
\nonumber\\ 
D(x^{\mu}=0,{\rm front})&=&\frac{2}{(2\pi)^4}\int dp_+dp_1dp_2dp_-\frac{1}{4p_+p_--(p_1)^2-(p_2)^2-m^2+i\epsilon},
\label{14.2}
\end{eqnarray}
which still contain all momenta. As shown in Fig. \ref{undressedtadpole}, the bubble can be represented as a circle with a cross on the circumference, with the cross representing a $\phi^2$ insertion. The tadpole graph would occur as a one loop graph in a $\lambda \phi^3$ theory, with an external $\phi$ field bringing zero momentum into the cross where two $\phi$ fields are created. It is thus the limit in which points are brought together in a coordinate space Wick expansion, and is thus a limit of a Feynman time-ordered propagator, with the $i\epsilon$ prescription in (\ref{14.2}) originating in the $i\epsilon$ in the time-ordered product. It is not the limit $x^{\mu}\rightarrow 0$ of a two-point function such as $-i\langle \Omega |\phi(x)\phi(0)|\Omega\rangle$. The tadpole graph is not a disconnected graph (a disconnected graph would be a circle without any insertion). The tadpole graph would also appear in a $g\phi\bar{\psi}\psi$ theory with the cross representing a fermion-antifermion insertion at the point where the scalar $\phi$ brings zero momentum into the loop. The tadpole graph appears in mass renormalization, in theories of symmetry breaking as described in the Appendix, and it can also appear in the matter energy-momentum tensor that is coupled to gravity (for constant $\phi$ in (\ref{1.2}) $T_{\mu\nu}=\tfrac{1}{2}g_{\mu\nu}m^2\phi^2$, while for the fermion (\ref{4.8a}) $g^{\mu\nu}T_{\mu\nu}=m\bar{\psi}\psi$), to thus be of relevance to cosmology and the cosmological constant problem.

Since we do obtain a closed vacuum bubble graph by letting $x^{\mu}\rightarrow 0$ in $D(x^{\mu})$, with either $x^0$ or $x^+$ now zero we would no longer be able to use either as a regulator. Vacuum and non-vacuum diagrams thus need to be treated separately from each other. Moreover, with the lower half plane circle at infinity contribution being suppressed in a complex $\omega$ plane contour integral when $\tau>0$ ($\tau$ being $x^0$ or $x^+$), by closing in the lower half plane we obtain a representation of the distribution $\theta(\tau>0)$  of the form 
\begin{align}
\theta(\tau>0)=-\frac{1}{2\pi i}\int d\omega \frac{e^{-i\omega \tau}}{\omega+i\epsilon}
=-\frac{1}{2\pi i}\left[\oint_{-\infty}^{\infty} d\omega \frac{e^{-i\omega \tau}}{\omega+i\epsilon}
-\int_{2\pi}^{\pi}\frac{iRe^{i\theta}d\theta e^{-i\omega \tau}}{Re^{i\theta}}\right]
=-\frac{1}{2\pi i}\left(-2\pi i-0\right)=1.
\label{14.3}
\end{align}
However if we set $\tau$ to zero we lose the regulator and now the circle does contribute. Closing below the real $\omega$ axis gives 
\begin{align}
\theta(0)=-\frac{1}{2\pi i}\int \frac{ d\omega}{\omega+i\epsilon}
=-\frac{1}{2\pi i}\left[\oint_{-\infty}^{\infty}  \frac{ d\omega}{\omega+i\epsilon}
-\int_{2\pi}^{\pi}\frac{iRe^{i\theta}d\theta }{Re^{i\theta}}\right]=-\frac{1}{2\pi i}(-2\pi i+\pi i)=\frac{1}{2},
\label{14.4}
\end{align}
 viz. the value we used for $\theta(0)$ in (\ref{1.23a}) and (\ref{14.1}), as achieved precisely because of the circle contribution. Thus unless we can suppress the circle contribution by some means in graphs with $\tau=0$, we will need to include it. While we actually can suppress the circle contribution in the instant-time vacuum case since in that case there are two powers of momenta in propagator denominators, to thus give a $1/(p_0)^2$ circle at infinity suppression, in the light-front case the circle contribution cannot be suppressed  since we only have a $1/p_+$ suppression coming from the propagator. The light-front vacuum sector is thus seen to be qualitatively different from the instant-time vacuum sector.
 
In passing we should note that one does not need to actually rely on suppression of the lower half circle at infinity, since even if one closes above the real $\omega$ axis, a choice of contour for which $\tau>0$ then causes  $\exp(-iR\cos \theta \tau+R\sin\theta \tau)$ to diverge point by point, it must still be the case that one gets the same answer as when one closes below, as one can evaluate $\int_{-\infty}^{\infty}dp_+$ for real $p_+$ by closing the contour any way one wants. What happens on the upper half plane circle when $\tau>0$ is that while  $\exp(R\sin\theta \tau)$ diverges point by point the divergent terms cancel when one does the integral along the upper half circle (the $\exp(-iR\cos \theta \tau)$ term oscillates the divergence away \cite{Mannheim2005}),  with the surviving contributions then coming from the endpoints \cite{Mannheim2005}, points at which $\sin\theta=0$, i.e., points at which there is no divergence. To illustrate the issue, one can evaluate  $I=\int_{-\infty}^{\infty}d\omega e^{-i\omega \tau}$ by closing above or below. On closing below or above one obtains 
 \begin{eqnarray}
 I({\rm below})&=&-\int_{2\pi}^\pi Rie^{i\theta}d\theta\exp(-i\tau Re^{i\theta})=-\frac{i}{\tau}\exp(-i\tau Re^{i\theta})\Big{|}_{2\pi}^{\pi}= \frac{2}{\tau}\sin R\tau, 
 \nonumber\\
 I({\rm above})&=&-\int_{0}^{\pi}Rie^{i\theta}d\theta\exp(-i\tau Re^{i\theta})=-\frac{i}{\tau}\exp(-i\tau Re^{i\theta})\Big{|}_{0}^{\pi}=\frac{2}{\tau}\sin R\tau. 
 \label{14.5}
 \end{eqnarray}
On taking the $R\rightarrow \infty$ limit in either case one obtains $I=2\pi\delta(\tau)$, just as one should. Then, since $d\theta(\tau)/d\tau=\delta(\tau)$, the same remarks apply to the contour integral representation of $\theta(\tau)$.

Since we have already shown the equivalence of the four-dimensional $D(x^0>0,{\rm instant})$ and $D(x^+>0,{\rm front})$ Feynman diagrams when $x^{\mu}$ is nonzero, their equivalence must persist even if we take $x^{\mu}$ to be zero. However, while the equivalence must persist, there are some subtleties associated with the circle at infinity contributions that we need to take into account since we have lost the use of $x^{0}$ and $x^+$ as circle at infinity regulators.  While we shall discuss these subtleties below, we note that technically it is not actually necessary to address the issue, since we can use the $x^{\mu}$ in $D(x^{\mu})$ as the regulator. Specifically, if we consider $D(x^{\mu})$ to be a point split version of $D(x^{\mu}=0)$, we can define $D(x^{\mu}=0)$ once and for all as the $x^{\mu}\rightarrow 0$ limit of $D(x^{\mu})$, and indeed that is how tadpole graphs occur in a field theoretic Feynman diagram Wick expansion. With such a definition the instant-time and light-front vacuum bubbles are automatically equal, and neither is zero. 

\subsection{Instant-time Vacuum Sector Diagrams}
\label{S14b}
 
Nonetheless, it is still of interest to evaluate $D(x^{\mu}=0)$ directly as a contour integral. Evaluating $D(x^{\mu}=0)$ in instant-time quantization as a contour integral is straightforward since there are two powers of $p_0$ in the denominator. Thus in the complex $p_0$ plane one can still drop the circle at infinity, close below and either only include poles below the real $p_0$ axis or evaluate the integral on the real $p_0$ axis by the regulator method. Given the mathematical equivalence of (\ref{10.2}) and (\ref{10.4}), for both procedures one thus obtains
\begin{eqnarray}
D(x^{\mu}=0,{\rm instant})&=&D(x^{\mu}=0,{\rm instant},{\rm regulator})=D(x^{\mu}=0,{\rm instant},{\rm pole})
\nonumber\\
&=&-\frac{i}{(2\pi)^3}\int_{-\infty}^{\infty} \frac{d^3p}{2E_p}
=-\frac{1}{16\pi^2}\int_0^{\infty}\frac{d\alpha}{\alpha^2}e^{-i\alpha m^2-\alpha\epsilon}.
\label{14.6}
\end{eqnarray}

\subsection{Light-front Vacuum Sector Diagrams}
\label{S14c}

However in the light-front case there is only one power of $p_+$ in the denominator,  and thus in the complex $p_+$ plane one cannot ignore the circle at infinity (in either the lower-half or upper-half plane). We can of course still use the exponential regulator on the real $p_+$ axis, and in analog to  (\ref{12.3}) this leads us directly to 
\begin{align}
D(x^{\mu}=0,{\rm front},{\rm regulator})
&=-\frac{2i}{(2\pi)^4}\int_{-\infty}^{\infty} dp_+dp_-dp_1dp_2
\int_0^{\infty}
d\alpha e^{i\alpha(4p_+p_--(p_1)^2-(p_2)^2-m^2+i\epsilon)}
\nonumber\\
&=-\frac{2i}{(2\pi)^3}\int_{-\infty}^{\infty} dp_-\int_0^{\infty}d\alpha e^{-i\alpha m^2-\alpha\epsilon}\frac{\pi}{i\alpha}\delta(4\alpha p_-)
\nonumber\\
&=-\frac{1}{16\pi^2}\int_0^{\infty}\frac{d\alpha}{\alpha^2}e^{-i\alpha m^2-\alpha\epsilon}
\label{14.7}
\end{align}
for the light-front vacuum bubble, to thus be both nonzero and in complete agreement with the instant-time vacuum bubble given in (\ref{14.6}). The presence of the $\delta(4\alpha p_-)$ term shows the centrality of the $p_-=0$ mode.

If we instead wish to evaluate $D(x^{\mu}=0,{\rm front})$ via contour integration,  for nonnegative $p_-$ we can enclose poles in the lower-half $p_+$ plane as per (\ref{12.2}), and symbolically set $\int_{-\infty}^{\infty}dp_+=-\int_{2\pi}^{\pi}iRe^{i\theta}d\theta+{\rm pole~contribution}$. Moreover, as well as needing to have to include the circle at infinity, just as with $D(x^{\mu}\neq 0,{\rm front})$ we need to take into account the fact that $p_-$ can vanish, to then make $F_p^2/4p_-=((p_1)^2+(p_2)^2+m^2)/4p_-$ infinite, and thus put a $p_+$ pole on the circle at infinity.  For the pole term we thus again have to introduce the $\delta$ cutoff at small $p_-$, and on doing so obtain
\begin{align}
D(x^{\mu}=0,{\rm front},{\rm pole})&=\frac{-2i}{(2\pi)^3}\int_{-\infty}^{\infty} dp_1\int_{-\infty}^{\infty}dp_2 \int_{\delta}^{\infty}\frac{dp_-}{4p_-}.
\label{14.8}
\end{align}
Then on setting $p_-=1/\alpha$ we obtain 
\begin{align}
D(x^{\mu}=0,{\rm front},{\rm pole})&=-\frac{i}{16\pi^3}\int_{-\infty}^{\infty} dp_1\int_{-\infty}^{\infty}dp_2 \int_{0}^{1/\delta}\frac{d\alpha}{\alpha}
\nonumber\\
&=-\frac{i}{16\pi^3}\int_{-\infty}^{\infty} dp_1\int_{-\infty}^{\infty}dp_2 \int_{0}^{\infty}\frac{d\alpha}{\alpha},
\label{14.9}
\end{align}
where, as in the nonzero $x^{\mu}$ case, we have let $\delta$ go to infinity at the end of the calculation. As we see, $D(x^{\mu}=0,{\rm front},{\rm pole})$ is not equal to $D(x^{\mu}=0,{\rm front},{\rm regulator})$, and thus the circle at infinity contribution $D(x^{\mu}=0,{\rm front},{\rm lower~circle})$ cannot be zero. Moreover, as written, (\ref{14.9}) is not just not equal to (\ref{14.7}), it is even independent of $m$. And in addition it is not equal to the $m$-dependent $x^{\mu}\rightarrow 0$ limit of $D(x^+>0,{\rm front},{\rm pole})$ as given in (\ref{12.6}), to thus show the role that $x^{\mu}$ can play as a regulator, with evaluating the pole graphs before or after taking the $x^{\mu}\rightarrow 0$ limit not coinciding. The reason why they do not coincide is because in the $x^+/4\delta$ upper limit in the integral in (\ref{12.5}) we let $\delta $ go to zero while keeping $x^+$ nonzero, while in (\ref{14.8}) we let $x^+$ go to zero before letting $\delta$ go to zero. Thus in the light-front case the $x^{\mu}\rightarrow 0$ limit is singular. Since the $x^{\mu}\rightarrow 0$ limit is not singular in the instant-time case, we see that there is a qualitative difference between the light-front and instant-time vacuum sectors

Moreover, since $D(x^{\mu}=0,{\rm front},{\rm lower~circle})$ has to be nonzero, it cannot be given as the $x^{\mu}\rightarrow 0$ limit of $D(x^+>0,{\rm front},{\rm lower~circle})$ since $D(x^+>0,{\rm front},{\rm lower~circle})$ is zero. To evaluate $D(x^{\mu}=0,{\rm front},{\rm lower~circle})$ we set $p_+=Re^{i\theta}$ and write the  circle contribution to $D(x^{\mu}=0,{\rm front})$ in the form
\begin{align}
D(x^{\mu}=0,{\rm front},{\rm lower~circle})&=-
\frac{2}{(2\pi)^4}\int_{-\infty}^{\infty}dp_1\int_{-\infty}^{\infty}dp_2\int_{-\infty}^{\infty}dp_-
\int_{2\pi}^{\pi}iRe^{i\theta}d\theta\frac{1}
{4Re^{i\theta}p_--F_p^2+i\epsilon},
\label{14.10}
\end{align}
where $F_p^2=(p_1)^2+(p_2)^2+m^2$.  With $D(x^{\mu}=0,{\rm front},{\rm lower~circle})$ being $m$ dependent, with $D(x^{\mu}=0,{\rm front},{\rm regulator})$ being $m$ dependent, and with $D(x^{\mu}=0,{\rm front},{\rm pole})$ being $m$ independent, since they are related by
\begin{eqnarray}
D(x^{\mu}=0,{\rm front},{\rm regulator})=D(x^{\mu}=0,{\rm front},{\rm lower~circle})+D(x^{\mu}=0,{\rm front},{\rm pole}),
\label{14.11}
\end{eqnarray}
we see that without needing to calculate $D(x^{\mu}=0,{\rm front},{\rm lower~circle})$ explicitly, it must be the case that the $m$-independent pole contribution is cancelled completely by the lower circle contribution, to yield a net $m$-dependent $D(x^{\mu}=0,{\rm front},{\rm lower~circle})+D(x^{\mu}=0,{\rm front},{\rm pole})$.

Since the pole term contribution is cancelled by the circle contribution, there must be an alternative way of calculating $D(x^{\mu}=0,{\rm front})$ that avoids having to deal with  pole terms altogether. It can thus only involve circle at infinity contributions, and we present it here since it is actually instructive in its own right. To formulate it we note that we can use the exponential regulator on the circle at infinity provided it is well-behaved. Thus we set  
\begin{align}
D(x^{\mu}=0,{\rm front},{\rm circle})=-\frac{2i}{(2\pi)^4}\int_{-\infty}^{\infty} dp_+\int_{-\infty}^{\infty}dp_-\int_{-\infty}^{\infty}dp_1\int_{-\infty}^{\infty}dp_2\int_0^{\infty}
d\alpha e^{i\alpha(4p_+p_--(p_1)^2-(p_2)^2-m^2+i\epsilon)}.
\label{14.12}
\end{align}
From (\ref{14.12}) we see that on setting $p_+=Re^{i\theta}$ on a circle at infinity of radius $R$ we can get convergence at $\alpha=\infty$ if $4i\alpha p_-R(\cos\theta+i\sin\theta)=4i\alpha p_-R\cos\theta-4\alpha p_-R\sin\theta$ converges, i.e., if $p_-\sin\theta$ is positive. With positive $p_-$ this would then require that $\sin\theta$ be positive, while negative $p_-$ would require that $\sin\theta$ be negative. Now $\sin\theta$ is positive for $0<\theta <\pi$, and negative for $\pi<\theta<2\pi$. Thus for positive $p_-$ we must close above the real $p_+$ axis (in contrast for $D(x^{\mu}=0,{\rm front},{\rm lower~circle})$ discussed above where we closed below the real axis), while for negative $p_-$ we must close below the real $p_+$ axis. However, for positive $p_-$ the poles in $p_+$ are below the real axis, while for negative $p_-$ the poles in $p_+$ are above the real axis. Thus in applying the exponential regulator on the circle at infinity we always have to close the contour so that we do not encounter any poles at all, to thereby show that the circle contribution cannot be ignored.  

Symbolically we can  set 
\begin{eqnarray}
\int_{-\infty}^{\infty} dp_+=\int_{-\infty}^{\infty} dp_+(p_->0)+\int_{-\infty}^{\infty} dp_+(p_-<0)=-\int_{0}^{\pi}iRe^{i\theta}d\theta(p_->0)-\int_{2\pi}^{\pi}iRe^{i\theta}d\theta(p_-<0).
\label{14.13}
\end{eqnarray}
And thus for $p_->0$ we obtain an upper circle contribution to $D(x^{\mu}=0,{\rm front})$ of the form
\begin{align}
&D(x^{\mu}=0,p_->0,{\rm front},{\rm upper~circle})
\nonumber\\
&=\frac{2i}{(2\pi)^4}\int_0^{\infty}dp_-\int_{-\infty}^{\infty} dp_1\int_{-\infty}^{\infty}dp_2\int _0^{\pi} iRe^{i\theta}d\theta\int_0^{\infty}d\alpha e^{i\alpha(4p_-Re^{i\theta}-(p_1)^2-(p_2)^2-m^2+i\epsilon)}
\nonumber\\
&=\frac{1}{8\pi^3}\int_0^{\infty}dp_-\int_0^{\infty}\frac{d\alpha}{\alpha}e^{-i\alpha m^2-\alpha \epsilon}
\int _0^{\pi} iRe^{i\theta}d\theta e^{4i\alpha p_-Re^{i\theta}}
\nonumber\\
&=\frac{1}{8\pi^3}\int_0^{\infty}dp_-\int_0^{\infty}\frac{d\alpha}{\alpha}e^{-i\alpha m^2-\alpha \epsilon}
\frac{e^{4i\alpha p_-Re^{i\theta}}}{4i\alpha p_-}\Big{|}_0^{\pi}
\nonumber\\
&=\frac{1}{8\pi^3}\int_0^{\infty}dp_-\int_0^{\infty}\frac{d\alpha}{\alpha}e^{-i\alpha m^2-\alpha \epsilon}\frac{(e^{-4i\alpha p_-R}-e^{4i\alpha p_-R})}{4i\alpha p_-}
\nonumber\\
&=-\frac{1}{4\pi^3}\int_0^{\infty}dp_-\int_0^{\infty}\frac{d\alpha}{\alpha}e^{-i\alpha m^2-\alpha \epsilon}
\frac{\sin(4\alpha p_-R)}{4\alpha p_-},  
\label{14.14}
\end{align}
and note in passing the role of the endpoints in the $\theta$ integral, as per (\ref{14.5}). Finally, on letting $R$ go to infinity we obtain
\begin{align}
&D(x^{\mu}=0,p_->0,{\rm front},{\rm upper~circle})=-\frac{1}{4\pi^2}\int_0^{\infty}dp_-\int_0^{\infty}\frac{d\alpha}{\alpha}e^{-i\alpha m^2-\alpha \epsilon}
\delta(4\alpha p_-) 
\nonumber\\
&=-\frac{1}{8\pi^2}\int_{-\infty}^{\infty}dp_-\int_{0}^{\infty}\frac{d\alpha}{\alpha}e^{-i\alpha m^2-\alpha \epsilon}\delta(4\alpha p_-) 
=-\frac{1}{32\pi^2}\int_{0}^{\infty}\frac{d\alpha}{\alpha^2}e^{-i\alpha m^2-\alpha \epsilon}.
\label{14.15}
\end{align}
Moreover, since $\exp(4i\alpha p_-Re^{i\theta})|_{0}^{\pi}=\exp(4i\alpha p_-Re^{i\theta})|_{2\pi}^{\pi}$, and since $\delta(4\alpha p_-)$ is even under $p_-\rightarrow -p_-$, it follows that $D(x^{\mu}=0,p_->0,{\rm front},{\rm upper~circle})$ and $D(x^{\mu}=0,p_-<0,{\rm front},{\rm lower~circle})$ are equal. The presence of the $\delta(4\alpha p_-)$ term again shows the centrality of $p_-=0$, with again only $p_-=0$ contributing. Thus finally we obtain
\begin{align}
D(x^{\mu}=0,{\rm front})&=D(x^{\mu}=0,p_->0,{\rm front},{\rm upper~circle})+D(x^{\mu}=0,p_-<0,{\rm front},{\rm lower~circle})
\nonumber\\
&=-\frac{1}{16\pi^2}\int_0^{\infty}\frac{d\alpha}{\alpha^2}e^{-i\alpha m^2-\alpha \epsilon}.
\label{14.16}
\end{align}
We recognize (\ref{14.16}) as (\ref{14.7}), to thus show that one can determine $D(x^{\mu}=0,{\rm front})$ entirely by circle at infinity contributions, with pole terms making no contribution. We also recognize (\ref{14.16}) as (\ref{14.6}), and thus by direct evaluation again confirm that the instant-time and light-front vacuum bubbles are equal, with both being nonzero.

As a final comment, we note that as written in (\ref{14.2}) the momentum integrals in $D(x^{\mu}=0,{\rm instant})$ and $D(x^{\mu}=0,{\rm front})$ can be transformed into each other by a change of variables of the form $p_0\rightarrow p_+$, $p_3\rightarrow p_-$. Thus their equivalence is not surprising. However,  under such a transformation a contour in the complex $p_0$ plane would transform into a contour in the complex $p_+$ plane, with $D(x^{\mu}=0,{\rm instant})$ and $D(x^{\mu}=0,{\rm front})$ thus being equivalent before one actually performs the contour integrations, with only the full pole plus circle contribution on an instant-time contour mapping into the full pole plus circle contribution on a light-front contour. There is no separate mapping of pole into pole or circle into circle. And with there being no circle contribution in the instant-time case as the circle in the complex instant-time  $p_0$ plane is suppressed (two powers of $p_0$ in the $(p_0)^2-(p_1)^2-(p_2)^2-(p_3)^2$ term in the denominator), in a contour that closes below the real $p_0$ axis $D(x^{\mu}=0,{\rm instant})$ is given entirely by pole terms. In contrast, in $D(x^{\mu}=0,{\rm front})$ there is a circle contribution as the circle in the complex light-front  $p_+$ plane is not suppressed (only one power of $p_+$ in the $4p_+p_--(p_1)^2-(p_2)^2$ term in the denominator), in a contour that closes below the real $p_+$ axis $D(x^{\mu}=0,{\rm front})$ is given by both circle and pole terms. It thus only by including the contribution of light-front circle terms that the equivalence of  $D(x^{\mu}=0,{\rm instant})$ and $D(x^{\mu}=0,{\rm front})$ that we seek can be established.

\section{Vacuum  Sector Fock Space Formalism}
\label{S15}

\subsection{Instant-time Vacuum  Sector Fock Space Formalism}
\label{S15a}

It is of interest to compare and contrast our result with the on-shell Fock space treatment of vacuum bubbles. For the instant-time case, the $x^{\mu}=0$ instant-time field expansion that follows from (\ref{11.1}) is of the form 
\begin{eqnarray}
\phi(x^{\mu}=0)=\int \frac{d^3p}{(2\pi)^{3/2}(2E_p)^{1/2}}[a(\vec{p})+a^{\dagger}(\vec{p})],
\label{15.1}
\end{eqnarray}
an expression that we note is still $m$ dependent through the $E_p=(p^2+m^2)^{1/2}$ factor. The 
insertion of (\ref{15.1}) into the vacuum $D(x^{\mu}=0)=-i\langle \Omega|\phi(0)\phi(0)|\Omega\rangle$ immediately leads to
\begin{align}
&D(x^{\mu}=0,{\rm instant},{\rm Fock})=-\frac{i}{(2\pi)^3}\int_{-\infty}^{\infty} \frac{d^3p}{2E_p}
=-\frac{1}{16\pi^2}\int_0^{\infty}\frac{d\alpha}{\alpha^2}e^{-i\alpha m^2-\alpha \epsilon}.
\label{15.2}
\end{align}
Comparing with (\ref{14.6}) we obtain 
\begin{align}
D(x^{\mu}=0,{\rm instant},{\rm Fock})=D(x^{\mu}=0,{\rm instant},{\rm pole})
\label{15.3}
\end{align}
just as we should, since only poles contribute to the instant-time vacuum Feynman diagram. And since only poles contribute we also obtain 
\begin{align}
D(x^{\mu}=0,{\rm instant})=D(x^{\mu}=0,{\rm instant},{\rm Fock})=D(x^{\mu}=0,{\rm instant},{\rm pole}),
\label{15.4}
\end{align}
to thus show the equivalence of the Feynman and Fock space prescriptions in the instant-time vacuum sector.

\subsection{Light-front Vacuum  Sector Fock Space Formalism}
\label{S15b}

In the light-front vacuum sector  the $x^{\mu}=0$ light-front field expansion that follows from (\ref{13.4}) is of the form  
\begin{align}
\phi(x^{\mu}=0)=\frac{2}{(2\pi)^{3/2}}\int_{-\infty}^{\infty}dp_1\int_{-\infty}^{\infty}dp_2\int_0^{\infty} \frac{dp_-}{(4p_-)^{1/2}}
[a(\vec{p})+a^{\dagger}(\vec{p})],
\label{15.5}
\end{align}
and it has no dependence on the mass $m$ of the field whatsoever. Its insertion into $D(x^{\mu}=0)=-i\langle \Omega|\phi(0)\phi(0)|\Omega\rangle$ yields
\begin{align}
D(x^{\mu}=0,{\rm front},{\rm Fock})&=\frac{-2i}{(2\pi)^3}\int_{-\infty}^{\infty} dp_1\int_{-\infty}^{\infty}dp_2 \int_{0}^{\infty}\frac{dp_-}{4p_-},
\label{15.6}
\end{align}
an expression that consequently also has no dependence on the mass $m$ of the field.
Comparing with (\ref{14.8}) and (\ref{14.9}) we thus obtain 
\begin{align}
D(x^{\mu}=0,{\rm front},{\rm Fock})=D(x^{\mu}=0,{\rm front},{\rm pole}).
\label{15.7}
\end{align}
As we see, this time we do not recover the full light-front vacuum regulator value given in (\ref{14.7}), and indeed we could not since (\ref{14.7}) depends on $m$ while (\ref{15.6}) does not.  The general rule then is that the on-shell evaluation always coincides with the pole term evaluation, and if the pole is not the only contributor to the Feynman contour then the Feynman and Fock space prescriptions cannot agree and one must use the Feynman diagram prescription, with the off-shell Feynman approach containing information that cannot be accessed in an on-shell Fock space approach. And thus for the light-front vacuum sector $D(x^{\mu}=0,{\rm front})$ one must use the full four-dimensional Feynman formalism, and when one does one obtains a nonzero mass-dependent expression for $D(x^{\mu}=0,{\rm front})$ that coincides with the instant-time evaluation $D(x^{\mu}=0,{\rm instant})$ of the vacuum Feynman diagram. 

As well as being a useful diagnostic for vacuum graphs, as we show below, a dependence of vacuum graphs on mass or a lack thereof has a reflection in the infinite momentum frame analysis of vacuum graphs that we present below (a $v=c$ limit in which mass becomes negligible). It is thus of interest to identify why it is that there is no mass dependence in the light-front $\phi(x^{\mu}=0)$ in the first place. To be specific, we note  in the instant-time case given in (\ref{11.1}) that even if we set $x^{\mu}=0$ there is still a dependence on mass due to the $1/(2E_p)^{1/2}$ term in the integration measure. However in the light-front case given in (\ref{13.4}) this factor is replaced by the mass-independent $1/(4p_-)^{1/2}$, to leave the light-front $\phi(x^{\mu}=0)$ as given in (\ref{15.5}) with no mass dependence at all. The reason for this difference originates in the canonical quantization procedure. Specifically, in the instant-time case the canonical conjugate of $\phi$ is $\partial_0\phi$, and when (\ref{11.1}) is inserted in $[\phi,\partial_0\phi]$ this generates an $E_p$ factor that needs to be cancelled by the integration measure in (\ref{11.1}) so as to generate a delta function for $[\phi,\partial_0\phi]$. However, in the light-front case the canonical conjugate is $\partial_-\phi$, and one thus needs a $p_-$ factor in the integration measure in (\ref{13.4}) so as to generate a delta function for $[\phi,\partial_-\phi]$, to thereby cause the light-front $\phi(x^{\mu}=0)$ to be mass independent.

Beyond the fact that the light-front on-shell prescription only corresponds to the pole contribution to the Feynman contour and does not account for the circle at infinity contribution, we in addition note that
in the light-front on-shell approach to scattering processes one only needs to consider forward in time processes, and thus in the scattering case one only needs Green's functions such as the non-vacuum $D(x^+>0,{\rm front})$. In this case all intermediate states have to have $p_+>0$, $p_->0$, and with the tadpole graph involving a $p_+=0$, $p_-=0$ insertion, if this same reasoning were to be applied to vacuum graphs, there would then be no $p_+>0$, $p_->0$ on-shell contribution at all, with the tadpole graph then vanishing \cite{Brodsky:1997de}. However, this reasoning does not apply to vacuum graphs, since as we see from (\ref{14.1}) the vacuum graph involves both forward and backward in time scattering. Thus in the circle diagram given in Fig. \ref{undressedtadpole} a propagator that leaves the cross with positive outgoing $p_-$ goes around the circle and then brings positive $p_-$ back to the cross, with bringing positive $p_-$ back to the cross being equivalent to leaving the cross with outgoing negative $p_-$. In a scattering amplitude in the on-shell Light-Front Hamiltonian formalism all intermediate states would be outgoing and have $p_-$ positive, and thus the on-shell approach does not encompass negative $p_-$ or zero $p_-$ contributions. Consequently, vacuum bubbles have to be treated differently than scattering amplitudes, with light-front vacuum bubble diagrams being nonzero. Moreover, we note that if we dress vacuum bubbles there will always be one line in the diagram that brings zero $p_-$ into the cross where the insertion is located (see e.g. Fig. \ref{vacuumtadpoledressed} below). Thus in the light-front case all vacuum bubbles are nonvanishing, just as they are in instant-time quantization. This of course must be the case since the path integral and Feynman diagram equivalence that we established above for instant-time  and light-front Green's functions holds for their vacuum graph limit as well.

\subsection{The Light-front Vacuum Sector Fock Space Formalism Puzzle}
\label{S15c}

While we have established that the light-front vacuum graph is given in (\ref{14.7}) by the $m$-dependent $D(x^{\mu}=0,{\rm front},{\rm regulator})$, there is a puzzle since there does not seem to be anything wrong with the on-shell derivation of the $m$-independent $D(x^{\mu}=0,{\rm front},{\rm Fock})$ given in (\ref{15.6}) as it uses the on-shell $\phi(x^{\mu}=0)$ given in (\ref{15.5}). And indeed, this is the correct expression for the on-shell expansion for $\phi(x^{\mu}=0)$, as it follows directly from the Fock expansion for $\phi(x^{\mu}\neq 0)$ given in (\ref{13.4}). And all we have done is insert $\phi(x^{\mu}=0)$ into the $-i\langle \Omega |\phi(0)\phi(0)|\Omega\rangle$ vacuum graph,  and yet there is a problem. Moreover, this procedure does work for instant-time vacuum graphs, it is just for light-front graphs that there is a problem. 

The resolution of the puzzle is that since the product of two fields at the same spacetime point is divergent we have to first point split and then derive $D(x^{\mu}=0)$ from $D(x^{\mu}\neq 0)$. In the light-front case, but not in the instant-time case, letting $x^{\mu}$ go to zero first and then evaluating the $-i\langle \Omega |\phi(0)\phi(0)|\Omega\rangle$ matrix element is different from first evaluating $D(x^{\mu}\neq 0)$ and then letting $x^{\mu}$ go to zero. With it being the latter prescription that coincides with the construction of tadpole graphs in a Wick expansion in the first place, this is the one we must use. 

To identify the specific difference between the two prescriptions we need to develop a single formalism in which we can realize both prescriptions simultaneously, so that we can then compare and see exactly where the Fock space approach breaks down. Since we know that there are circle at infinity contributions in the Feynman diagram prescription but not in the Fock space prescription, we need to find a single formalism in which we bypass circle at infinity issues altogether.  To this end we need to represent time-ordering theta functions in a form that does not involve closing a  Feynman contour. We thus employ the real frequency axis exponential regulator, and set 
\begin{eqnarray}
\theta(x^+)&=&\frac{1}{2\pi }\int_{-\infty}^{\infty}  d\omega \int_0^{\infty}d\alpha e^{-i\omega x^+}e^{i\alpha(\omega+i\epsilon)}=\int_0^{\infty}d\alpha e^{-\alpha \epsilon}\delta(\alpha-x^+)
\nonumber\\
&=&\int_0^{\infty}d\alpha e^{-x^+ \epsilon}\delta(\alpha-x^+)=\int_0^{\infty}d\alpha \delta(\alpha-x^+),
\label{15.8}
\end{eqnarray}
with the $i\epsilon$ again providing convergence at $\alpha=\infty$. That $\int_0^{\infty}d\alpha \delta(\alpha-x^+)$ indeed is $\theta(x^+)$ follows since $\int_0^{\infty}d\alpha \delta(\alpha-x^+)=1$ if $x^+$ is positive, and $\int_0^{\infty}d\alpha \delta(\alpha-x^+)=0$ if $x^+$ is negative.

In evaluating $D(x^{\mu},{\rm front},{\rm regulator})$ with $x^{\mu}\neq 0$ and with no restriction on the sign of  $x^+$, we  follow (\ref{12.3}) and obtain

\begin{eqnarray}
&&D(x^{\mu},{\rm front},{\rm regulator})
\nonumber\\
&&=-\frac{2i}{(2\pi)^4}\int_{-\infty}^{\infty} dp_+\int_{-\infty}^{\infty} dp_1\int_{-\infty}^{\infty} dp_2\int_{-\infty}^{\infty} dp_-e^{-i(p_+x^++p_-x^-+p_1x^1+p_2x^2)}
\int_0^{\infty}
d\alpha e^{i\alpha(4p_+p_--(p_1)^2-(p_2)^2-m^2+i\epsilon)}
\nonumber\\
&&=-\frac{2i}{(2\pi)^3}\int_{-\infty}^{\infty} dp_1\int_{-\infty}^{\infty} dp_2\int _{0}^{\infty}dp_-e^{-i(p_-x^-+p_1x^1+p_2x^2)}
\int_0^{\infty}d\alpha e^{i\alpha(-(p_1)^2-(p_2)^2-m^2+i\epsilon)}\delta(4\alpha p_--x^+)
\nonumber\\
&&~~-\frac{2i}{(2\pi)^3}\int_{-\infty}^{\infty} dp_1\int_{-\infty}^{\infty} dp_2\int _{-\infty}^{0}dp_-e^{-i(p_-x^-+p_1x^1+p_2x^2)}
\int_0^{\infty}d\alpha e^{i\alpha(-(p_1)^2-(p_2)^2-m^2+i\epsilon)}\delta(4\alpha p_--x^+).
\label{15.9}
\end{eqnarray}
On changing the signs of $p_-$, $p_1$ and $p_2$ in the last integral, recalling that $F_p^2=(p_1)^2+(p_2)^2+m^2$, and substituting for $\alpha$ using the delta functions we obtain
\begin{eqnarray}
&&D(x^{\mu},{\rm front},{\rm regulator})
\nonumber\\
&&=-\frac{2i}{(2\pi)^3}\int_{-\infty}^{\infty} dp_1\int_{-\infty}^{\infty} dp_2\int _{0}^{\infty}\frac{dp_-}{4p_-}e^{-i(p_-x^-+p_1x^1+p_2x^2)}
\int_0^{\infty}d\alpha e^{ix^+(-F_p^2+i\epsilon)/4p_-}\delta(\alpha -x^+/4p_-)
\nonumber\\
&&~~-\frac{2i}{(2\pi)^3}\int_{-\infty}^{\infty} dp_1\int_{-\infty}^{\infty} dp_2\int _{0}^{\infty}\frac{dp_-}{4p_-}e^{i(p_-x^-+p_1x^1+p_2x^2)}
\int_0^{\infty}d\alpha e^{ix^+(F_p^2-i\epsilon)/4p_-}\delta(\alpha +x^+/4p_-).
\label{15.10}
\end{eqnarray}
Then, using (\ref{15.8}), and with the sign of $p_-$ not being negative  we obtain
\begin{eqnarray}
&&D(x^{\mu},{\rm front},{\rm regulator})
\nonumber\\
&&=-\frac{2i\theta(x^+)}{(2\pi)^3}\int_{-\infty}^{\infty} dp_1\int_{-\infty}^{\infty} dp_2\int _{0}^{\infty}\frac{dp_-}{4p_-}e^{-i(F_p^2x^+/4p_-+p_-x^-+p_1x^1+p_2x^2)}
\nonumber\\
&&~~-\frac{2i\theta(-x^+)}{(2\pi)^3}\int_{-\infty}^{\infty} dp_1\int_{-\infty}^{\infty} dp_2\int _{0}^{\infty}\frac{dp_-}{4p_-}e^{i(F_p^2x^+/4p_-+p_-x^-+p_1x^1+p_2x^2)}.
\label{15.11}
\end{eqnarray}
We recognize (\ref{15.11}) as being the $m$-dependent (\ref{13.3}), a result we had obtained both via pole contributions and via use of the Fock space expansion for $\phi(x^{\mu})$ given in (\ref{13.4}). 

Now if we set $x^{\mu}=0$ in (\ref{15.11}) we would appear to obtain the $m$-independent $D(x^{\mu}=0,{\rm front},{\rm Fock})$ given in (\ref{15.6}). However, we cannot take the $x^+\rightarrow 0$ limit since the quantity $x^+/4p_-$ is undefined if $p_-$ is zero, and $p_-=0$ is included in the integration range. Hence the limit is singular. 

To obtain a limit that is not singular we note that we can set $x^{\mu}$ to zero in (\ref{15.9}) as there the limit is well-defined, and this leads to 
\begin{eqnarray}
&&D(x^{\mu}=0,{\rm front},{\rm regulator})
\nonumber\\
&&=-\frac{2i}{(2\pi)^3}\int_{-\infty}^{\infty} dp_1\int_{-\infty}^{\infty} dp_2\int _{0}^{\infty}dp_-
\int_0^{\infty}d\alpha e^{i\alpha(-(p_1)^2-(p_2)^2-m^2+i\epsilon)}\delta(4\alpha p_-)
\nonumber\\
&&~~-\frac{2i}{(2\pi)^3}\int_{-\infty}^{\infty} dp_1\int_{-\infty}^{\infty} dp_2\int _{-\infty}^{0}dp_-
\int_0^{\infty}d\alpha e^{i\alpha(-(p_1)^2-(p_2)^2-m^2+i\epsilon)}\delta(4\alpha p_-)
\nonumber\\
&&=-\frac{2i}{(2\pi)^3}\int_{-\infty}^{\infty} dp_1\int_{-\infty}^{\infty} dp_2\int _{-\infty}^{\infty}dp_-
\int_0^{\infty}\frac{d\alpha}{4\alpha} e^{i\alpha(-(p_1)^2-(p_2)^2-m^2+i\epsilon)}\delta(p_-).
\label{15.12}
\end{eqnarray}
If we do the momentum integrations we obtain the $m$-dependent
\begin{eqnarray}
&&D(x^{\mu}=0,{\rm front},{\rm regulator})=-\frac{1}{16\pi^2}\int_0^{\infty}\frac{d\alpha}{\alpha^2}e^{-i\alpha m^2-\alpha\epsilon}.
\label{15.13}
\end{eqnarray}
We recognize (\ref{15.13}) as being of the same form as the $m$-dependent $D(x^{\mu}=0,{\rm front},{\rm regulator})$ given in (\ref{14.7}). We thus have to conclude that the limit $x^{\mu}\rightarrow 0$ of (\ref{15.11}) is not (\ref{15.6}) but is (\ref{15.13}) instead. The technical difference between (\ref{15.13}) and (\ref{15.11}) is that to obtain (\ref{15.11}) we did the $\alpha$ integration first, while to obtain (\ref{15.13}) we did the $p_-$ integration first. Only the latter procedure takes care of the $p_-=0$ contribution. Thus to conclude, we note that because of singularities we first have to point split, and when we do so we find that it is the $m$-dependent (\ref{15.13}) that is the correct value for the light-front vacuum graph.

\subsection{General Comments}
\label{S15d}

We should note that since the vacuum matrix elements that we have calculated here are divergent (for boson or fermion loops), they need to be renormalized. For flat spacetime instant-time vacuum graphs one can cancel divergences by normal ordering in the standard way. However, since normal ordering involves moving all annihilation operators to the right and all creation operators to the left in a vacuum matrix element, it does not encompass the circle at infinity contributions that occur in light-front vacuum graphs. For light-front vacuum graphs we thus need to deal with circle at infinity contributions and such contributions are foreign to standard renormalization techniques, and indeed in their presence one may not be able to effect a Wick rotation to Euclidean momenta. To get round this we note that for renormalization one does not actually need to consider circle at infinity contributions per se since one can evaluate Feynman diagrams using the real frequency axis exponential regulator, just as was done in the light-front (\ref{14.7}). Then one can introduce a second field with a regulator mass $M$, the Pauli-Villars prescription, and evaluate its contribution using the same exponential regulator and then take the difference. Moreover, beyond this, even if one were to normal order in the standard way, when a symmetry is broken dynamically by a fermion bilinear condensate one is actually interested in $\langle S|\colon\bar{\psi}\psi\colon|S\rangle$ where $|S\rangle$ is a spontaneously broken vacuum and the dots mean normal ordering with respect to the normal vacuum $|N\rangle$, with $\langle S|\colon\bar{\psi}\psi\colon|S\rangle$ being equal to $\langle S|\bar{\psi}\psi|S\rangle-\langle N|\bar{\psi}\psi|N\rangle$. I.e., one is interested in the change in the vacuum matrix element as one changes the vacuum. Finally, for curved spacetime one cannot normal order at all anyway since gravity couples to energy and not to energy difference. The one place left to eliminate matter source divergences is gravity itself. However then one would need a quantum gravitational theory. With conformal gravity being a consistent quantum gravity theory, in conformal gravity one is able to cancel vacuum divergences via an interplay between the gravity and matter source sectors, and this is discussed in \cite{Mannheim2017}.

\section{Infinite Momentum Frame Considerations}
\label{S16}

In the instant-time vacuum sector we had found that  $D(x^{\mu}=0,{\rm instant},{\rm Fock})$ and $D(x^{\mu}=0,{\rm instant},{\rm pole})$ are equal, with both being given by (\ref{15.2}). On transforming (\ref{15.2}) to the infinite momentum frame and comparing with (\ref{15.6}) and (\ref{14.8}) or (\ref{14.9}) we obtain 
\begin{eqnarray} 
&&D(x^{\mu}=0,{\rm instant},{\rm Fock})=D(x^{\mu}=0,{\rm instant},{\rm pole})=-\frac{i}{(2\pi)^3}\int_{-\infty}^{\infty} \frac{d^3p}{2E_p}
\nonumber\\
&&\rightarrow -\frac{i}{(2\pi)^3}\int _{-\infty}^{\infty} dp_1\int _{-\infty}^{\infty}dp_2\int _0^{\infty}\frac{dp_-}{2p_-}
=D(x^{\mu}=0,{\rm front},{\rm Fock})=D(x^{\mu}=0,{\rm front},{\rm pole}).
\label{16.1}
\end{eqnarray}
As such, the infinite momentum frame is doing what it is supposed to do, namely it is transforming an instant-time on-shell graph into a light-front on-shell graph. However, we have seen that the light-front mass-independent on-shell evaluation of the vacuum graph does not agree with correct mass-dependent value provided by the off-shell light-front vacuum Feynman diagram.  Thus in this respect not only is the on-shell prescription failing for light-front vacuum graphs, so is the infinite momentum frame prescription.

There is an oddity in (\ref{16.1}), one peculiar to the infinite momentum frame. Since the mass-dependent quantity $dp_3/2E_p$ is Lorentz invariant, under a Lorentz transformation with a velocity less than the velocity of light it must transform into itself and thus must remain mass dependent. However, in the infinite momentum frame it transforms into a quantity $dp_-/2p_-$ that is mass independent. This is because velocity less than the velocity of light and velocity equal to the velocity of light are inequivalent, since an observer that is able to travel at less than the velocity of light is not able to travel at the velocity of light. Lorentz transformations at the velocity of light are different than those at less than the velocity of light, and at the velocity of light observers (viz. observers on the light cone) can lose any trace of mass. 

Moreover, (\ref{16.1}) also raises a puzzle. Specifically,  while the instant-time on-shell evaluation of the vacuum $D(x^{\mu}=0,{\rm instant},{\rm Fock})$ does coincide with the instant-time evaluation of the vacuum off-shell Feynman diagram  $D(x^{\mu}=0,{\rm instant},{\rm regulator})$, and while the instant-time evaluation of the off-shell Feynman diagram  $D(x^{\mu}=0,{\rm instant},{\rm regulator})$ does coincide with the light-front evaluation of the off-shell Feynman diagram  $D(x^{\mu}=0,{\rm front},{\rm regulator})$, nonetheless, the light-front on-shell evaluation of the vacuum $D(x^{\mu}=0,{\rm front},{\rm Fock})$ does not coincide with the light-front evaluation of the off-shell Feynman diagram $D(x^{\mu}=0,{\rm front},{\rm regulator})$.

The resolution of this puzzle lies in the contribution of the circle at infinity to the Feynman contour. In the instant-time case the integral $\int dp_0dp_3/[(p_0)^2-(p_3)^2-(p_1)^2-(p_2)^2-m^2+i\epsilon]$ is suppressed on the circle at infinity in the complex $p_0$ plane ($p_3$ being finite), and only poles contribute. However, when one goes to the infinite momentum frame in the instant-time case $dp_3$ also becomes infinite ($p^3=mv/(1-v^2)^{1/2}$) and the circle contribution is no longer suppressed. Specifically,  on the instant-time circle at infinity, the term that is of relevance behaves as $\int Rie^{i\theta}d\theta dp_3/(R^2e^{2i\theta}-(p_3)^2)$, and on setting $\epsilon=1/R$ in the infinite momentum frame limit,  as per (\ref{9.32}) the circle term behaves as the unsuppressed $\int Rie^{i\theta}d\theta Rdp_-/(R^2e^{2i\theta}-R^2p_-^2)=\int ie^{i\theta}d\theta dp_-/(e^{2i\theta}-p_-^2)$. Thus in the instant-time case one cannot ignore the circle at infinity in the infinite momentum frame even though one can ignore it for observers moving with finite momentum. Consequently, the initial reduction from the instant-time Feynman diagram to the on-shell instant-time Hamiltonian prescription is not valid in the infinite momentum frame, and one has to do the full four-dimensional Feynman contour integral instead. 

We had noted earlier a general rule that the on-shell evaluation always coincides with the pole term evaluation, and that if the pole is not the only contributor to the Feynman contour then the Feynman and Hamiltonian prescriptions cannot agree and one must use the Feynman prescription. We can now add that if we ignore the effect of an infinite Lorentz boost on the instant-time circle at infinity, the instant-time infinite momentum frame evaluation always coincides with the light-front pole term evaluation, and if the light-front pole is not the only contributor to the light-front Feynman contour then the Feynman and infinite momentum frame evaluations cannot agree and one must use the light-front Feynman contour or exponential regulator prescription.

\section{Dressing the Vacuum Graph}
\label{S17}
\begin{figure}[H]
\begin{center}
\includegraphics[scale=0.15]{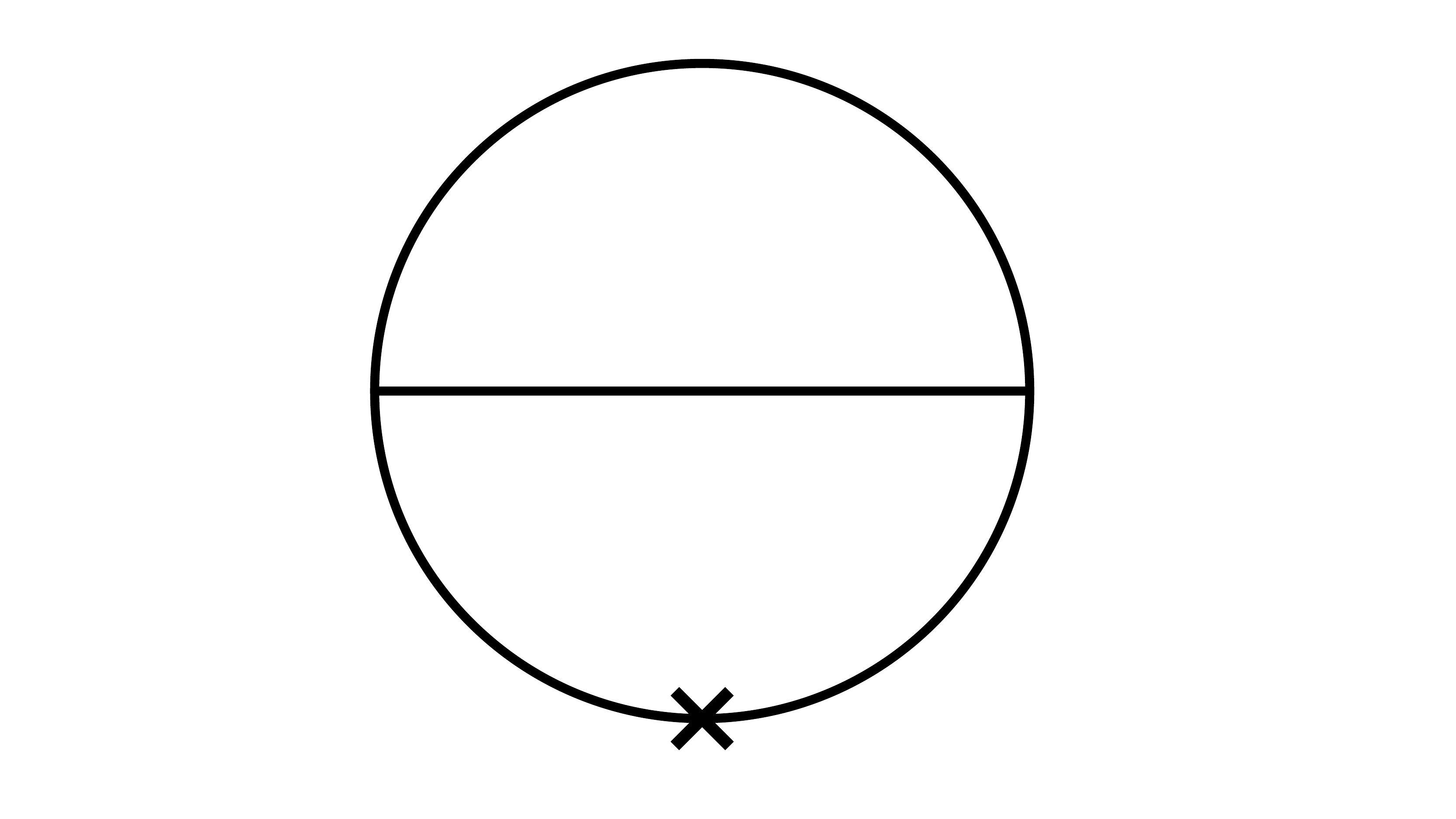}
\caption{Dressed $\langle\Omega|\phi(0)\phi(0)|\Omega\rangle$}
\label{vacuumtadpoledressed}
\end{center}
\end{figure}

While we have already established the equivalence of instant-time quantization and light-front quantization to all orders in perturbation theory, for completeness we discuss what explicitly happens to circle at infinity contributions in dressings to light-front tadpole graphs. The first dressing to the graph in Fig. \ref {undressedtadpole} is shown in Fig. \ref{vacuumtadpoledressed}. This graph is  actually a self-energy graph within a vacuum loop. To see this for the tadpole graph for instance, momentarily separate the lines at the cross. This then becomes a $\Sigma(p)$ self-energy renormalization graph. However, this renormalization comes with a $\delta m$ and a $Z$, and the graph can be replaced by a dressed propagator. To calculate 
\begin{eqnarray}
I=\int \frac{d^4kd^4p}{(4p_+p_--F^2_p+i\epsilon)(4k_+k_--F^2_k+i\epsilon)[4(k_++p_+)(k_-+p_-)-F^2_{p+k}+i\epsilon]},
\label{17.1}
\end{eqnarray}
viz. the Fig. \ref{vacuumtadpoledressed} tadpole with $F_p^2=(p_1)^2+(p_2)^2+m^2$, we first do the $d^4k$ integration with $p_{\mu}$ held fixed, i.e., up to irrelevant factors we evaluate
\begin{eqnarray}
\Sigma(p)=\int \frac{dk_+dk_1dk_2dk_-}{(4k_+k_--(k_1)^2-(k_2)^2-m^2+i\epsilon)[4(k_++p_+)(k_-+p_-)-(k_1+p_1)^2-(k_2+p_2)^2-m^2+i\epsilon]}.
\label{17.2}
\end{eqnarray}
There is no circle at infinity contribution as the denominator has two powers of $k_+$. The graph diverges as a single logarithm, i.e., as ${\rm log}(\Lambda^2/p^2)$ at large $p^2$. Introducing a mass renormalization counter term $\delta m=-{\rm log}(\Lambda^2/m^2)$ gives  ${\rm log}(m^2/p^2)$. The original graph is thus 
\begin{eqnarray}
I=\int dp_+dp_1dp_2dp_-\frac{{\rm log}(m^2/p^2)}{(4p_+p_--(p_1)^2-(p_2)^2-m^2+i\epsilon)},
\label{17.3}
\end{eqnarray}
and the circle at infinity is not suppressed. The concerns raised in this paper thus carry over to dressed light-front vacuum graphs as well and cannot be ignored.

\section{Conclusions}
\label{S18}

In this paper we have examined some general aspects of light-front quantization, and have obtained results, some that are known to the literature and some that are not. In particular by imposing Poincare invariance in the form $[P_{\mu},\phi]=-i\partial_{\mu}\phi$ we have been able to identify the equal light-front time canonical commutators or anticommutators needed for light-front quantization in the scalar, fermion and both Abelian and non-Abelian gauge field cases, and in so doing have found some new commutation and anticommutation relations involving $\partial_+$ derivatives of the fermion and gauge fields. Additionally, we have shown that equal instant-time commutation and anticommutation relations cannot be transformed into equal light-front time commutation and anticommutation relations by a coordinate transformation, with the two quantization procedures seemingly being intrinsically different. Nonetheless, we have also shown that one actually can derive equal light-front time commutators and anticommutators starting not from equal-time instant-time commutators and anticommutators but from unequal-time instant-time commutators and anticommutators. Thus, equal light-front time commutator and anticommutator relations do not need to be independently postulated, with instant-time and light-front commutators and anticommutators being completely equivalent. Moreover, we have also shown that c-number instant-time and light-front Feynman graphs are actually equal in value, with the equal light-front time commutators actually readjusting in order to ensure that this be the case. Thus whether one quantizes using instant-time or light-front canonical commutators one obtains precisely the same values for Feynman diagrams.

Once one has obtained a light-front commutation relation such as $[\phi(x^+,x^1,x^2,x^-),2\partial_-\phi(x^+,y^1,y^2,y^-)]=i\delta(x^1-y^1)\delta(x^2-y^2)\delta(x^--y^-)$, one can realize it by an on-shell three-dimensional Fock space field expansion of the form  $\phi(x)=f_p(x)a(p)+f_p^{\dagger}(x)a^{\dagger}_p$ where $p$ is a three-vector, where the modes $f_p(x)$ are wave function solutions to the equations of motion, and where the energy of each mode is fixed by $p_{\mu}p^{\mu}=m^2$. As such, this on-shell field description contains the same information as the commutation relations that the fields satisfy. 

However, for quantum field theory that is not sufficient information. Specifically, while one could use the Fock space expansion to evaluate quantities such as the $\langle \Omega|\phi(x)\phi(0)|\Omega\rangle$ two-point function, more information is contained in a  Feynman propagator such as $-i\langle \Omega|(\theta(x^+)\phi(x)\phi(0)+\theta(-x^+)\phi(0)\phi(x))|\Omega\rangle$ because of its time ordering theta functions. These theta functions are not ordinary functions but are distributions, and one way of realizing them is as contour integrals in a complex frequency plane with the $i\epsilon$ factor indicating in which way to close the contour. This leads to an off-shell four-dimensional contour integral Feynman diagram approach in which  all four components of $p_{\mu}$ are free to vary and to not be constrained to obey $p_{\mu}p^{\mu}=m^2$, with theta functions being intrinsically off-shell quantities. 

When one then closes a Feynman contour in a complex energy plane one obtains pole contributions and circle at infinity contributions, and these can have different consequences in off-shell four-dimensional instant-time and light-front  Feynman diagrams. To actually obtain suppression of circle at infinity contributions, in non-vacuum Feynman diagrams one can use the external time coordinate $x^0$ or $x^+$ as a regulator (this shows the value of studying Green's functions in coordinate space rather than momentum space), while for vacuum graphs one needs sufficient powers of the energy in propagator denominators. Then, if circle at infinity contributions are suppressed and only pole terms contribute, then because poles are on shell one finds that the off-shell Feynman diagram prescription is equivalent to the on-shell Fock space Hamiltonian description in which one expands a quantum field in terms of on-shell wave functions and creators and annihilators of on-shell particle states according to the generic $\phi(x)=f_p(x)a(p)+f_p^{\dagger}(x)a^{\dagger}_p$ and inserts it into $-i\langle \Omega|(\theta(x^+)\phi(x)\phi(0)+\theta(-x^+)\phi(0)\phi(x))|\Omega\rangle$. For instant-time quantization this equivalence holds  for both non-vacuum and vacuum Feynman diagrams. Similarly, we have shown that for light-front non-vacuum graphs this also turns out to be the case. For light-front non-vacuum graphs this can thus be regarded as a first principles derivation of the on-shell Light-Front Hamiltonian prescription for scattering amplitudes as obtained by starting from the off-shell Feynman prescription. 

However, in the light-front vacuum sector this equivalence does not hold, as there the circle at infinity contribution is not suppressed as there is neither an external $x^+$ regulator or sufficient powers of the energy in propagator denominators, and circle at infinity contributions must thus be included. Circle at infinity contributions have no Fock space counterpart, with the on-shell Light-Front Hamiltonian prescription and thus the closely related infinite momentum frame prescription then failing to correctly describe light-front vacuum graphs. Thus for vacuum graphs one cannot use the on-shell Fock space expansion and one must instead use the full off-shell Feynman diagram prescription. And when one does so, one finds that in light-front quantization vacuum graphs are not just nonzero, they are identical in value to their instant-time counterparts. Thus to summarize, one can obtain the light-front three-dimensional on-shell Fock space description when light-front four-dimensional off-shell Feynman diagram contours are given purely by pole terms, but not otherwise. If circle at infinity contributions are not suppressed one must use the full four-dimensional off-shell Feynman prescription, with this being the case for light-front vacuum graphs. 

In constructing Green's functions for fermions there are tip of the light cone singularities that are not present in the bosonic case. Nonetheless, we have shown that these singularities do not prevent us from establishing the equivalence of fermion Green's functions in the light-front quantization and instant-time quantization cases. One of the fundamental issues in quantum field theory is whether different quantization procedures produce different theories or the same theory. Our analysis shows that as long as Feynman diagrams or path integrals  are general coordinate invariant, any two quantization procedures that are themselves related by a general coordinate transformation will necessarily lead to the same Green's functions for scattering (i.e., non-vacuum) processes. However the vacuum sectors have to be considered in their own right as there could be tip of the light cone singularities, and for light-front quantization we have shown by direct evaluation  that they do not prevent us from establishing the equivalence of the light-front and instant-time quantization procedures. 

As long as we stay away from $x^+=0$ we only need to deal with poles in light-front Feynman diagrams and everything can be described by on-shell physics, just as is done in the Light-Front Hamiltonian approach. However at $x^+=0$ circle at infinity contributions are present and an on-shell description (just poles) is inadequate. Thus for the light-front vacuum sector there is information in off-shell Feynman diagrams that is not accessible to, and thus goes beyond, the on-shell Light-Front Hamiltonian approach. 

Since in general in quantum theory invariance under translations entails  invariance under unitary transformations, we thus establish not only that instant-time quantization and light-front quantization are equivalent procedures, they are unitarily equivalent procedures.  And while we have shown the equivalence of light-front quantization and instant-time quantization, as a practical matter it is typically the case that  calculations done with light-front quantization are simpler than calculations done with instant-time quantization with fewer diagrams typically being required. It is in this sense then that the light-front approach is favored.

\begin{acknowledgments}
The authors would like to thank Dr. K. Y.-J. Chiu for her helpful insights. PDM would like to thank Dr. T. G. Rizzo for the kind hospitality of the Theory Group at the Stanford Linear Accelerator Center where part of this work was performed. The work of SJB and PL were supported in part by the Department of Energy under contract DE-AC02-76SF00515. SLAC-PUB-17390. The work of PL was also supported by the Agence Nationale de la Recherche under project No. ANR--18--ERC1--0002. 
PDM also wishes to acknowledge useful conversations with Dr. L. Jin and Dr. J. Dressel. SLAC-PUB-17390.
\end{acknowledgments}

\appendix
\numberwithin{equation}{section}
\setcounter{equation}{0}
\section{Light-front Variables}
\label{SA}

\subsection{Standard Form}

Introducing the light-front coordinates $x^{\pm}=x^0\pm x^3$, so that $s^2=g_{\mu\nu}x^{\mu}x^{\nu}=(x^0)^2-(x^3)^2-(x^1)^2-(x^2)^2=x^+x^--(x^1)^2-(x^2)^2$ allows us to define the instant-time and light-front metrics as
\begin{eqnarray}
g_{\mu\nu}({\rm instant})=
\begin{pmatrix}
1 &  0 & 0&0 \cr 
0& -1&0&0\cr
0&0&-1&0\cr
0&0&0&-1
\end{pmatrix},\quad
g_{\mu\nu}({\rm front})=
\begin{pmatrix}
0&  \tfrac{1}{2} & 0&0 \cr 
\tfrac{1}{2}& 0&0&0\cr
0&0&-1&0\cr
0&0&0&-1
\end{pmatrix},
\label{A1}
\end{eqnarray}
with determinants $-g$ given by  $-{\rm det}[g_{\mu\nu}({\rm instant})]=1$, $-{\rm det}[g_{\mu\nu}({\rm front})]=1/4$. Raising and lowering in the light-front case is done with $g_{\mu\nu}({\rm front})$ and gives $x_-=x^+/2$, $x_+=x^-/2$.

For derivatives we have to identify what is held fixed. If we want first to hold $x^0$ or $x^3$ fixed, we need to express $x^+$ and $x^-$ in terms of $x^0$ and $x^3$, and thus use $x^{\pm}=x^0\pm x^3$. Then with $x^3$ held fixed we obtain 
\begin{eqnarray}
\partial_0=\frac{\partial}{\partial x^0}=\frac{\partial x^+}{\partial x^0}\frac{\partial}{\partial x^+}+\frac{\partial x^-}{\partial x^0}\frac{\partial}{\partial x^-}=\frac{\partial}{\partial x^+}+\frac{\partial}{\partial x^-}=\partial_++\partial_-.
\label{A2}
\end{eqnarray} 
And with $x^0$ held fixed we obtain 
\begin{eqnarray}
\partial_3=\frac{\partial}{\partial x^3}=\frac{\partial x^+}{\partial x^3}\frac{\partial}{\partial x^+}+\frac{\partial x^-}{\partial x^3}\frac{\partial}{\partial x^-}=\frac{\partial}{\partial x^+}-\frac{\partial}{\partial x^-}=\partial_+-\partial_-,
\label{A3}
\end{eqnarray} 
and thus
\begin{eqnarray}
\Box=g^{\mu\nu}\partial_{\mu}\partial_{\nu}=\partial_0^2-\partial_3^2-\partial_1^2-\partial_2^2
=4\partial_+\partial_--\partial_1^2-\partial_2^2.
\label{A4}
\end{eqnarray}
Similarly, if we want to hold $x^+$ or $x^-$ fixed, we use $x^0=(x^++x^-)/2$, $x^3=(x^+-x^-)/2$. Then with $x^-$ held fixed we obtain 
\begin{eqnarray}
\partial_+=\frac{\partial}{\partial x^+}=\frac{\partial x^0}{\partial x^+}\frac{\partial}{\partial x^0}+\frac{\partial x^3}{\partial x^+}\frac{\partial}{\partial x^3}=\frac{1}{2}\frac{\partial}{\partial x^0}+\frac{1}{2}\frac{\partial}{\partial x^3}=\frac{1}{2}\partial_0+\frac{1}{2}\partial_3,
\label{A5}
\end{eqnarray} 
\begin{eqnarray}
\partial_-=\frac{\partial}{\partial x^-}=\frac{\partial x^0}{\partial x^-}\frac{\partial}{\partial x^0}+\frac{\partial x^3}{\partial x^-}\frac{\partial}{\partial x^3}=\frac{1}{2}\frac{\partial}{\partial x^0}-\frac{1}{2}\frac{\partial}{\partial x^3}=\frac{1}{2}\partial_0-\frac{1}{2}\partial_3.
\label{A6}
\end{eqnarray} 
Thus again we obtain (\ref{A4}).

For integrals, with $(-g)^{1/2}=1/2$ for the light-front determinant we have
\begin{eqnarray}
\int dx^0dx^1dx^2dx^3=\tfrac{1}{2}\int dx^+dx^-dx^1dx^2=2\int dx_+dx_-dx_1dx_2.
\label{A7}
\end{eqnarray}
Applying (\ref{A7})  to $\delta(x^0+x^3)=\delta(x^+)$ we obtain
\begin{eqnarray}
\int dx^1dx^2dx^3=\tfrac{1}{2}\int dx^-dx^1dx^2=\int dx_+dx_1dx_2,
\label{A8}
\end{eqnarray}
as used in the text.

For the four-momentum we define the energy $p_+$ as the conjugate of $x^+$ and the momentum $p_-$ as the conjugate of $x^-$ so that $p_{\mu}x^{\mu}=g_{\mu\nu}p^{\mu}x^{\nu}=p_{+}x^{+}+p_{-}x^{-}+p_{1}x^{1}+p_{2}x^{2}$. With the light-front metric given in (\ref{A1}) we can identify covariant components as $p_{\mu}=g_{\mu\nu}p^{\nu}$, so that $p_+=p^-/2$ and $p_-=p^+/2$.  With $x^0=\tfrac{1}{2}(x^++x^-)$ and $x^3=\tfrac{1}{2}(x^+-x^-)$ we obtain $p_0x^0+p_3x^3 =\tfrac{1}{2}(p_0+p_3)x^++\tfrac{1}{2}(p_0-p_3)x^-$, and can thus identify $p_+=\tfrac{1}{2}(p_0+p_3)$, $p_-=\tfrac{1}{2}(p_0-p_3)$, $4p_+p_-=(p_0)^2-(p_3)^2$, $p_0=p_++p_-$, $p_3=p_+-p_-$. Similarly, we have  $p^-=2p_+=p_0+p_3$, $p^+=2p_-=p_0-p_3$, i.e, $p^-=p^0-p^3$, $p^+=p^0+p^3$.

\subsection{Normalized Form}

Define a normalized light-front coordinate basis as $x^{+}=(x^0+ x^3)/\surd{2}$, $x^{-}=(x^0- x^3)/\surd{2}$, $x^{0}=(x^++ x^-)/\surd{2}$, $x^{3}=(x^+- x^-)/\surd{2}$,
so that $(x^0)^2-(x^3)^2-(x^1)^2-(x^2)^2=2x^+x^--(x^1)^2-(x^2)^2$. The light-front metric is given by 
\begin{eqnarray}
g_{\mu\nu}({\rm front})=
\begin{pmatrix}
0&  1 & 0&0 \cr 
1& 0&0&0\cr
0&0&-1&0\cr
0&0&0&-1
\end{pmatrix},
\label{A9}
\end{eqnarray}
with determinant $-{\rm det}[g_{\mu\nu}({\rm front})]=1$ of the same normalization as $-{\rm det}[g_{\mu\nu}({\rm instant})]=1$, with the normalized basis conveniently leaving the determinant of the metric unchanged. Derivatives are given by 
\begin{eqnarray}
\partial_0&=&\frac{1}{\surd{2}}(\partial_++\partial_-),\quad \partial_3=\frac{1}{\surd{2}}(\partial_+-\partial_-),
\quad \partial_+=\frac{1}{\surd{2}}(\partial_0+\partial_3),\quad \partial_-=\frac{1}{\surd{2}}(\partial_0-\partial_3),
\nonumber \\
\Box &=&\partial_+\partial_-+\partial_-\partial_+-\partial_1^2-\partial_2^2=2\partial_+\partial_--\partial_1^2-\partial_2^2,
\label{A10}
\end{eqnarray} 
with all the derivatives being given the same weight in $\Box =\partial_+\partial_-+\partial_-\partial_+-\partial_1^2-\partial_2^2$.

\section{Fermion Considerations}
\label{SB}

\subsection{A Light-front Dirac Equation Oddity }

In writing down the Dirac equation Dirac did not start with the covariant
\begin{eqnarray}
(i\gamma^{0}\partial_t+i\gamma^k\partial_k -m)\psi=0,
\label{B1a}
\end{eqnarray}
but instead started with 
\begin{eqnarray}
i\partial_{t}\psi+i\alpha^k\partial_k\psi-\beta m\psi=0.
\label{B2a}
\end{eqnarray}
These equations are equivalent since  (\ref{B1a}) and (\ref{B2a}) can be derived from each other by multiplying through by $\beta=\gamma^0$ and setting $\gamma^k=\beta\alpha^k$.  In the Dirac basis  for the gamma matrices one takes $\gamma_{\rm D}^{0}$ to be diagonal and has
\begin{eqnarray}
\gamma_{\rm D}^{0}=\beta_{\rm D}=\begin{pmatrix} I&0\\ 0&-I \end{pmatrix},\quad
\alpha_{\rm D}^{k}=\begin{pmatrix}0&\sigma_k \\ \sigma_k &0\end{pmatrix},\quad
\gamma_{\rm D}^{k}=\beta_{\rm D}\alpha_{\rm D}^{k}=\begin{pmatrix}0&\sigma_k\\ -\sigma_k&0\end{pmatrix}, \quad \gamma_{\rm D}^{5}=\begin{pmatrix}0&I\\ I&0\end{pmatrix}.
\label{B3a}
\end{eqnarray}

In light-front coordinates the covariant Dirac equation takes the form  
\begin{eqnarray}
(i\gamma^{+}\partial_++i\gamma^-\partial_-+i\gamma^1\partial_1 +i\gamma^2\partial_2-m)\psi=0,
\label{B4a}
\end{eqnarray}
where $\gamma^{\pm}=\gamma^0\pm \gamma^3$. However there is no light-front analog of (\ref{B2a}), since $(\gamma^+)^2$ and $(\gamma^-)^2$ are zero, to thus not be invertible. Thus even though the invertible $\gamma^0$ and $\gamma^3$ obey $(\gamma^0)^2=1$, $(\gamma^3)^2=-1$, the non-invertible $\gamma^+$ and $\gamma^-$ are divisors of zero. Consequently, one cannot multiply the light-front (\ref{B4a}) by $\gamma^+$ and obtain a light-front analog of (\ref{B2a}). Since $\gamma^+$ and $\gamma^-$ are divisors of zero, there is no similarity transformation that can effect $S\gamma^0S^{-1}=\gamma^+$, $S\gamma^3S^{-1}=\gamma^-$, and there thus are intrinsic differences between light-front fermions and instant-time fermions. Moreover, it  is because $\gamma^+$ and $\gamma^-$ are divisors of zero that (\ref{B4a}) breaks up into good and fermions, with $\Lambda^{\pm}=(1/2)\gamma^0\gamma^{\pm}$ being projection operators, i.e., being operators that also are not invertible. Despite this intrinsic difference between light-front and instant-time fermions, we have nonetheless been able to show that their Green's functions are identical.

Finally, we note that even though  $\gamma^+$ and $\gamma^-$ are themselves divisors of zero, the products $\gamma^+\gamma^-$ and $\gamma^-\gamma^+$ are not, with  combination $\gamma^+\gamma^-+\gamma^-\gamma^+$ evaluating to $4$. In consequence, $i\gamma^+\partial_++i\gamma^-\partial_-$ squares to $-4\partial_+\partial_-$, with the Klein-Gordon equation in the form $[4\partial_+\partial_--(\partial_1)^2-(\partial_2)^2+m^2]\psi=0$ then following from (\ref{B4a}).

\subsection{Weyl Basis for the Dirac Gamma Matrices}

In working with the $\Lambda^+$ and $\Lambda^-$ projection operators it would be very convenient if we could find a basis for the gamma matrices in which $\Lambda^+$ and $\Lambda^-$ are diagonal. It turns out that there is such a basis, the one Weyl used to diagonalize $\gamma^5$. The Weyl basis $\gamma^{\mu}_W$ is constructed from the Dirac basis $\gamma^{\mu}_{\rm D}$ via the similarity transform (see e.g.  \cite{Mannheim1984})
\begin{eqnarray}
\gamma^{\mu}_{\rm W}=\tfrac{1}{\surd{2}}(1-\gamma^5_{\rm D}\gamma^0_{\rm D})\gamma^{\mu}_{\rm D}\tfrac{1}{\surd{2}}(1+\gamma^5_{\rm D}\gamma^0_{\rm D}).
\label{B5a}
\end{eqnarray}
This yields 
\begin{align}
&\gamma_{\rm W}^{0}=\begin{pmatrix} 0&-1\\ -1&0 \end{pmatrix},\quad
\gamma_{\rm W}^{k}=\begin{pmatrix}0&\sigma_k\\ -\sigma_k&0 \end{pmatrix},\quad
\alpha_{\rm W}^{k}=\begin{pmatrix}\sigma_k&0 \\ 0&-\sigma_k \end{pmatrix},\quad
\gamma_{\rm W}^{5}=\begin{pmatrix}I&0 \\ 0&-I \end{pmatrix},
\label{B6a}
\end{align}
\begin{align}
&\Lambda^{\pm}_{\rm W}=\tfrac{1}{\surd{2}}(1-\gamma^5_{\rm D}\gamma^0_{\rm D})\left[ \tfrac{1}{2}\left(1\pm\gamma^0_{\rm D}\gamma^3_{\rm D}\right)\right]\tfrac{1}{\surd{2}}(1+\gamma^5_{\rm D}\gamma^0_{\rm D})=\tfrac{1}{2}\left(1\pm\gamma^0_{\rm W}\gamma^3_{\rm W}\right).
\label{B7a}
\end{align}
In this basis we find that  not only is $\gamma^5_{\rm w}$ diagonal, $\Lambda^+_{\rm W}$ and $\Lambda^-_{\rm W}$ are diagonal too. They take the form 
\begin{eqnarray}
\Lambda^+_{\rm W}=\begin{pmatrix}1&0&0&0\\ 0&0&0&0\\ 0&0&0&0\\ 0&0&0&1\end{pmatrix},\quad \Lambda^-_{\rm W}=\begin{pmatrix}0&0&0&0\\ 0&1&0&0\\ 0&0&1&0\\ 0&0&0&0 \end{pmatrix},
\label{B8a}
\end{eqnarray}
and effect
\begin{eqnarray}
\Lambda^+_{\rm W}\begin{pmatrix} \psi_1\\ \psi_2\\ \psi_3\\ \psi_4\end{pmatrix}=
\begin{pmatrix} \psi_1\\ 0\\ 0\\ \psi_4\end{pmatrix},\quad \Lambda^-_{\rm W}\begin{pmatrix}\psi_1\\ \psi_2\\ \psi_3\\\psi_4\end{pmatrix}
=\begin{pmatrix} 0\\ \psi_2\\ \psi_3\\  0 \end{pmatrix}.
\label{B9a}
\end{eqnarray}
Hence in the Weyl basis we can treat the good and bad fermions as two-component spinors.

\subsection{Good and Bad Fermion Bilinears}

It is of interest to decompose fermion bilinear operators into good and bad fermions $\psi_{(\pm)}=\Lambda^{\pm}\psi$ where $\Lambda^{\pm}=(1/2)\gamma^0(\gamma^0\pm\gamma^3)=(1/2)\gamma^0\gamma^{\pm}$. On recalling that $\gamma^0\gamma^{\mu}\gamma^0=(\gamma^{\mu})^{\dagger}$, we see that  $\Lambda^{\pm}=(\Lambda^{\pm})^{\dagger}$. We can thus set $(\psi_{(\pm)})^{\dagger}=(\Lambda^{\pm}\psi)^{\dagger}=\psi^{\dagger}\Lambda^{\pm}=\psi_{(\pm)}^{\dagger}$. For the fermion vector current $V^{\mu}=\bar{\psi}\gamma^{\mu}\psi$, we find that  in the light-front $V^{+}=\psi^{\dagger}\gamma^0\gamma^{+}\psi=2\psi^{\dagger}_{(+)}\psi_{(+)}$, to thus consist of good fermions alone. Similarly $V^{-}=\psi^{\dagger}\gamma^0\gamma^{-}\psi=2\psi^{\dagger}_{(-)}\psi_{(-)}$ consists of bad fermions alone, while  $V^{1}=\psi^{\dagger}\gamma^0\gamma^{1}\psi$ and $V^{2}=\psi^{\dagger}\gamma^0\gamma^{2}\psi$ contain both $\psi_{(+)}$ and $\psi_{(-)}$. In consequence, the conservation condition $\partial_{\mu}V^{\mu}=0$ involves both good and bad fermions, and while the charge  $Q=\tfrac{1}{2}\int dx^-dx^1dx^2 V^{+}=\int dx^-dx^1dx^2 \psi^{\dagger}_{(+)}\psi_{(+)}$ only consists of good fermions, its light-front time derivative $\partial_+Q=-\tfrac{1}{2}\int dx^-dx^1dx^2(\partial_-V^{-}+\partial_1V^1+\partial_2V^2)$ involves bad fermions. Since $\psi_{(-)}$ is subject to the nonlocal constraint given in (\ref{4.16a}), viz. $\psi_{(-)}=-\frac{i}{2}(\partial_-)^{-1}\left[-i\gamma^0(\gamma^1\partial_1+\gamma^2\partial_2)+m\gamma^0\right]\psi_{(+)}$, to secure the light-front time independence of $Q$ requires that the fields be more convergent asymptotically than in the instant-time case. 

These remarks apply equally to the axial vector current  $A^{\mu}=\bar{\psi}\gamma^{\mu}\gamma^5\psi$. In the light-front case $A^{+}=\psi^{\dagger}\gamma^0\gamma^{+}\gamma^5\psi=2\psi^{\dagger}_{(+)}\gamma^5\psi_{(+)}$ is composed of good fermions alone. Similarly $A^{-}=\psi^{\dagger}\gamma^0\gamma^{-}\gamma^5\psi=2\psi^{\dagger}_{(-)}\gamma^5\psi_{(-)}$ is composed of bad fermions alone, while $A^{1}=\psi^{\dagger}\gamma^0\gamma^{1}\gamma^5\psi$ and $A^{2}=\psi^{\dagger}\gamma^0\gamma^{2}\gamma^5\psi$ contain both $\psi_{(+)}$ and $\psi_{(-)}$. Moreover, while $Q^5=\tfrac{1}{2}\int dx^-dx^1dx^2 A^{+}=\int dx^-dx^1x^2 \psi^{\dagger}_{(+)}\gamma^5\psi_{(+)}$ is composed of good fermions alone, even if $\partial_{\mu}A^{\mu }=0$, the light  front time derivative of $Q^5$ will involve bad fermions.

The mass operator $\bar{\psi}\psi=\psi^{\dagger}\gamma^0\psi$ contains both good and bad fermions. Similarly, the pseudoscalar $\bar{\psi}i\gamma^5\psi=i\psi^{\dagger}\gamma^0\gamma^5\psi$ also contains both good and bad fermions. Writing them out in full we have
\begin{eqnarray}
\bar{\psi}\psi=\psi_{(+)}^{\dagger}\gamma^0\psi_{(-)}+\psi_{(-)}^{\dagger}\gamma^0\psi_{(+)},\quad i\bar{\psi}\gamma^5\psi=i\psi_{(+)}^{\dagger}\gamma^0\gamma^5\psi_{(-)}+i\psi_{(-)}^{\dagger}\gamma^0\gamma^5\psi_{(+)}.
\label{B10a}
\end{eqnarray}

With the energy-momentum tensor given in (\ref{4.5}) only differing from the canonical energy-momentum tensor $T^{\mu\nu}=i\bar{\psi}\gamma^{\mu}\partial^{\nu}\psi$ by a total divergence,  and with this total divergence not affecting the momentum generators as they are integrals of $T^{\mu\nu}$, we can use the canonical energy-momentum tensor.  For it we find that $T^{++}$, $T^{+-}$, $T^{+1}$ and $T^{+2}$ are composed of good fermions alone, $T^{-+}$, $T^{--}$, $T^{-1}$, $T^{-2}$ are composed of bad fermions alone, and  $T^{11}$, $T^{12}$ and $T^{22}$ contain both good and bad fermions. The conservation condition $\partial_{\mu}T^{\mu\nu}=0$ involves both good and bad fermions. Consequently, while the all four of the momentum operators $P^{\mu}=\tfrac{1}{2}\int dx^-dx^1dx^2T^{+\mu}$ only involve good fermions, their light-front time derivatives involve both good and bad fermions.

For the light-front angular momentum generators $M^{\mu\nu}=\tfrac{1}{2}\int dx^-dx^1dx^2(x^{\mu}T^{+\nu}-x^{\nu}T^{+\mu})$ we see that all six components of $M^{\mu\nu}$ contain good fermions alone.  
Armed with this information we now see how it impacts on Ward identities.

\section{Ward Identities}
\label{SC}

Consider the operator $T[j^{\mu}(x)B(0)]$, where $j^{\mu}$ is a vector or axial-vector current and $B(0)$ is a string of fields all at $x^{\mu}=0$. Consider first instant-time quantization. Differentiating with respect to $x^{\mu}$ we obtain the operator identity (which holds independent of in which states we take matrix elements)
\begin{eqnarray}
\partial_{\mu}T^{(0)}[j^{\mu}(x)B(0)]
=\delta(x^0)[j^0(x),B(0)]+T^{(0)}[\partial_{\mu}j^{\mu}(x)B(0)].
\label{C1}
\end{eqnarray}
We now restrict to  the case where $\partial_{\mu}j^{\mu}(x)=0$, and  take matrix elements in the vacuum (normal or spontaneously broken). Since there is only one four-momentum $p^{\mu}$ in Fourier space we can set
\begin{eqnarray}
\langle \Omega|T^{(0)}[j^{\mu}(x)B(0)]|\Omega\rangle=\frac{1}{(2\pi)^4}\int d^4p e^{ip\cdot x}p^{\mu}F(p^2),
\label{C2}
\end{eqnarray}
where $F(p^2)$ is a scalar function. With $Q(x^0)=\int d^3x j^0(x)$, we integrate both sides of (\ref{C1}) with $\int d^4x$. This gives
\begin{eqnarray}
i\int d^4p\delta^4(p)p^2F(p)=\langle \Omega|[Q(x^0=0),B(0)]|\Omega\rangle.
\label{C3}
\end{eqnarray}
Should the right-hand side of (\ref{C3}) not vanish (i.e., $Q(x^0=0)|\Omega\rangle \neq0$), there would then have to be  a massless pole at $p^2=0$ on the left-hand side. This then is the Goldstone theorem.

In light-front quantization everything goes through the same way and we have 
\begin{eqnarray}
i\int d^4p\delta^4(p)p^2F(p)=\langle \Omega|[Q(x^+=0),B(0)]|\Omega\rangle.
\label{C4}
\end{eqnarray}
For the axial-vector Ward identity we set $B(0)=\bar{\psi}i\gamma^5\psi$ in (\ref{C4}), set $j^{\mu}=A^{\mu}$, and need to evaluate $[Q^5(x^+=0),\bar{\psi}i\gamma^5\psi]$. To this end we note that generically 
\begin{eqnarray}
A^{\dagger}BC^{\dagger}D-C^{\dagger}DA^{\dagger}B=
A^{\dagger}(BC^{\dagger}+C^{\dagger}B)D-C^{\dagger}(DA^{\dagger}+A^{\dagger}D)B
-(A^{\dagger}C^{\dagger}+C^{\dagger}A^{\dagger})BD+C^{\dagger}A^{\dagger}(BD+DB).
\label{C5}
\end{eqnarray}
Provided that $Q^5$ is independent of $x^+$, on setting $A^{\dagger}=\psi^{\dagger}_{(+)}(y)$, $B=\psi_{(+)}(y)$, $C^{\dagger}=\psi_{(+)}^{\dagger}(x)$, $D=\psi_{(-)}(x)$ and then $C^{\dagger}=\psi_{(-)}^{\dagger}(x)$, $D=\psi_{(+)}(x)$, and on recalling (\ref{4.13a}), (\ref{4.16a}),   and (\ref{4.23a}),  we obtain the needed axial-vector Ward identity commutator in the light-front case:
\begin{eqnarray}
[Q^5,\bar{\psi}(x)i\gamma^5\psi(x)]&=&\int dy^-dy^1dy^2[\psi^{\dagger}_{(+)}(y)\gamma^5\psi_{(+)}(y),
i\psi_{(+)}^{\dagger}(x)\gamma^0\gamma^5\psi_{(-)}(x)+i\psi_{(-)}^{\dagger}(x)\gamma^0\gamma^5\psi_{(+)}(x)]
\nonumber\\
&=&i\psi_{(+)}^{\dagger}(x)\gamma^0\psi_{(-)}(x)+i\psi_{(-)}^{\dagger}(x)\gamma^0\psi_{(+)}(x)=i\bar{\psi}(x)\psi(x).
\label{C6}
\end{eqnarray}
As we see, both sides of (\ref{C6}) contain bad fermions even though $Q^5$ does not. However, while we do need the good fermion bad  fermion anticommutator given in (\ref{4.23a}), we do not need the troublesome bad fermion bad fermion anticommutator given in (\ref{4.22a}). Moreover, despite the fact that (\ref{4.23a}) contains spatial derivatives of delta functions, all such terms conveniently drop out in (\ref{C6}), leaving (\ref{C6}) with exactly the same form that it has in the instant-time case. The bad fermions are thus needed to maintain the axial-vector Goldstone theorem, with the axial-vector Ward identity 
\begin{eqnarray}
\int d^4p\delta^4(p)p^2F(p)=\langle \Omega|\bar{\psi}(x)\psi(x)|\Omega\rangle
\label{C7}
\end{eqnarray}
receiving nonlocal contributions in the light-front case.

With the vacuum being translation invariant we can set $\langle \Omega|\bar{\psi}(x)\psi(x)|\Omega\rangle=\langle \Omega|\bar{\psi}(0)\psi(0)|\Omega\rangle$. Spontaneous breaking in the axial vector sector thus requires that the vacuum be such that the expectation value $\langle \Omega|\bar{\psi}(0)\psi(0)|\Omega\rangle$ be nonzero, and this expectation value will be nonvanishing if the fermion has a mass. For instance, evaluating the expectation value for a free massive fermion (the mean field approximation to a chiral invariant four-Fermi theory \cite{Mannheim2017}) yields
\begin{eqnarray}
\langle \Omega|\bar{\psi}(0)\psi(0)|\Omega\rangle=-\frac{2i}{(2\pi)^4}\int dp_+dp_1dp_2dp_-{\rm Tr}\frac{1}{\slashed{p}-m+i\epsilon}=-\frac{8i}{(2\pi)^4}\int dp_+dp_1dp_2dp_-\frac{m}{p^2-m^2+i\epsilon},
\label{C8}
\end{eqnarray}
an expression that is nonzero if $m$ is nonzero.

The nonvanishing of $\langle \Omega|\bar{\psi}(0)\psi(0)|\Omega\rangle$ is central to dynamical symmetry breaking by fermion bilinear condensates. Since we have shown that instant-time and light-front Green's functions are equal, $\langle \Omega|\bar{\psi}(0)\psi(0)|\Omega\rangle$ will be nonzero in both instant-time and light-front quantization if the fermion has a mass. As we showed in (\ref{15.5}) the light-front  on-shell $\phi(0)$, and thus the light-front on-shell $\psi(0)$ and $\psi^{\dagger}(0)$, are mass independent. Inserting the light-front  $\psi(0)$ and $\psi^{\dagger}(0)$ into $\langle \Omega|\bar{\psi}(0)\psi(0)|\Omega\rangle$ thus could not lead to the mass dependent (\ref{C8}). We thus see that one cannot use the three-dimensional on-shell light-front expressions for fields in vacuum matrix elements that have no spacetime dependence. Rather, one must use the full four-dimensional Feynman formalism, with $\langle \Omega|\bar{\psi}(0)\psi(0)|\Omega\rangle$ as evaluated in (\ref{C8}) and (\ref{9.25a}) being both mass dependent and nonzero.

\end{document}